\documentclass[lettersize,10pt]{article}
\usepackage{latexsym,amssymb,amsmath,epsfig}
\usepackage[margin=0.7in]{geometry}
\usepackage{setspace,color}
\usepackage{tikz}
\usepackage{floatrow}
\usepackage{parskip}
\usepackage{grffile}
\usepackage{graphicx}
\usepackage{caption, subcaption}
\usepackage{upgreek}
\usepackage[super]{nth}
\usepackage{bm}
\usepackage[shortlabels]{enumitem}
\usepackage{caption}
\usepackage{titlesec}
\usepackage{hyperref}

\hypersetup{
colorlinks,
citecolor=black,
filecolor=black,
linkcolor=black,
urlcolor=black
}

\numberwithin{equation}{section}

\newcommand{\eps}{\varepsilon}
\newcommand{\lap}{\Delta}
\renewcommand{\v}[1]{\mathbf{#1}}
\newcommand{\mrm}[1]{\mathrm{#1}}
\newcommand{\bigo}{\mathcal{O}}
\newcommand{\mc}[1]{\mathcal{#1}}

\providecommand{\keywords}[1]{\textbf{keywords ---} #1}

\newcommand{\Gmat}{\mathcal{G}}

\newcommand{\chivec}{\bm{\chi}}

\newcommand{\ubar}{\bar{u}}

\newcommand{\Emat}{\mathcal{E}}

\newcommand{\hole}{\Omega_\eps}
\newcommand{\pdedomain}{\bar{\Omega}}

\newcommand{\Rljj}{R_{\lambda_{j,j}}}
\newcommand{\Glji}{G_{\lambda{j,i}}}
\newcommand{\Glij}{G_{\lambda{i,j}}}
\newcommand{\Rlhole}{R_{\lambda_{0,0}}}
\newcommand{\Glhole}{G_{\lambda_{0,i}}}
\newcommand{\thetal}{\theta_\lambda}
\newcommand{\gl}{\mathbf{g}_\lambda}
\newcommand{\Gmatl}{\mathcal{G}_\lambda}

\newcommand{\Imat}{\mathcal{I}}
\newcommand{\Bmat}{\tilde{\mathcal{B}}}
\newcommand{\Mmat}{\mathcal{M}}

\newcommand{\Bhat}{\tilde{B}}

\newcommand{\tol}{\mrm{TOL}}

\begin{document}

\title{Spot Patterns in the 2-D Schnakenberg Model with Localized
  Heterogeneities}
\author{ Tony Wong\thanks{Dept. of Mathematics, Univ. of British
		Columbia, Vancouver, B.C., Canada.} \and Michael
	J. Ward\footnotemark[1]\,\, \thanks{corresponding author,
		\texttt{ward@math.ubc.ca}}}

\maketitle

\begin{abstract}
  A hybrid asymptotic-numerical theory is developed to analyze the
  effect of different types of localized heterogeneities on the
  existence, linear stability, and slow dynamics of localized spot
  patterns for the two-component Schnakenberg reaction-diffusion model
  in a 2-D domain.  Two distinct types of localized heterogeneities
  are considered: a strong localized perturbation of a spatially
  uniform feed rate and the effect of removing a small hole in the
  domain, through which the chemical species can leak out. Our hybrid
  theory reveals a wide range of novel phenomena such as, saddle-node
  bifurcations for quasi-equilibrium spot patterns that otherwise
  would not occur for a homogeneous medium, a new type of spot
  solution pinned at the concentration point of the feed rate, spot
  self-replication behavior leading to the creation of more than two
  new spots, and the existence of a creation-annihilation attractor
  with at most three spots. Depending on the type of localized
  heterogeneity introduced, localized spots are either repelled or
  attracted towards the localized defect on asymptotically long time
  scales. Results for slow spot dynamics and detailed predictions of
  various instabilities of quasi-equilibrium spot patterns, all based
  on our hybrid asymptotic-numerical theory, are illustrated and
  confirmed through extensive full PDE numerical simulations.
\end{abstract}



\section{Introduction}\label{sec:intro}

Localized spot patterns, in which a solution component becomes
spatially localized near certain time-varying discrete points within a
bounded multi-dimensional domain, is a well-known
``far-from-equilibrium'' spatial pattern that occurs for certain
two-component reaction-diffusion (RD) systems in the singular limit of
a large diffusivity ratio. This class of localized pattern is observed
in many chemical and biological systems, such as the
chlorine-dioxide-malonic acid reaction \cite{bddd98}, the
ferrocyanide-iodate-sulphite reaction \cite{lmps94, ls95}, and the
initiation of plant root hair cells mediated by the plant hormone
auxin \cite{abw18}, among others (see \cite{vanag} and \cite{knob} for
surveys). In a spatially homogeneous 2-D medium, and for various
specific RD systems, the slow dynamical behavior of quasi-equilibrium
spot patterns, together with their various types of bifurcations that
trigger a range of different instabilities of the pattern such as
spot-annihilation, spot-replication, and temporal oscillations of the
spot amplitude, have been well-studied \cite{cw12, kww09, rrw14, tw16,
  tw18, ww03, ww03b, ww08, xk17}. The primary focus of this article is
to investigate, for one prototypical RD system, how certain spatial
heterogeneities in the model affect the dynamics and instabilities of
quasi-equilibrium spot patterns, and lead to new dynamical phenomena
that would otherwise not occur in a medium free of defects. For
tractability of our analysis, and as we describe below, we will focus
only on certain types of spatially localized heterogeneities.

There is a growing literature, primarily in a 1-D setting, of
analyzing the effect of a spatial heterogeneity in either the
diffusivity or reaction kinetics on pattern-formation behavior for
two-component RD systems with regards to both small amplitude patterns
(see \cite{page1}, \cite{page2}, \cite{krause2} and the references
therein) and for localized far-from-equilibrium spike-type patterns
(cf.~\cite{abw18}, \cite{bast}, \cite{bast_comp}, \cite{brena1d},
\cite{pvh_dirac}, \cite{kw_siamrev}, \cite{kwei}, \cite{kx},
\cite{precur_kpw}, \cite{tw18}, \cite{ward2}, \cite{wei-win_2},
\cite{wei-win_1}, \cite{wei_cluster}). In particular, the analysis in
\cite{precur_kpw} and \cite{wei-win_2} has revealed that a precursor
gradient in the reaction kinetics can lead to the existence of stable
asymmetric spike patterns for the Gierer-Meinhardt (GM) model, which
would otherwise not occur in a homogeneous medium. A precursor field
for the GM model can also lead to stable steady-states consisting of
spike clusters near critical points of the precursor.  In
\cite{kkwg18} it was shown that a different type of smooth
heterogeneity in the 1-D GM model can lead to the formation of a
creation-annihilation attractor, which consists of periodically
repeating cycles of spike formation, propagation, and annihilation
against a domain boundary. In the limit of a large number of spikes
that are confined by a spatial heterogeneity, a mean field equation
for the spike density was derived in \cite{kx} and \cite{kw_siamrev}
for the 1-D GM and Schnakenberg models, respectively, and in
\cite{kwei} for 2-D spot clusters for the GM model. For the 1-D
Schnakenburg model, the mean field limiting equation in
\cite{kw_siamrev} revealed the existence of a creation-annihilation
attractor in which spikes undergo self-replication in the interior of
the spike cluster, while other spikes are annihilated at the edges of
the cluster. For an extended Klausmeir RD model of spatial ecology,
coarsening and pinning behavior of 1-D spike patterns with various
spatial and temporal heterogeneities were analyzed in \cite{bast} and
\cite{bast_comp}. Spike dynamics and pinning effects for a 1-D RD
model where the nonlinearities have small spatial support, as is
typical for catalytic reactions, was studied in \cite{pvh_dirac}.

For a different class of localized pattern consisting of either a
propagating pulse-type or a transition-layer solution, there has been
much effort at analyzing the effect of a small step function barrier
on pulse propagation properties for the three-component
Fitzhugh-Nagumo RD system (cf.~\cite{chen_nagumo}, \cite{pvh_fitz1},
\cite{pvh_fitz2}, \cite{nishiura2}) and for two-component bistable RD
systems (cf.~\cite{ikeda}, \cite{nishi}).  For this type of jump-type
spatial heterogeneity, the focus has been to analytically determine
parameter ranges where a 1-D propagating pulse will either be
reflected, transmitted, or pinned by the barrier. In a 2-D setting,
\cite{nish3} provides a numerical study of similar propagation and
collision properties for a single localized spot in the presence of a
1-D step-function line barrier.

In contrast to the simpler 1-D case, there are relatively few
analytical studies of the effect of spatial heterogeneities for RD
systems in higher spatial dimensions.  For a generalized
Schnakenberg-type RD system modeling the initiation of root hair
profusion in plant cells, a spatially inhomogeneous auxin gradient in
a 2-D rectangular domain was shown to lead to the alignment of
localized spots in the direction of the gradient (cf.~\cite{abw18}).
In a 2-D rectangular domain, it was shown in \cite{ktww18} for a
generalized Klausmeir RD system, modeling vegetation patterns in
semi-arid environments, that an anisotropic diffusivity can stabilize
a localized stripe pattern to transverse perturbations. With isotropic
diffusion, the homoclinic stripe would be unstable to either breakup
into spots or zigzag deformations. Spot-pinning behavior for RD
systems on closed manifolds of non-constant curvature, which can be
viewed as intrinsic spatial heterogeneities, has been analyzed for the
Schnakenberg model in \cite{tt19}. In \cite{tw18}, which is most
closely related to our study, the role of Robin boundary conditions
and boundary fluxes on the slow dynamics and instabilities of
quasi-equilibrium spot patterns for the Brusselator RD model was
analyzed.
  
The goal of this paper is to analyze the effect of various types of
{\em localized} heterogeneities for the singularly perturbed
Schnakenberg model in a bounded 2-D domain $\Omega$, formulated as
\begin{equation}\label{proto:model}
  v_t = \eps^2 \Delta v - v + uv^2\,, \quad \tau u_t = D \Delta u + a -
  \eps^{-2} uv^2\,, \quad \v{x}\in \Omega \,; \qquad
  \partial_n v = \partial_n u = 0 \,, \quad \v{x} \in \partial\Omega \,,
\end{equation}
where $0<\eps\ll 1$, while $D>0$ and $\tau>0$ are ${\mathcal O}(1)$
constants.  One heterogeneity will be introduced through strong, but
local, perturbations in the feed rate $a=a(\v{x})$, which
characterizes the amount of material that is introduced from the
substrate. Another localized heterogeneity that we will consider is to
analyze the effect of perturbing \eqref{proto:model} by removing a
small hole in the domain, which thereby allows leakage of the chemical
species out of the domain.

For these types of localized heterogeneities we will
extend the hybrid asymptotic-numerical framework of \cite{kww09} and
\cite{tw18} to analyze the existence, linear stability, and slow
dynamics of quasi-equilibrium spot patterns. Depending on the type of
localized heterogeneity introduced, spot patterns are either repelled
or attracted towards the defect on a long time scale of order
${\mathcal O}(\eps^{-2})$. By formulating and analyzing various
spectral problems arising from the linear stability analysis for
instabilities of the quasi-equilibrium pattern on short
${\mathcal O}(1)$ time-scales, we will show how peanut-splitting and
competition instabilities that trigger either spot self-replication or
spot-annihilation events, respectively, are affected by the type of
localized heterogeneity. For a localized heterogeneity where there is
a slowly moving localized source of feed in the domain, we will
combine our linear stability theory for quasi-equilibrium spot
patterns with our derived ODE system for slow spot dynamics to
construct a novel attractor consisting of spot-replication and
spot-annihilation events that has a maximum of three spots in the
domain at any time.

To both illustrate and validate our asymptotic theory for various
types of localized heterogeneities, throughout this paper we will
compare our predictions for slow spot dynamics and spot amplitude
instabilities with full PDE simulations of \eqref{proto:model}. The
full simulations are done using the open source finite element
software FEniCS \cite{alnaes2015fenics}, which automates the mesh
generation and finite element assembly from user inputs. Our choice of
node sizes range, approximately, from $20000$ to $80000$. For
time-stepping we used either a Backward-Euler time stepping scheme or
a BDF-2 (backward differentiation formula), the latter of which is
preferable for computing spot amplitude temporal oscillations
due to a Hopf bifurcation.

The outline of this paper is as follows.  In \S \ref{sec:proto} we
summarize the theoretical framework, largely based on \cite{kww09},
for analyzing the existence, linear stability, and slow dynamics of
quasi-equilibrium spot patterns for \eqref{proto:model} for the case
where the feed rate is spatially homogeneous. In providing this
background material, in subsequent sections we can expedite the
analysis of the effect of various types of localized heterogeneities
by simply highlighting the modifications that are needed to the
theoretical framework in \S \ref{sec:proto}. In \S
\ref{sec:ring_ex_two} we show the new result that quasi-equilibrium
two-spot patterns in the unit disk undergo a spot-annihilation
instability as the feed-rate decreases below a saddle-node point
associated with two-spot quasi-equilibria (see
Fig.~\ref{fig:GCEP_two_all} below). The bifurcation structure and
imperfect sensitivity of two-spot quasi-equilibria in the unit disk
are illustrated by using the continuation software COCO \cite{ds13}.

In \S \ref{sec:leakage} we consider the effect on the existence,
linear stability, and slow spot dynamics for \eqref{proto:model} when
the localized defect consists of removing a small hole of radius
${\mathcal O}(\eps)$ from the domain, while imposing a homogeneous
Dirichlet condition on the boundary of the small hole. With this type
of localized heterogeneity, which allows for the possibility of both
chemical species to leak out of the 2-D domain, we show
that a one-spot quasi-equilibrium solution exists only if the feed
rate is large enough or if the spot is sufficiently far enough away
from the hole. More specifically, in contrast to the scenario for a
homogeneous medium, spot quasi-equilibria are shown to exhibit a novel
saddle-node bifurcation structure in the unit disk in terms of either
the feed rate or the distance from the hole. In addition, from the
derivation of a modified system for slow spot dynamics, we show that
localized spots are dynamically repelled from the small hole (see
Fig.~\ref{fig:leak_2} and Fig.~\ref{fig:leak_3} below). Moreover, we show
that a significantly larger threshold value for the feed rate, as compared
to the case for a homogeneous medium, is needed to initiate spot
self-replication events (see Fig.~\ref{fig:leak_split} below). Although
perforated domains have been well-studied in the context of narrow capture
mean first passage time problems for Brownian particles (see
\cite{Venu} and the references therein), the effect of a perforated
domain, resulting in an {\em open} reaction-diffusion system
(cf.~\cite{tw18}), on localized pattern formation problems has to our
knowledge not been analyzed previously. The analysis in \S
\ref{sec:leakage} of the effect of a hole is done by combining the
strong localized perturbation theory approach for perforated domains
(cf.~\cite{whk93}, \cite{wk93}) with the theoretical framework of
\cite{kww09} for the analysis of localized spot patterns.

In \S \ref{sec:het} we extend the asymptotic theory in \S
\ref{sec:proto} to allow for a localized spatially heterogeneous feed
rate that consists of a spatially uniform feed that is augmented by a
large, but concentrated, source of feed. The concentrated source of
feed is modeled by a Gaussian of small variance centered within the
domain, and corresponds to a typical regularization of a Dirac
singularity. By deriving a modified ODE system for slow spot dynamics
for this type of defect, we show that depending on the initial spot
location and the relative magnitude of the concentrated feed to the
background feed level, a one-spot pattern in the unit disk can either
become pinned to the concentration point of the localized feed in
finite time or else reach a new equilibrium location that is biased
towards this concentration point. The results are encapsulated in the
saddle-node bifurcation diagram for one-spot quasi-equilibria shown
below in Fig.~\ref{one:equil}.  In this case, the localized
heterogeneity has an attractive effect on spot dynamics. For a
two-spot quasi-equilibrium ring pattern in the unit disk, and with a
concentration of the feed rate centered at the origin, we show the
qualitatively new result that the two-spot pattern will be linearly
stable to competition instabilities in parameter regimes that would
otherwise would lead to instabilities with a spatially uniform feed
rate. For this pattern, the equilibrium ring radius is shown to
represent a balance between the attractive interaction towards the
concentration point of the feed rate and the well-known repulsive
inter-spot interaction.

Motivated by the finite-time pinning behavior predicted in \S
\ref{sec:het}, in \S \ref{sec:pinned_one} we construct a new type of
spot solution where the spot is pinned at the point of concentration
of the spatially localized feed rate. The amplitude of this spot is
shown to depend on the maximum value of the concentrated feed. By
analyzing instabilities of this new type of spot profile to locally
non-radially symmetric perturbations, we show in
Fig.~\ref{pinned_lambda0} below the qualitatively new result that the
usual peanut-splitting mode is not necessarily the first angular mode
to go unstable as parameters are varied. This theoretical prediction
is confirmed with full PDE numerical simulations where it is shown
that a localized spot, pinned at the concentration point of the feed,
can undergo a spot self-replication process leading to either two or
three new spots (see
Fig.~\ref{pinned_split_exp1}--\ref{pinned_split_exp3} below).
Finally, full PDE simulations show that a localized spot can remain
pinned at the concentration point of the feed even when this
concentration point is evolving dynamically in the domain.

For a heterogeneous substrate with a concentrated source of feed, in
\S \ref{sec:pinned_quasi} we analyze the existence, linear stability,
and slow spot dynamics for quasi-equilibrium $N+1$ spot patterns that
consist of $N$ unpinned spots together with an additional spot
centered at the concentration point of the feed rate. By deriving a
globally coupled eigenvalue problem, we formulate a criterion for
which this pattern undergoes a competition instability, triggering a
spot-annihilation event, that is due to a zero-eigenvalue crossing of
the linearization. Finally, in \S \ref{sec:loop}, by allowing the
concentration point of the feed to evolve dynamically on a ring
concentric within the unit disk, we combine our linear stability
theory for the onset of spot self-replication or spot-annihilation
together with our result for slow spot dynamics to predict the existence
of a creation-annihilation loop, or attractor, that has a maximum of
three spots in the disk at any one time. This attractor is modeled by
augmenting the ODE's for slow spot dynamics with a procedure to create
two new spots after the peanut-splitting linear stability threshold is
exceeded. In our algorithm, a second procedure is used to remove a
spot once a competition instability, due to a zero-eigenvalue
crossing, is detected from the globally coupled eigenvalue
problem. Quantitative results obtained from this hybrid algorithm over
three cycles of the creation-annihilation loop are favorably compared
with full numerical PDE simulation results in
Fig.~\ref{pinned_split_comp_exp1_part1}--\ref{pinned_split_comp_exp1_part6}
below.

Finally, in \S \ref{sec:discussion} we discuss a few related problems with
spatial heterogeneities that warrant further investigation.

\section{Spot patterns in the Schnakenberg model with
a spatially uniform feed rate}\label{sec:proto}

In \S \ref{sec:proto_qe_slow} we briefly summarize some results
of \cite{kww09} for the construction of quasi-equilibrium $N$-spot
patterns for \eqref{proto:model} and to characterize heir slow dynamics.

\subsection{Quasi-equilibria and slow spot dynamics}\label{sec:proto_qe_slow}

In the limit $\eps\to 0$ we first construct an $N$-spot
quasi-equilibrium solution for \eqref{proto:model} with spots centered
at $\v{x}_1,\ldots,\v{x}_N$. We assume that the spots are
well-separated in the sense that $|\v{x}_i - \v{x}_j| = \bigo(1)$ for
$i\neq j$ and $\mbox{dist}(\v{x}_j,\partial\Omega)=\bigo(1)$ for
$j=1,\ldots,N$.  We assume that the quasi-equilibrium pattern is
linearly stable on ${\mathcal O}(1)$ time intervals.

In the inner region near the $j^{\text{th}}$ spot, we let
$\v{x}_j = \v{x}_j(\sigma)$ where $\sigma=\eps^2 t$ is the slow time
scale (cf.~\cite{kww09}). We introduce the inner variables
\begin{subequations}\label{proto:inner_expansion}
\begin{equation}
  v=\sqrt{D} V_{j}(\v{y}) \,, \quad u = \frac{1}{\sqrt{D}} U_j(\v{y}) \,,
  \qquad \mbox{where} \qquad \v{y} \equiv \eps^{-1} \left(\v{x} -
    \v{x}_j(\sigma)\right) \,, \quad \mbox{and} \quad \rho = |\v{y}| \,,
\end{equation}
together with the inner expansion
\begin{equation}
  V_j = V_{j0}(\rho) +  V_{j1} + \cdots \,, \qquad
  U_j = U_{j0}(\rho)  + U_{j1} + \cdots \,.
\end{equation}
\end{subequations}
Upon substituting \eqref{proto:inner_expansion} into
\eqref{proto:model}, we collect powers of $\eps$ to obtain,
at leading order, the radially symmetric {\em core problem}
\begin{subequations}\label{proto:core_full}
\begin{align}
  \Delta_\rho V_{j0} - V_{j0} + U_{j0} V_{j0}^2 &= 0 \,, \quad
  \Delta_\rho U_{j0} - U_{j0} V_{j0}^2 = 0 \,, \quad 0<\rho<\infty \,,
                      \label{proto:core_problem}\\
  V_{j0}^{\prime}(0)=U_{j0}^{\prime}(0)=0 \,; \qquad &V_{j0} \to 0 \,,
   \quad U_{j0} \sim S_j \log \rho + \chi(S_j) \,, \quad \mbox{as} \quad
  \rho \to \infty \,, \label{proto:core_problem_far_field}
\end{align}
\end{subequations}
where
$\Delta_\rho \equiv \partial_{\rho\rho} + \rho^{-1} \partial_\rho$ and
$S_j$ is called the spot source strength.  At next order, we find
that $\v{v}_1 \equiv (V_{j1},U_{j1})^T$ satisfies
\begin{subequations}\label{proto:v1all}
\begin{equation}\label{proto:v1}
  \Delta_\v{y} \v{v}_1 + \Mmat_{j} \v{v}_1 = \v{f}_j \,, \qquad
  \v{y} \in \mathbb{R}^2 \,,
\end{equation}
where $\Delta_{\v{y}}$ denote derivatives in $\v{y}$, and where we have
defined
\begin{equation}\label{proto:v1mat}
\Mmat_{j} \equiv 
\begin{pmatrix} 
-1 + 2U_{j0}V_{j0} & V_{j0}^2 \\
-2U_{j0}V_{j0} & -V_{j0}^2 
\end{pmatrix}\,,
\qquad \v{f}_j \equiv \begin{pmatrix}
  0 \\
  -V_{j0}^{\prime}(\v{e}_\phi \cdot \v{\dot{x}_j})
  \end{pmatrix} \,.
\end{equation}
\end{subequations}
Here $\v{e}_\phi\equiv (\cos\theta,\sin\theta)^T$ and
$\v{\dot{x}_j}\equiv {d\v{x}_j/d\sigma}$. For \eqref{proto:v1} we
can impose that $V_{j1} \to 0$ as $|\v{y}|\to \infty$. However, the
far-field condition of $U_{j1}$ is determined only after asymptotic
matching to an appropriate outer solution. 

\begin{figure}[htbp]
\begin{subfigure}[b]{0.35\textwidth}
\includegraphics[width=\textwidth,height=4.2cm]{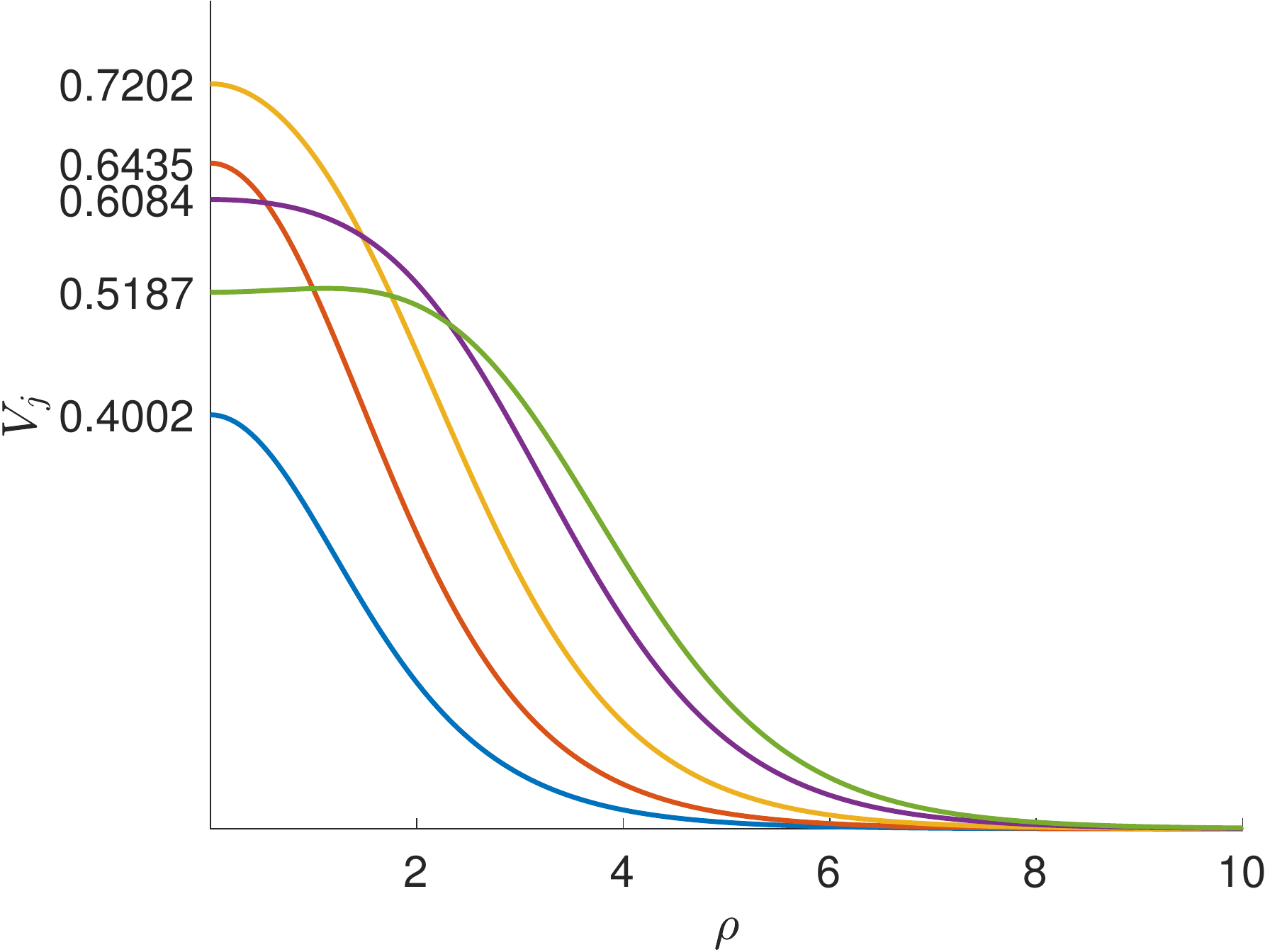}
\caption{$V_j$ with different $S$}
\end{subfigure}
\hfill
\begin{subfigure}[b]{0.3\textwidth}
\includegraphics[width=\textwidth,height=4.2cm]{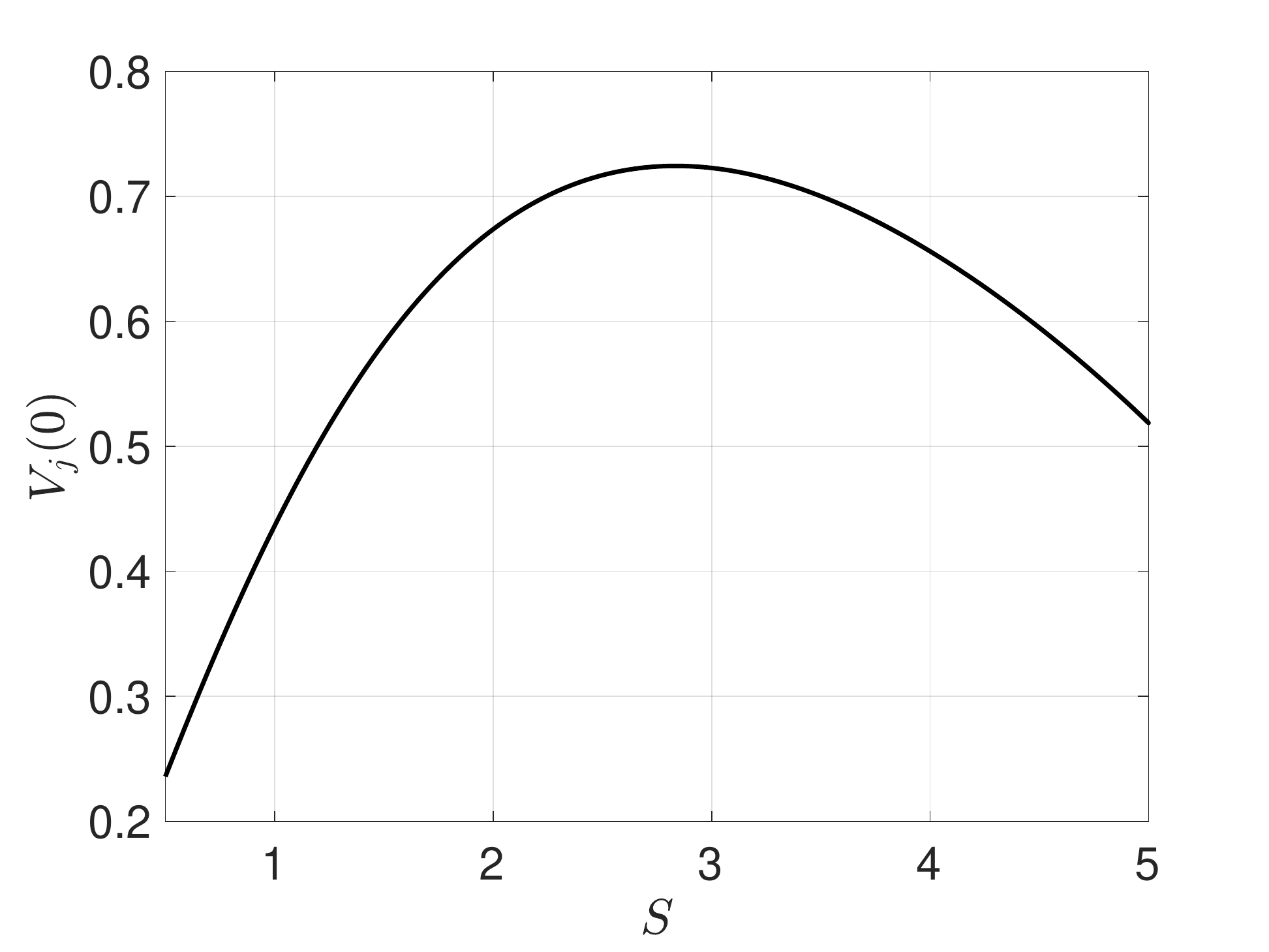}
\caption{Spot height versus $S$}
\label{demo_spot_height}
\end{subfigure}
\hfill
\begin{subfigure}[b]{0.3\textwidth}
\includegraphics[width=\textwidth,height=4.2cm]{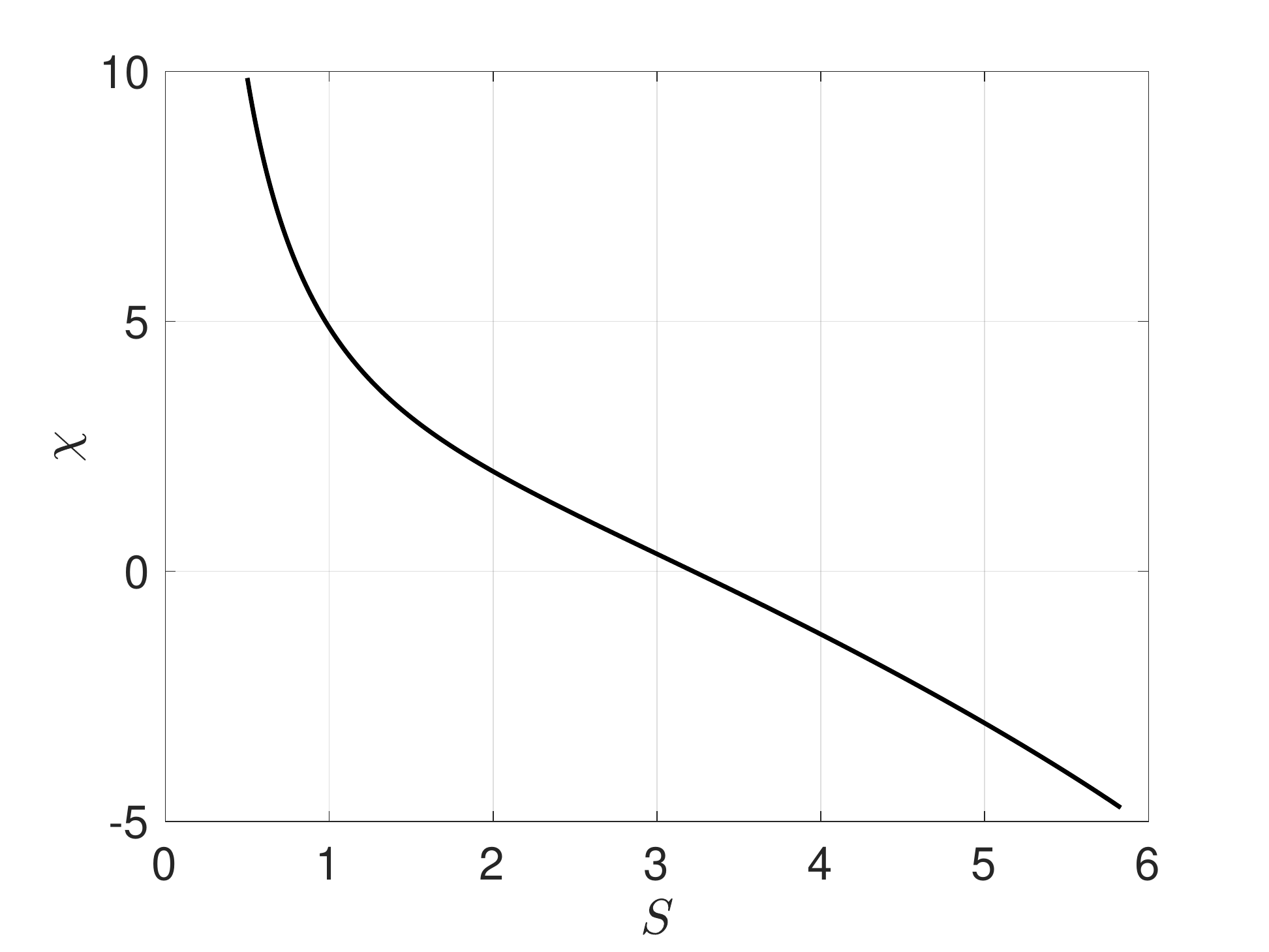}
\caption{$\chi(S)$}
\end{subfigure}
\caption{Left panel: The core solution component $V_j(\rho)$, computed
  from \eqref{proto:core_full}, for $S=0.9 \,, 1.8 \,, 3.1 \,, 4.4 $
  and $5$ with corresponding spot heights $V_j(0)$ given approximately
  by $0.4002 \,, 0.6435 \,, 0.7202 \,, 0.6435 \,, 0.5187$,
  respectively. Middle panel: Spot height versus $S$, showing a
  monotone increase on $S<S^{\star}\approx 2.83$ and a unique spot
  height for each $S>0$. Right panel: $\chi$ versus
  $S$.}\label{fig:core_old}
\end{figure}

In Fig.~\ref{fig:core_old} we plot the numerical solution
$V_{j0}(\rho)$ to the core problem \eqref{proto:core_full} for several
source strengths. We show that there is a unique spot height
$V_j(0)$ and a unique $\chi$ for each source strength.

To derive the outer problem for $u$, we integrate the equation
for $U_{j0}$ in \eqref{proto:core_problem} over $0<\rho<\infty$ to obtain
the identity
\begin{equation}\label{proto:Sj}
S_j = \int_0^\infty U_{j0} V_{j0}^2 \, \rho \, d\rho.
\end{equation}
Then, in the limit $\eps\to 0$, we use \eqref{proto:Sj} to 
obtain, in the sense of distributions, that
\begin{equation}\label{proto:correspondence_rule}
  \eps^{-2} uv^2 \to \sum\limits_{i=1}^N\left(\int_{\mathbb{R}^2} (D^{-1/2}U_{i0})
    (D^{1/2}V_{i0})^2 d\v{y} \right) \delta(\v{x}-\v{x}_i) =
  2\pi \sqrt{D} \sum\limits_{i=1}^N S_i \, \delta(\v{x}-\v{x}_i) \,.
\end{equation}
By using this distributional limit in \eqref{proto:model}, we obtain
that the outer problem, defined away from the spots, is 
\begin{equation}\label{proto:outer_problem}
  \Delta u + \frac{a}{D} - \frac{2\pi}{\sqrt{D}}\sum\limits_{i=1}^N S_i \,
  \delta(\v{x}-\v{x}_i) = 0 \quad \mbox{in} \quad \Omega\,, \qquad
  \partial_n u = 0 \quad \mbox{on} \quad \partial\Omega \,.
\end{equation}
By integrating \eqref{proto:outer_problem} over $\Omega$ and using the
divergence theorem, we get
\begin{equation}\label{proto:sum_Sj}
\sum\limits_{i=1}^N S_i = \frac{a|\Omega|}{2\pi\sqrt{D}} \equiv p_a \,,
\end{equation}
where $|\Omega|$ denotes the area of $\Omega$. The solution to
\eqref{proto:outer_problem} is represented as
\begin{equation}\label{proto:outer_solution}
  u(\v{x}) = -\frac{2\pi}{\sqrt{D}} \sum\limits_{i=1}^N S_i \, G(\v{x};\v{x}_i) +
  \ubar \,,
\end{equation}
where $\ubar$ is an undetermined additive constant and $G(\v{x};\v{z})$ is
the unique Neumann Green's function satisfying 
\begin{equation}\label{proto:neu_green}
\begin{aligned}
  &\Delta G = \frac{1}{|\Omega|} - \delta(\v{x}-\v{z}) \quad \mbox{in} \quad
  \Omega \,, \quad \partial_n G = 0 \quad \mbox{on} \quad\partial\Omega \,;
  \qquad \int_{\Omega} G(\v{x};\v{z}) \, d\v{x}=0\,, \\
  & G \sim -\frac{1}{2\pi} \log |\v{x}-\v{z}| + R(\v{z};\v{z}) +
  \nabla_{\v{x}} R(\v{x},\v{z}) \vert_{\v{x}=\v{z}} \cdot (\v{x}-\v{z}) +
  {\mathcal O}(|\v{x}-\v{z}|^2) \,, \quad \mbox{as} \quad \v{x} \to \v{z} \,,
\end{aligned}
\end{equation}
where $R(\v{z};\v{z})$ is called the regular part of $G$.

To determine a nonlinear algebraic system for the source strengths,
and a DAE system for slow spot dynamics, we must match the near-field
behavior as $\v{x}\to\v{x}_j$ of the outer solution
\eqref{proto:outer_solution} to the far-field behavior of the two-term
inner solution, which is given from
\eqref{proto:core_problem_far_field} and \eqref{proto:inner_expansion}
by
\begin{equation}\label{proto:matching_inner}
  u \sim \frac{1}{\sqrt{D}} \left( S_j \log \rho + \chi(S_j) +
    \eps U_{j1} + \cdots \right) = \frac{1}{\sqrt{D}} \left(S_j
    \log|\v{x}-\v{x}_j| + \frac{S_j}{\nu} + \chi(S_j) \right) +
  \frac{\eps U_{j1}}{\sqrt{D}} + \cdots \,,
\end{equation}
as $\rho=\eps^{-1}|\v{x}-\v{x}_j|\to \infty$, where we have defined
$\nu \equiv {-1/\log\eps}$. Then, by Taylor-expanding
\eqref{proto:outer_solution} as $\v{x}\to\v{x}_j$, and replacing
$\v{x}-\v{x}_j=\eps \v{y}$, we obtain after some algebra that
\begin{equation}\label{proto:matching_outer}
  u \sim \frac{S_j}{\sqrt{D}} \log|\v{x}-\v{x}_j| - \frac{2\pi}{\sqrt{D}}
  \left(S_j R_{j,j} + \sum\limits_{i\neq j}^N S_i G_{j,i}\right) +
  \ubar - \frac{\eps}{\sqrt{D}} \, (\pmb{\beta}_j \cdot \v{y}) + \cdots \,,
\end{equation}
where $\pmb{\beta}_j$ is defined by
\begin{equation}\label{proto:betaj}
  \pmb{\beta}_j \equiv 2\pi \left(S_j\nabla_{\v{x}} R_{j,j} +
    \sum\limits_{i \neq j}^N S_i \nabla_{\v{x}} G_{j,i} \right) \,.
\end{equation}
Here we have labeled $R_{j,j} \equiv R(\v{x}_j;\v{x}_j)$,
$G_{j,i} \equiv G(\v{x}_j;\v{x}_i)$, $\nabla_{\v{x}}R_{j,j}\equiv
\nabla_{\v{x}}R(\v{x};\v{x}_j)\vert_{\v{x}=\v{x}_j}$, and
$\nabla_{\v{x}}G_{j,i}\equiv
\nabla_{\v{x}}G(\v{x};\v{x}_i)\vert_{\v{x}=\v{x}_j}$.

By comparing the leading terms in \eqref{proto:matching_inner}
and \eqref{proto:matching_outer}, and recalling
\eqref{proto:sum_Sj}, we obtain in matrix form that
\begin{subequations}\label{proto:nas_orig}
\begin{equation}\label{proto:source_strength_2}
  \v{s} + 2 \pi \nu \, \Gmat \, \v{s} + \nu \chivec = \nu \ubar \sqrt{D} \,
  \v{e} \,, \qquad \v{e}^T \v{s} = p_a \equiv \frac{a|\Omega|}{2\pi\sqrt{D}}\,,
\end{equation}
where we have defined $\v{s}$, $\chivec$, $\v{e}$, and the Neumann
Green's matrix $\Gmat \in \mathbb{R}^{N\times N}$ for
$\v{x}_1, \ldots, \v{x}_N$ by
\begin{equation}\label{proto:def_nas}
  \v{s} = (S_1, \ldots, S_N)^T \,, \qquad \chivec \equiv
  (\chi(S_1), \ldots, \chi(S_N))^T \,, \qquad
  \v{e} \equiv (1,\ldots, 1)^T \in \mathbb{R}^N \,, \qquad
  (\Gmat)_{i\, j} = 
\begin{cases}
	R_{j,j} \quad \text{if } i = j \,, \\
	G_{i,j} \quad \text{if } i \neq j \,.
\end{cases}
\end{equation}
\end{subequations}
By left-multiplying \eqref{proto:source_strength_2} by $\v{e}^T$, and
by using $\v{e}^T\v{s}=p_a$, we can isolate $\ubar$. Then, by
substituting $\ubar$ back into \eqref{proto:source_strength_2}, we can
decouple \eqref{proto:source_strength_2} to obtain that $\v{s}$
satisfies the nonlinear algebraic system (NAS)
\begin{equation}\label{proto:source_system}
  \v{s} + 2 \pi \nu \, (\Imat - \Emat) \, \Gmat \v{s} + \nu \,
  (\Imat - \Emat) \, \chivec = \frac{p_a}{N} \v{e} \,, \qquad \mbox{with}
  \qquad \ubar = \frac{p_a + 2 \pi \nu \, \v{e}^T \Gmat \, \v{s} + \nu \,
    \v{e}^T \chivec}{\nu \sqrt{D} N} \,.
\end{equation}
Here $\Emat\equiv N^{-1}\v{e}\v{e}^T\in \mathbb{R}^{N\times N}$ and
$\Imat \in \mathbb{R}^{N\times N}$ is the identity matrix.

To determine the slow dynamics, we proceed to next order and match the
$\bigo(\eps)$ terms in \eqref{proto:matching_inner} and
\eqref{proto:matching_outer}. This yields that the far-field behavior
for the solution $U_{j1}$ to \eqref{proto:v1all} is
\begin{equation}\label{proto:Uj1_far_field}
  U_{j1} \sim -\pmb{\beta}_j \cdot \v{y}\,, \quad \mbox{as} \quad
  \rho = |\v{y}| \to \infty \,,
\end{equation}
where $\pmb{\beta}_j$ is defined in \eqref{proto:betaj}.  The ODE
system for the spot locations is obtained by imposing a solvability
condition on the solution to \eqref{proto:v1all} with far-field
behavior \eqref{proto:Uj1_far_field}. By differentiating the core
problem \eqref{proto:core_full} with respect to $y_1$ and $y_2$, it
follows that the homogeneous problem
$\lap_\v{y} \v{\Phi} + M_j \v{\Phi} = \v{0}$ has two non-trivial
solutions. As such, there are two solutions to the corresponding
homogeneous adjoint problem
$\lap_\v{y} \v{\Psi} + M_j^T \v{\Psi} = \v{0}$. These two solutions
have the form
\begin{equation}
  \v{\Psi}_c = \v{P}(\rho)\cos\phi \,, \qquad \v{\Psi}_s = \v{P}(\rho) \sin\phi
  \,,
\end{equation}
where $\v{P}(\rho) \equiv \left( P_1(\rho), P_2(\rho) \right)^T$ is the
normalized nontrivial solution to 
\begin{equation}\label{proto:adjoint_problem_radial}
  \lap_\rho \v{P} - \frac{1}{\rho^2} \, \v{P} + M_j^T \v{P} = \v{0} \,,
  \quad \mbox{with} \quad \v{P} \sim \begin{pmatrix}
  0 \\
 {1/\rho}
\end{pmatrix} \quad \mbox{as} \quad \rho \to \infty \,.
\end{equation}

In \cite{kww09} the solvability condition is obtained by multiplying
\eqref{proto:v1} by $\v{\Psi}_c$ and $\v{\Psi}_s$ and applying Green's
second identity on a sufficiently large circle where the far-field
conditions \eqref{proto:Uj1_far_field} and
\eqref{proto:adjoint_problem_radial} are imposed. This yields the
following ODE system for $\v{x}_j(\sigma)$, for $j = 1,\ldots,N$, with
$\sigma=\eps^2 t$, that characterize the slow spot dynamics:
\begin{equation}\label{proto:DAE}
  \frac{d\v{x}_j}{d\sigma} = - \gamma(S_j) \, \pmb{\beta}_j \,, \qquad
  \gamma(S_j) \equiv -\frac{2}{\int_0^\infty P_1 V_{j0}^{\prime} \, \rho \,
    d\rho} \,.
\end{equation}
Here $\pmb{\beta}_j$ is defined in \eqref{proto:betaj}, while
$S_1,\ldots,S_N$ satisfies the NAS \eqref{proto:source_system}. The
plot in Fig.~3 of \cite{kww09} of the numerically computed $\gamma(S_j)$
shows that $\gamma(S_j)>0$.
 
\subsection{Linear stability analysis}\label{sec:proto_linstab}

The slow spot dynamics \eqref{proto:DAE} is valid only when the
quasi-equilibrium solution is linearly stable on ${\mathcal O}(1)$
time-scales. In this subsection we analyze the linear stability of the
quasi-equilibrium solution, denoted by $v=v_e$ and $u=u_e$.  To do so,
we introduce the perturbation $v = v_e + e^{\lambda t} \phi$ and
$u = u_e + e^{\lambda t} \eta$ into \eqref{proto:model}, and upon
linearizing we obtain
\begin{equation}\label{proto:linstab_linearization}
  \eps^2 \lap \phi - \phi + 2u_ev_e\phi + v_e^2\eta = \lambda \phi\,, \qquad
  D\lap\eta + a - \eps^{-2} \left(2u_ev_e\phi + v_e^2\eta\right) = \tau
  \lambda\eta\,, \quad \mbox{in} \quad \Omega \,,
\end{equation}
with $\partial_n \phi = \partial_n \eta = 0$ on $\partial\Omega$.

In the inner region near the $j^{\text{th}}$ spot we have to leading
order that $v_e \sim \sqrt{D} \,V_{j0}(\rho)$ and
$u_e \sim U_{j0}(\rho)/\sqrt{D}$, where
$\rho = \eps^{-1}|\v{x}-\v{x}_j|$, with $V_{j0}$ and $U_{j0}$
satisfying the core problem \eqref{proto:core_full}. By letting
$\phi \sim e^{im\theta}\Phi_j(\rho)$ and
$\eta \sim e^{im\theta}N_j(\rho)/D$, for the integer angular mode
$m\geq 0$, we obtain the leading order inner eigenvalue problem
\begin{equation}\label{proto:lin_stab_eigenproblem}
  \lap_\rho \Phi_j - \frac{m^2}{\rho^2}\Phi_j - \Phi_j + 2 \, U_{j0} V_{j0} \Phi_j
  + V_{j0}^2 N_j = \lambda \Phi_j \,, \qquad \lap_\rho N_j -
  \frac{m^2}{\rho^2}N_j - 2 U_{j0} V_{j0} \Phi_j - V_{j0}^2 N_j = 0\,.
\end{equation}

We first consider non-radially symmetric perturbations for which
$m > 0$. The case $m=1$ corresponds, trivially, to the translation
mode $(\Phi_j, N_j) = (U_{j0}^{\prime}, V_{j0}^{\prime})$ with
$\lambda = 0$. For angular modes with $m \geq 2$, we impose
$\Phi_j \to 0$ exponentially as $\rho \to 0$. In addition, owing to
the ${m^2 N_j/\rho^2}$ term in \eqref{proto:lin_stab_eigenproblem}, we
impose the far-field decay condition $N_j \sim \bigo(\rho^{-m})$ as
$\rho \to \infty$. The eigenvalue $\lambda_\text{max}$ in
\eqref{proto:lin_stab_eigenproblem} with the largest real part has
been numerically calculated in \cite{kww09} for a range of $S_j$. For
each $m\geq 2$, it was found that $\lambda_\text{max}$ is real and
negative (positive) when $S_j < \Sigma_m$ ($S_j > \Sigma_m$) (see
Fig.~4 of \cite{kww09}). Moreover, as shown numerically in
\cite{kww09}, the ordering principle $\Sigma_2 < \Sigma_3 < \ldots$
holds for the stability thresholds for non-radially symmetric
perturbations. As such, the mode $m=2$, referred as to the
peanut-splitting mode, is the first to lose stability when $S_j$ is
increased. The critical threshold for this mode is
$\Sigma_2\approx 4.302$.  In \cite{tw20} it was shown that this
symmetry-breaking bifurcation is subcritical and, for a steady-state
spot, it triggers a nonlinear spot self-replication process.

In contrast to the local analysis of instabilities associated with
non-radially symmetric perturbations, the eigenvalue problem for
radially symmetric perturbations with $m=0$ is derived by globally
coupling local problems near each spot. To derive this {\em globally
  coupled eigenvalue problem} (GCEP), we set $m=0$ in
\eqref{proto:lin_stab_eigenproblem} and impose that $N_j\sim c_j\log\rho$
as $\rho\to\infty$, where $c_j$ is an unknown constant. We then write
\begin{equation}\label{proto:m=0_scale}
    \Phi_j=c_j\tilde{\Phi}_j \,, \qquad  N_j=c_j\tilde{N}_j\,,
\end{equation}
so as to obtain from \eqref{proto:lin_stab_eigenproblem} that
\begin{subequations}\label{proto:m=0_eigenproblem}
\begin{align}
\lap_\rho \tilde{\Phi}_j - &\tilde{\Phi}_j + 2 \,
  U_{j0} V_{j0} \, \tilde{\Phi}_j + V_{j0}^2 \, \tilde{N}_j = \lambda
 \tilde{\Phi}_j\,, \qquad \lap_\rho \tilde{N}_j - 2 \, U_{j0} V_{j0} \,
 \tilde{\Phi}_j - V_{j0}^2 \, \tilde{N}_j = 0\,, \quad \rho > 0 \,, \\
    &\tilde{\Phi}_j^{\prime}(0)=\tilde{N}_j^{\prime}(0)=0 \,; \qquad
      \tilde{\Phi}_j \to 0\,, \quad \tilde{N}_j \sim \log\rho +
     \tilde{B}(S_j;\lambda) + o(1)\,, \quad \mbox{as} \quad \rho \to \infty\,,
\end{align}
\end{subequations}
where $\tilde{B}(S_j;\lambda)$ must be calculated numerically from
\eqref{proto:m=0_eigenproblem}. However, by differentiating the core
problem \eqref{proto:core_full} with respect to $S_j$, we observe that
$\partial_S V_{j0}$ and $\partial_S U_{j0}$ satisfy
\eqref{proto:m=0_eigenproblem} when $\lambda = 0$. As a result, we
have the identity that $\tilde{B}(S_j;0) = \chi^{\prime}(S_j)$.

By integrating the $\tilde{N}_j$ equation in \eqref{proto:m=0_eigenproblem},
and using \eqref{proto:m=0_scale}, we obtain the identity
\begin{equation}\label{proto:linstab_cj}
  c_j = \int_0^{\infty} \left( 2 U_{j0} V_{j0} \Phi_j + V_{j0}^2 N_j \right) \rho
 \, d\rho \,.
\end{equation}
Then, in the limit $\eps\to 0$, we use \eqref{proto:linstab_cj} to 
derive, in the sense of distributions, that
\begin{equation}\label{proto:linstab_corr_rule}
  \eps^{-2} \left(2 u_e v_e \phi + v_e^2 \eta\right) \to 2\pi
  \sum\limits_{i=1}^N c_i \, \delta(\v{x}-\v{x}_i) \,.
\end{equation}
We use \eqref{proto:linstab_corr_rule}, together with the asymptotic
matching condition $\eta\sim {c_j \tilde{N_j}/D}$, where $\tilde{N}_j$
has the far-field behavior as $\rho\to\infty$ in
\eqref{proto:m=0_eigenproblem}, to obtain the following outer problem
for $\eta$, defined away from the spots:
\begin{subequations}\label{proto:linstab_all_outer}
\begin{align}
  \lap \eta - \frac{\tau \lambda}{D} \eta &= \frac{2\pi}{D}\sum\limits^N_{i=1}
    c_i \delta(\v{x}-\v{x}_i) \quad \mbox{in} \quad \Omega \,, \qquad
    \partial_n \eta=0 \quad \mbox{in}\quad \partial\Omega                                    \label{proto:linstab_outer_problem}\\
  \eta & \sim \frac{c_j}{D} \left[\log|\v{x}-\v{x}_j| + \frac{1}{\nu} +
    \tilde{B}(S_j;\lambda)\right]\,, \quad \mbox{as} \quad
  \v{x} \to \v{x}_j \,, \qquad \mbox{for} \quad j=1,\ldots,N \,.
  \label{proto:linear_stability_matching_inner_1}
\end{align}
\end{subequations}

For $\lambda\neq 0$, we represent the solution to
\eqref{proto:linstab_outer_problem} as
\begin{equation}\label{proto:linstab:outer_solution}
\eta = -\frac{2\pi}{D}\sum\limits_{i=1}^N c_i \, G_\lambda(\v{x};\v{x}_i) \,,
\end{equation}
where $G_\lambda(\v{x},\v{z})$ is the eigenvalue-dependent Green's function
satisfying
\begin{subequations}\label{proto:eig_green}
\begin{align}
  \lap &G_\lambda - \frac{\tau \lambda}{D} G_\lambda = -\delta(\v{x}-\v{z}) \quad
   \mbox{in} \quad \Omega\,, \qquad \partial_n G_\lambda = 0 \quad \mbox{on}
   \quad \partial\Omega\,, \\
  &G_\lambda \sim -\frac{1}{2\pi}\log|\v{x}-\v{z}| + R_\lambda(\v{z};\v{z}) + o(1)
      \quad \mbox{as} \quad \v{x} \to \v{z} \,.
\end{align}
\end{subequations}
By Taylor-expanding $\eta$ in \eqref{proto:linstab:outer_solution} as
$\v{x}\to \v{x}_j$, and then equating the resulting expression with
\eqref{proto:linear_stability_matching_inner_1}, we conclude that
\begin{equation}\label{proto:linear_stability_matching_spot}
  c_j + 2 \pi \nu \left( c_j \Rljj + \sum\limits_{i\neq j}^N c_i \,\Glji \right) +
  \nu \, c_j \tilde{B}(S_j;\lambda)  = 0 \,, \qquad j=1\,,\ldots,N\,,
\end{equation}
where $\Rljj \equiv R_\lambda(\v{x}_j;\v{x}_j)$ and
$\Glji \equiv G_\lambda(\v{x}_j;\v{x}_i)$. In matrix form, and
with  $\v{c}\equiv (c_1,\ldots,c_N)^T$,
\eqref{proto:linear_stability_matching_spot} is equivalent to
\begin{subequations}\label{proto:GCEP_all}
\begin{equation}\label{proto:GCEP}
  \Mmat(\lambda) \v{c} = \v{0}\,, \qquad \mbox{where} \qquad
  \Mmat(\lambda) \equiv \Imat + 2\pi\nu\Gmatl + \nu\Bmat \,.
\end{equation}
Here $\Imat \in \mathbb{R}^{N\times N}$ is the identity matrix, while the
symmetric Green's matrix $\Gmat_\lambda$ and diagonal matrix $\Bmat$
are defined by
\begin{equation}\label{proto:Gmatl_Bmat}
\big(\Gmatl\big)_{ij} \equiv
\begin{cases}
\Rljj &\quad \mbox{if } i = j \,, \\
\Glij &\quad \mbox{if } i \neq j \,,
\end{cases}
\qquad
\big(\Bmat\big)_{ij} =
\begin{cases}
\tilde{B}(S_j;\lambda) &\quad \mbox{if } i = j \,, \\
0 &\quad \mbox{if } i \neq j \,.
\end{cases}
\end{equation}
The homogeneous matrix system \eqref{proto:GCEP} for $\v{c}$, referred
to as the GCEP, has a nontrivial solution if and only if 
\begin{equation}\label{proto:det_m}
  \det\Mmat(\lambda)=0 \,.
\end{equation}
\end{subequations}
A discrete root $\lambda$ to \eqref{proto:det_m} for which
$\mbox{Re}(\lambda)>0$ corresponds to a locally radially symmetric
instability near the spots, while the corresponding eigenvector $\v{c}$
characterizes the small-scale perturbation of the spot amplitudes.

In this way, the linear stability properties associated with locally
radially symmetric perturbations near the spots is reduced to the
problem of determining the number ${\mathcal N}$ of roots of
$\det \Mmat(\lambda)=0$ in the right-half $\mbox{Re}(\lambda)>0$ of
the spectral plane. To do so, we formulate and numerically implement a
winding number procedure over the counterclockwise contour
${\mathcal C}_\zeta$ that consists of the semi-circle
$|\lambda| = \zeta > 0$, for
$-\pi/2 \leq \mbox{arg}\lambda \leq \pi/2$, and the imaginary segment
$\{\lambda = i\lambda_I : -\zeta \leq \lambda_I \leq \zeta\}$. However,
since $\Mmat(\lambda)$ is undefined at $\lambda=0$, we need to first
find the behavior of $\det\Mmat(\lambda)$ as $\lambda\to 0$ so as to
remove this singularity. To do so, we let $\lambda\to 0$ in
\eqref{proto:eig_green} and readily calculate that
\begin{equation}\label{proto:green_zero}
  \mc{G}_\lambda = \frac{D}{|\Omega|\tau\lambda} \v{e}\v{e}^T + \mc{Q} \,,
  \qquad \mbox{where} \qquad \mc{Q} \equiv \mc{G} + \bigo(\tau \lambda)
  \quad \mbox{as} \quad \lambda \to 0 \,.
\end{equation}
Here $\mc{G}$ is the Neumann Green's matrix and
$\v{e}\equiv (1,\ldots,1)^T$.  Since $\v{e}\v{e}^T$ is a rank one matrix, we
substitute \eqref{proto:green_zero} into \eqref{proto:GCEP} for
$\Mmat$ and, by using the well-known matrix determinant lemma, we obtain
\begin{equation}\label{proto:M_lambda_singularity}
  \det \Mmat(\lambda) = \det\left(\Imat + 2 \pi \nu \mc{Q} + \nu \Bmat \right) +
  \frac{2\pi\nu D}{|\Omega|\tau\lambda}
  \left[ \v{e}^T \mrm{adj}\left(\Imat + 2 \pi \nu \mc{Q} +
  \nu \Bmat\right) \v{e} \right] \,,
\end{equation}
where $\mrm{adj}(\mc{A})$ denotes the adjugate of a matrix
$\mc{A}$. From \eqref{proto:M_lambda_singularity} it follows that
$\det\Mmat(\lambda)$ has a simple pole at $\lambda=0$. As a result, it
is convenient to introduce the function ${\mathcal T}(\lambda)$
defined by ${\mathcal T}(\lambda) \equiv \lambda \det \Mmat(\lambda)$,
which has a removable singularity at $\lambda=0$ and has the same
number ${\mathcal N}$ of zeroes in $\mbox{Re}(\lambda)>0$ as does
$\det \Mmat(\lambda)$. The argument principle for ${\mathcal T}$ yields
that
\begin{equation}\label{proto:wind}
  {\mathcal N} = {\mathcal P} +
  \frac{1}{2\pi} \lim_{\zeta\to \infty} \left[ \mbox{arg}
    {\mathcal T}(\lambda) \right]_{{\mathcal C}_\zeta} \,, \qquad
  \mbox{where} \qquad {\mathcal T}(\lambda) \equiv \lambda \det \Mmat(\lambda)
  \,.
\end{equation}
Here ${\mathcal P}$ is the number of poles of ${\mathcal T}(\lambda)$
in $\mbox{Re}(\lambda)>0$. Since ${\mathcal G}_{\lambda}$ is analytic
in $\mbox{Re}(\lambda)>0$, any such pole can only arise from the
diagonal matrix $\Bmat$ as defined by
\eqref{proto:Gmatl_Bmat}. However, from a numerical computation of the
local problem \eqref{proto:m=0_eigenproblem}, we find that $\Bmat$ is
analytic in $\mbox{Re}(\lambda)>0$ and so ${\mathcal P}=0$ in
\eqref{proto:wind}. To determine ${\mathcal N}$ in the examples below,
the change
$\left[ \mbox{arg} {\mathcal T}(\lambda) \right]_{{\mathcal C}_\zeta}$
in the argument of ${\mathcal T}$ over the contour ${\mathcal C}_\zeta$
is computed numerically.

Next, we study zero-eigenvalue crossings. Since
$\tilde{B}(S_j;0)=\chi^{\prime}(S_j)$, the outer problem
\eqref{proto:linstab_outer_problem} when $\lambda=0$ becomes
\begin{subequations}\label{proto:linstab_all_outer0}
\begin{align}
  \lap \eta &= \frac{2\pi}{D}\sum\limits^N_{i=1}
              c_i \delta(\v{x}-\v{x}_i) \quad \mbox{in} \quad \Omega \,, \qquad
              \partial_n \eta=0 \quad \mbox{in}\quad \partial\Omega                                    \label{proto:linstab_outer_problem0}\\
  \eta & \sim \frac{c_j}{D} \left[\log|\v{x}-\v{x}_j| + \frac{1}{\nu} +
         \chi^{\prime}(S_j)\right]\,, \quad \mbox{as} \quad
         \v{x} \to \v{x}_j \,, \quad j=1,\ldots,N \,.
  \label{proto:linear_stability_matching_inner_10}
\end{align}
\end{subequations}
From the divergence theorem we conclude that
$\sum_{i=1}^{N}c_i=0$. With this constraint, we represent the solution
to \eqref{proto:linstab_outer_problem0} in terms of the Neumann
Green's function $G$ of \eqref{proto:neu_green} as
\begin{equation}\label{proto:linstab_zero}
  \eta = -\frac{2\pi}{D}\sum\limits_{i=1}^N c_i \, G(\v{x};\v{x}_i) +
  \bar{\eta} \,,
\end{equation}
where $\bar{\eta}$ is an additive constant to be determined. Then, we
Taylor-expand \eqref{proto:linstab_zero} as $\v{x}\to \v{x}_j$ by
recalling the local behavior of $G$ in \eqref{proto:neu_green}. By
equating the resulting expression with the required singularity
condition \eqref{proto:linear_stability_matching_inner_10}, we obtain a
matrix system for $\v{c}=(c_1,\ldots,c_N)^T$ and $\bar{\eta}$ of the
form
\begin{equation}\label{proto:c_system}
  (\Imat + 2\pi\nu \, \Gmat + \nu \Bmat_0) \v{c} = \nu \, \bar{\eta} \,
  \v{e} \,,   \qquad  \v{e}^T\v{c}=0 \,,
\end{equation}
where $\Gmat$ is the Neumann Green's matrix and where the
diagonal matrix $\Bmat_0$ is defined by
\begin{equation}\label{zero:bmat}
\left( \Bmat_0 \right)_{i\, j} =
\begin{cases}
\chi^{\prime}(S_j) &\quad \mbox{if } i = j \,, \\
0 &\quad \mbox{if } i \neq j \,,  
\end{cases} \qquad \longrightarrow \qquad \Bmat_0 \equiv
\mbox{diag}\left(\chi^{\prime}(S_1),\ldots, \chi^{\prime}(S_N) \right)\,.
\end{equation}

By left-multiplying \eqref{proto:c_system} by $\v{e}^T$, and using
$\v{e}^T\v{c}=0$, we find that
$\bar{\eta} = N^{-1}\left(2\pi\v{e}^T\Gmat\v{c} + \v{e}^T\Bmat_0 \v{c}\right)$.
By substituting this expression back into the first equation in
\eqref{proto:c_system} we derive that
\begin{equation}\label{proto:GCEP_zero_crossing}
\Mmat_0 \v{c} = \v{0} \,, \qquad \mbox{where} \qquad \Mmat_0 \equiv
\Imat + 2\pi\nu(\Imat-\Emat)\Gmat + \nu(\Imat-\Emat)\Bmat_0 \,,
\end{equation}
where $\Emat\equiv N^{-1}\v{e}\v{e}^T$. We conclude that a
zero-eigenvalue crossing associated with locally radially symmetric
perturbations near the spots occurs if and only if $\det\Mmat_0=0$.
Since the corresponding nontrivial eigenmode $\v{c}$ satisfies
$\v{e}^T \v{c} = 0$, it is referred to as a {\em competition mode} as
it locally preserves the sum of all the spot amplitudes.

Finally, we relate the zero-eigenvalue crossing condition
$\det\Mmat_0=0$ to the local solvability of the NAS
\eqref{proto:source_system}. Suppose, for a particular fixed parameter
set, that $\v{s}=\v{s}_e$ is a non-degenerate solution to the NAS
\eqref{proto:source_system} in the sense that the Jacobian matrix of
the NAS is invertible at $\v{s}=\v{s}_e$. Upon introducing the
perturbation $\v{s}=\v{s}_e+\v{c}$ into \eqref{proto:source_system}
where $|\v{c}|\ll 1$, we linearize the NAS to readily determine that
this Jacobian matrix is in fact the GCEP matrix $\Mmat_0$ of
\eqref{proto:GCEP_zero_crossing}, in which
$\Bmat_0 \equiv \mbox{diag}\left(\chi^{\prime}(S_{1e}),\ldots,
  \chi^{\prime}(S_{Ne}) \right)$. As a result, if $\v{s}_e$ is a
non-degenerate solution to the NAS \eqref{proto:source_system} we must
have $\det\Mmat_0\neq 0$, and so $\lambda=0$ is not an eigenvalue of
the GCEP. Therefore, it is only at a bifurcation point of the NAS
\eqref{proto:source_system} where a zero-eigenvalue crossing of the
GCEP can occur. This correspondence is summarized as
\begin{equation}\label{proto:cross_nas}
  \det\Mmat_0=0  \qquad \Longleftrightarrow \qquad \v{s}_e
  \mbox{ is at a bifurcation point of the NAS } \eqref{proto:source_system} \,.
\end{equation}
	
\subsection{An $N$-spot ring pattern}\label{sec:basic_ring}

An $N$-spot ring pattern is a pattern of $N$ equally-spaced spots
located on a ring of radius $r_0$, with $0<r_0<1$, that is concentric
within the unit disk $\Omega$. For $j=1,\ldots,N$, the locations of the
spots on the ring can be taken as
\begin{equation}\label{proto:ring}
  \v{x}_j = r_0 \, \v{e}_{\theta_j} \,, \qquad \v{e}_{\theta_j} \equiv
  \left(\cos\theta_j \,, \, \sin\theta_j \right)^T \,, \qquad \theta_j
  \equiv \frac{2\pi (j-1)}{N} \,, \quad \mbox{for} \quad j=1,\ldots,N\,.
\end{equation}
For a ring pattern, the symmetric Neumann Green's matrix $\Gmat$ is
also circulant, and so it has the eigenvector
$\v{e}=(1,\ldots,1)^T$. As a result, the NAS
\eqref{proto:source_system} admits a symmetric solution where the
spots have the common source strength $S_j=S_{c}\equiv {p_a/N}$, for
$j=1,\ldots,N$, where $p_a$ is given in \eqref{proto:sum_Sj}.

As shown in Appendix \ref{app:neum}, with
$\v{x}_j = r_0(\sigma) \v{e}_{\theta_j}$ the spot dynamics given in
\eqref{proto:DAE} can be reduced to the scalar ODE 
\begin{equation}\label{proto:ring_scalar_ode}
  \frac{dr_0}{d\sigma} = \gamma(S_c) S_c \left( \frac{N-1}{2r_0} -
    \frac{N r_0^{2N-1}}{1 - r_0^{2N}} - N r_0 \right) \,, \qquad
  \mbox{with} \quad S_c = \frac{a|\Omega|}{2\pi N \sqrt{D}}\,,
\end{equation}
for the ring radius, where $\sigma=\eps^2 t$. On $0<r_0<1$, this ODE
\eqref{proto:ring_scalar_ode} has a globally stable equilibrium point
$r_{0e}$, given by the unique root to
\begin{equation}\label{proto:equilibrium_ring_radius}
 \frac{N-1}{2N} - r_0^2 = \frac{r_0^{2N}}{1 - r_0^{2N}} \,.
\end{equation}

From \S \ref{sec:proto_linstab}, the $N$-spot ring pattern is linear
stable to locally non-radially symmetric perturbations near the spots
only when $S_c<\Sigma_2\approx 4.302$, where
$S_{c}={a|\Omega|/[2\pi N\sqrt{D}]}$ with $|\Omega|=\pi$. In terms of
the feed rate $a$, this stability condition holds when
$a<a_f\equiv 2\Sigma_2 \sqrt{D} N\approx 8.6 D N$.

Next, we study the linear stability associated with radially-symmetric
perturbations near the spots. For a ring pattern, the GCEP
\eqref{proto:GCEP} becomes
\begin{equation}\label{proto:GCEP_ring0}
  \Mmat \v{c} = \v{0} \,, \qquad \mbox{where} \qquad \Mmat = (1 + \nu \Bhat_c)
  \Imat + 2 \pi \nu \, \Gmatl \,.
\end{equation}
Here $\Bhat_c \equiv \Bhat(S_c;\lambda)$ is to be calculated from the
inner problem \eqref{proto:m=0_eigenproblem} with $S_j=S_c$. Owing to
the cyclic structure of the ring pattern, the symmetric Green's matrix
$\Gmatl$ is also a circulant matrix and, as a result, it has the matrix
spectrum (see Appendix \ref{appendix:circulant})
\begin{equation}\label{proto:gmat_circ}
  \Gmatl \v{e}=\omega_1 \v{e} \,, \qquad
  \Gmatl\v{q}_j=\omega_{j}\v{q}_j \,, \quad j=2,\ldots,N \,;
  \qquad \v{e}^T \v{q}_j=0 \,, \quad \v{q}_i^T \v{q}_j=0 \,, \quad
  i\neq j \,,
\end{equation}
where $\v{q}_j$ for $j=2,\ldots,N$ are given in
\eqref{circulant_spectrum2}.  The matrix eigenvalues $\omega_j$ are
given in terms of the first row of $\Gmatl$ by
\eqref{circulant_spectrum2}, while the entries in $\Gmatl$ can be
evaluated numerically from the infinite series result in
\eqref{proto:eig_green_series_all} of Appendix \ref{app:neum} for the
eigenvalue-dependent Green's function of \eqref{proto:eig_green}.

Since $\Mmat$ in \eqref{proto:GCEP_ring0} represents an update to
$\Gmatl$ by a multiple of the identity matrix, the eigenspace of
$\Mmat$ is the same as $\Gmatl$. As a result, we simply substitute
$\v{c}_1=\v{e}$ and $\v{c}_j=\v{q}_j$ into \eqref{proto:GCEP_ring0} to
obtain the root finding problems ${\mathcal F}_j=0$, which are defined
in terms of $\omega_j$ in \eqref{proto:gmat_circ} by
\begin{equation}\label{proto:GCEP_ring}
  {\mathcal F}_j \equiv 1 + \nu \Bhat(S_c;\lambda)
  + 2 \pi \nu \, \omega_j \,, \qquad
   j = 1,\ldots, N \,. 
\end{equation}
We refer to $\v{c}_1=\v{e}$ and $\v{c}_j=\v{q}_j$, for $j=2,\ldots,N$,
as the synchronous mode and asynchronous modes, respectively. 

\subsubsection{Example: instabilities associated with a two-spot
  ring pattern}\label{sec:ring_ex_two}

We begin by analyzing the zero-eigenvalue crossing in the GCEP for an
$N$-spot ring pattern. The criterion \eqref{proto:GCEP_zero_crossing}
becomes
\begin{equation}\label{proto:GCEP_zero_crossing_ring}
  \Mmat_0 \v{c} = \v{0} \,, \qquad \mbox{where} \qquad
  \Mmat_0 = \left(1 + \nu \chi^{\prime}(S_c)
  \right) \Imat + 2 \pi \nu (\Imat - \Emat) \Gmat 
    -\nu \chi^{\prime}(S_c) \Emat\,.
\end{equation}
The matrix $\Mmat_0$ shares the same eigenspace as the symmetric and
circulant matrix $\Gmat$, and so has eigenvectors
$\v{e},\v{q}_2,\ldots,\v{q}_N$ as in \eqref{proto:gmat_circ}. Since
$\Gmat \v{e}=\sigma_1\v{e}$, we use $\Emat \v{e}=\v{e}$ to calculate
$\Mmat_0\v{e}=\v{e}$. Therefore, the synchronous mode $\v{c}=\v{e}$ can
never be a nullvector for $\Mmat_0$. In contrast, with $\v{c}=\v{q}_j$ for
$j=2,\ldots,N$, we use $\Emat\v{q}_j=0$ to obtain that
$\Mmat_0\v{c}=\v{0}$ if and only if
\begin{equation}\label{proto:GCEP_competition_threshold} 
  1 + \nu \chi^{\prime}(S_c) + 2 \pi \nu \sigma_j = 0 \,, \qquad \mbox{where}
  \qquad \Gmat \v{q}_j = \sigma_j \v{q}_j\,, \quad j=2,\ldots,N \,.
\end{equation}
From \eqref{proto:cross_nas}, $\det\Mmat_0=0$ can only occur at a
bifurcation point for the NAS \eqref{proto:source_system}.

As an example, we investigate competition instabilities for a two-spot
equilibrium ring pattern in the unit disk with $\eps=0.02$, $D=1$, and
ring radius $r_0=0.4536$ determined from
\eqref{proto:equilibrium_ring_radius}. Since $\v{q}_2=(1,-1)^T$ is the
only competition mode, we use $S_c={a/[2N\sqrt{D}]}={a/4}$ with
$j=N=2$ to write \eqref{proto:GCEP_competition_threshold} as a
nonlinear algebraic equation in the feed rate $a$. This equation is
solved numerically to obtain the competition threshold
$a_{comp} \approx 4.45$. To interpret this bifurcation point, we
determine asymmetric branches of two-spot ring patterns from the NAS
\eqref{proto:source_system} with $N=2$. Labeling $S_1$ and $S_2$ as
the source strengths of the two spots, we set $S_2 = p_a - S_1$ in
\eqref{proto:source_system} with $N=2$ to obtain a scalar nonlinear
algebraic equation for $S_1$ given by
\begin{equation}\label{proto:S1}
  \left(1 + 2 \pi \nu (R_{11} - G_{12}) \right) S_1 +
  \frac{\nu}{2} \left[\chi(S_1) - \chi(p_a-S_1)\right] = \frac{p_a}{2}
\left( 1 + 2\pi\nu(R_{11} - G_{12}) \right) \,, \quad \mbox{where}
 \quad p_a \equiv \frac{a}{2}\,. 
\end{equation}

\begin{figure}[htbp]
\begin{subfigure}{0.45\textwidth}
 \includegraphics[width=\textwidth,height=4.4cm]{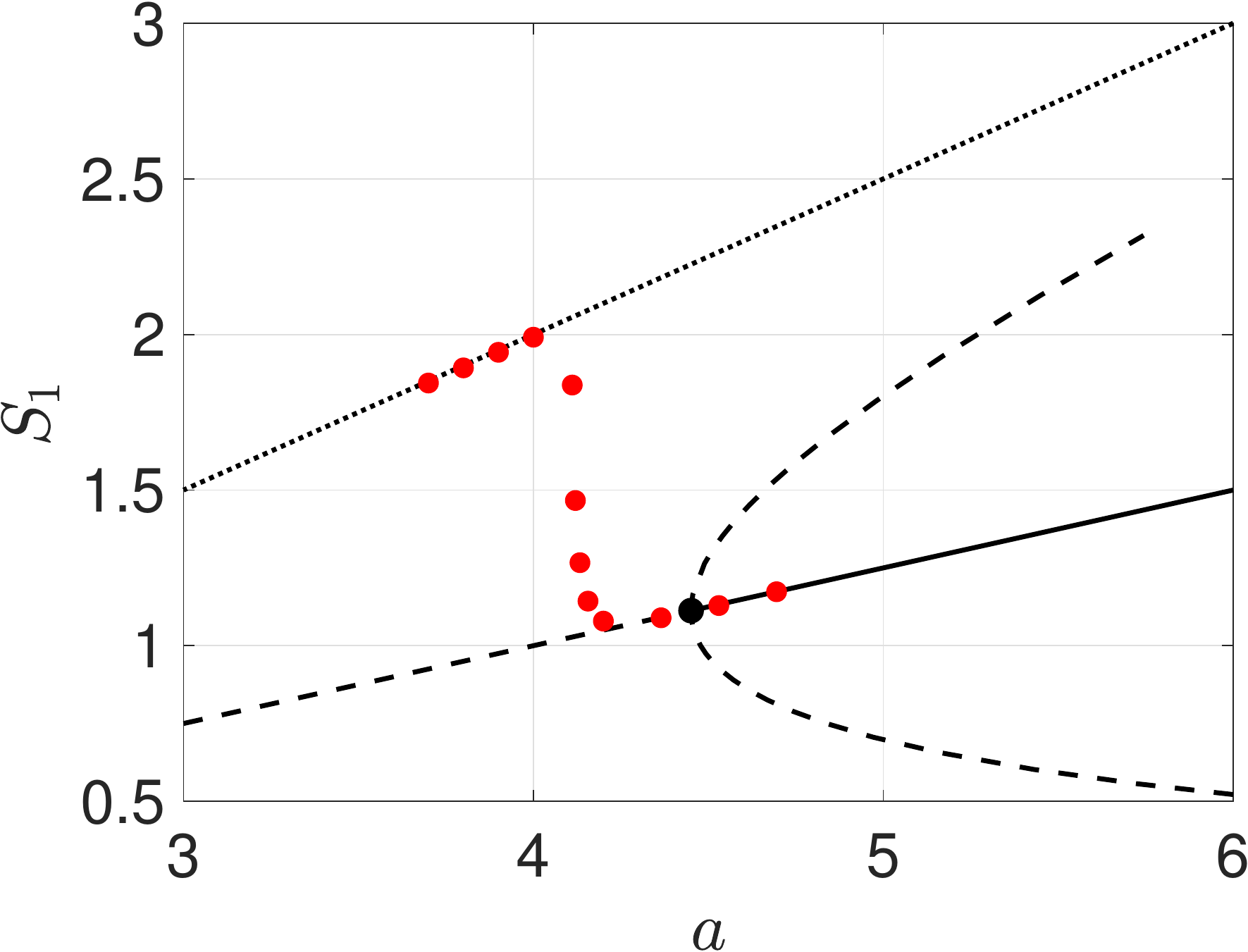}
\caption{Pitchfork bifurcation: two-spot ring pattern}
\label{GCEP_two_spots_sym}
\end{subfigure}
\begin{subfigure}{0.45\textwidth}
\includegraphics[width=\textwidth,height=4.4cm]{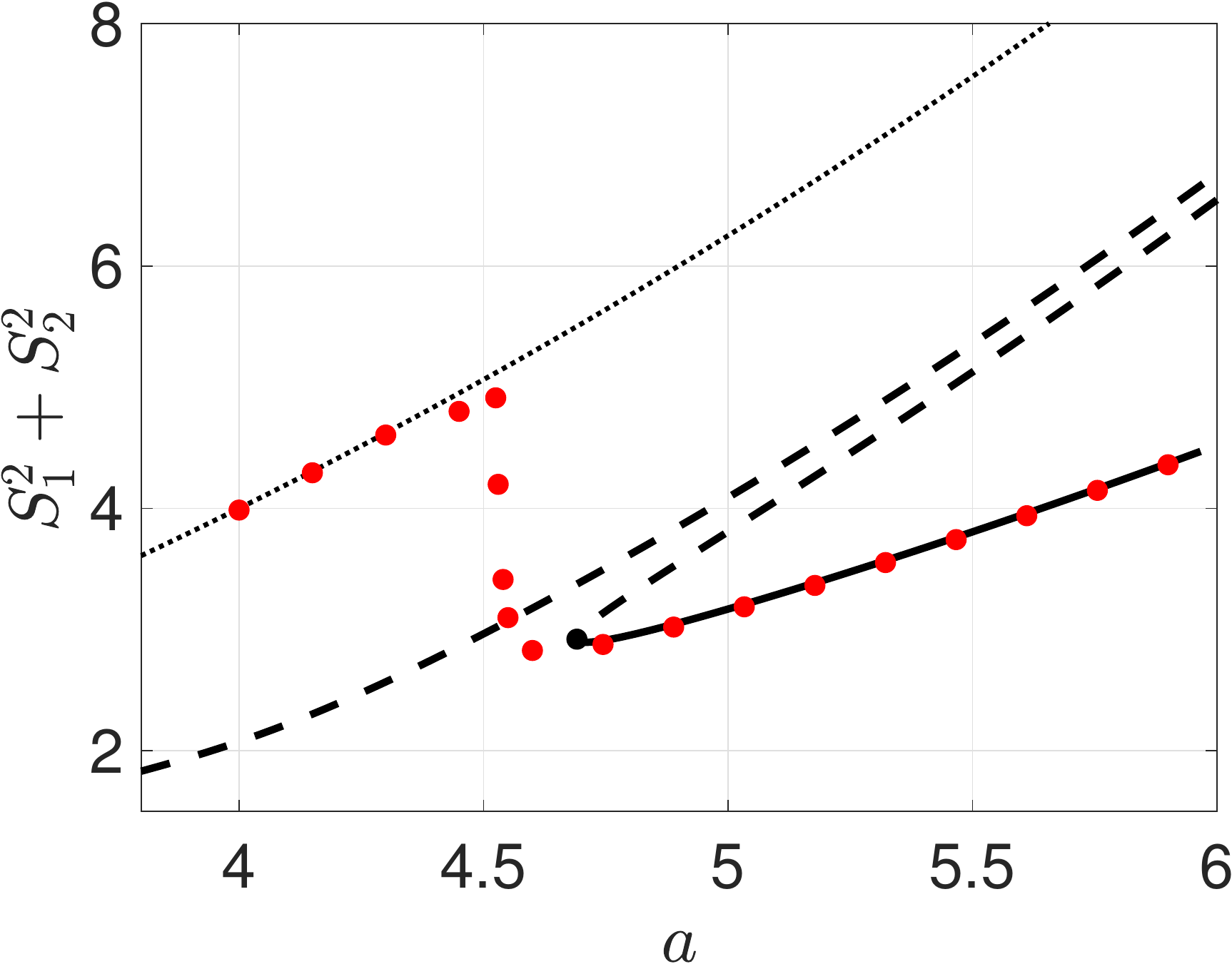}
\caption{Imperfect bifurcation: perturbed two-spot pattern}
\label{fig:GCEP_two_spots_asym}
\end{subfigure}
\caption{Left panel: Source strength $S_1$ versus $a$ for a two-spot
  equilibrium ring pattern with $\eps=0.02$, $D=1$, and ring radius
  $r_0=0.4536$. The solid (dashed) portion in the pitchfork structure
  has zero (one) unstable eigenvalue for the GCEP
  \eqref{proto:GCEP_ring0}. The red dots represent the $S_1$
  data interpolated from the PDE simulation with slowly decreasing
  feed rate $a = \max(4.7-0.005\,t \,,3.5)$. We observe that $S_1$
  jumps to the one-spot branch where $S_1={a/2}$ (dotted line). Right
  panel: Same parameters except that now spots are at
  $\v{x}_1 = (0.5,0)$ and $\v{x}_2=(-0.6,0)$. The thick solid and
  dashed curves are the stable and unstable branches of two-spot
  quasi-equilibria. The source strength from PDE data (red dots) is
  obtained by mapping from the the spot height. As $a$ is swept with
  $a = \max(4, 6 - (\eps/2)t)$ below the saddle-node point, only one
  spot survives. The sum of squares of the source strength jumps to
  the one-spot branch $S_1^2 = a^2/4$ (dotted line).}\label{fig:GCEP_two_all}
\end{figure}

By solving \eqref{proto:S1} numerically, in
Fig.~\ref{GCEP_two_spots_sym} we show the bifurcation structure of
$S_1$ versus $a$. The symmetric branch corresponds to the common
source strength $S_1=S_2\equiv S_c = {a/4}$. It undergoes a pitchfork
bifurcation at $a=a_{\text{comp}}\approx 4.45$, for which from
\eqref{proto:cross_nas} a zero-eigenvalue crossing for the GCEP must
occur. Moreover, asymmetric branches of quasi-equilibria with
$S_1\neq S_2$ exist for $a>a_{\text{comp}}$. In the same figure, we
superimpose PDE simulation data computed from \eqref{proto:model} with
a slowly decreasing feed rate $a = \max(4.7-0.005\,t \,,3.5)$. As the
feed rate drops below $a_{\text{comp}}$, only one spot survives and
there is a fast transition to the one-spot branch where $S_1 = {a/2}$
(dotted line in Fig.~\ref{GCEP_two_spots_sym}). 

To illustrate an imperfection sensitivity in the bifurcation structure
of two-spot quasi-equilibria, we consider a two-spot pattern with
spots located at $\v{x}_1 = (0.5,0)$ and $\v{x}_2=(-0.6,0)$ in the
unit disk with $\eps=0.02$ and $D=1$.  Through numerical continuation
of the NAS \eqref{proto:source_system} with bifurcation parameter $a$
using {\em COCO} \cite{ds13}, in Fig.~\ref{fig:GCEP_two_spots_asym} we
observe two isolated branches of $S_1^2 + S_2^2$, with one branch
having a saddle-node bifurcation at $a\approx 4.609$, which must
correspond to a zero-eigenvalue of the GCEP. The linear stability
properties of these branches, as indicated in the caption of
Fig.~\ref{fig:GCEP_two_spots_asym}, was obtained from a numerical
computation of the winding number in \eqref{proto:wind}. From the
results of a full PDE computation of \eqref{proto:model} with a slowly
decreasing feed-rate $a = \max(4, 6 - (\eps/2)t)$ with
$\eps=0.02$, in Fig.~\ref{fig:GCEP_two_spots_asym} we show that as $a$
sweeps below the saddle-node point for two-spot quasi-equilibria, one
spot gets annihilated while the remaining spot jumps to the stable
one-spot branch.

\begin{figure}[htbp]
\begin{subfigure}{0.45\textwidth}
\includegraphics[width=\textwidth,height=4.3cm]{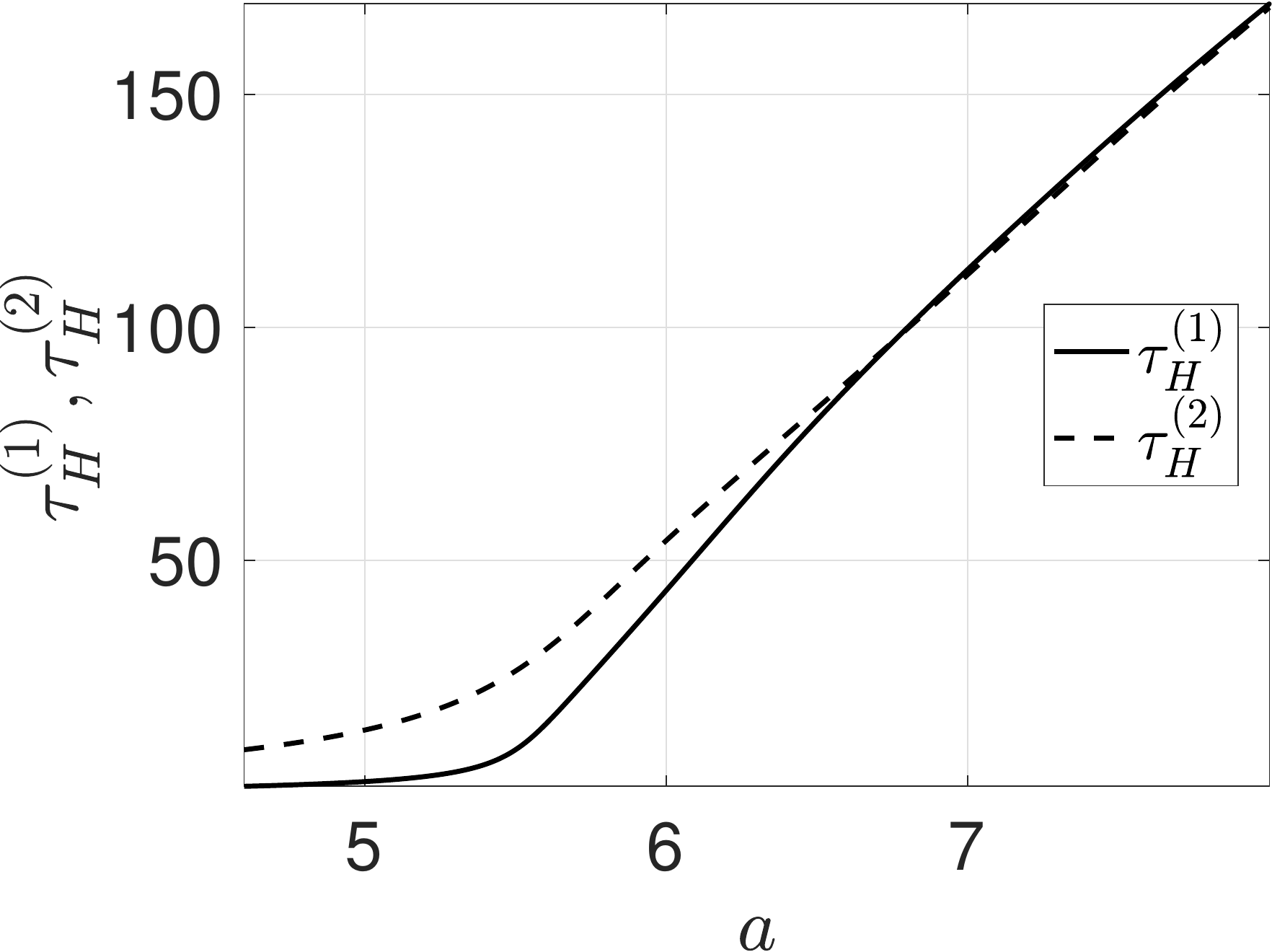}
\caption{Hopf bifurcation threshold for $\tau$}
\end{subfigure} 
\hfill
\begin{subfigure}{0.45\textwidth}
\includegraphics[width=\textwidth,height=4.3cm]{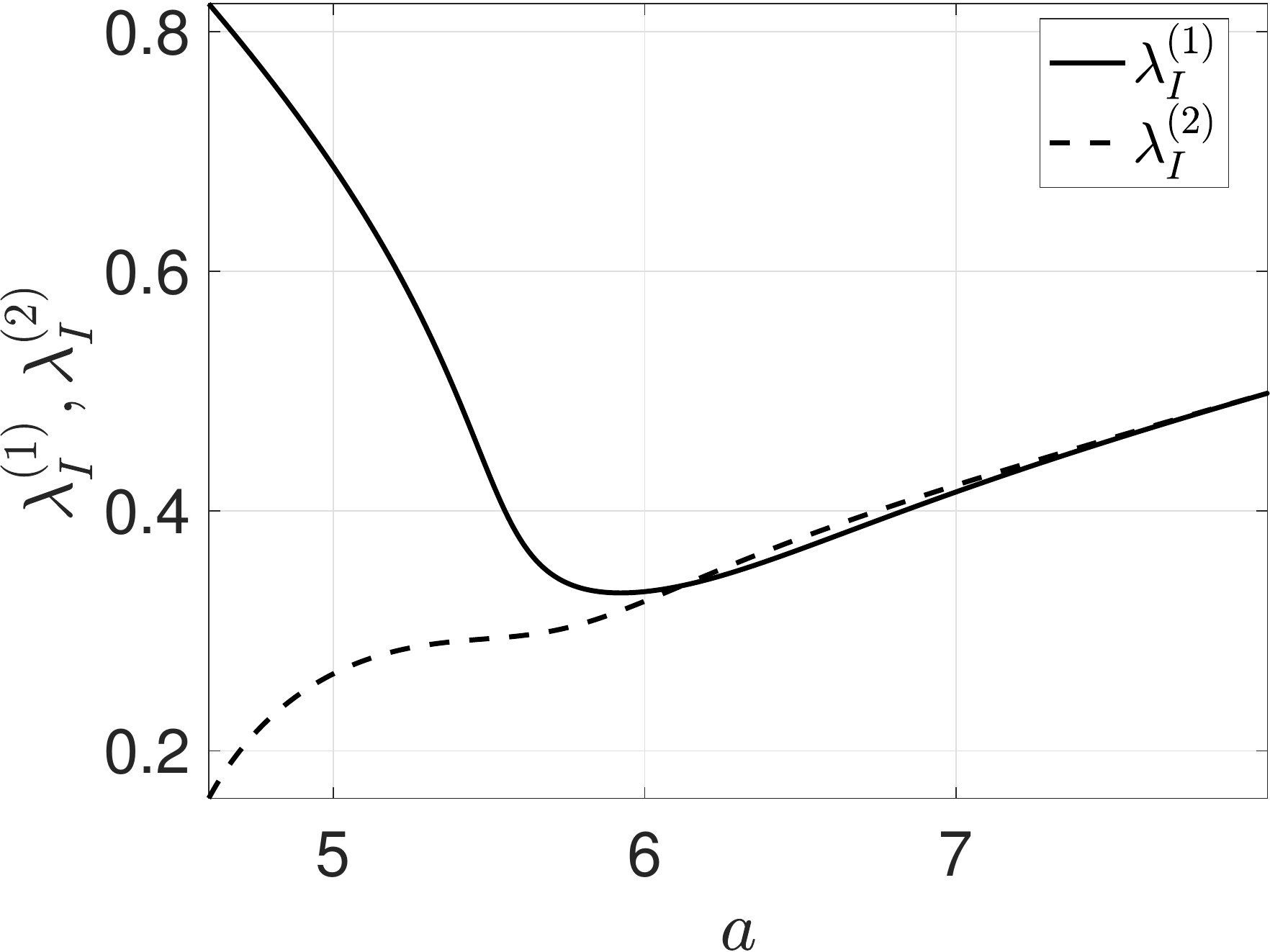}
\caption{Imaginary eigenvalue at the Hopf bifurcation}
\end{subfigure}
\caption{Left panel: The Hopf bifurcation value of $\tau$ for the
  synchronous $(j=1)$ and asynchronous mode $(j=2)$, as computed from
  \eqref{proto:hopf_example}, for the linearization of a two-spot ring
  steady-state solution with $\eps=0.02$, $D=1$, and ring radius
  $r_0=0.4536$. The thresholds become almost indistinguishable as the
  feed rate $a$ is increased. Right panel: The corresponding imaginary
  eigenvalue for the two modes.}\label{fig:hopf_ring}
\end{figure}

Next, we illustrate how a pair of unstable eigenvalues emerge from a
Hopf bifurcation as $\tau$ is increased.  We consider a two-spot
equilibrium ring pattern in the unit disk with $\eps=0.02$ and
$D=1$. The two spots are centered at $(\pm r_0,0)$, where
$r_0 \approx 0.4536$ is the steady-state two-spot ring radius, as
calculated from \eqref{proto:equilibrium_ring_radius} when $N=2$. By
varying the feed rate $a$, on the range
$4.45 \approx a_{\text{comp}}<a<a_f\approx 17.2$ (heavy solid curve in
Fig.~\ref{GCEP_two_spots_sym}), we use \eqref{proto:GCEP_ring} to
numerically compute the Hopf bifurcation thresholds for $\tau$ for the
synchronous mode $(j=1)$ and the asynchronous mode $(j=2)$. This is
done by using Newton's method to solve for
$(\lambda_I^{(j)}, \tau_H^{(j)})$ in
\begin{equation}\label{proto:hopf_example}
  \mbox{Re}\left[ F_j(i\lambda_I^{(j)}, \tau_H^{(j)}) \right] = 0 \,, \qquad
  \mbox{Im}\left[ F_j(i\lambda_I^{(j)}, \tau_H^{(j)}) \right] = 0 \,, \quad
  \mbox{for} \quad j = 1\,, 2\,.
\end{equation}
The results for $\tau_{H}^{(j)}$ and $\lambda_I^{(j)}$ for $j=1,2$
versus the feed rate $a$ are shown in the left and right panels of
Fig.~\ref{fig:hopf_ring}, respectively. From this figure, we observe
that the mode that synchronizes the temporal oscillations in the spot
amplitudes is the first to go unstable as $\tau$ is increased. A
numerical implementation of the winding number criterion in
\eqref{proto:wind} yields that the two-spot ring pattern is linearly
stable when $\tau<\min(\tau_H^{(1)}, \tau_H^{(2)})$.

\begin{figure}[htbp]
\begin{subfigure}{0.45\textwidth}
\includegraphics[width=\textwidth,height=4.3cm]{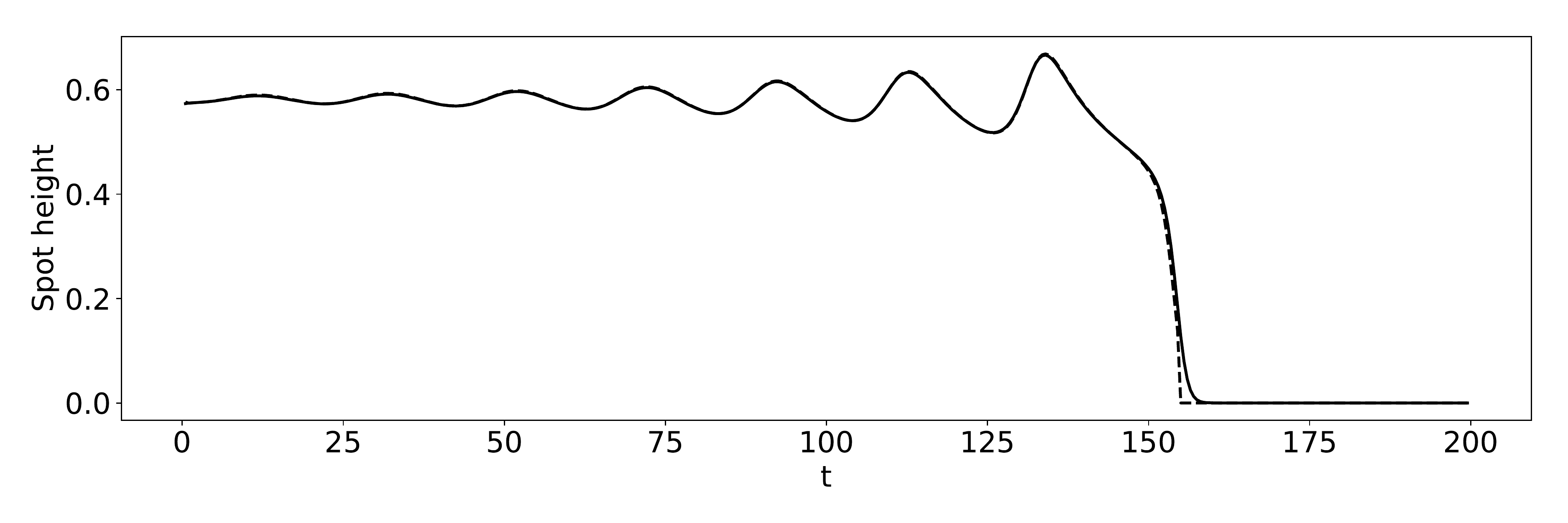}
\caption{$a=6$ and $\tau_{H}^{(1)} < \tau < \tau_{H}^{(2)}$.}
\label{GCEP_hopf_exp1_set4}
\end{subfigure}
\begin{subfigure}{0.45\textwidth}
\includegraphics[width=\textwidth,height=4.3cm]{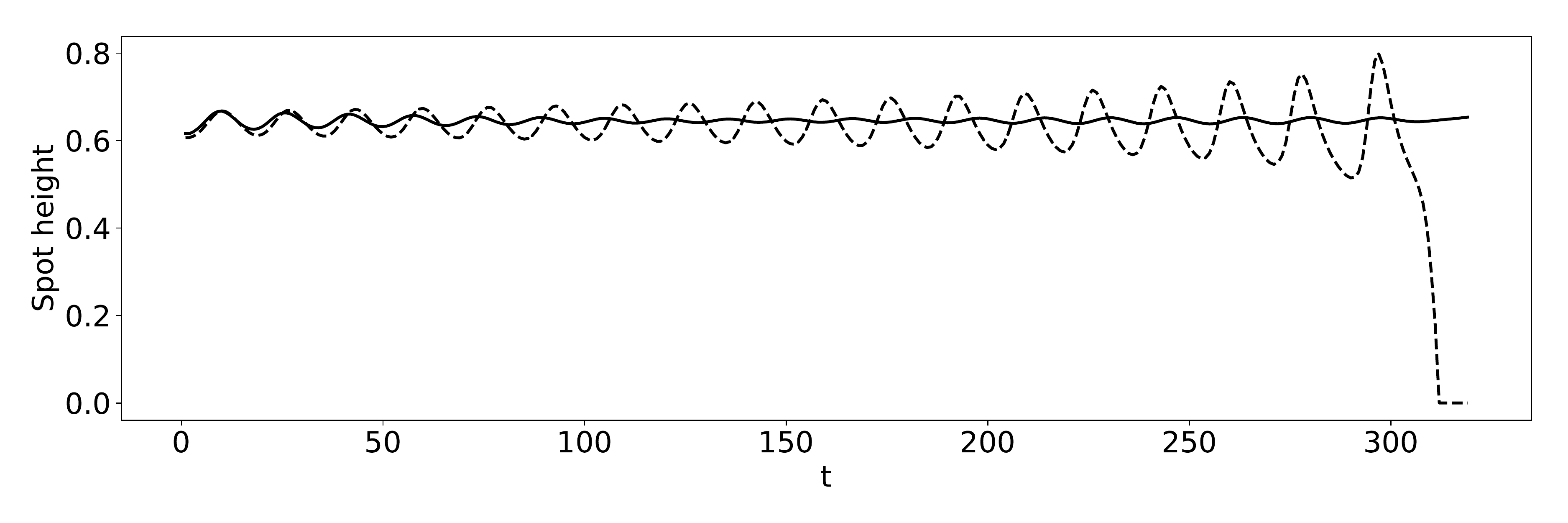}
\caption{$a=7.2$ and $\tau_{H}^{(2)} < \tau_{H}^{(1)} < \tau$.}
\label{GCEP_hopf_exp1_set10}
\end{subfigure}
\caption{PDE simulation results of \eqref{proto:model} for the spot
  amplitudes versus time starting from a two-spot steady-state ring
  pattern with $\eps=0.02$, $D=1$, and ring radius $r_0=0.4536$.  Left
  panel: $a=6$ and $\tau=54$. Synchronous oscillations occur, leading to
  the annihilation of both spots. Right panel: $a=7.2$ and $\tau=137$.
  Eventually asynchronous spot amplitude oscillations occur, leading to the
  annihilation of only one spot.}
\end{figure}

To confirm the Hopf bifurcation threshold, as calculated from
\eqref{proto:hopf_example}, we compute full numerical solution to the
PDE \eqref{proto:model} for $\eps=0.02$, $D=1$, using as an initial
condition a two-spot ring pattern with ring radius $r_0=0.4536$. For
$a=6$, we have $\tau_H^{(1)} \approx 43.56$ and
$\tau_H^{(2)} \approx 54.28$ from \eqref{proto:hopf_example}. With the
choice $\tau = 54$, for which
$\tau_{H}^{(1)} < \tau < \tau_{H}^{(2)}$, we predict from the GCEP
that the amplitudes of the two spots will oscillate in phase.  In the
PDE simulation results of Fig.~\ref{GCEP_hopf_exp1_set4} we show that
there are synchronous oscillations of the spot amplitudes, which
eventually leads to the disappearance of both spots. By increasing the
feed rate to $a=7.2$, we have $\tau_H^{(1)} \approx 124.56$ and
$\tau_H^{(2)} \approx 123.11$ from \eqref{proto:hopf_example}. With
the choice $\tau = 137$, we predict that the two spots will be
unstable to both synchronous and asynchronous perturbations in the
spot amplitudes.  In the PDE simulation results of
Fig.~\ref{GCEP_hopf_exp1_set10} we show that, although initially the
spot amplitudes oscillate synchronously. as time increases these
oscillations become asynchronous, and eventually one of the two
spots is annihilated.

\section{A perforation of the domain as a localized
  defect}\label{sec:leakage}

In this section we analyze how the existence, linear stability, and slow
dynamics of quasi-equilibrium spot patterns are affected by
removing a small circular hole of radius ${\mathcal O}(\eps)$ from $\Omega$,
given by
\begin{equation*}
\hole = \big\{\v{x} \in \Omega \, : \, |\v{x}-\v{x}_0| \leq C\eps \, \big\} \,,
\end{equation*}
where $C>0$ is an $\mc{O}(1)$ parameter controlling the size of the hole. 
In the perforated domain, the Schnakenberg model is
\begin{subequations}\label{leakage:full}
\begin{align}
  v_t &= \eps^2 \lap v - v + uv^2 \,, \qquad \tau u_t = D\lap u + a -
        \frac{uv^2}{\eps^2} \,, \quad \mbox{in} \quad \pdedomain
\equiv \Omega \setminus \hole  \,,  \label{leakage:pde}\\
  \partial_n v &= \partial_n u = 0  \quad \mbox{on} \quad \partial\Omega \,;
  \qquad v = u = 0 \quad \mbox{on} \quad \partial\hole \,. \label{leakage:bc}
\end{align}
\end{subequations}
The homogeneous Dirichlet boundary conditions on $\partial\hole$ models
the leakage of the activator $v$ and the substrate $u$ through the
boundary of the small hole.

\subsection{Quasi-equilibrium $N$-spot pattern and slow
  dynamics}\label{sec:leakage_qe}

We begin by constructing a quasi-equilibrium $N$-spot pattern with
spots located at $\v{x}_1, \ldots, \v{x}_N$ in the perforated
domain. We assume, initially, that this pattern is linearly stable on
$\mc{O}(1)$ time intervals. In our analysis below, we assume that
$|\v{x}_i-\v{x}_j|=\bigo(1)$ for $i\neq j$, and that
$\mbox{dist}(\v{x}_i,\partial\Omega)=\bigo(1)$ and
$\mbox{dist}(\v{x}_i,\partial\hole)=\bigo(1)$ for $i=1,\ldots,N$.

Following the derivation in \S \ref{sec:proto_qe_slow}, the outer
problem for the inhibitor field, defined away from the spots, is
\begin{equation}\label{leakage:outer_problem0}
  \lap u + \frac{a}{D} - \frac{2\pi}{\sqrt{D}} \sum\limits_{i=1}^N S_i \,
  \delta(\v{x}-\v{x}_i) = 0 \quad \mbox{in} \quad \pdedomain \,, \quad
  \partial_n u= 0  \quad \mbox{on} \quad \partial\Omega \,; \qquad u = 0
  \quad \mbox{on} \quad \partial\hole \,,
\end{equation}
where $S_1,\ldots,S_N$ denote the spot source strengths.  However,
this outer problem is of singular perturbation type since $u$ must
satisfy the extra conditon $u=0$ on $\partial\hole$.  To proceed, we
will use strong localized perturbation theory to replace the effect of
the hole with a Dirac singularity. To do so, near the hole centered at
$\v{x}_0$ we introduce local coordinates
$\v{y} = \eps^{-1}(\v{x}-\v{x}_0)$ and $u \sim
U_0(\v{y})/\sqrt{D}$. From \eqref{leakage:outer_problem0}, we obtain
to leading order that
\begin{equation}\label{leakage:U0}
  \lap_{\v{y}} U_0 = 0 \,, \quad |\v{y}| \geq C \,; \qquad U_0 = 0 \,,
  \quad \mbox{on} \quad  |\v{y}| = C \,,
\end{equation}
which has the solution $U_0 = S_0 \log\left(|\v{y}|/C\right)$, where $S_0$
is to be determined. This yields the matching condition
\begin{equation}\label{leakage:matching_hole1}
  u \sim \frac{U_0}{\sqrt{D}} \sim \frac{S_0}{\sqrt{D}}\left(\log|\v{x}-\v{x}_0|
 + \frac{1}{\nu} - \log C \right) \,, \quad \mbox{as}\quad \v{x} \to \v{x}_0
  \,,
\end{equation}
where $\nu={-1/\log\eps}$. Owing to the identity
\begin{equation}\label{leak:iden}
  \int_{\partial\hole} - D \partial_n u \vert_{\partial\hole} 
  \, ds \sim 2\pi S_0 \sqrt{D} \,,
\end{equation}
where $\partial_n$ denotes the outward normal derivative to
$\pdedomain$, the constant $S_0$ is proportional to the diffusive flux
of inhibitor through the hole. The strength of this leakage term,
mediated by $S_0$, is calculated below in a self-consistent way.

By superimposing the Dirac singularity
$\dfrac{2\pi S_0}{\sqrt{D}} \, \delta(\v{x}-\v{x}_0)$ on the outer
problem to account for the logarithmic singularity in
\eqref{leakage:matching_hole1}, we replace
\eqref{leakage:outer_problem0} with the modified outer problem
\begin{equation}\label{leakage:outer_problem}
  \lap u + \frac{a}{D} - \frac{2\pi}{\sqrt{D}} \sum\limits_{i=0}^{N} S_i \,
  \delta(\v{x}-\v{x}_i) = 0 \quad \mbox{in} \quad  \Omega \,; \qquad
  \partial_n u= 0 \quad \mbox{on} \quad \partial\Omega \,,
\end{equation}
which is defined at ${\mathcal O}(1)$ distances from 
the spot locations and from the center of the hole.

The solution to \eqref{leakage:outer_problem} is represented in terms
of the Neumann Green's function of \eqref{proto:neu_green} as
\begin{equation}\label{leakage:outer_solution}
  u(\v{x}) = -\frac{2\pi}{\sqrt{D}}\sum\limits_{i=0}^N S_i \, G(\v{x};\v{x}_i) +
  \ubar\,,
\end{equation}
where $\ubar$ is a constant to be determined. By applying the divergence
theorem to \eqref{leakage:outer_problem} we get
\begin{equation}\label{leakage:source_strength_sum}
\sum\limits_{i=0}^{N} S_i = \frac{a|\Omega|}{2\pi\sqrt{D}} \equiv p_a\,.
\end{equation}

We let $\v{x}\to\v{x}_0$ in \eqref{leakage:outer_solution} in order to
asymptotically match the local behavior of $u$ with the far-field behavior
\eqref{leakage:matching_hole1} for the solution near the hole. This
matching yields the algebraic equation
\begin{equation}\label{leakage:matching_hole}
  S_0 + 2\pi\nu\left(S_0 R_{0,0} + \sum\limits_{i=1}^N S_i \, G_{0,i}\right) -
  \nu S_0 \log C  = \nu\sqrt{D} \, \ubar \,,
\end{equation}
where $R_{0,0} \equiv R(\v{x}_0;\v{x}_0)$ and
$G_{0,i} \equiv G(\v{x}_0;\v{x}_i)$.

Next, we match the local behavior of the outer solution in
\eqref{leakage:outer_solution} near each spot with the far-field
behavior \eqref{proto:matching_inner} of the corresponding inner
solution.  Letting $\v{x}\to\v{x}_j$ in \eqref{leakage:outer_solution}
we obtain that 
\begin{equation}\label{leakage:matching_spot_outer}
\begin{split}
  u &\sim \frac{S_j}{\sqrt{D}} \log |\v{x}-\v{x}_j| -
  \frac{2\pi}{\sqrt{D}} \left( S_j R_{j,j} + \sum\limits_{i \neq j}^N S_i \,
    G_{j,i} + S_0 \, G_{j,0} \right) + \ubar \\
  &\qquad - 2 \pi \left( S_j \nabla_\v{x} R_{j,j} + \sum\limits_{i \neq j}^N
    S_i \nabla_\v{x} G_{j,i} + S_0 \nabla_\v{x} G_{j,0} \right) \cdot
  (\v{x}-\v{x}_j) + {\mathcal O}(|\v{x}-\v{x}_j|^2) \,, \qquad j=1,\ldots, N \,.
\end{split}
\end{equation}
By matching the $\mc{O}(1)$ terms in \eqref{proto:matching_inner} and
\eqref{leakage:matching_spot_outer}, we obtain that
\begin{equation}\label{leakage:matching_spot}
  S_j + 2\pi\nu\left(S_j R_{j,j} + \sum\limits_{i\neq j}^N S_i \, G_{j,i} + S_0 \,
    G_{j,0} \right) + \nu\chi(S_j) = \nu\sqrt{D}\ubar \,, \qquad j=1,\ldots,N\,.
\end{equation}

We write the nonlinear algebraic system \eqref{leakage:source_strength_sum},
\eqref{leakage:matching_hole}, and \eqref{leakage:matching_spot} for
$S_0,\ldots,S_N$ and $\bar{u}$ in matrix form as
\begin{subequations}
\begin{equation}\label{leakage:nas}
  S_0 = p_a - \v{e}^T \v{s} \,, \qquad
  \v{s} + 2\pi \nu \left(\Gmat\v{s} + S_0\v{g} \right) + \nu\chivec =
  \left(\nu \sqrt{D} \ubar \right) \v{e} \,, \qquad
 \theta S_0 = \nu \sqrt{D} \ubar - 2 \pi \nu \v{g}^T \v{s} \,,
\end{equation}
where we have defined
\begin{equation}\label{leakage:nas_def}
  \begin{split}
    &\v{s} \equiv (S_1, \ldots, S_N)^T \,, \qquad \v{g} \equiv
    (G_{0,1}, \ldots, G_{0,N})^T \,, \qquad \v{e} \equiv (1,\ldots,1)^T \in
    \mathbb{R}^N \,, \\
    & \chivec \equiv (\chi(S_1), \ldots, \chi(S_N))^T \,, \qquad \theta \equiv
    1 + 2 \pi \nu R_{0,0} - \nu \log C \,.
\end{split}
\end{equation}
\end{subequations}
Here $\Gmat \in \mathbb{R}^{N\times N}$ is the Neumann Green's matrix
characterizing inter-spot interactions for spots centered at
$\v{x}_1, \ldots, \v{x}_N$. By eliminating $S_0$ between the first and
third equations in \eqref{leakage:nas}, we can solve for $\ubar$ as
\begin{equation}\label{leakage:temp3}
  \ubar = \frac{\theta p_a + \v{s}^T (2\pi\nu\v{g} - \theta \v{e})}
  {\nu\sqrt{D}} \,.
\end{equation}
By substituting \eqref{leakage:temp3} together with
$S_0=p_a-\v{e}^T\v{s}$ into the middle equation of \eqref{leakage:nas}
we obtain the following nonlinear algebraic system for the vector
$\v{s}$ of spot strengths:
\begin{equation}\label{leakage:source_strength}
  \v{s} + 2\pi\nu\Gmat\v{s} + (\v{e}^T \v{s}) \left(\theta\v{e} -
    4\pi\nu\v{g}\right) + \nu\chivec = p_a \left(\theta \v{e} -
    2\pi\nu\v{g}\right)\,.
\end{equation} 

Next, to derive the DAE system for slow spot dynamics, we match
\eqref{proto:matching_inner} with \eqref{leakage:matching_spot_outer}
for the $\mc{O}(\eps)$ gradient terms. Denoting
$\v{y} = \eps^{-1}(\v{x}-\v{x}_j)$, and using $S_0=p_a-\v{e}^T\v{s}$,
this yields the following far-field behavior for the correction
$U_{j1}$ to the leading order core solution, as defined in
\eqref{proto:inner_expansion}:
\begin{equation}
  U_{j1} \sim - \left( \pmb{\beta}_j + 2\pi S_0 \nabla_\v{x} G_{j,0} \right)
  \cdot \v{y} = - \left[ \pmb{\beta}_j + 2 \pi \left(p_a - \sum\limits_{i=1}^N S_i \right) \nabla_\v{x} G_{j,0} \right] \cdot \v{y}  \,, \quad \mbox{as}
  \quad |\v{y}|\to \infty \,.
\end{equation}
Here $\pmb{\beta}_j$ is defined in \eqref{proto:betaj}. Following the
derivation in \S \ref{sec:proto_qe_slow}, we conclude that the DAE
system for slow spot dynamics is given by
\begin{equation}\label{leakage:DAE}
  \frac{d\v{x}_j}{d\sigma} = - \gamma(S_j) \left[\pmb{\beta}_j + 2 \pi
    \left(p_a - \sum\limits_{i=1}^N S_i \right) \nabla_\v{x} G_{j,0}\right] \,,
  \qquad j=1,\ldots,N \,,
\end{equation}
where $\sigma=\eps^2 t$ and $\v{s}\equiv (S_1,\ldots,S_N)^T$ satisfies
the nonlinear algebraic system \eqref{leakage:source_strength}. Here
$\gamma(S_j)$ is defined in \eqref{proto:DAE}.

\begin{figure}[htbp]
\begin{subfigure}[b]{0.19\textwidth}
\includegraphics[width=\textwidth]{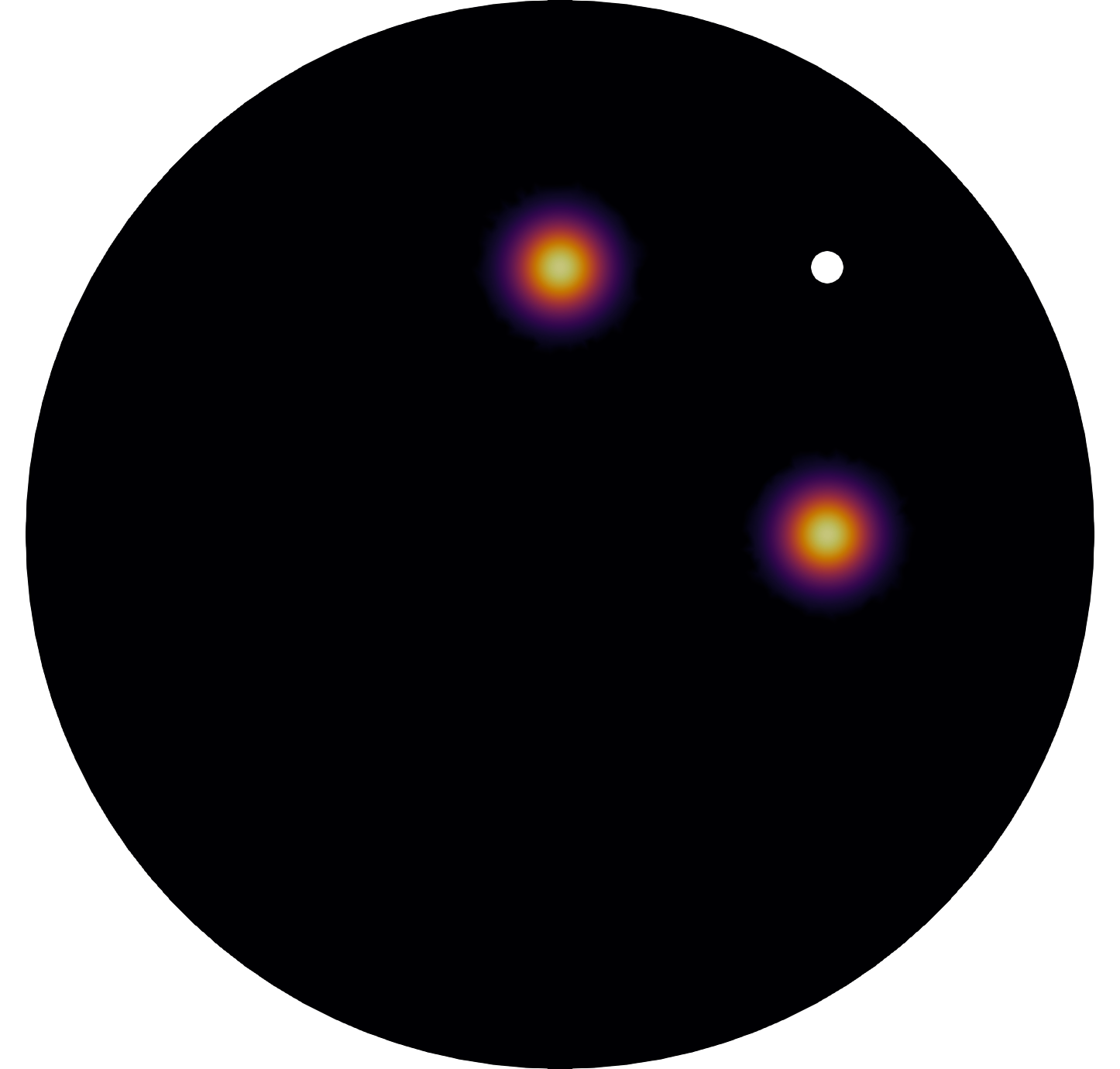}	
\caption{$t=0$}
\end{subfigure}
\hfill
\begin{subfigure}[b]{0.19\textwidth}
\includegraphics[width=\textwidth]{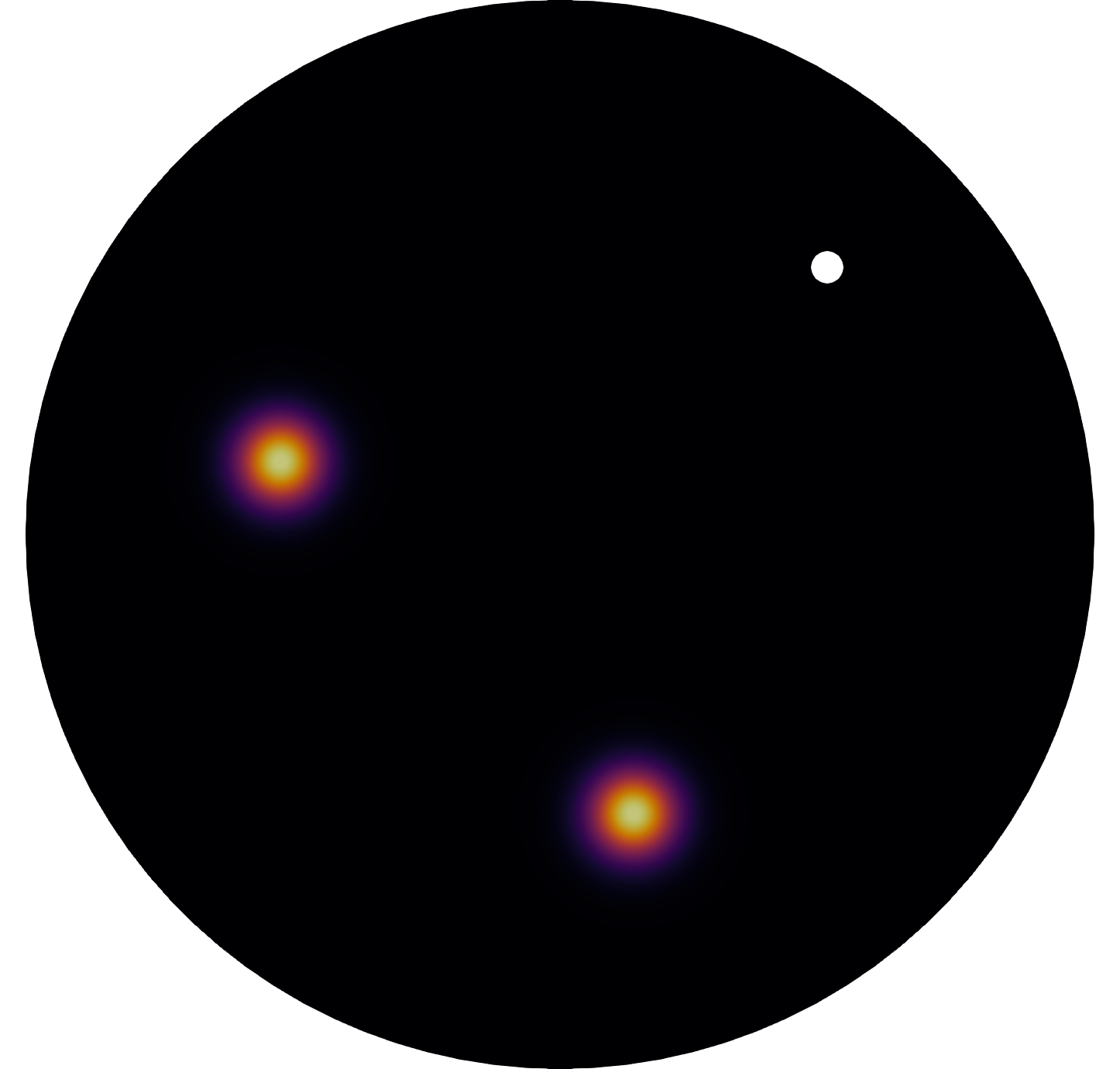}	
\caption{$t=999$}
\end{subfigure}
\begin{subfigure}[b]{0.29\textwidth}
\includegraphics[width=\textwidth]{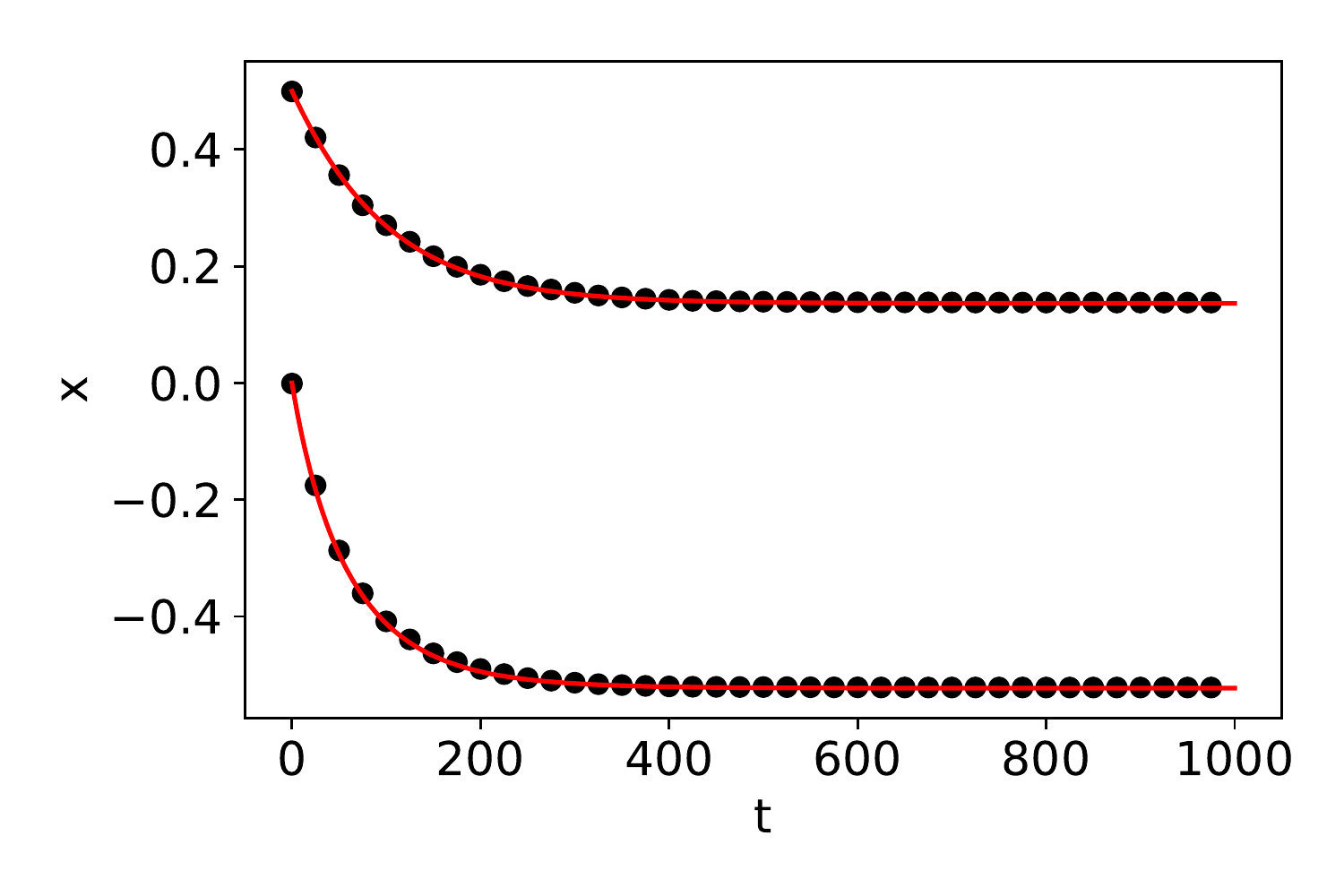}
\caption{$x$-coordinate of spots}
\end{subfigure}
\begin{subfigure}[b]{0.29\textwidth}
\includegraphics[width=\textwidth]{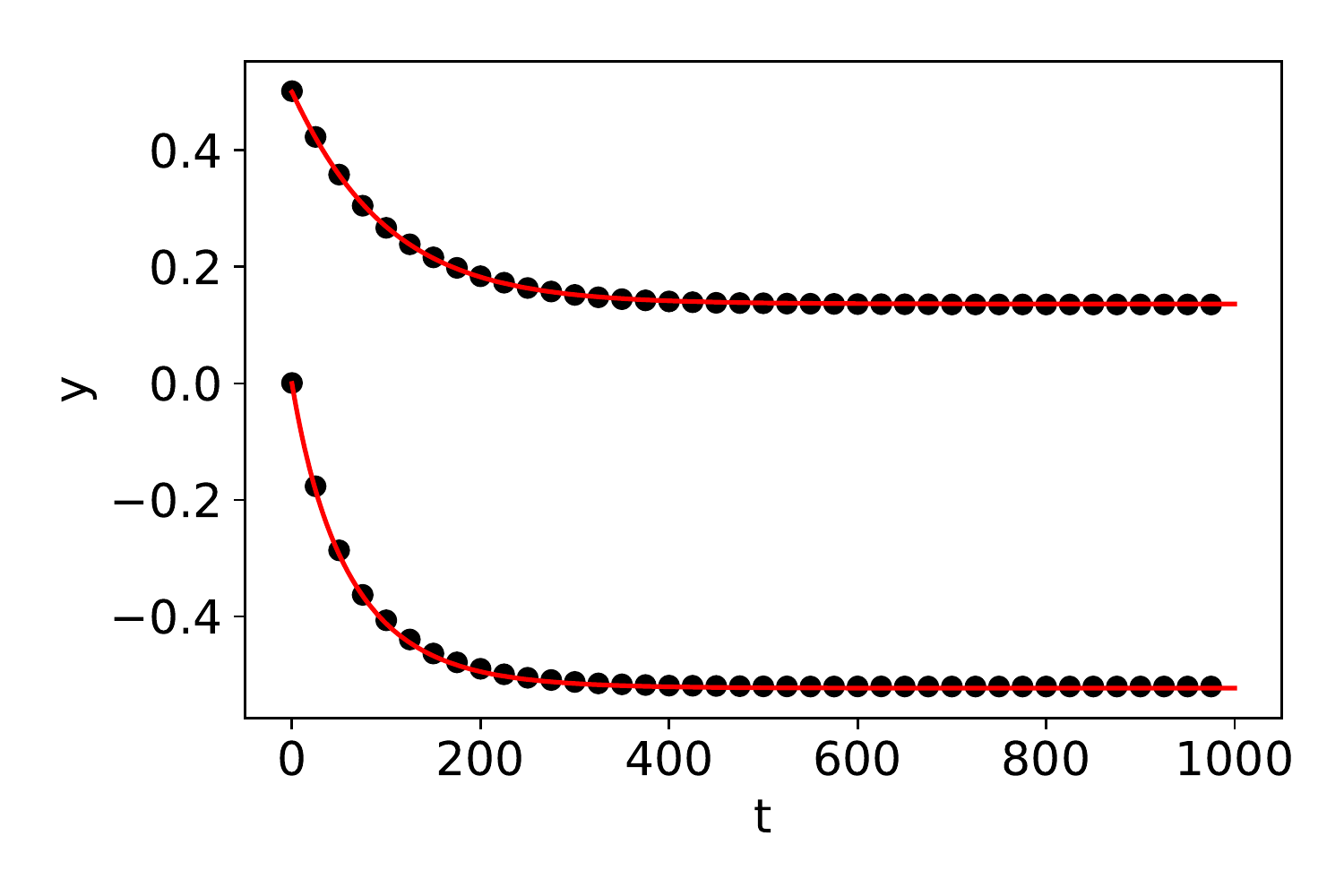}
\caption{$y$-coordinate of spots}
\end{subfigure} \\
\caption{For $\eps=0.03, \, D = \tau = 1, \, a = 16$ and a hole at
  $\v{x}_0 = (0.5,0.5)$ with radius $\eps$ ($C=1$), two spots
  initially located at $(0.5,0)$ and $(0,0.5)$, respectively, share
  the same source strength $S \approx 3.0599$. In (a) and (b), we show
  the numerical PDE solution of $v$ at $t=0$ and $t=999$,
  respectively. In (c) and (d), we show the very close agreement of
  spot trajectories obtained by the PDE simulation (black dots) and
  the DAE \eqref{leakage:DAE} and \eqref{leakage:source_strength} (red
  solid line).}\label{fig:leak_2}
\end{figure}

\begin{figure}[htbp]
\begin{subfigure}[b]{0.19\textwidth}
\includegraphics[width=\textwidth]{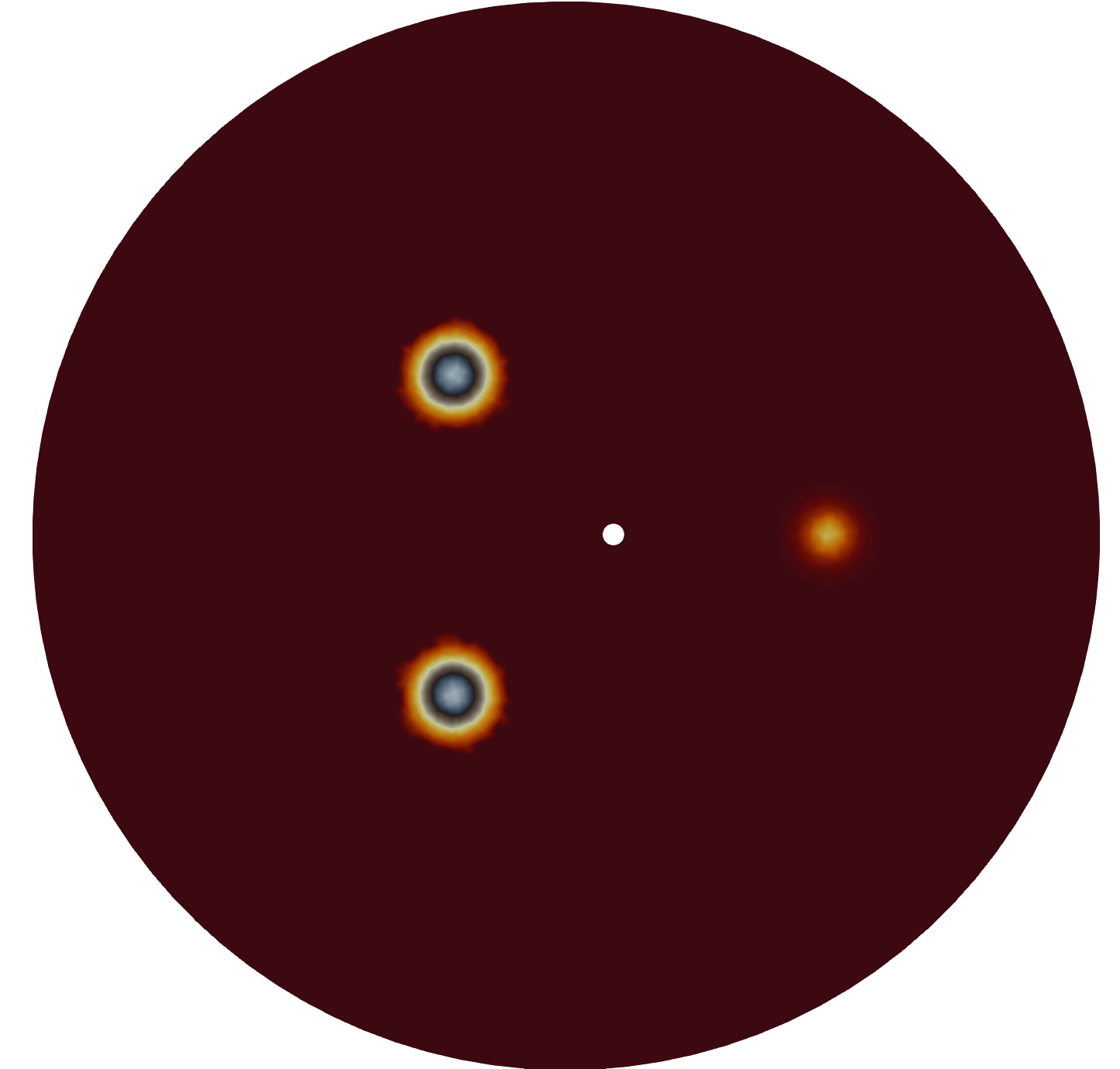}	
\caption{$t=0$}
\end{subfigure}
\hfill
\begin{subfigure}[b]{0.19\textwidth}
\includegraphics[width=\textwidth]{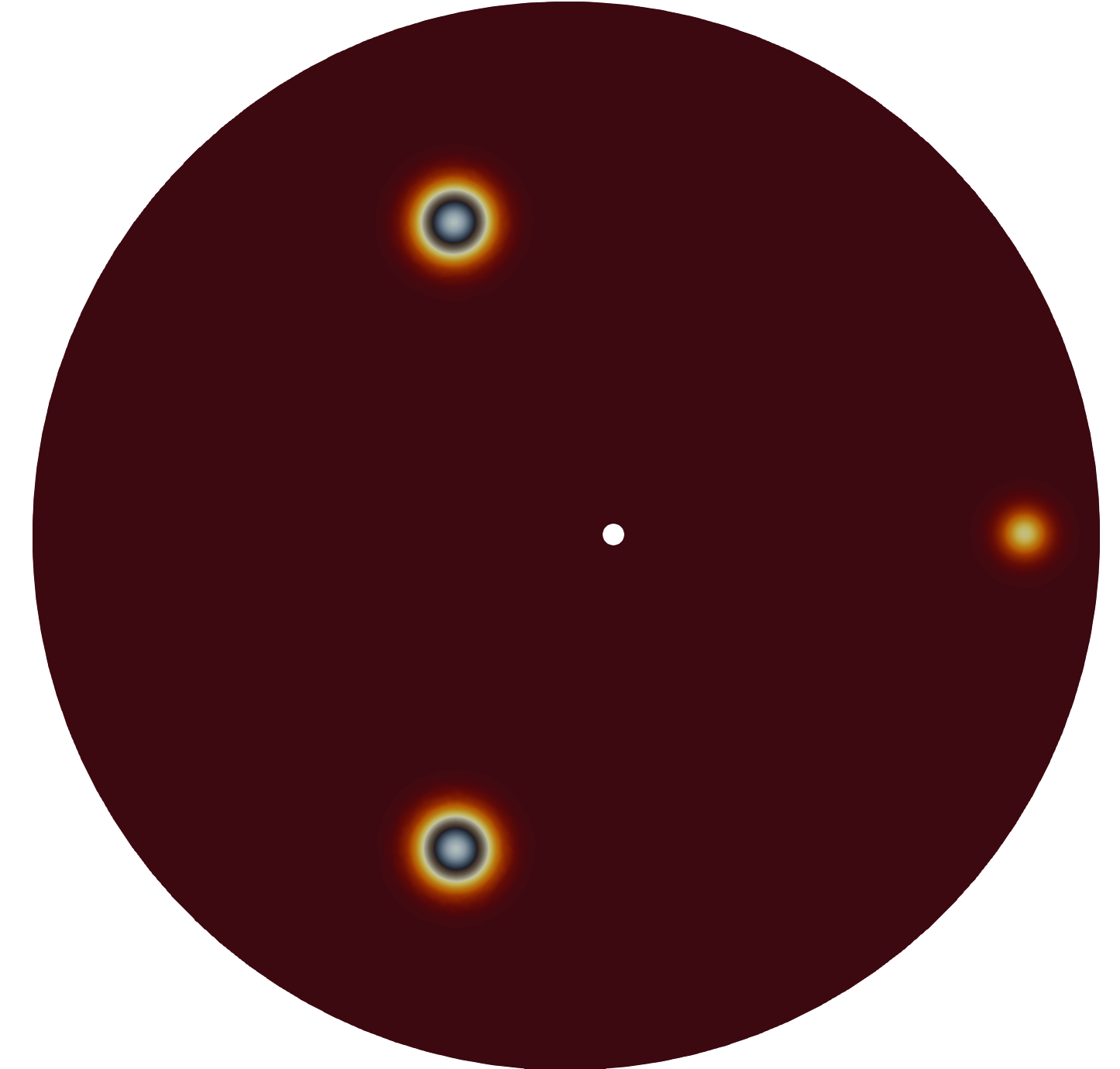}
\caption{$t=1998$}
\end{subfigure}
\begin{subfigure}[b]{0.29\textwidth}
\includegraphics[width=\textwidth]{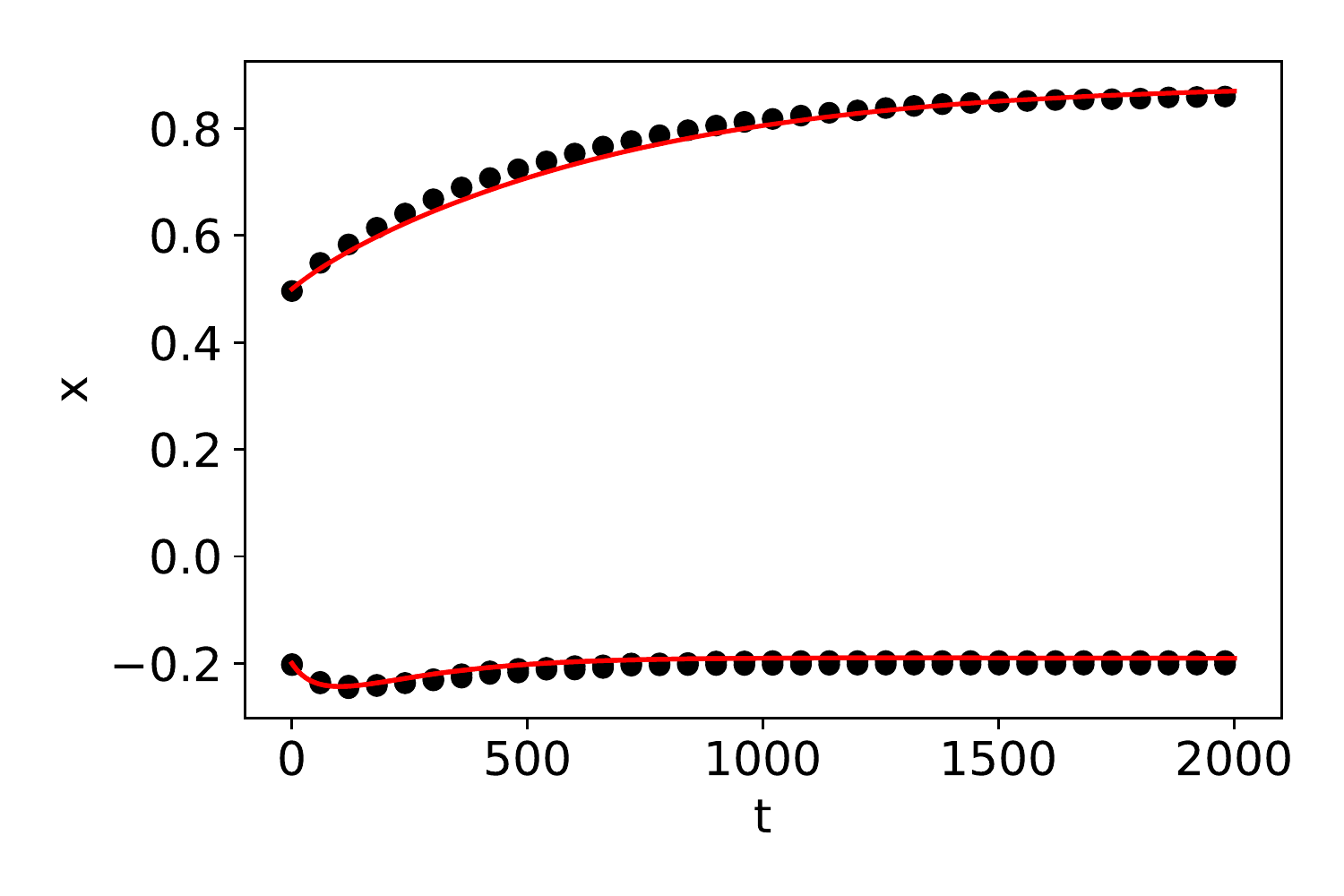}
\caption{$x$-coordinate of spots}
\end{subfigure}
\begin{subfigure}[b]{0.29\textwidth}
\includegraphics[width=\textwidth]{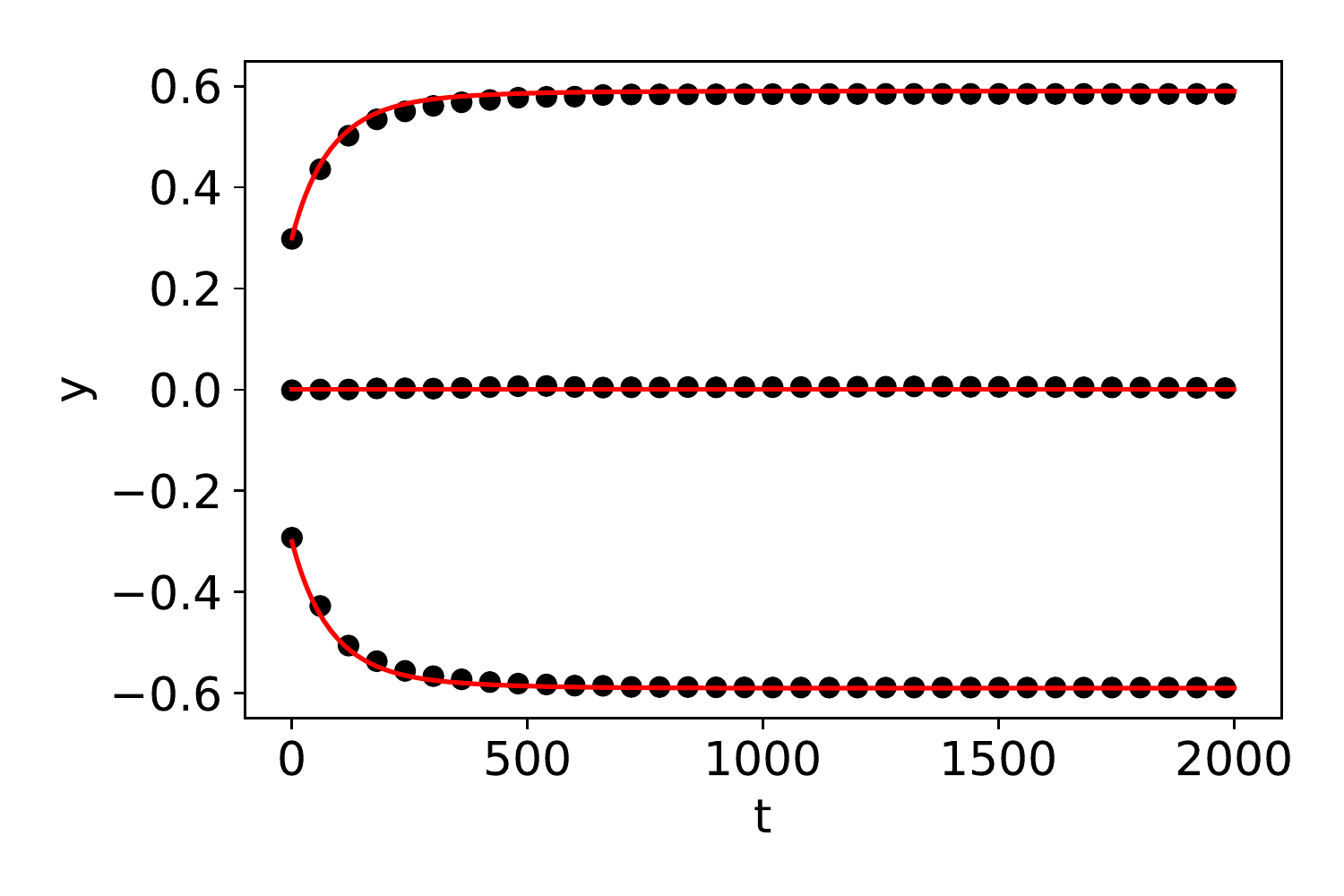}
\caption{$y$-coordinate of spots}
\end{subfigure}
\caption{For $\eps=0.02, \, D = \tau = 1, \, a = 20$ and a hole at
  $\v{x}_0 = (0.1,0)$ with radius $\eps$ ($C=1$), three spots are
  initially located at $(0.5,0)$ ($S\approx 0.4043$) and
  $(-0.2,\pm 0.3)$ ($S \approx 3.2549$), respectively. In (a) and (b),
  we show the numerical PDE solution of $v$ at $t=0$ and $t=1998$,
  respectively. In (c) and (d), we show the very close agreement of
  spot trajectories obtained by the PDE simulation (black dots) and
  the DAE \eqref{leakage:DAE} and \eqref{leakage:source_strength} (red
  solid line). We note that the $x$-coordinates of two spots on the
  left of the hole almost coincide and their trajectories in the
  $x$-direction are indistinguishable.}\label{fig:leak_3}
\end{figure}

\begin{figure}[htbp]
\begin{subfigure}[b]{0.25\textwidth}
\includegraphics[width=\textwidth]{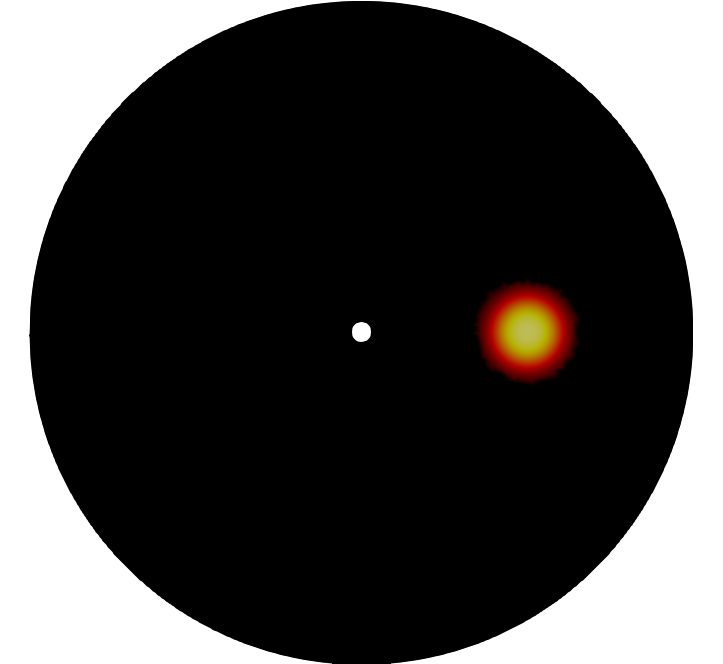}
\caption{$t=0$}
\end{subfigure}
\hfill
\begin{subfigure}[b]{0.25\textwidth}
\includegraphics[width=\textwidth]{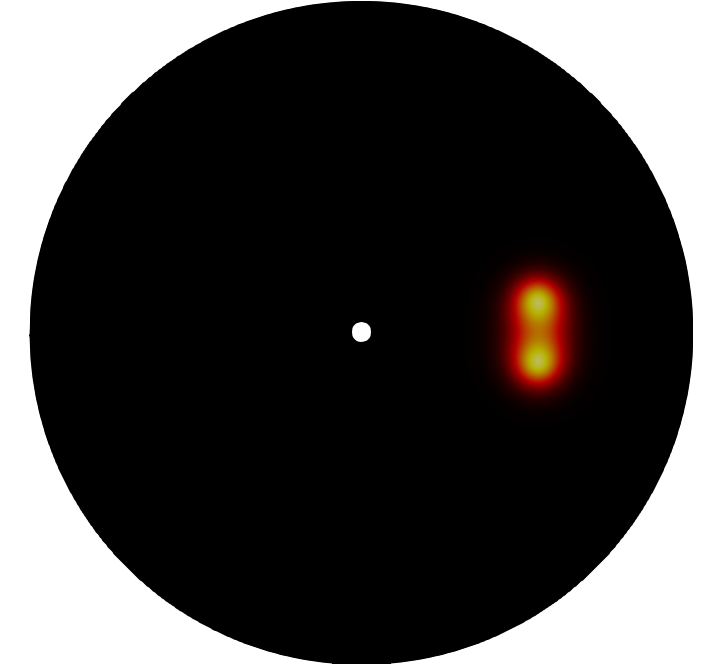}
\caption{$t=28$}
\end{subfigure}
\hfill
\begin{subfigure}[b]{0.25\textwidth}
\includegraphics[width=\textwidth]{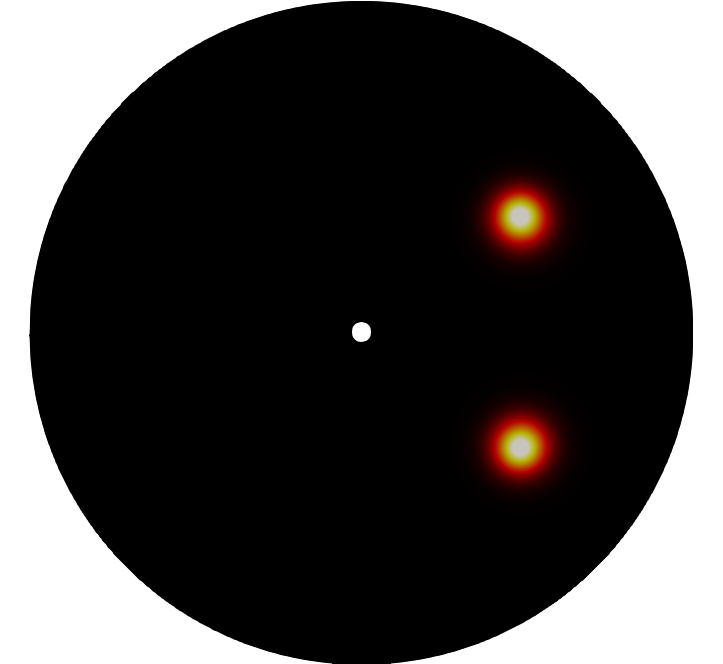}
\caption{$t=80$}
\end{subfigure}
\caption{For $\eps=0.03 \,, \tau = D = 1 \,, a=18 $ and a hole at the
  center with radius $\eps \,(C=1)$, a spot located at $\v{x}=(0.5,0)$
  initially has source strength $S\approx 4.42779$, which is greater
  than the peanut-splitting threshsold $\Sigma_2\approx 4.302$ in
  \S \ref{sec:proto_linstab}. We confirm the predicted
  spot-splitting event with this PDE simulation.}\label{fig:leak_split}
\end{figure}

In the unit disk, in Fig.~\ref{fig:leak_2} and Fig.~\ref{fig:leak_3}
we show a very favorable comparison between the spot trajectories as
computed from the DAE system \eqref{leakage:DAE} and
\eqref{leakage:source_strength} and from the full PDE system
\eqref{leakage:full} for the case of two or three spots,
respectively. The hole location and radius, and the other parameter
values, are given in the figure captions. From Fig.~\ref{fig:leak_2}
and Fig.~\ref{fig:leak_3}, we observe that there is a repulsive
interaction between the spots and the small hole. By increasing the
feed-rate parameter $a$, in Fig.~\ref{fig:leak_split} we show that a
one-spot solution will exhibit spot self-replication when the spot
source strength exceeds the peanut-splitting threshold
$\Sigma_2\approx 4.302$. However, in contrast to the case of a
hole-free unit disk where the critical feed-rate parameter for the
onset of a peanut-instability of a spot is $a_c =2\Sigma_2\approx 8.6$,
and is independent of the spot location, we observe from
Fig.~\ref{fig:leak_split} that a much larger feed rate is needed to
trigger a peanut-splitting instability when the domain contains a
hole.  Moreover, the required threshold of the feed rate depends on
the relative locations of the spot and the center of the hole.

\subsection{Linear stability analysis}\label{sec:leakage_linstab}

In this subsection we analyze the linear stability on an ${\mathcal O}(1)$
time-scale of the quasi-equilibria, denoted by $v_e$ and $u_e$, as
constructed in \S \ref{sec:leakage_qe}. We substitute
$v = v_e + e^{\lambda t} \phi$ and $u = u_e + e^{\lambda t} \eta$ into
\eqref{leakage:pde} and \eqref{leakage:bc}, and linearize to obtain
\begin{equation}\label{leakage:linstab_linearization}
\begin{split}
  \eps^2 \lap \phi - \phi + 2u_ev_e\phi + v_e^2\eta = \lambda \phi \,, &\qquad
  D\lap\eta + a - \eps^{-2}\left(2u_ev_e\phi + v_e^2\eta\right) =
  \tau\lambda\eta\,, \quad \mbox{in} \quad \pdedomain \,, \\
  \partial_n \phi = \partial_n \eta = 0 \quad \mbox{on} \quad \partial \Omega
  \,; &\qquad \phi = \eta = 0 \quad \mbox{on} \quad \partial \Omega_\eps \,.
\end{split}
\end{equation}

Following the analysis in \S \ref{sec:proto_linstab}, we obtain the
local eigenvalue problem \eqref{proto:lin_stab_eigenproblem}. The
analysis of instabilities associated with non-radially symmetric
perturbations near a spot is the same as given in \S \ref{sec:proto_linstab}
and the criterion is based on the source strengths. We conclude that
the $j^{\mbox{th}}$ spot is linearly unstable to the peanut-splitting
mode when $S_j>\Sigma_2\approx 4.302$, where $S_j$ is obtained from the
nonlinear algebraic system \eqref{leakage:source_strength} that depends
on the location of the hole.

We focus on deriving a GCEP associated with radially symmetric
perturbation near a spot, in which $m=0$ in the local problem
\eqref{proto:lin_stab_eigenproblem}.  Using the distributional limit
\eqref{proto:linstab_corr_rule}, we obtain for $\lambda\neq 0$ that
the outer problem for $\eta$ away from the spots is
\begin{equation}\label{leakage:linstab_outer_problem0}
  \lap \eta - \frac{\tau \lambda}{D} \eta -
  \frac{2\pi}{D}\sum\limits^N_{i=1} c_j \delta(\v{x}-\v{x}_i) = 0 \quad
  \mbox{in} \quad \pdedomain \,, \qquad \partial_n \eta = 0 \quad \mbox{on}
  \quad \partial\Omega \,; \qquad \eta = 0 \quad \mbox{on} \quad
  \partial\Omega_\eps \,.
\end{equation}

Similar to the derivation of outer problem
\eqref{leakage:outer_problem}, we approximate the zero Dirichlet
boundary condition for $\eta$ on the hole boundary by a Dirac Delta
forcing of undetermined strength
$2\pi c_0 D^{-1}\, \delta(\v{x}-\v{x}_0)$. In this way, the modified
outer problem for $\eta$ defined at ${\mathcal O}(1)$ distances from
the spots and the hole is
\begin{equation}\label{leakage:linstab_outer_problem}
  \lap \eta - \frac{\tau \lambda}{D} \eta -
  \frac{2\pi}{D}\sum\limits^N_{i=0} c_i \, \delta(\v{x}-\v{x}_i) = 0 \quad
  \mbox{in} \quad \Omega \,, \qquad \partial_n \eta = 0 \quad \mbox{on}
  \quad \partial\Omega \,,
\end{equation}
which is subject to the matching condition 
\begin{equation}\label{leakage:linstab_matching_hole1}
  \eta \sim \frac{c_0}{D}\left(\log|\v{x}-\v{x}_0| + \frac{1}{\nu} - \log C
  \right) \,, \quad \mbox{as} \quad \v{x} \to \v{x}_0 \,.
\end{equation}

The solution to \eqref{leakage:linstab_outer_problem} is represented in
terms of the eigenvalue-dependent Green's function
$G_\lambda$ of \eqref{proto:eig_green} by
\begin{equation}\label{leakage:linstab_outer_solution}
\eta = -\frac{2\pi}{D}\sum\limits_{i=0}^N c_i G_\lambda(\v{x};\v{x}_i) \,.
\end{equation}
We let $\v{x}\to \v{x}_0$ in \eqref{leakage:linstab_outer_solution}
and equate the resulting ${\mathcal O}(1)$ limiting behavior with
\eqref{leakage:linstab_matching_hole1}. This matching condition yields
that
\begin{equation}\label{leakage:linstab_matching_hole}
  c_0 + 2 \pi \nu \left( c_0 \Rlhole + \sum\limits_{i=1}^N c_i \, \Glhole
  \right) - \nu c_0 \log C  = 0 \,,
\end{equation}
where $\Rlhole \equiv R_{\lambda}(\v{x}_0;\v{x}_0)$ and
$\Glhole \equiv G_{\lambda}(\v{x}_0;\v{x}_i)$.

Next, by expanding \eqref{leakage:linstab_outer_solution} as
$\v{x} \to \v{x}_j$, we have for each $j=1,\ldots,N$ that
\begin{equation}\label{leakage:linstab_matching}
  \eta \sim \frac{c_j}{D}\log|\v{x}-\v{x}_j| - \frac{2\pi}{D}\left(
    c_j R_\lambda(\v{x}_j;\v{x}_j) + \sum\limits_{i\neq j}^N c_i \,
    G_\lambda(\v{x}_j;\v{x}_i) + c_0 \, G_\lambda(\v{x}_j; \v{x}_0) \right)\,.
\end{equation}
Upon matching \eqref{leakage:linstab_matching} with the far-field
behavior \eqref{proto:linear_stability_matching_inner_1} of the inner
problem we obtain
\begin{equation}\label{leakage:linstab_matching_spot}
  c_j + 2 \pi \nu \left( c_j \Rljj + \sum\limits_{i=1,i\neq j}^N c_i \,
    \Glji + c_0 \, G_{\lambda_{j,0}}\right) + \nu c_j \tilde{B}(S_j;\lambda)  = 0\,,
\end{equation}
where $\Rljj \equiv R_\lambda(\v{x}_j;\v{x}_j)$ and
$\Glji \equiv G_\lambda(\v{x}_j;\v{x}_i)$.

We write \eqref{leakage:linstab_matching_hole} and
\eqref{leakage:linstab_matching_spot} in matrix form as
\begin{subequations}\label{leakage:linstab_temp}
\begin{equation}\label{leakage:linstab_temp_1}
 \thetal c_0 + 2 \pi \nu\,\gl^T\v{c}=\v{0} \,, \qquad
 \v{c} \,+ \, 2 \pi \nu\left(\mc{G}_\lambda \v{c} + c_0 \gl\right) +
 \nu \Bmat \v{c} = \v{0} \,,
\end{equation}
where the matrices $\Gmatl$ and $\Bmat$ are defined in
\eqref{proto:Gmatl_Bmat}. In \eqref{leakage:linstab_temp_1} we
have defined
\begin{equation}
  \v{c} \equiv \left( c_1, \ldots, c_N\right)^T \,, \quad \gl \equiv
  \left( G_{\lambda_{0,1}}, \ldots, G_{\lambda_{0,N}} \right)^T \,, \quad
  \v{e} \equiv (1,\ldots,1)^T \in \mathbb{R}^N \,, \quad
  \thetal \equiv 1 + 2\pi\nu\Rlhole - \nu\log C \,.
\end{equation}
\end{subequations}
The GCEP is obtained by eliminating $c_0$ in
\eqref{leakage:linstab_temp_1}.  In this way, we conclude that a
discrete eigenvalue $\lambda$ of the linearization must be such that
\begin{subequations}\label{leakage:gcep}
\begin{equation}\label{leakage:gcep1}
  \Mmat \v{c} = 0\,,\qquad \mbox{where} \qquad \Mmat(\lambda) \equiv
  \theta_\lambda \left(\Imat + 2\pi\nu\Gmatl + \nu\Bmat \right) - 4\pi^2\nu^2\,
  \gl\gl^T \,,
\end{equation}
has a nontrivial solution $\v{c}\neq \v{0}$. Here $\Imat$ is the $N\times N$
identity matrix. Any such $\lambda\neq 0$ satisfying
\begin{equation}\label{leakage:gcep2}
\det \Mmat(\lambda) = 0\,,
\end{equation}
\end{subequations}
for which $\mbox{Re}(\lambda)>0$, corresponds to an instability
associated with locally radially symmetric perturbations near the
spots.

As similar to the analysis in \S \ref{sec:proto_linstab}, we must
consider separately the special case of a zero-eigenvalue crossing
where $\lambda = 0$. When $\lambda=0$, the solution to the
modified outer problem \eqref{leakage:linstab_outer_problem} is
\begin{equation}\label{leakage:linstab_outer_solution_reduced}
  \eta = -\frac{2\pi}{D} \sum\limits_{i=0}^N c_i \, G(\v{x};\v{x}_i) + \bar{\eta}
  \,, \qquad \mbox{where} \qquad \sum_{i=0}^{N} c_i=0 \,,
\end{equation}
and where $\bar{\eta}$ is an additive constant to be found. Here, $G$
is the Neumann Green's function satisfying \eqref{proto:neu_green}.
By matching the local behavior of $\eta$ to the far-field behavior
\eqref{leakage:linstab_matching_hole1} near the hole as well as to the
far field behavior \eqref{proto:linear_stability_matching_inner_1}
near the spots, we obtain in matrix form that
\begin{subequations}\label{leakage0:sys}
\begin{equation}\label{leakage0:sys_1}
  \theta c_0 + 2 \pi \nu \v{g}^T \v{c} = \nu D \bar{\eta} \,, \qquad
  \v{c} + 2 \pi \nu \left( \Gmat \v{c} + c_0 \v{g} \right) + \nu \Bmat_0 \v{c}
  = \nu D \, \bar{\eta} \, \v{e} \,,
\end{equation}
where $\Gmat\in \mathbb{R}^{N\times N}$ is the Neumann Green's matrix and
$\Bmat_0=\mbox{diag}\left(\chi^{\prime}(S_1),\ldots,\chi^{\prime}(S_N)\right)$,
as is given in \eqref{zero:bmat}. In \eqref{leakage0:sys_1} we have
defined
\begin{equation}
  \v{c} \equiv \left( c_1, \ldots, c_N\right)^T \,, \quad \v{g} \equiv
  \left( G_{0,1}, \ldots, G_{0,N} \right)^T \,, \quad
  \v{e} \equiv (1,\ldots,1)^T \in \mathbb{R}^N \,, \quad
  \theta \equiv 1 + 2\pi\nu R(\v{x}_0;\v{x}_0) - \nu\log C \,,
\end{equation}
\end{subequations}
where $G_{0,i}\equiv G(\v{x}_0;\v{x}_i)$. Since $\sum_{i=0}^{N} c_i=0$, we
can write $c_0=-\v{e}^T\v{c}$. Upon eliminating $c_0$ in
\eqref{leakage0:sys_1}, we conclude that $\lambda=0$ is an eigenvalue
of the linearization if and only if
\begin{equation}\label{leakage0:gcep}
  \Mmat_0 \v{c} = 0\,,\qquad \mbox{where} \qquad \Mmat_0 \equiv
  \Imat + \theta N\Emat + 2\pi\nu\,\Gmat + \nu \Bmat_0 - 2\pi\nu
  \left(\v{g}\v{e}^T + \v{e}\v{g}^T\right)\,,
\end{equation}
has a nontrivial solution $\v{c}\neq \v{0}$. Parameter values corresponding
to zero-eigenvalue crossings are where $\det\Mmat_0=0$.

\subsection{A ring pattern of $N$-spots with leakage at the
  center}\label{sec:leakage:ring}

We consider a ring pattern of $N$-spots, with spots centered at
\eqref{proto:ring}, in the perforated unit disk $\pdedomain$ that has
a hole of radius $C\eps$ at the origin. Since the $N$ spots have a common
source strength $S_c$, we let $\v{s}=S_c\,\v{e}$ in
\eqref{leakage:source_strength}. Upon using ${\mathcal G}\v{e}=
{p(r_0)\v{e}/N}$ from \eqref{ring:gmat_eig}, where $r_0$ is the ring
radius, together with
\begin{equation}\label{leakage:ring_int}
  \theta = 1 + 2\pi\nu R(\v{0};\v{0}) - \nu \log C = 1 -
  \nu\left( \log C + \frac{3}{4} \right) \,, \qquad
  \v{g} = G(\v{x}_j;\v{0})\, \v{e} = \frac{1}{2\pi}\left(
    -\log r_0 + \frac{r_0^2}{2} - \frac{3}{4} \right) \v{e} \,,
\end{equation}
as calculated from \eqref{ring:gcomplex}, we obtain
from \eqref{leakage:source_strength} and
\eqref{leakage:source_strength_sum} that $S_c$ satisfies the scalar
nonlinear equation
\begin{equation}\label{leakage:common_source_strength}
  S_c + \frac{\nu S_c}{N+1} \log\left[ \frac{r_0^{N+1}}{NC^N(1-r_0^{2N})} \right]
  + \frac{\nu\chi(S_c)}{N+1} = \frac{p_a}{N+1} \left[ 1 + \nu
    \left( \log\left(\frac{r_0}{C}\right) - \frac{r_0^2}{2} \right) \right]\,,
  \qquad p_a= \frac{a}{2\sqrt{D}} \,.
\end{equation}

Next, by using \eqref{ring:gcomplex_der}, we calculate for a ring
pattern that
\begin{equation*}
  2 \pi \left( p_a - \sum\limits_{i=1}^N S_i \right) \nabla_\v{x}
  G_{j,0} = (p_a - N S_c) \left(r_0 - \frac{1}{r_0}\right) \v{e}_{\theta_j} \,,
\end{equation*}
where $\v{e}_{\theta_j}$ is defined in \eqref{proto:ring}. Upon using
this result, together with the expression \eqref{proto:betaj_ring} for
$\beta_j$ for a ring pattern, the ODE system \eqref{leakage:DAE} for
slow spot dynamics reduces to the following scalar ODE for the ring radius
$r_0$:
\begin{equation}\label{leakage:ring_dynamics}
  \frac{d r_0}{d \sigma} = \gamma(S_c) \left[p_a\left(\frac{1}{r_0}-r_0\right) -
    S_c \left(\frac{N+1}{2r_0} + \frac{Nr_0^{2N-1}}{1-r_0^{2N}}\right)\right]\,,
\end{equation}
where $\sigma=\eps^2 t$. Here $S_c=S_c(r_0)$ is determined from the
nonlinear constraint \eqref{leakage:common_source_strength}. It
follows that the equilibrium ring radius $r_0=r_{0e}$ of
\eqref{leakage:ring_dynamics} with common source strength $S_c$ is a
root of
\begin{equation}\label{leakage:equilibrium_ring}
  S_c \left(\frac{N+1}{2r_{0e}} + \frac{Nr_{0e}^{2N-1}}{1-r_{0e}^{2N}}\right) =
  p_a\left(\frac{1}{r_{0e}}-r_{0e}\right)\,,
\end{equation}
where $S_c=S_c(r_{0e})$ satisfies \eqref{leakage:common_source_strength}.

Next, the GCEP \eqref{leakage:gcep1} for a ring pattern reduces to finding
values of $\lambda$ for which there are nontrivial solutions to
\begin{equation}\label{leakage:ring_gcep}
  \Mmat \v{c} = \v{0} \,, \qquad \mbox{with} \qquad
  \Mmat \equiv \thetal \left( 1 + \nu \Bhat_c  \, \Imat + 2\pi\nu \,\Gmatl
  \right) - 4 N \pi^2 \nu^2 \beta_\lambda^2 \Emat \,, \qquad \Emat \equiv
  \frac{1}{N} \v{e}\v{e}^T \,,
\end{equation}
where $\Bhat_c \equiv \Bhat(S_c;\lambda)$ is calculated from
\eqref{proto:m=0_eigenproblem} and where $\beta_\lambda \equiv
G_\lambda(\v{x}_1;\v{0})= \ldots = G_\lambda(\v{x}_N;\v{0})$. Since
$\Gmatl$ is a cyclic symmetric matrix, it has the eigenspace
$\v{c} = \v{e}$ and $\v{c} = \v{q}_j$, where $\v{e}^T\v{q}_{j}=0$ and
$\v{q}_{j}^T\v{q}_i=0$ for $i\neq j$ and $i,j=2,\ldots,N$. In this way,
from \eqref{leakage:ring_gcep}, the discrete eigenvalues $\lambda$ for
the synchronous ($\v{c}=\v{e}$) mode and competition modes
($\v{c}=\v{q}_j$, $j=2,\ldots,N$) are the roots of
\begin{subequations}\label{leakage:ring_gcep_modes}
\begin{align}
&F_1 \equiv \thetal(1 + \nu\Bhat_c + 2 \pi \nu \, \omega_1) - 4 N \pi^2 \nu^2 \beta_\lambda = 0 \,, \label{leakage:GCEP_ring_syn}\\[5pt]
  &F_j \equiv \thetal(1 + \nu \Bhat_c + 2 \pi \nu \, \omega_j) = 0 \,, \quad
    j=2 \,, \ldots \,, N \,, \label{leakage:GCEP_ring_asyn}
\end{align}
\end{subequations}
where the matrix eigenvalues $\omega_i=\omega_i(\lambda)$ of $\Gmatl$
are defined by $\Gmatl \v{e} = \omega_1 \, \v{e}$ and
$\Gmatl \v{q}_i = \omega_i \, \v{q}_i$ for $i=2,\ldots,N$.

Next, we derive the threshold condition on the parameters for
which there is a zero-eigenvalue crossing in the GCEP. We use
$\Bmat_0 = \Bhat(S_c;0) \, \Imat \equiv \chi^{\prime}(S_c) \, \Imat$,
together with \eqref{leakage:ring_int}, to obtain 
that \eqref{leakage0:gcep} reduces to
\begin{equation}\label{leakage:GCEP_ring_reduced}
  \Mmat_0 \v{c} = 0 \,, \qquad \mbox{where} \qquad
  \Mmat_0 \equiv \left[1 + \nu \chi^{\prime}(S_c)\right] \, \Imat +
  N\left[ 1 - \nu \left(\log\frac{C}{r_0^2} + r_0^2 - \frac{3}{4}\right)
  \right] \Emat + 2\pi\nu\,\Gmat \,.
\end{equation}
By using \eqref{ring:gmat_eig}, we conclude from
\eqref{leakage:GCEP_ring_reduced} that a zero-eigenvalue crossing for
the mode $\v{c}=\v{e}$ occurs if and only if $S_c$ satisfies
\begin{equation}\label{leakage:GCEP_ring_synchronous_threshold}
N+1 + \nu \chi^{\prime}(S_c) + \nu \log 
\left(\frac{r_0^{N+1}}{N C^N (1-r_0^{2N})}\right) = 0 \,.
\end{equation}

We now show that this zero-eigenvalue threshold condition
\eqref{leakage:GCEP_ring_synchronous_threshold} occurs precisely at
the value of $S_c$ for which the root $S_c=S_c(r_0)$ to
\eqref{leakage:common_source_strength} has a saddle-node
bifurcation. To see this, we differentiate
\eqref{leakage:common_source_strength} with respects to $S_c$ to
obtain
\begin{equation}\label{leakage:dr0/dSc}
  \frac{\nu}{r_0} \left( \frac{p_a(C - r_0^2)}{r_0} - \frac{S_c (N+1 +
      (N-1)r_0^{2N})}{r_0(1-r_0^{2N})}\right) \frac{d r_0}{d S_c} = N + 1 +
  \nu  \chi^{\prime}(S_c) + \nu \log \left(\frac{r_0^{N+1}}{N C^N (1 -
      r_0^{2N})}\right) \,.
\end{equation} 
At a saddle-point point $(r_{0f},S_{cf})$ we have ${dr_0/dS_c} = 0$,
and so the right-hand side of \eqref{leakage:dr0/dSc} must vanish at
that point, which yields
\eqref{leakage:GCEP_ring_synchronous_threshold}.  We conclude that a
zero-eigenvalue crossing of the GCEP can only occur at the location of
a saddle-node bifurcation point for a quasi-equilibrium ring pattern.

Next, to determine the threshold condition on the parameters for a
zero-eigenvalue crossing for the competition modes, we substitute
$\v{c}=\v{q}_i$ for $i=2,\ldots,N$ into \eqref{leakage:GCEP_ring_reduced},
and use $\Emat \v{q}_i=0$ to obtain
\begin{equation}\label{leakage:GCEP_ring_asynchronous_threshold}
1 + \nu\chi^{\prime}(S_c) + 2 \pi \nu \sigma_i = 0 \,,  \qquad i=2,\ldots, N\,.
\end{equation}
Here $\sigma_i$ are eigenvalues of the Neumann Green's matrix for
which ${\mathcal G}\v{q}_i=\sigma_i\v{q}_i$ for $i=2,\ldots,N$. Roots
of the coupled problem
\eqref{leakage:GCEP_ring_asynchronous_threshold} and
\eqref{leakage:ring_int} correspond to the threshold values
$(S_c^{(i)}, r_0^{(i)})$, for $i=2,\ldots,N$, where a zero-eigenvalue
crossing of the GCEP occurs.

\subsubsection{A one-spot quasi-equilibrium}

\begin{figure}[htbp]
\begin{subfigure}{0.45\textwidth}
\includegraphics[width=\textwidth,height=4.3cm]{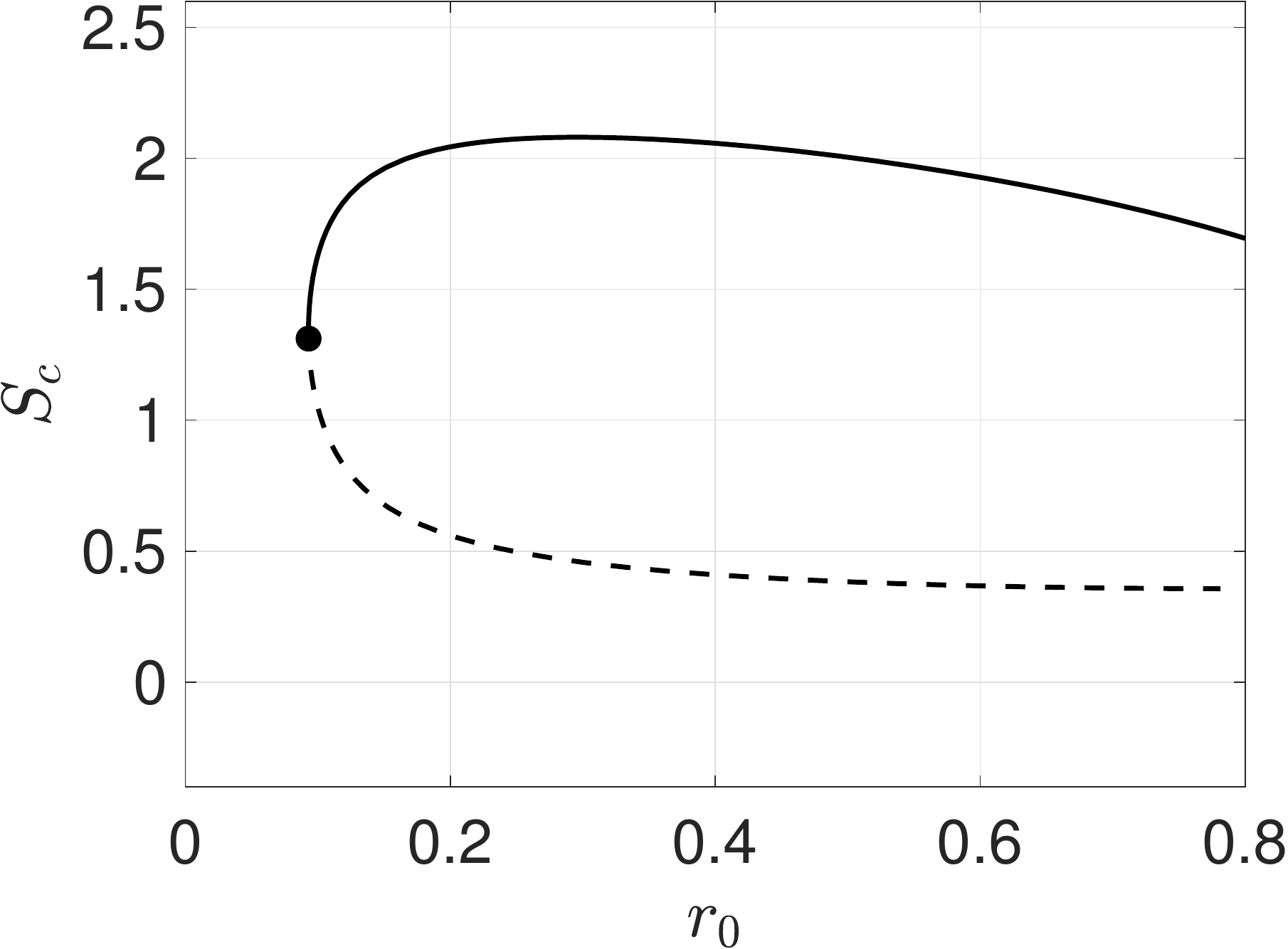}
\caption{Bifurcation diagram}
\label{leakage:N=1_exp1_Sc_vs_r0}
\end{subfigure}
\hfill
\begin{subfigure}{0.45\textwidth}
\includegraphics[width=\textwidth,height=4.8cm]{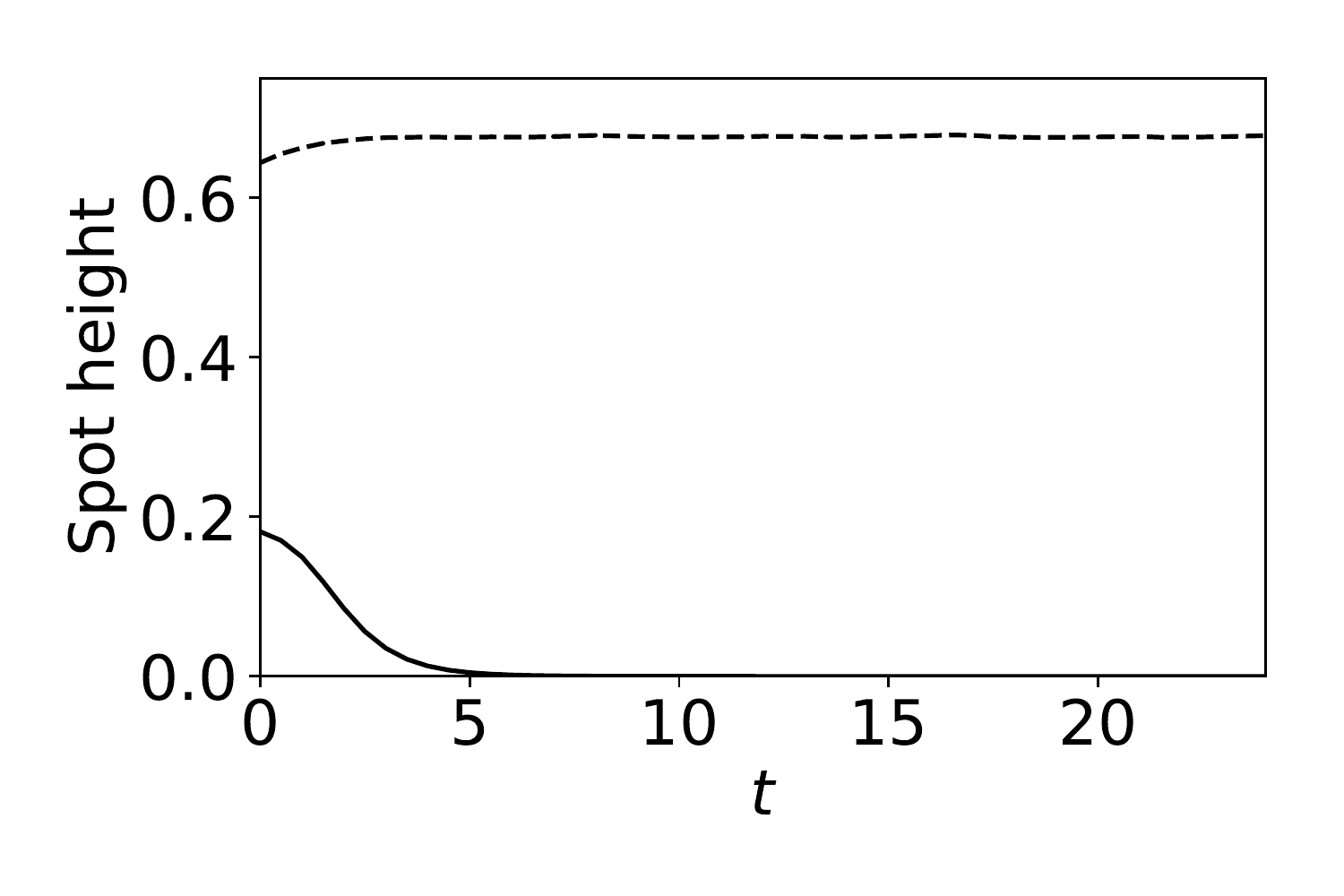}
\caption{Spot height}
\label{leakage_ring_exp3and4}
\end{subfigure}
\caption{We fix $C=1$, $D=\tau=1$, $\eps=0.02$, and $a=10$. Left
  panel: $S_c$ versus $r_0$ for a one-spot quasi-equilibrium solution,
  as computed from \eqref{leakage:common_source_strength}. The
  saddle-node bifurcation is at
  $(r_{0f}, S_{cf}) \approx (0.0930, 1.3114)$.  As indicated by (a), a
  one-spot pattern with $r_0 = 0.4$ has two possible source strengths,
  which are $S \approx 0.4094$ (lower branch) and $S \approx 2.0576$
  (upper branch). Right panel: short-time evolution of the spot
  amplitude, defined as the maximum of $v$, with these two initial
  source strengths, as computed from the full PDE
  \eqref{leakage:full}. The bottom (solid) curve shows that the spot
  on the lower branch is rapidly annihilated.}
\end{figure}

We first consider a one-spot quasi-equilibrium solution in the
perforated unit disk $\pdedomain$. In this subsection, we fix
$\eps=0.02$ and $D=\tau=1$. By taking the ring radius $r_0$ as a
bifurcation parameter, in Fig.~\ref{leakage:N=1_exp1_Sc_vs_r0} we show
that \eqref{leakage:common_source_strength} has a fold bifurcation
structure for the source strength of the spot.  From this figure, we
observe that a one-spot quasi-equilibrium solution does not exist when
the spot is too close to the center of the hole located at the origin.
In contrast, when there is no hole, a one-spot quasi-equilibrium
solution exists for all $r_0\geq 0$ in the unit disk. We have
numerically verified that along the lower branch in
Fig.~\ref{leakage:N=1_exp1_Sc_vs_r0} the GCEP
\eqref{leakage:ring_gcep} has an unstable eigenvalue, while along the
upper branch it has no unstable eigenvalues. To verify these linear
stability predictions of the GCEP, for a one-spot quasi-equilibrium
solution with $r_0 = 0.4$ we performed full PDE simulations on
\eqref{leakage:full} with two source strengths, as indicated in the
bifurcation diagram in Fig.~\ref{leakage:N=1_exp1_Sc_vs_r0}. The
short-time evolution of the spot amplitude presented in
Fig.~\ref{leakage_ring_exp3and4} shows that the one-spot solution on
the lower branch is quickly annihilated, while the amplitude of the
spot on the upper branch is stabilized at a nearby value. These full
PDE results are in agreement with the linear stability predictions
based on the GCEP.

\begin{figure}[htbp]
\begin{subfigure}[b]{0.45\textwidth}
\includegraphics[width=\textwidth,height=4.3cm]{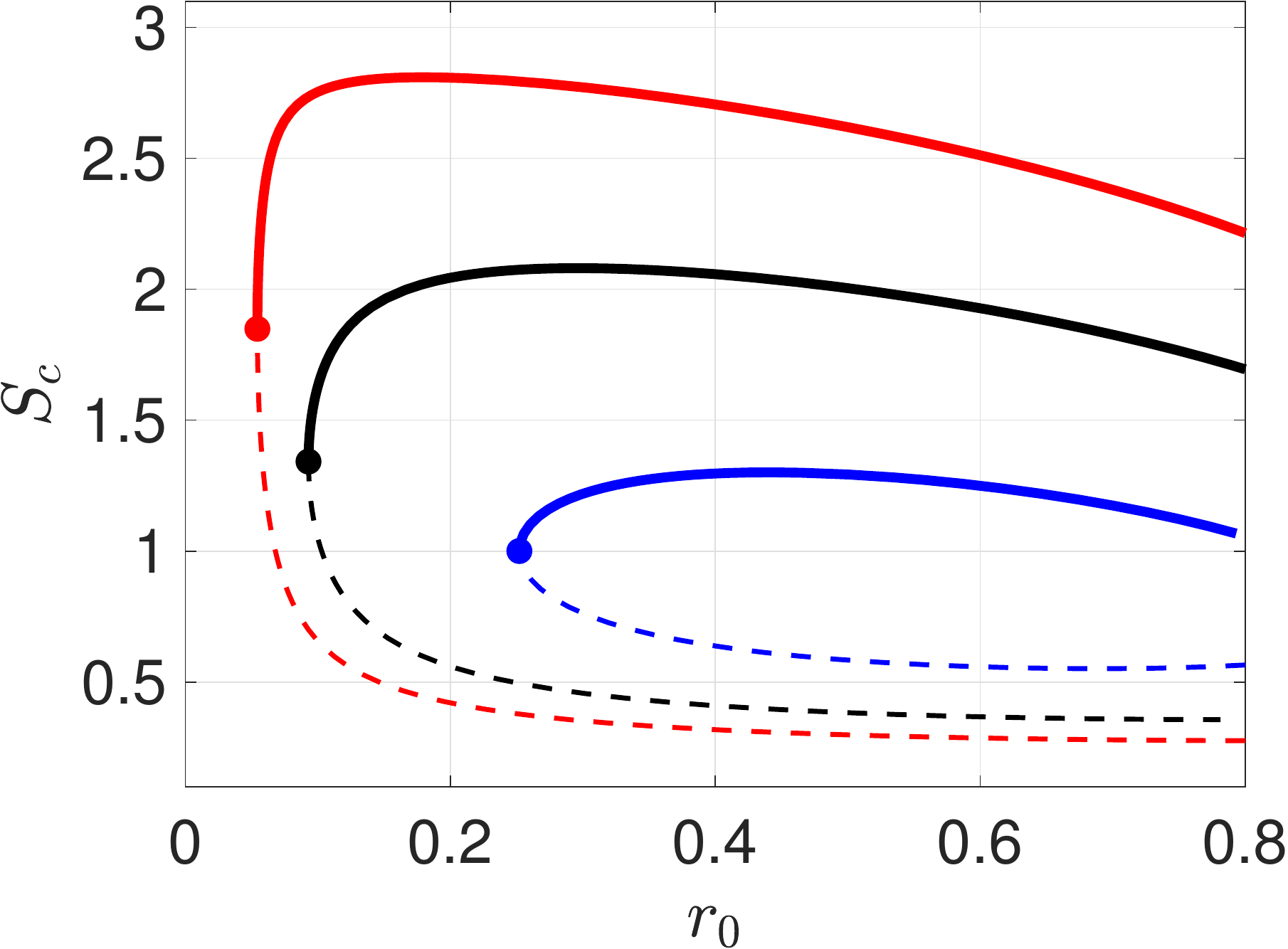}
\caption{Outside to inside : $a=12, \, 10, \, 8$.}
\label{leakage_N=1_a}
\end{subfigure}
\hfill
\begin{subfigure}[b]{0.45\textwidth}
\includegraphics[width=\textwidth,height=4.3cm]{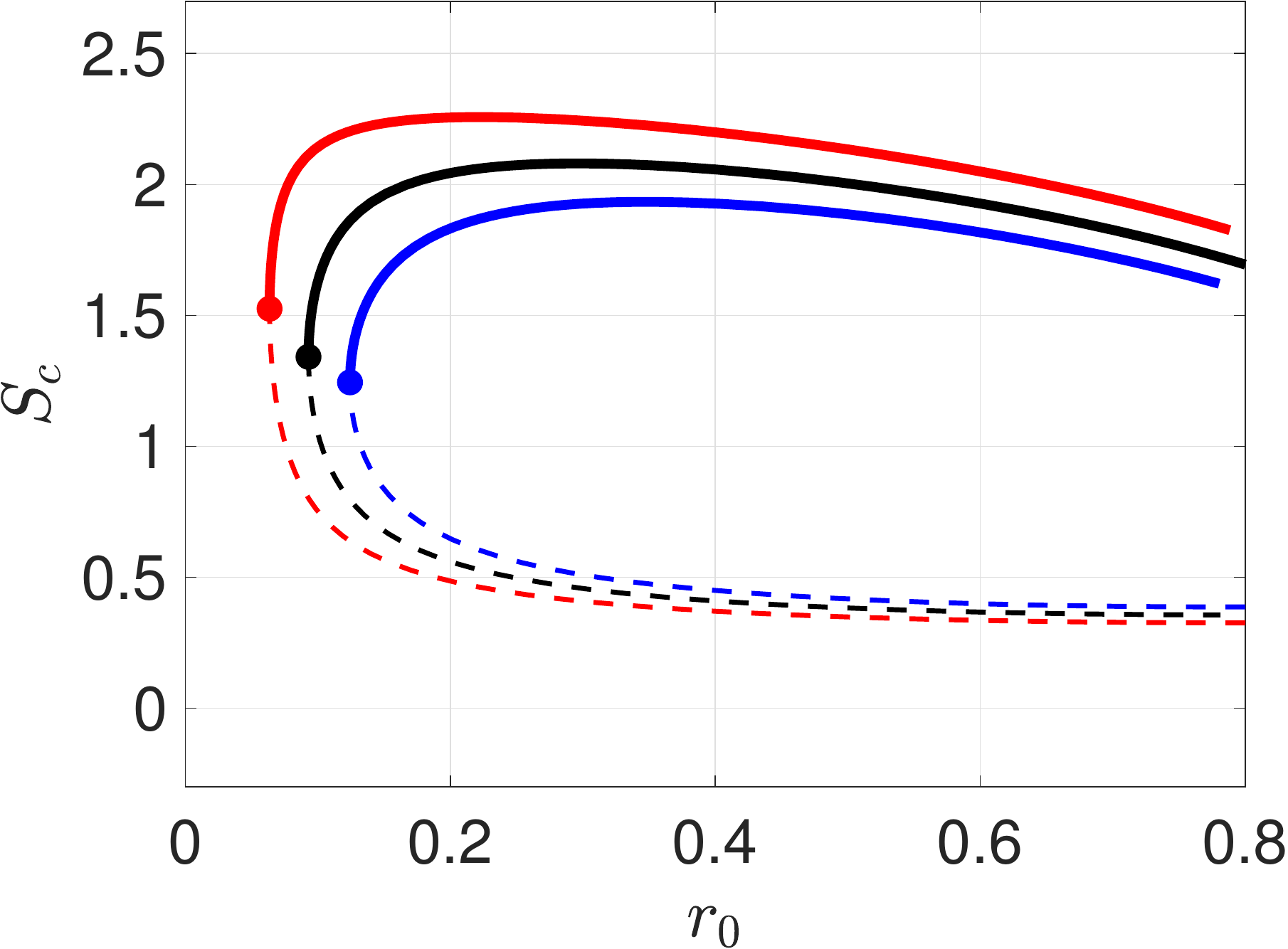}
\caption{Outside to inside : $C=0.8, \, 1, \, 1.2$.}
\label{leakage_N=1_C}
\end{subfigure}
\hfill
\caption{In both panels, the middle curve is the same as in
  Fig.~\ref{leakage:N=1_exp1_Sc_vs_r0}, which corresponds to
  $\eps=0.02, \,C=1$ and $a=10$. All lower branches have an unstable
  eigenvalue for the GCEP \eqref{leakage:ring_gcep}. Left panel: We
  fix $C=1$. The saddle-node bifurcation for $a=8$ and $a=12$ occurs
  at $r_{0f}\approx 0.2521$ and $r_{0f}\approx 0.0544$,
  respectively. Right panel: We fix $a=10$. The saddle-node
  bifurcation for $C=0.8$ and $C=1.2$ occurs at $r_{0f}\approx 0.0636$
  and $r_{0f}\approx 0.1243$, respectively. As either the
  hole radius decreases or the feed rate increases, a one-spot
  quasi-equilibrium solution can exist closer to the
  hole.}
\end{figure}

In Fig.~\ref{leakage_N=1_a} and Fig.~\ref{leakage_N=1_C}, we show how
the $S_c$ versus $r_0$ bifurcation diagram, computed from
\eqref{leakage:common_source_strength}, changes with respect to the
feed-rate parameter $a$ and the parameter $C>0$ that controls the
radius $\eps C$ of the hole.  We observe that as either $a$ increases
or $C$ decreases (smaller hole radius), a one-spot quasi-equilibrium
solution can exist closer to the hole.

\begin{figure}[htbp]
\begin{subfigure}{0.45\textwidth}
\includegraphics[width=\textwidth,height=4.3cm]{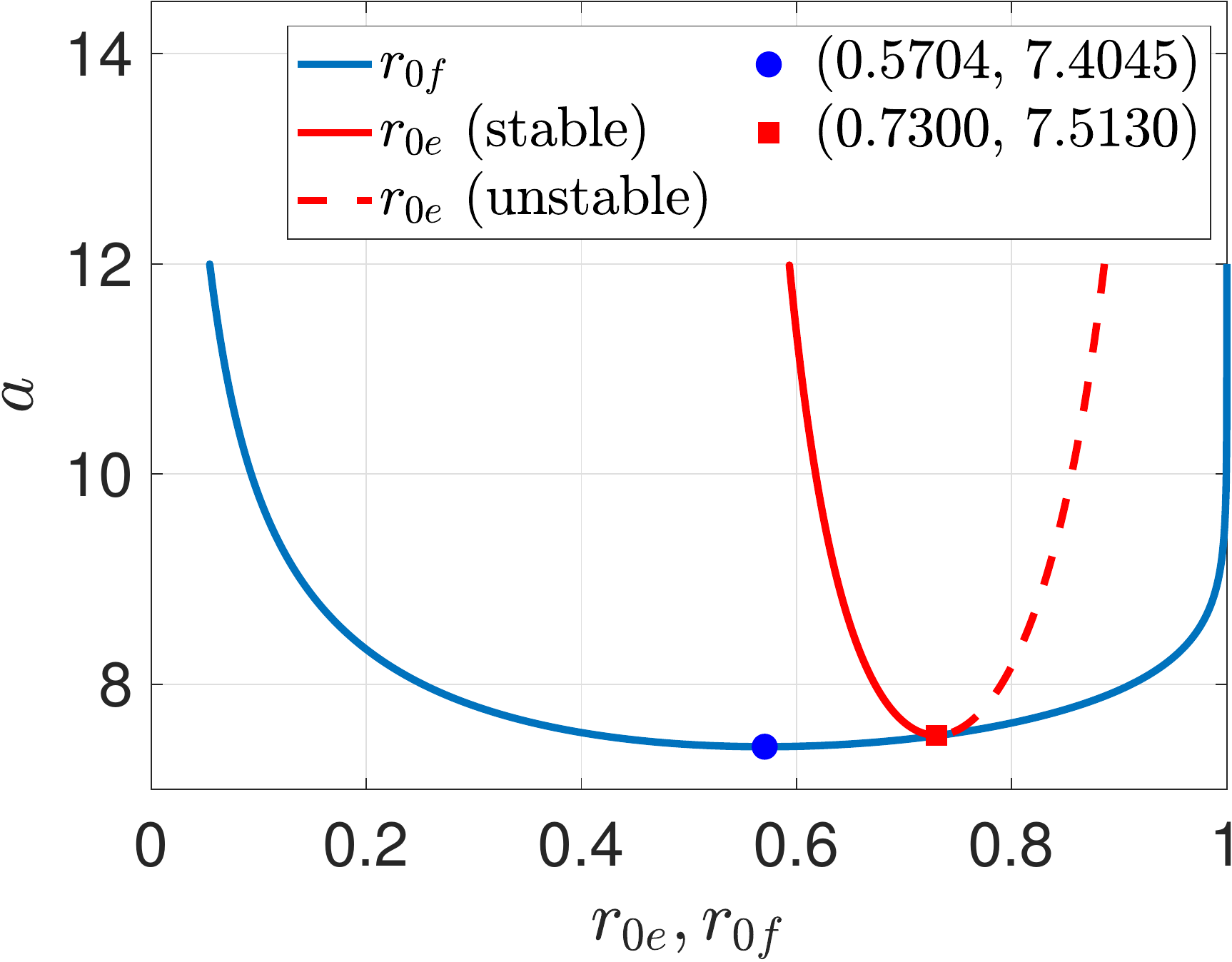}
\caption{Fix $C=1$}
\label{leakage_N=1_r0e_r0f_exp1}
\end{subfigure}
\hfill
\begin{subfigure}{0.45\textwidth}
\includegraphics[width=\textwidth,height=4.3cm]{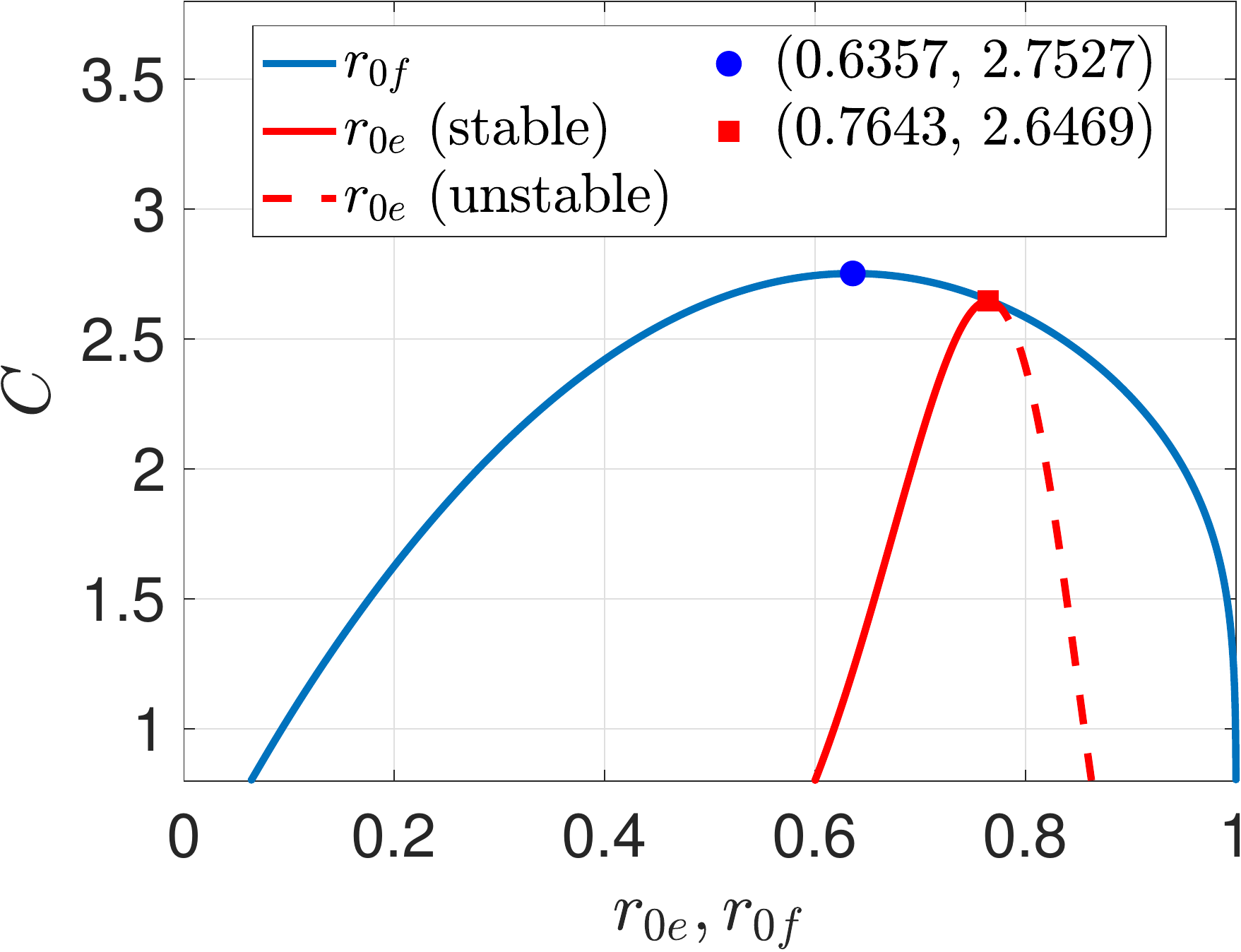}
\caption{Fix $a=10$}
\label{leakage_N=1_r0e_r0f_exp3}
\end{subfigure}
\hfill
\caption{The saddle-node structures of $r_{0f}$ (bigger U-shape) and
  the equilibrium ring radius $r_{0e}$ (smaller U-shape) with respect
  to the feed rate $a$ (left panel) and the hole radius parameter $C$
  (right panel) for a one-spot solution. Along the dashed portion of
  the $r_{0e}$ branch, the GCEP \eqref{leakage:ring_gcep} has an
  unstable eigenvalue.  For each feed-rate $a$ exceeding a threshold,
  there is only one stable equilibrium location for the one-spot
  solution.}
\end{figure}

Next, we use numerical continuation on
\eqref{leakage:common_source_strength} and the saddle-node condition
\eqref{leakage:GCEP_ring_synchronous_threshold} to determine how the
saddle-node point $r_{0f}$ for the ring radius depends on the
feed-rate parameter $a$ when $C=1$. A similar numerical continuation
of \eqref{leakage:common_source_strength} and the steady-state ring
radius condition \eqref{leakage:equilibrium_ring}, also reveals a
saddle-node bifurcation structure of $r_{0e}$. These results,
presented in Fig.~\ref{leakage_N=1_r0e_r0f_exp1}, show that a one-spot
quasi-equilibrium solution exists only when $a$ is greater than the
saddle-node value $a_f \approx 7.4045$. For each $a>a_f$, there are
two fold-point values of $r_{0f}$ for quasi-equilibria: one near the
boundary of the unit disk (not shown in
Fig.~\ref{leakage:N=1_exp1_Sc_vs_r0}) while the other is closer to the
hole.  For each $a>7.513$, there are two steady-state equilibrium ring
radii, with only one of these being linearly stable for the GCEP
\eqref{leakage:ring_gcep}. In Fig.~\ref{leakage_N=1_r0e_r0f_exp3},
where we fixed $a=10$, we show a similar saddle-node bifurcation
structure for $r_{0f}$ and $r_{0e}$ versus the parameter $C$, which
controls the radius of the hole. We observe that there is no
quasi-equilibrium one-spot solution if the hole radius exceeds a
certain threshold.

\begin{figure}[htbp]
\begin{subfigure}[b]{0.45\textwidth}
\includegraphics[width=\textwidth,height=4.3cm]{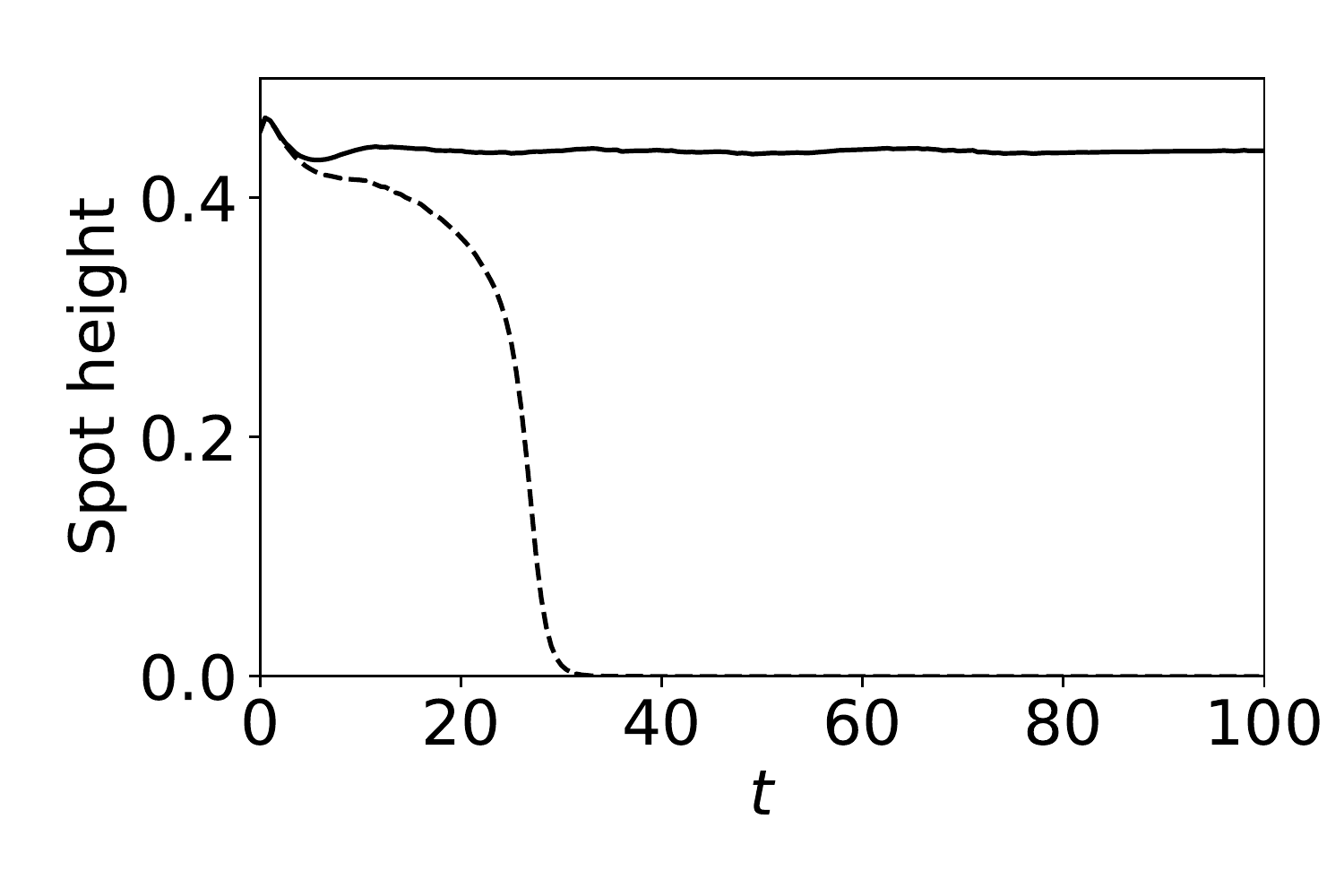}
\caption{$a=7.6$ v.s. $a=\max(7.6-0.01\,t,\,7.4)$}
\label{leakage_ring_exp5and6}
\end{subfigure}
\hfill
\begin{subfigure}[b]{0.45\textwidth}
\includegraphics[width=\textwidth,height=4.3cm]{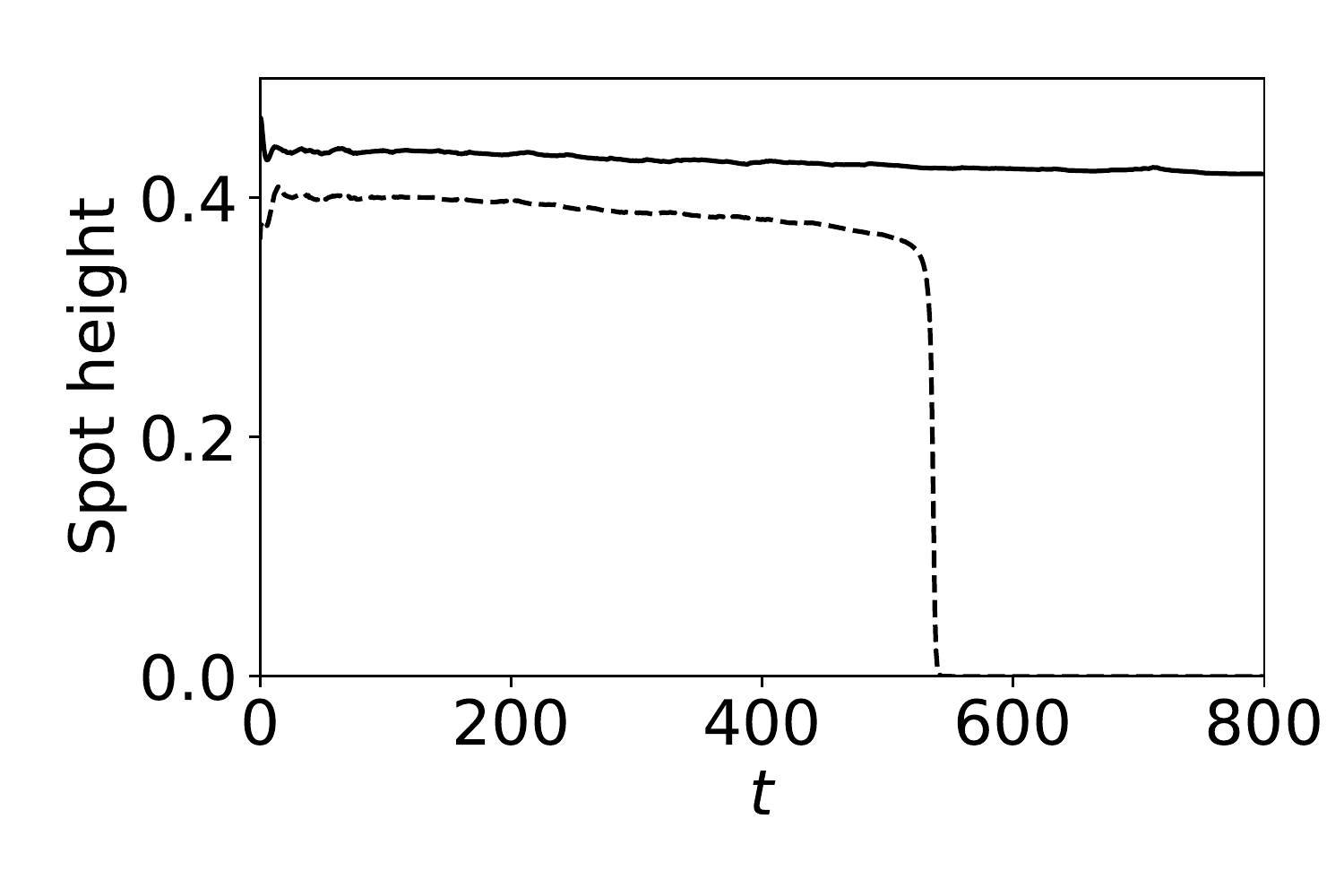}
\caption{$a=7.6$ v.s. $a=7.48$}
\label{leakage_ring_exp5and7}
\end{subfigure}
\caption{We fix $C=1$. Left panel: Short-time evolution of the
  amplitude of a one-spot quasi-equilibrium solution for a constant
  feed rate $a=7.6$ (solid line) and for a slowly decreasing feed rate
  $a=\max(7.6-0.01\,t,\,7.4)$ (dashed line).  Right panel: Longer time
  evolution of the spot amplitude for $a\equiv7.6$ (solid line) and
  $a=7.48$ (dashed line). When $a=7.6$, the one-spot solution has
  become close to its equilibrium value when $t \approx 2000$ (not
  shown). For both panels the initial spot location was at $r_0=0.57$,
  and the numerical results were computed from the full PDE
  \eqref{leakage:full}.}
\end{figure}

In Fig.~\ref{leakage_ring_exp5and6}, we show full PDE results computed
from \eqref{leakage:full} for a one-spot quasi-equilibrium solution,
initially located at $r_0 = 0.57$, in which the feed-rate parameter is
slowly decreased in time according to
$a = \max(7.6 - 0.01\,t, \,7.4)$. From this figure, we observe that
the spot amplitude collapses to zero, leading to spot annihilation, at
a time $t\approx20$. This rapid decay of the spot amplitude is due to
the non-existence of one-spot quasi-equilibria for $r_0=0.57$ when $a$
decreases below the saddle-node value $a_f$. Alternatively, in
Fig.~\ref{leakage_ring_exp5and6}, the full PDE simulation results
shows that the one-spot quasi-equilibrium persists when the feed rate
is fixed at $a=7.6 > a_f$. To motivate a further, but more delicate,
PDE simulation result, we observe from
Fig.~\ref{leakage_N=1_r0e_r0f_exp1} that the saddle-node value for
$r_{0e}$ occurs at $a_e \approx 7.5130$, which is greater than
$a_f\approx 7.4045$. For any feed rate $a$ between $a_f$ and $a_e$, a
quasi-equilibrium one-spot solution exists for some range of $r_0$,
but there is no steady-state equilibrium value $r_{0e}$.  In
Fig.~\ref{leakage_ring_exp5and7} we show results from a full PDE
simulation of \eqref{leakage:full} for a one-spot quasi-equilibrium
initially located at $r_0 = 0.57$ and with feed-rate $a=7.48$, which
satisfies $a_f<a<a_e$. We observe that the one-spot quasi-equilibrium
survives only until $t \approx 540$, when the slowly drifting
  spot is repelled sufficiently from the hole that it crosses the
  quasi-equilibrium existence threshold. In contrast, the
corresponding PDE simulation with $a=7.6 > a_e$ shows that the
one-spot quasi-equilibrium solution persists, and slowly drifts
away from the hole towards its stable equilibrium location
at around $t \approx 2000$ (not shown). 

\subsubsection{Hopf bifurcation of a one-spot quasi-equilibrium
solution}

\begin{figure}[htbp]
  \includegraphics[width=0.45\textwidth,height=4.2cm]{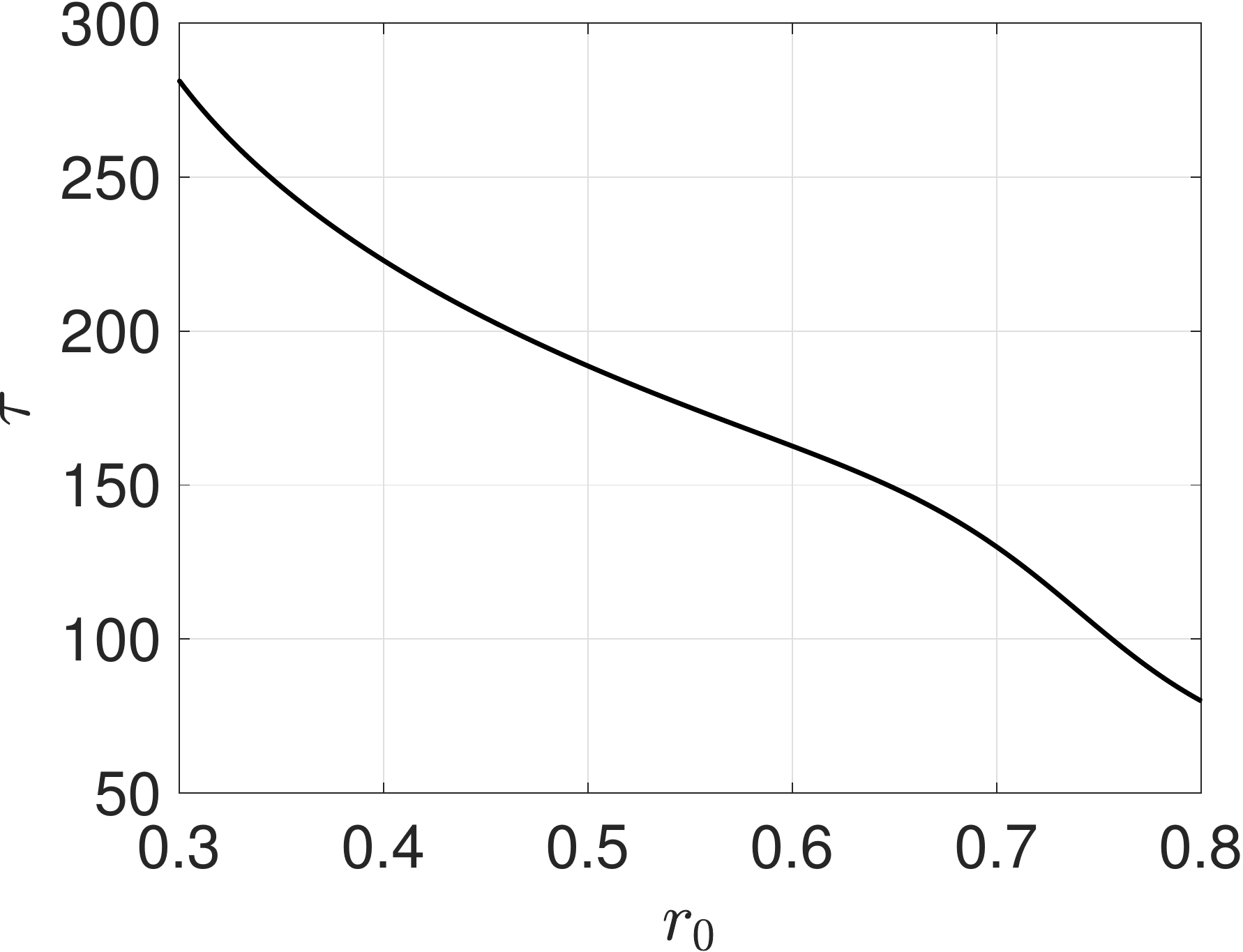}
\caption{Hopf bifurcation threshold $\tau_H$ versus $r_0$ for a
  one-spot quasi-equilibrium solution, as computed from
  \eqref{leakage:one_hopf}, for $a=10$, $D=C=1$, and
  $\eps=0.02$.}
\label{leakage_N=1_hopf_exp1}
\end{figure}

\begin{figure}[htbp]
\begin{subfigure}{0.32\textwidth}
\includegraphics[width=\textwidth,height=4.2cm]{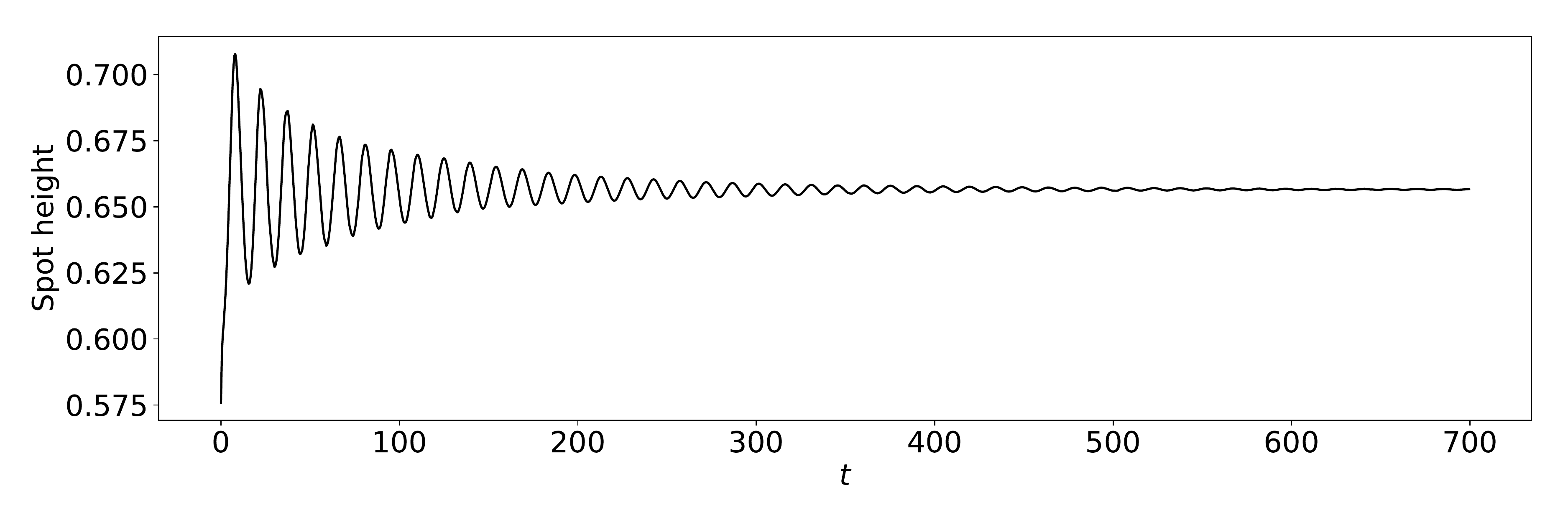}
\caption{$\tau = 162$}
\label{leakage_hopf_exp1}
\end{subfigure}
\begin{subfigure}{0.32\textwidth}
\includegraphics[width=\textwidth,height=4.2cm]{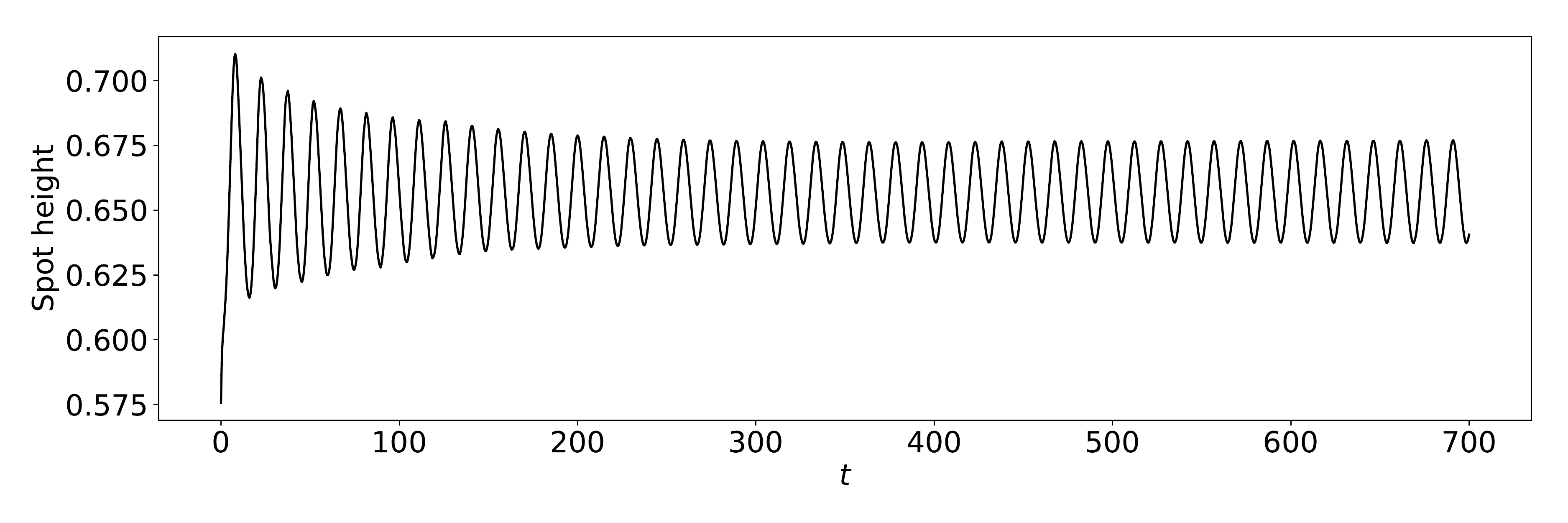}
\caption{$\tau = 168$}
\label{leakage_hopf_exp2}
\end{subfigure}
\begin{subfigure}{0.32\textwidth}
\includegraphics[width=\textwidth,height=4.2cm]{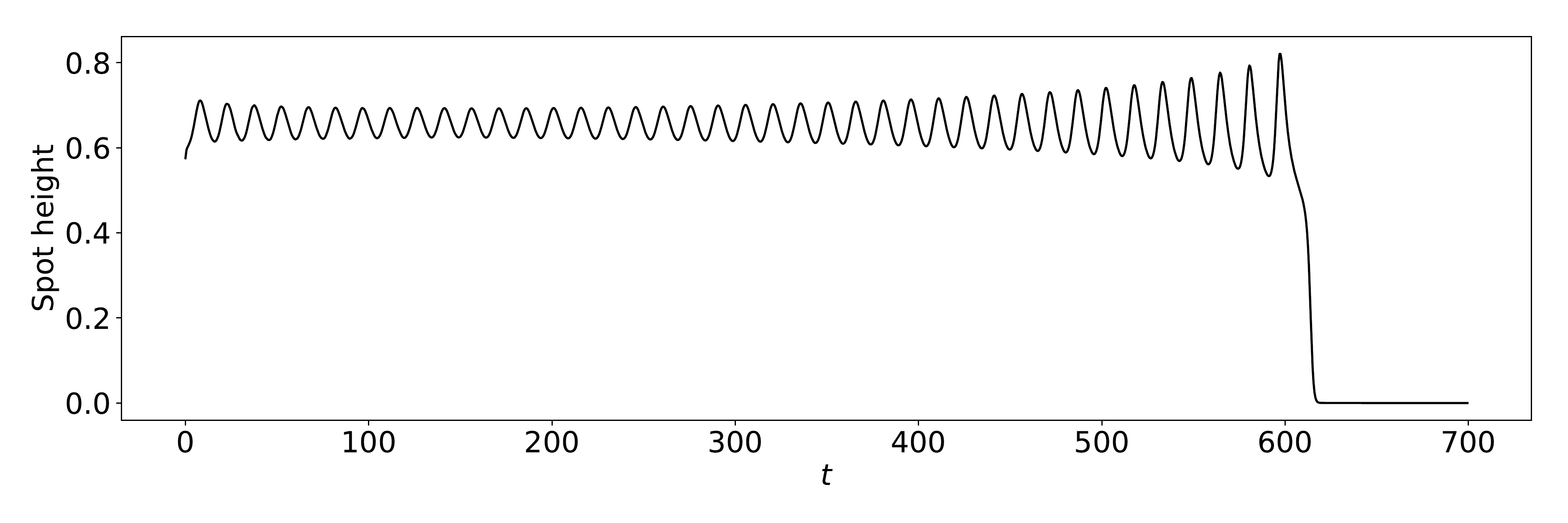}
\caption{$\tau = 170$}
\label{leakage_hopf_exp3}
\end{subfigure}
\caption{For $\eps=0.02 \,, a=10\,, C=D=1$, we choose three values of
  $\tau$ near the Hopf bifurcation threshold $\tau_H \approx 162.6$
  for a one-spot quasi-equilibrium solution centered at $r_0=0.6$.
  (a) $\tau=162 < \tau_H$: the spot amplitude has decaying
  oscillations. (b) $\tau=168 > \tau_H$: small amplitude oscillations
  indicating a supercritical Hopf bifurcation. (c) $\tau=170$: spot
  amplitude oscillations grow and trigger an oscillatory collapse of the
  spot.}
\label{leakage_hopf_exp1_2_3}
\end{figure}

Next, we demonstrate the occurrence of a Hopf bifurcation in the spot
amplitude for a one-spot quasi-equilibrium solution in the perforated
unit disk. By fixing $\eps=0.02\,, a = 10$, and $C = D = 1$, in
Fig.~\,\ref{leakage_N=1_hopf_exp1} we plot the Hopf bifurcation
threshold value $\tau = \tau_H$ on the range $r_0 \in [0.3,0.8]$, as
obtained by numerically solving for the pair $(\tau_H, \lambda_I)$ from
\begin{equation}\label{leakage:one_hopf}
  \mrm{Re}\left[ F_1(\tau_H, i \lambda_I) \right] = 0 \,, \qquad
  \mrm{Im}\left[ F_1(\tau_H, i \lambda_I)\right] = 0 \,,
\end{equation}
where $F_1$ is defined in \eqref{leakage:GCEP_ring_syn}. In
particular, when $r_0=0.6$, we compute that $\tau_H \approx 162.6$. To
confirm this threshold value, in Fig.~\ref{leakage_hopf_exp1_2_3} we
plot the spot amplitude for a one-spot quasi-equilibrium solution with
$r_0=0.6$ for $\tau=162<\tau_H$, $\tau=168>\tau_H$, and for
$\tau=170$, as computed from a full PDE simulation of
\eqref{leakage:full}. For $\tau=168$ we observe a small-scale periodic
oscillation of the spot amplitude, suggesting that the Hopf
bifurcation is supercritical. However, for the larger value
$\tau=170$, we observe that the temporal oscillation in the spot
amplitude can grow and lead to spot annihilation.

\subsubsection{Competition instability of a two-spot pattern}

\begin{figure}[htbp]
\includegraphics[width=0.45\textwidth,height=4.3cm]{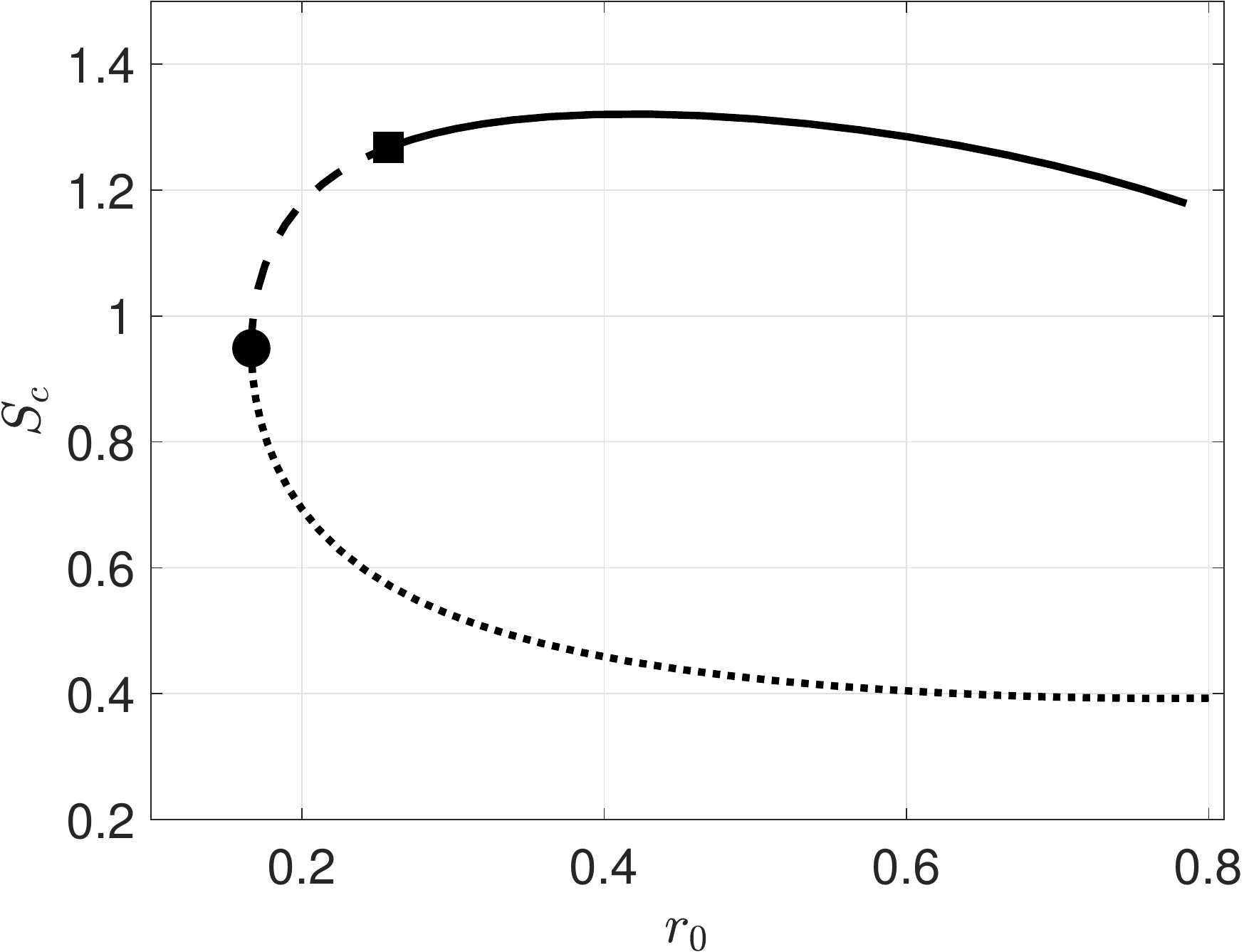}
\caption{The saddle-node bifurcation point and the competition
  threshold for a two-spot ring solution are shown as black circle and
  square markers, respectively. The dashed (solid) segment of upper
  branch corresponds where $r_0 < r_0^{(2)}$ ($r_0 > r_0^{(2)}$). Here
  $r_{0}^{(2)}$ is where there is a zero-eigenvalue crossing of the
  GCEP \eqref{leakage:GCEP_ring_reduced} for the competition mode.
  The parameters are $C=D=\tau=1$, $a=10$, and $\eps=0.02$.}
\label{leakage_N=2_exp1_Sc_vs_r0}
\end{figure}

\begin{figure}[htbp]
\begin{subfigure}[b]{0.32\textwidth}
\includegraphics[width=\textwidth,height=4.2cm]{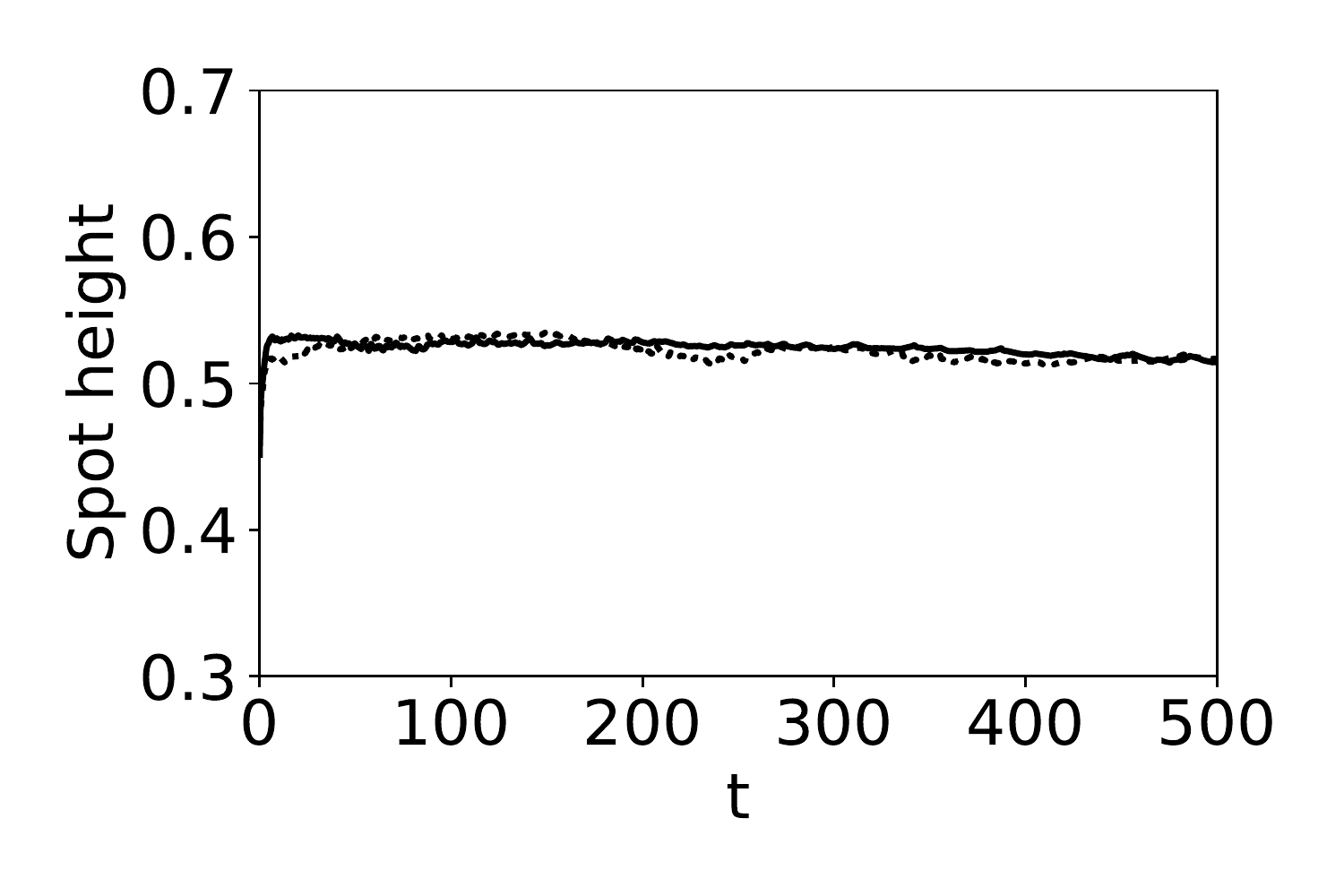}
\caption{Two spots survive.}
\end{subfigure} 
\hfill
\begin{subfigure}[b]{0.32\textwidth}
\includegraphics[width=\textwidth,height=4.2cm]{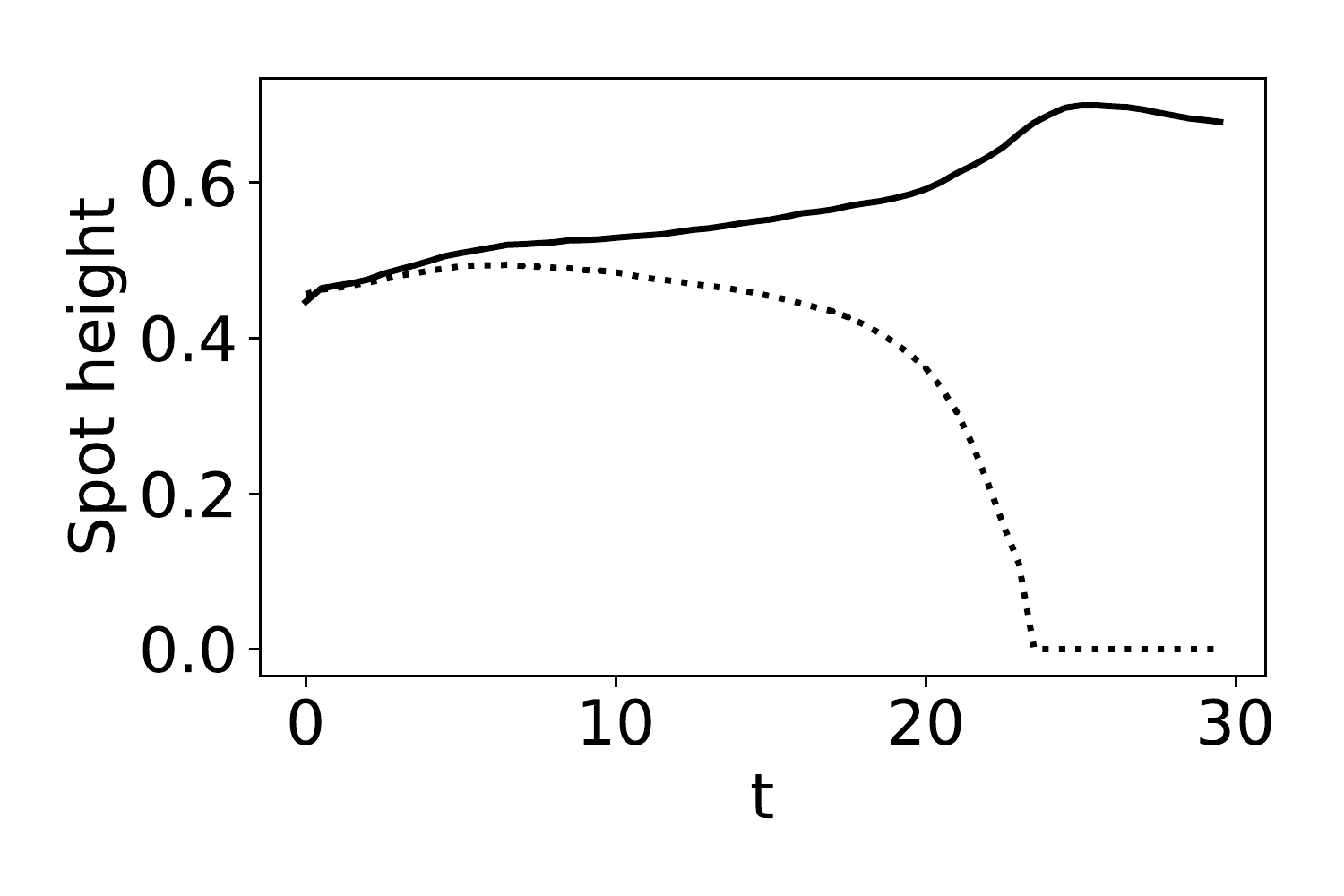}
\caption{One spot survives.}
\end{subfigure} 
\hfill
\begin{subfigure}[b]{0.32\textwidth}
\includegraphics[width=\textwidth,height=4.2cm]{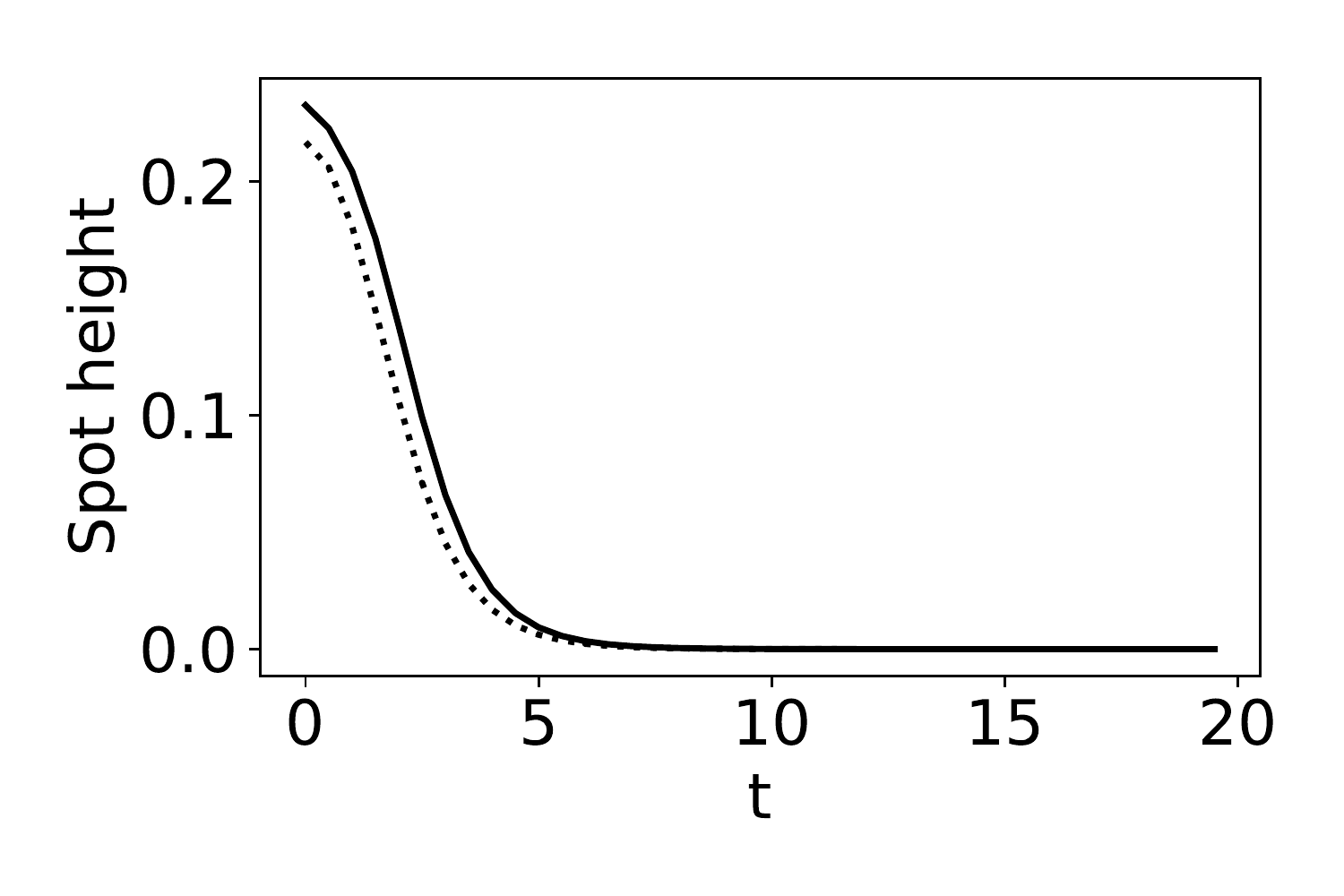}
\caption{Both spots die.}
\end{subfigure} 
\caption{The evolution of the spot amplitudes for a two-spot
  quasi-equilibrium solution, as computed from the full PDE
  \eqref{leakage:full}.  In (a) and (b), the two spots have initial
  condition on the upper branch of
  Fig.\,\ref{leakage_N=2_exp1_Sc_vs_r0}. Their initial locations are
  at $(\pm r_0,0)$, where $r_0=0.28 > r_0^{(2)}$ and
  $r_0=0.2 < r_0^{(2)}$ in (a) and (b), respectively. In (c), the
  two-spots have initial condition on the lower branch, with
  $r_0=0.29$. The parameters are $C=D=\tau=1$, $a=10$, and $\eps=0.02$.
  The PDE results are in agreement with linear stability predictions.}
\label{leakage_ring_8_9_10}
\end{figure}

Here we consider a two-spot quasi-equilibrium pattern in the
perforated unit disk, with parameters $\eps=0.02, \,C=D=\tau=1$, and
$a=10$. In Fig.\,\ref{leakage_N=2_exp1_Sc_vs_r0}, we plot the
bifurcation diagram of $S_c$ versus $r_0$ for $N=2$ spots, as computed
from \eqref{leakage:common_source_strength}, showing a saddle-node
bifurcation behavior. We calculate that the saddle-node point occurs
at $r_0=r_0^{(1)} \approx 0.1665$ and that the zero-eigenvalue
crossing for the competition mode, as computed from
\eqref{leakage:GCEP_ring_asynchronous_threshold}, occurs at
$r_0=r_0^{(2)} \approx 0.2573$.  This naturally divides the
bifurcation diagram into three segments with different stability
properties: the lower branch, the upper branch on
$r_{0}^{(1)}<r_0<r_{0}^{(2)}$, and the upper branch on
$r_0>r_{0}^{(2)}$. On the lower branch, we compute that there is a
root to $F_1=0$ to \eqref{leakage:GCEP_ring_syn} with
$\mbox{Re}(\lambda)>0$, and so the GCEP \eqref{leakage:ring_gcep} has
an unstable eigenvalue. This indicates that, on the lower branch, the
two-spot pattern is unstable to synchronous locally radially-symmetric
perturbations near the spots. Along the upper branch with
$r_0 < r_0^{(2)}$ there is a root to $F_2=0$ in
\eqref{leakage:GCEP_ring_asyn} with $\mbox{Re}(\lambda)>0$, and so
this segment of the bifurcation diagram is unstable to asynchronous
locally radially-symmetric perturbations. Finally, on the upper branch
with $r_0>r_0^{(2)}$, there is no root to
\eqref{leakage:GCEP_ring_asyn} in $\mbox{Re}(\lambda)>0$, and so this
segment is linearly stable.  These linear stability predictions are
validated in Fig.~\ref{leakage_ring_8_9_10} from full PDE simulations
of \eqref{leakage:full} with initial conditions chosen in these three
segments of the bifurcation diagram in
Fig.~\ref{leakage_N=2_exp1_Sc_vs_r0}.

\section{Pinning effects from a spatially localized feed-rate}\label{sec:het}

In this section, we analyze slow spot dynamics for the case where the
localized heterogeneity consists of a localized source of feed from
the substrate of the form
\begin{equation}\label{het:localized_feed}
  a(\v{x}) = a_0 + \eps^{-2} a_1 \Phi\big(|\v{x}-\pmb{\xi}|/\eps\big) \,, \qquad
  \Phi(r) \equiv \exp(-r^2/2)/(2\pi) \,,
\end{equation}
where $a_0>0$ and $a_1>0$ are constants. Here $\pmb{\xi}\in\Omega$ is the
location of the concentration of the feed.

\subsection{Quasi-equilibria and slow spot dynamics}
\label{sec:het_qe_slow}

We first modify our asymptotic construction of $N$-spot quasi-equilibria
given in \S \ref{sec:proto_qe_slow} to include the heterogeneous
feed rate of \eqref{het:localized_feed}. The asymptotic analysis for the
inner region near a spot is exactly the same as in \S \ref{sec:proto_qe_slow}.
Following the derivation in \S \ref{sec:proto_qe_slow}, the outer problem
for the inhibitor field, defined away from the spots, is
\begin{equation}\label{het:outer_problem}
  \lap u + \frac{a(\v{x})}{D} - \frac{2\pi}{\sqrt{D}}\sum\limits_{i=1}^N S_i \,
  \delta(\v{x}-\v{x}_i) = 0 \quad \mbox{in} \quad \Omega \,, \qquad
  \partial_n u = 0  \quad \mbox{on} \quad \partial\Omega \,,
\end{equation}
where $S_1, \ldots, S_N$ are the source strengths of the $N$ spots. By
applying the divergence theorem to \eqref{het:outer_problem} we get
\begin{equation}\label{het:sum_Sj}
  \sum\limits_{i=1}^N S_i = \frac{\int_\Omega a \,d\v{x} }{2\pi\sqrt{D}}
  \equiv p_a \,.
\end{equation}

We decompose the solution to \eqref{het:outer_problem} as
\begin{equation}\label{het:outer_solution}
  u(\v{x}) = \frac{u_2(\v{x})}{D} - \frac{2\pi}{\sqrt{D}}\sum\limits_{i=1}^N
  S_i G(\v{x} ; \v{x}_i) + \bar{u} \,,
\end{equation}
where $\ubar$ is a constant and $G$ is the Neumann Green's function of
\eqref{proto:neu_green}. Here $u_2(\v{x})$ is the unique solution to
\begin{equation}\label{het:u2}
  \lap u_2 = -a(\v{x}) + \frac{\int_\Omega a(\v{x}) d\v{x}}{|\Omega|}
  \quad \mbox{in} \quad \Omega \,, \quad \partial_n u_2 = 0 \quad \mbox{on}
  \quad \partial\Omega \,; \qquad \int_\Omega u_2 \, dx = 0 \,,
\end{equation}
which is given in terms of $G$ by
\begin{equation}\label{het:u2_closed_form}
u_2(\v{x}) = \int_\Omega a(\v{z}) \, G(\v{z} ; \v{x}) \, d\v{z} \,.
\end{equation}

As in \S \ref{sec:proto_qe_slow} we can perform an asymptotic matching
as $\v{x}\to\v{x}_j$ for $j=1,\ldots,N$ between the outer solution and
inner solutions to derive a nonlinear algebraic system for $\ubar$ and
the source strengths. Letting $\v{x}\to\v{x}_j$ in
\eqref{het:outer_solution}, we obtain that
\begin{equation}\label{het:outer_expansion}
\begin{split}
  u & \sim \frac{u_2(\v{x}_j)}{D}  + \frac{S_j}{\sqrt{D}}
  \log|\v{x}-\v{x}_j| - \frac{2\pi}{\sqrt{D}} \left(
S_j R_{j,j} + \sum\limits_{i \neq j}^N S_i G_{j,i} \right) + \ubar \\
& \quad + \left( \frac{1}{D} \, \nabla_{\v{x}} u_2(\v{x}_j) -
  \frac{2\pi}{\sqrt{D}} \left( S_j \nabla_{\v{x}} R_{j,j} +
    \sum\limits_{i \neq j}^N S_i \nabla_{\v{x}} G_{j,i} \right) \right)
\cdot (\v{x}-\v{x}_j) + {\mathcal O}(|\v{x}-\v{x}_j|^2) \,, \qquad
j=1,\ldots, N \,,
\end{split}
\end{equation}
where $R_{j,j} \equiv R(\v{x}_j; \v{x}_j)$ and $G_{j,i} \equiv G(\v{x}_j;\v{x}_i)$.

Upon matching \eqref{het:outer_expansion} with \eqref{proto:matching_inner}
for the $\bigo(1)$ terms, we write the resulting equations in matrix
form as
\begin{subequations}\label{het:source_init}
\begin{equation}\label{het:source_strength2}
  \v{s} + 2 \pi \nu \, \Gmat \v{s} + \nu \chivec = \nu \left(
    \frac{1}{\sqrt{D}}\,\v{u_2}  + \ubar \sqrt{D}  \, \v{e} \right)  \,,
  \qquad \v{e}^T\v{s}=p_a \,,
\end{equation}
where $\Gmat$ is the Neumann Green's matrix, and where we have defined 
\begin{equation}\label{het:source_def}
  \v{s} \equiv (S_1, \ldots, S_N)^T \,, \quad \chivec \equiv
  (\chi(S_1), \ldots, \chi(S_N))^T \,, \quad
  \v{e} \equiv (1,\ldots,1)^T \in \mathbb{R}^N \,, \quad
  \v{u}_2 \equiv (u_2(\v{x}_1), \ldots, u_2(\v{x}_N))^T \,.
\end{equation}
\end{subequations}

Upon left-multiplying \eqref{het:source_strength2} by $\v{e}^T$, we can
isolate $\ubar$ as
\begin{equation}\label{het:ubar}
  \ubar = \frac{p_a + 2 \pi \nu \, \v{e}^T \Gmat \v{s} + \nu \,
    \v{e}^T \chivec}{\nu N \sqrt{D}} - \frac{\v{e}^T \v{u}_2}{N D} \,.
\end{equation}
By using \eqref{het:ubar} to eliminate $\ubar$ in
\eqref{het:source_strength2}, we obtain a nonlinear
algebraic system for the vector of source strengths $\v{s}$,
\begin{equation}\label{het:source_system}
  \v{s} + 2\pi\nu (I - \Emat) \, \Gmat\v{s} + \nu(I - \Emat)\chivec =
  \frac{\nu}{\sqrt{D}}\,(I - \Emat)\v{u}_2 + \frac{p_a}{N} \v{e} \,, \qquad
  \mbox{where} \qquad \Emat = \frac{1}{N} \v{e}\v{e}^T\,,
\end{equation}
and $p_a$ is defined in \eqref{het:sum_Sj}.

To derive the DAE system for slow spot dynamics we must match
\eqref{proto:matching_inner} with \eqref{het:outer_expansion} for the
$\bigo(\eps)$ gradient terms. This matching yields the far-field
behavior for the inner correction term $U_{j1}$, as defined in
\eqref{proto:inner_expansion}, given by
\begin{equation}
  U_{j1} \sim \left( \frac{1}{\sqrt{D}} \, \nabla_\v{x} u_2(\v{x}_j) -
    \pmb{\beta}_j \right) \cdot \, \v{y} \, \quad \mbox{as} \quad
  |\v{y}|\to \infty \,,
\end{equation}
where $\v{y} = \eps^{-1} (\v{x}-\v{x}_j)$ and $\pmb{\beta}_j$ is
defined in \eqref{proto:betaj}.  Following the derivation in
\S \ref{sec:proto_qe_slow}, we conclude that the DAE system for slow
spot dynamics is given by
\begin{equation}\label{het:DAE}
  \frac{d\v{x}_j}{d\sigma} = \gamma(S_j) \left( \frac{1}{\sqrt{D}} \,
    \nabla_{\v{x}} u_2(\v{x}_j) - \pmb{\beta}_j  \right) \,, \qquad
    j=1,\ldots,N \,,
\end{equation}
where $\sigma=\eps^2 t$ and $\v{s}\equiv (S_1,\ldots,S_N)^T$ satisfies
the nonlinear algebraic system \eqref{het:source_system}. Here
$\gamma(S_j)$ is defined in \eqref{proto:DAE}.

As $\eps \to 0$, we can approximate, in the sense of distributions,
the heterogeneous feed rate in \eqref{het:localized_feed} as
\begin{equation}\label{het:aloc}
a(\v{x}) \to a_0  + a_1 \, \delta(\v{x} - \pmb{\xi}) \,.
\end{equation}
In this way, $u_2$ in \eqref{het:u2_closed_form} can be calculated
explicitly, by using Green's reciprocity and
$\int_{\Omega} G(\v{z};\v{x})\,d\v{x}=0$, as
\begin{equation}\label{het:u2_localized}
u_2(\v{x}) = \int_\Omega a(\v{z})G(\v{z};\v{x}) d\v{x} = a_1 G(\v{x};\bm{\xi})\,.
\end{equation}

\subsubsection{One-spot dynamics in the unit disk}\label{sec:het_one_spot}

For a one-spot solution, we use \eqref{het:aloc} in \eqref{het:sum_Sj}
to calculate $S_1$. Then, by using \eqref{het:u2_localized} in
\eqref{het:DAE}, together with the explicit expressions
\eqref{ring:gcomplex_der} for the gradients of the Neumann
Green's function for the unit disk, we obtain from \eqref{het:DAE}
that the slow dynamics of a one-spot quasi-equilibrium solution is
\begin{subequations}\label{het:one_spot}
\begin{equation}
  \frac{d\v{x}_1}{d\sigma} = -\frac{a_0\gamma(S_1)}{2\pi\sqrt{D}} \,
  \mathcal{H}(\v{x}_1) \,, \qquad \mbox{with} \qquad S_1 = \frac{a_0\pi + a_1}
  {2\pi \sqrt{D}} \,,
\end{equation}
where $\sigma=\eps^2 t$ and $\mathcal{H}$ is defined by
\begin{equation}\label{het:DAE_one_spot0}
  \mathcal{H}(\v{x}_1) \equiv \frac{a_1}{a_0}\left[\frac{\v{x}_1 -
      \bm{\xi}}{|\v{x}_1 - \bm{\xi}|^2} + \frac{\v{x}_1 |\bm{\xi}|^2 -
      \bm{\xi}}{|\v{x}_1|^2 |\bm{\xi}|^2 - 2\v{x}_1 \cdot\bm{\xi} + 1}
  \right] + \frac{\v{x}_1}{1 - |\v{x}_1|^2}\left[ \frac{a_1}{a_0}+
    \pi(2 - |\v{x}_1|^2) \right] \,.
\end{equation}
\end{subequations}

Without loss of generality we let $\pmb{\xi}=(\xi,0)$ with
$0<\xi<1$. By symmetry, any equilibrium to \eqref{het:one_spot} lies on
the line that connects the origin and $\bm{\xi}$. As such, we let
$\v{x}_1 = (r_0,0)$ and obtain from \eqref{het:one_spot} that
$r_0=r_0(\sigma)$ satisfies the scalar ODE
\begin{equation}\label{het:DAE_one_spot}
  \frac{dr_0}{d\sigma} = -\frac{a_0\gamma(S_1)}{2\pi\sqrt{D}} \, \mc{K}(r_0) \,,
  \qquad \mbox{where} \qquad \mc{K}(r_0) \equiv
  \frac{a_1}{a_0}\left(\frac{1}{r_0 - \xi} + \frac{r_0-\xi}{(1-r_0^2)(1-\xi r_0)
    }\right) + \frac{\pi r_0(2-r_0^2)}{1-r_0^2}\,.
\end{equation}
Since $\gamma(S_1)>0$ and $\mc{K}(r_0)>0$ on $\xi<r_0<1$, it follows
that ${dr_0/d\sigma}<0$ on the range $\xi<r_0<1$.

\begin{figure}[htbp]
\begin{subfigure}[b]{0.45\textwidth}
\includegraphics[width=\textwidth,height=4.3cm]{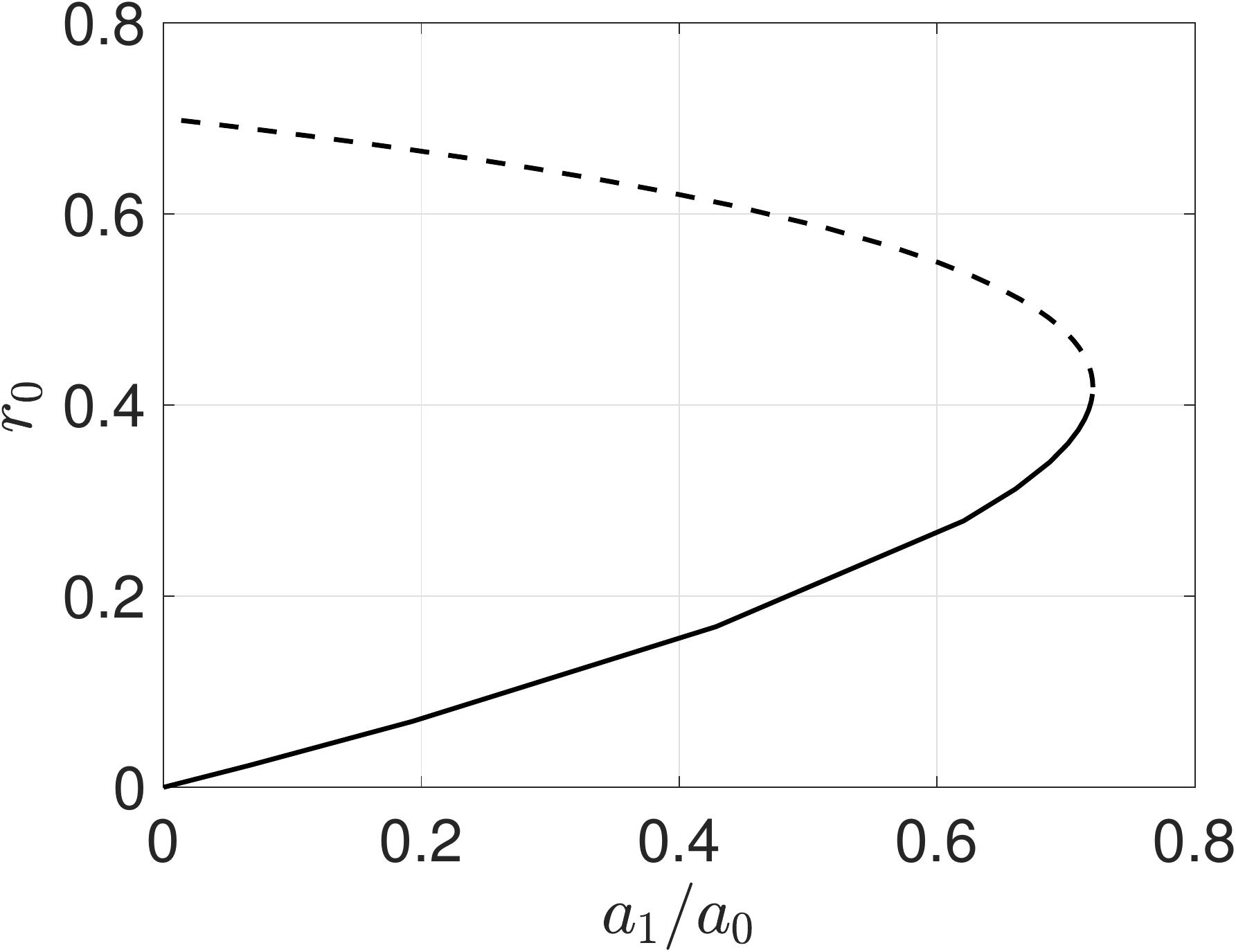}
\caption{Bifurcation diagram}
\label{het_one_spot_r0_vs_k}
\end{subfigure}
\hfill
\begin{subfigure}[b]{0.45\textwidth}
\includegraphics[width=\textwidth,height=4.3cm]{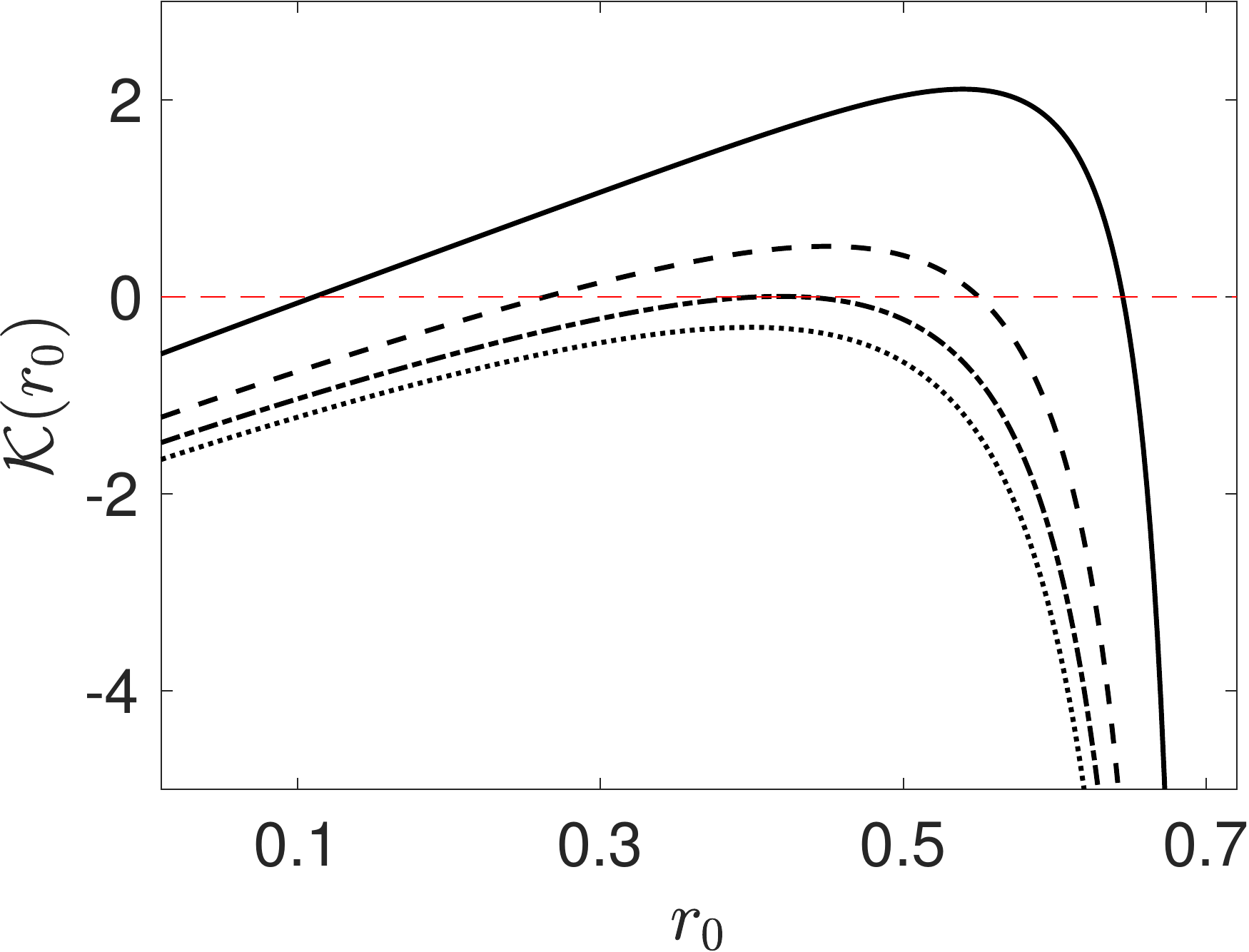} 
\caption{$\mc{K}$}
\label{het_K}
\end{subfigure}
\caption{The concentration point for the feed rate is
  $\bm{\xi} = (0.7,0)$. Left panel: The bifurcation diagram of the
  equilibria $r_{0e}$ of $\mc{K}(r_0)=0$, as defined in
  \eqref{het:DAE_one_spot}, versus ${a_1/a_0}$. A saddle-node
  bifurcation occurs at ${a_1/a_0} \approx 0.7208$. Right panel: From
  top to bottom, plots of $\mc{K}(r_0)$ for
  $a_1/a_0 = 0.3 \,, 0.6 \,, 0.72$ and $0.8$, respectively.}\label{one:equil}
\end{figure}

As such, any equilibrium $r_{0e}$ for \eqref{het:DAE_one_spot},
satisfying $\mc{K}(r_{0e})=0$, must be on the range $0<r_0<\xi$. The
effect of the relative magnitude of the localized feed to the
background feed appears in \eqref{het:DAE_one_spot} in the form of
their ratio $a_1/a_0$. Taking this ratio as a bifurcation parameter,
in Fig.~\ref{het_one_spot_r0_vs_k} we plot the bifurcation diagram of
the roots to $\mc{K}(r_0)=0$ for $\xi=0.7$.  We observe that there are
two equilibria $r_{0e}^{(1)} < r_{0e}^{(2)}$ provided that
$a_1/a_0 < 0.7208$, and none if ${a_1/a_0}>0.7208$. Since
$\mc{K}^{\prime}(r_{0e}^{(1)})>0$, we conclude that $r_{0e}^{1}$ is a
stable equilibrium point of \eqref{het:DAE_one_spot}, while
$r_{0e}^{2}$ is an unstable equilibrium.  To further demonstrate the
saddle-node bifurcation value of ${a_1/a_0}$, in Fig.\,\ref{het_K} we
plot $\mc{K}(r_0)$ on $0<r_0<\xi$ for the four values
${a_1/a_0} = 0.3 \,, 0.6 \,, 0.72$ and $0.8$. For ${a_1/a_0}<0.7208$,
we have that ${d r_0/d\sigma} > 0$ on the range $r_{0e}^{(2)}<r_0<\xi$
and ${d r_0/d\sigma}<0$ for $\xi<r_0<1$. Moreover, since
${dr_0/d\sigma} = \mc{O}\left[1/(r_0-\xi)\right]$ as $r_0 \to \xi$,
this implies that a spot initially located at some $r_0(0)$ with
$r_0(0) > r_{0e}^{(2)}$ will get pinned at the concentration point
$\xi$ of the feed rate at a finite time. Moreover, if ${a_1/a_0}>0.7208$,
this finite-time pinning will occur for {\em any} initial point
$r_0(0)$ in $0<r_0(0)<1$. 

We summarize the fate of a one-spot quasi-equilibrium solution with
slow dynamics \eqref{het:DAE_one_spot} as follows: The spot drifts to
the equilibrium $r_0=r_{0e}^{(1)}$ for any $r_0(0) < r_{0e}^{(2)}$
when ${a_1/a_0} < 0.7208$. The spot gets pinned at $r_0 = \xi$ if
$r_0(0) > r_{0e}^{(2)}$ and ${a_1/a_0}<0.7208$. The spot gets pinned
at $r_0=\xi$ for any $r_0(0)$ in $0<r_0(0)<1$ if ${a_1/a_0} > 0.7208$.
We emphasize that this saddle node threshold value for ${a_1/a_0}$ is
independent of the inhibitor diffusivity $D$.  Although our asymptotic
analysis, leading to the ODE \eqref{het:one_spot}, is only valid when
the spot is well-separated from the concentration point the feed rate,
i.e.~when $|\v{x}-\bm{\xi}|\gg {\mathcal O}(\eps)$, the prediction of
finite-time pinning phenomena provides a motivation for the analysis
in \S \ref{sec:pinned} of constructing a new type of spot solution
where the spot is pinned at the point of concentration of the feed rate.

\begin{figure}[htbp]
\begin{subfigure}[b]{0.32\textwidth}
\includegraphics[width=\textwidth,height=4.2cm]{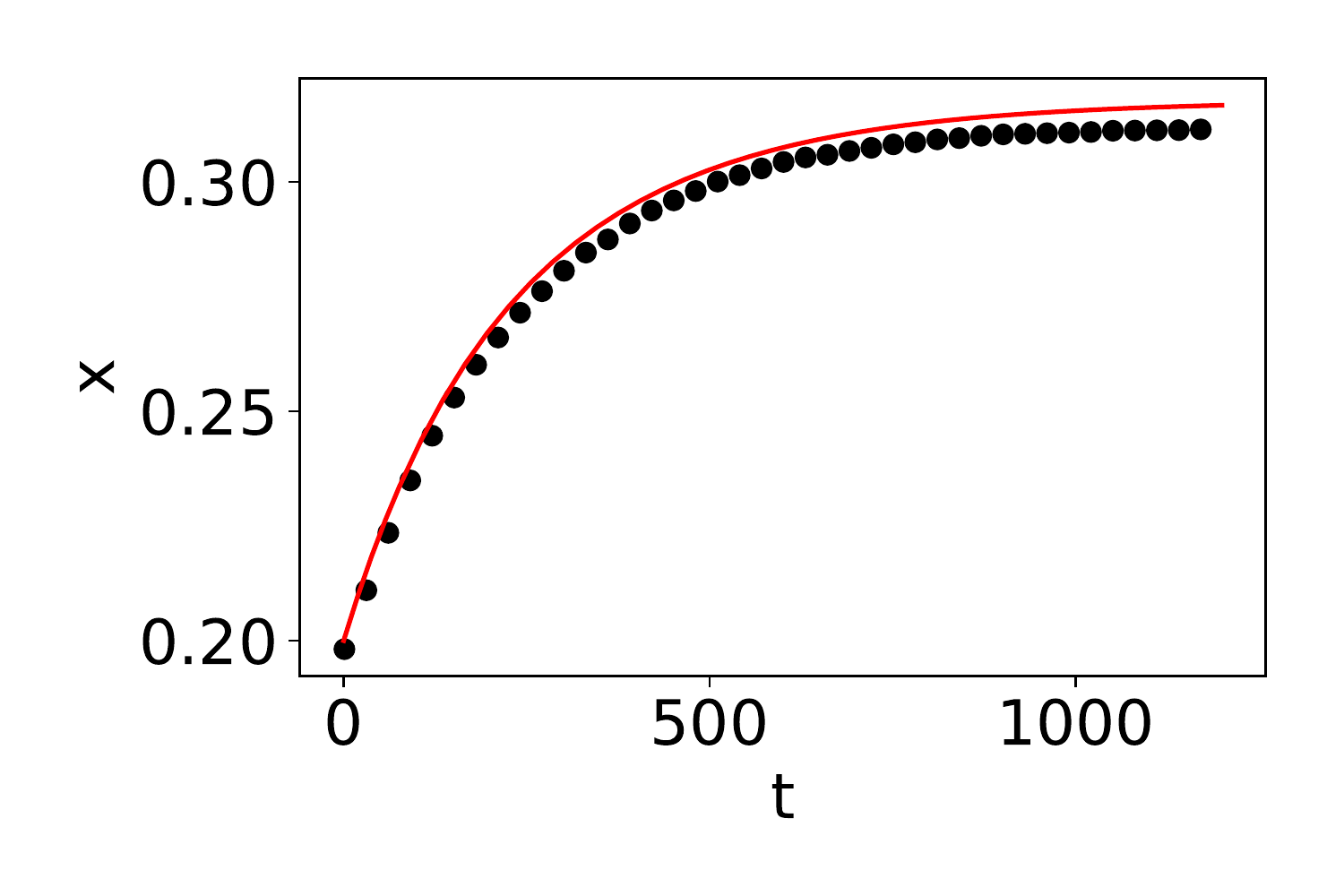}
\caption{Drift to $r_{0e}^{(1)}$}
\label{het_one_spot_exp1}
\end{subfigure}
\hfill
\begin{subfigure}[b]{0.32\textwidth}
\includegraphics[width=\textwidth,height=4.2cm]{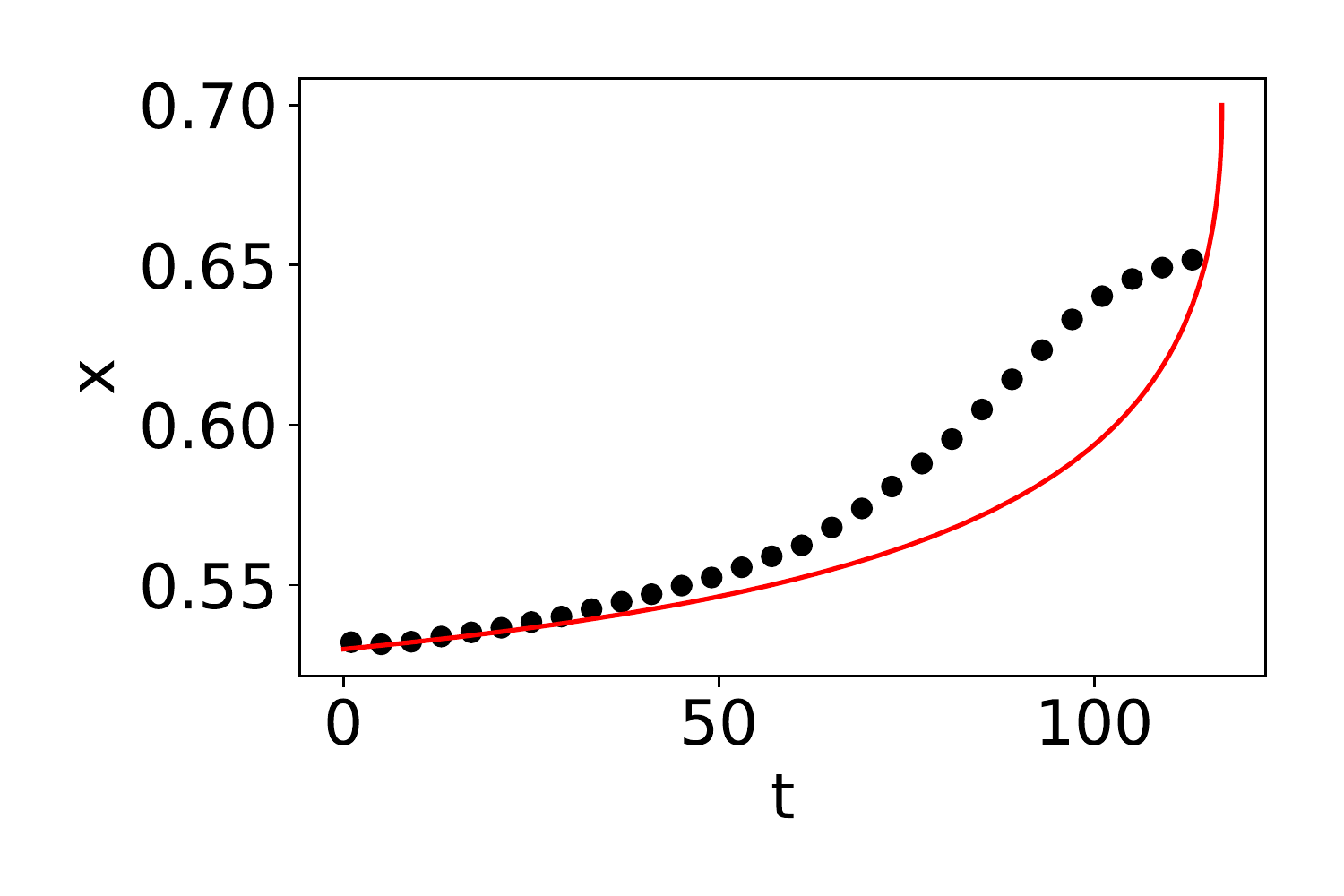}
\caption{Pinned at $\xi$}
\label{het_one_spot_exp2}
\end{subfigure}
\hfill
\begin{subfigure}[b]{0.32\textwidth}
\includegraphics[width=\textwidth,height=4.2cm]{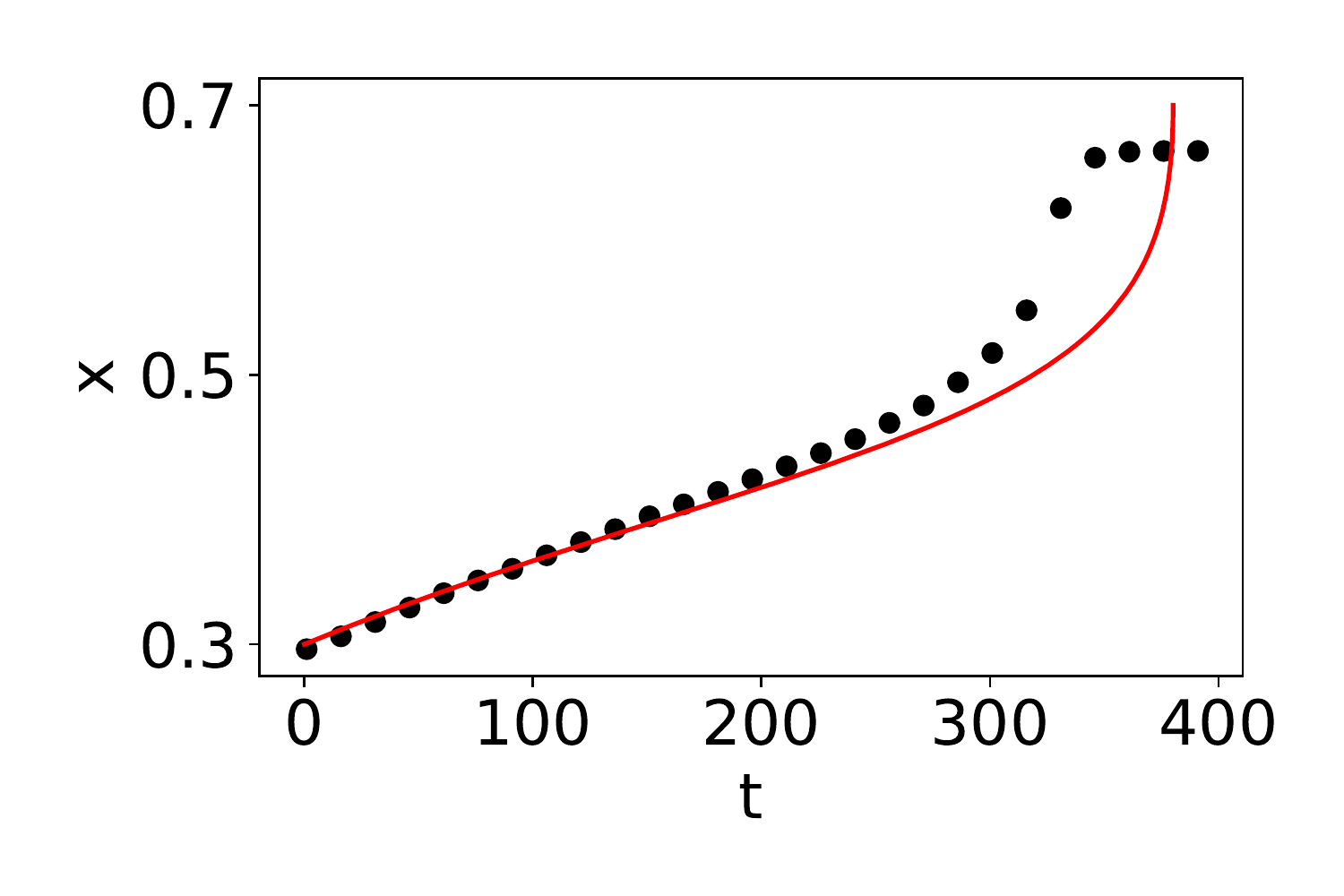}
\caption{Pinned at $\xi$}
\label{het_one_spot_exp3}
\end{subfigure}
\caption{The $x$-coordinates of the spot trajectory computed from the
  full PDE \eqref{proto:model} with \eqref{het:localized_feed} (black
  dots) and the scalar ODE \eqref{het:DAE_one_spot} with $\xi=0.7$. Left panel:
  $a_0=6$, $a_1=4$, and $r_0(0) = 0.2$. Middle panel:
  $a_0=6$, $a_1=4$, and $r_0(0) = 0.53$. Right panel: 
  $a_0=6$, $a_1=5$, and $r_0(0) = 0.3$.}
\end{figure}

To illustrate these results we compare predictions based on the scalar
ODE \eqref{het:DAE_one_spot} with full PDE simulations of
\eqref{proto:model} with the feed rate \eqref{het:localized_feed} in
the unit disk with $D=\tau=1$, and $\eps=0.03$. We set $\xi=0.7$ and
with the choice $a_0=6$ and $a_1=4$, for which ${a_1/a_0} < 0.7208$,
the two equilibrium locations are $r_{0e}^{(1)} \approx 0.3178$ and
$r_{0e}^{(2)} \approx 0.5090$. In Fig.~\ref{het_one_spot_exp1}, where
we compare results from full PDE simulations and the ODE
\eqref{het:DAE_one_spot}, we verify that a spot initially located at
$r_0(0) = 0.2 < r_{0e}^{(1)}$ slowly drifts to $r_{0e}^{(1)}$. In
contrast, for the same $a_0$ and $a_1$, but with initial value
$r_0(0) = 0.53 > r_{0e}^{(2)}$, we observe from
Fig.~\ref{het_one_spot_exp2} that the spot approaches $\xi = 0.7$. The
full PDE and ODE results are found to agree well until the spot is
near $\xi=0.7$. We remark that the velocity field in the ODE becomes
singular as $r_0 \to \xi$ owing to the Dirac delta function
approximation of the localized feed rate. Finally, if we increase the
relative strength of the concentration of the feed rate so that
$a_0=6$ and $a_1=5$, for which ${a_1/a_0}>0.7208$, we confirm from
Fig.~\ref{het_one_spot_exp3} that with $r_0(0)=0.3$ the spot gets
pinned at $\xi$ owing to the absence of any equilibrium for this ratio
${a_1/a_0}$.

\subsubsection{Two-spot dynamics in the unit disk}

Next, we consider a ring pattern of $N$-spots in the unit disk with localized
feed rate concentrated at the origin, so that $\pmb{\xi}=\v{0}$. By using
$\nabla_{\v{x}}u_2=a_1\nabla_{\v{x}}G(\v{x};\v{0})$, together with
\eqref{ring:gcomplex_der} and \eqref{proto:betaj_ring} for
$\nabla_{\v{x}}G(\v{x};\v{0})$ and $\beta_j$, respectively, we obtain from
\eqref{het:DAE} that the slow dynamics of the ring radius $r_0$ satisfies
the scalar ODE
\begin{equation}\label{het:ode_N_spots}
  \frac{dr_0}{d\sigma} = - \frac{a_0\gamma(S_c)}{2\pi r_0 \sqrt{D}} \,
  \mc{D}(r_0) \,, \qquad \mbox{where} \qquad
  \mc{D}(r_0) \equiv \frac{N+1}{2N} \left[ \frac{a_1}{a_0} - \pi
    \left( \frac{N-1}{N+1} \right)\right]  + \pi r_0^2 + \left(\pi +
    \frac{a_1}{a_0}\right) \frac{r_0^{2N}}{1-r_0^{2N}} \,.
\end{equation}
From \eqref{het:sum_Sj} and \eqref{het:aloc}, the common spot source
strength is $S_c={(a_0\pi + a_1)/[2\pi N\sqrt{D}]}$.
  
\begin{figure}[htbp]
  \includegraphics[width=0.45\textwidth,height=4.3cm]{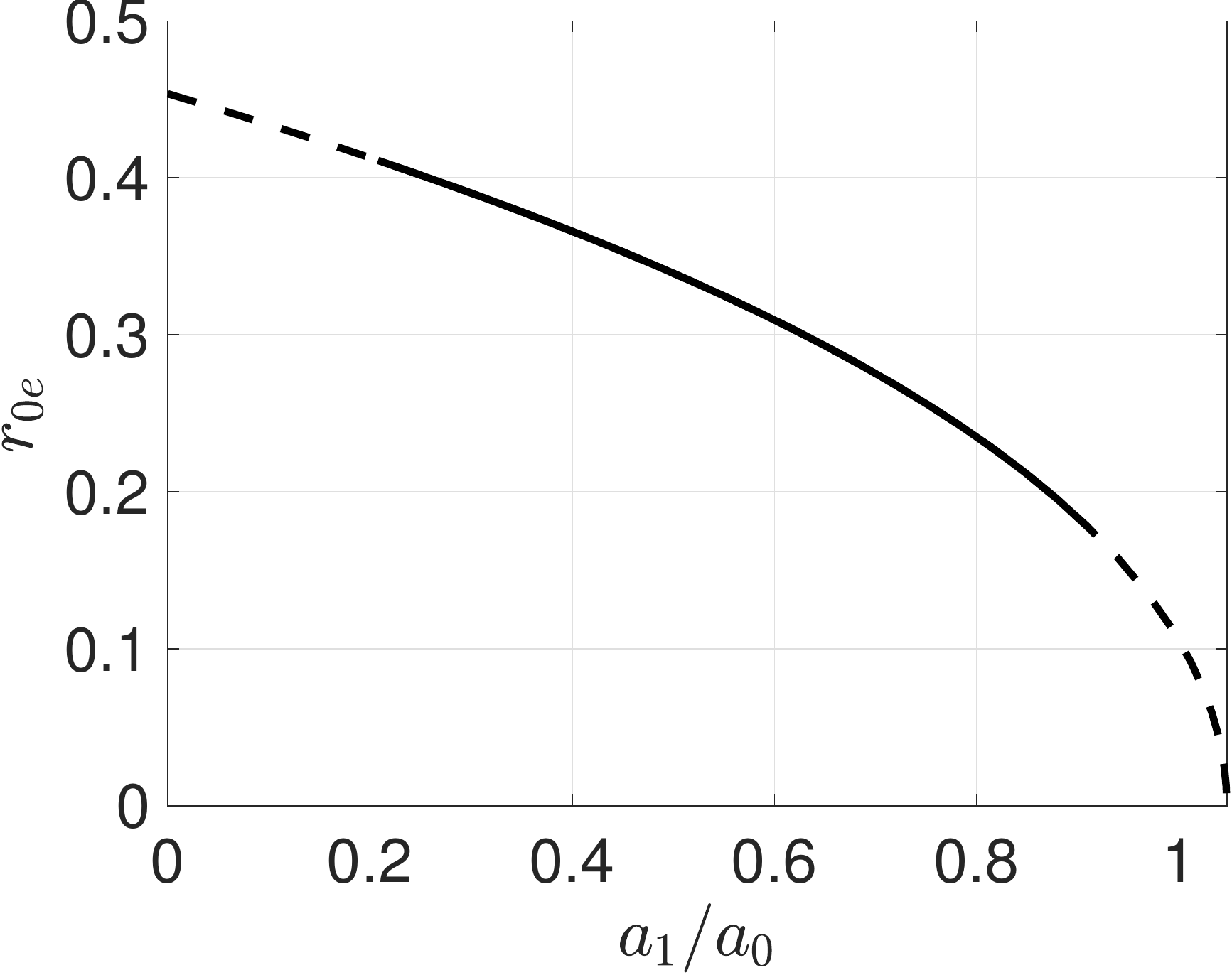}
  \caption{Bifurcation diagram of equilibrium ring radius versus the
    ratio ${a_1/a_0}$, as computed from setting $\mc{D}(r_0)=0$ in
    \eqref{het:ode_N_spots}, for a two-spot pattern in the unit disk
    with feed rate concentration at the origin. Fixing $a_0=4.3$ and
    $D=\tau=1$, on the range  $0.1777<r_{0e}<0.4111$ (solid portion) the
    concentration of feed at the origin renders the two-spot equilibrium
    solution linearly stable to a competition instability. The dashed
    portions are where the solution is unstable to competition.}
\label{het_two_spots_r0e_vs_k}
\end{figure}

The equilibrium ring radius $r_{0e}$ is a root to $\mc{D}(r_0)=0$. Since
$\mc{D}^{\prime}(r_0) > 0$ and $\mc{D}\to +\infty$ as $r_0\to 1^{-}$, the
ODE \eqref{het:ode_N_spots} must have an equilibrium point in $0<r_0<1$
when
\begin{equation}\label{het:ode_N_spots_eq_criterion}
  \mc{D}(0) =  \frac{N+1}{2N} \left[ \frac{a_1}{a_0} - \pi
    \left( \frac{N-1}{N+1} \right)\right] < 0 \,, \qquad
  \mbox{which implies} \qquad \frac{a_1}{a_0} < \pi
  \left( \frac{N-1}{N+1} \right) \,.
\end{equation}
For $N=2$, in Fig.~\ref{het_two_spots_r0e_vs_k} we plot the
bifurcation diagram of the equilibrium ring radius $r_{0e}$ versus the
ratio ${a_1/a_0}$. On the range
$0\leq {a_1/a_0} <{\pi/3}\approx 1.047$, we observe that there is a
unique equilibrium radius. We note that $r_{0e}\to 0$ when
${a_1/a_0} \to \pi/3 \approx 1.0472$, which is the upper bound for
${a_1/a_0}$ in \eqref{het:ode_N_spots_eq_criterion} for $N=2$.

Next, we fix $a_0=4.3$, and $D=\tau=1$. The analysis of competition
instabilities and the derivation of the GCEP for two-spot equilibria
with feed concentration at the origin is exactly the same as in \S
\ref{sec:basic_ring} provided that we use
$S_c={(a_0\pi + a_1)/[4\pi \sqrt{D}]}$ with $D=1$ for the common
source spot strength. This leads to the root finding criterion
\eqref{proto:GCEP_ring} with $j=N=2$ for the GCEP
\eqref{proto:GCEP_ring0} and the zero-eigenvalue crossing condition
\eqref{proto:GCEP_competition_threshold} with $j=N=2$.  When $a_1=0$
(no feed concentration), Fig.~\ref{GCEP_two_spots_sym} showed that
there is a competition instability for a steady-state two-spot ring
pattern if $a_0<4.45$. From a numerical computation of the winding
number \eqref{proto:wind} and the zero-eigenvalue crossing condition
\eqref{proto:GCEP_competition_threshold} with $j=N=2$, we obtain that
the dashed portions in the bifurcation diagram in
Fig.~\ref{het_two_spots_r0e_vs_k} for the equilibrium ring radius
correspond to where the two-spot equilibrium solution is unstable to a
competition instability. As expected, since $a_0=4.3<4.45$, we observe
that the two-spot equilibrium is unstable if $a_1$ is sufficiently
small.  Moreover, the two-spot equilibrium is unstable near
${a_1/a_0}\approx {\pi/3}$ since the spots become too closely spaced
(i.e. $r_{0e}$ is too small). However, the key new qualitative feature
of Fig.~\ref{het_two_spots_r0e_vs_k} is that there is a range of
${a_1/a_0}$ where a concentration of feed at the origin {\em
  stabilizes} a two-spot equilibrium solution, which without the
concentration of feed would be unstable to a competition stability.

\begin{figure}[htbp]
\begin{subfigure}[b]{0.32\textwidth}
\includegraphics[width=\textwidth,height=4.2cm]{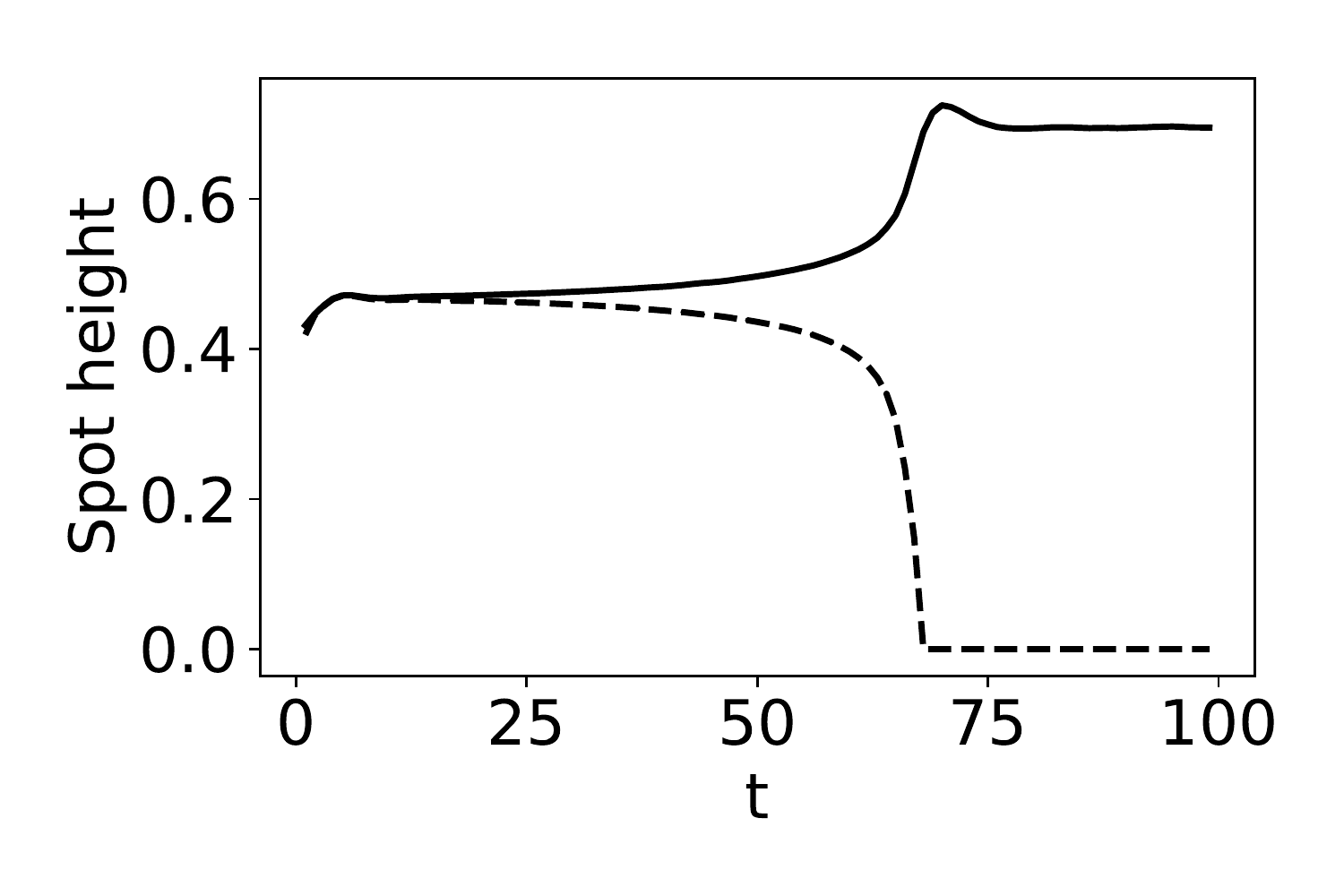}
\caption{$r_{0e} = 0.4330\,,$ $a_1/a_0 = 0.1040$.}
\label{het_two_spots_exp1}
\end{subfigure}
\hfill
\begin{subfigure}[b]{0.32\textwidth}
\includegraphics[width=\textwidth,height=4.2cm]{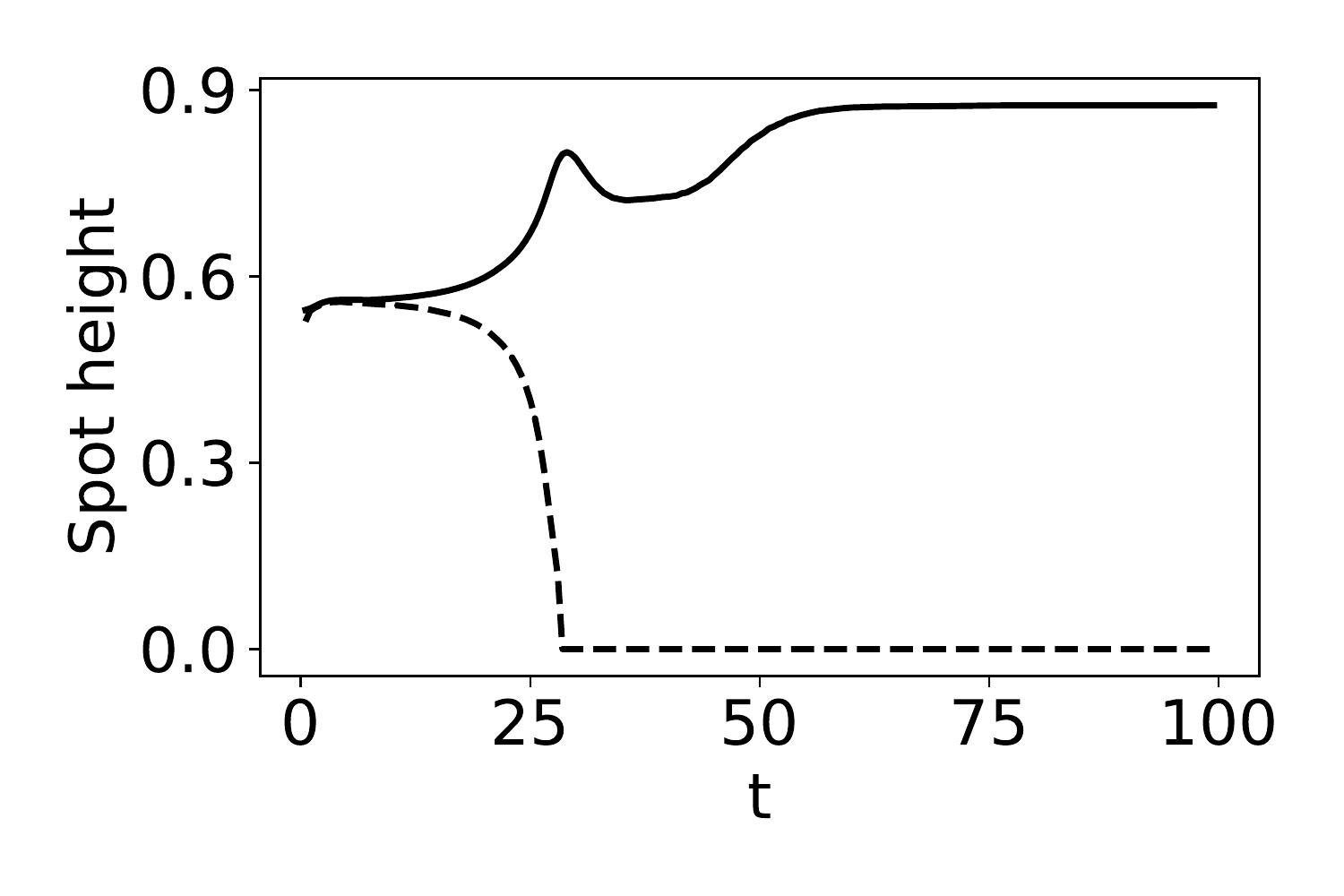}
\caption{$r_{0e} = 0.1127\,,$ $a_1/a_0 = 0.9932$.}
\label{het_two_spots_exp2}
\end{subfigure}
\hfill
\begin{subfigure}[b]{0.32\textwidth}
\includegraphics[width=\textwidth,height=4.2cm]{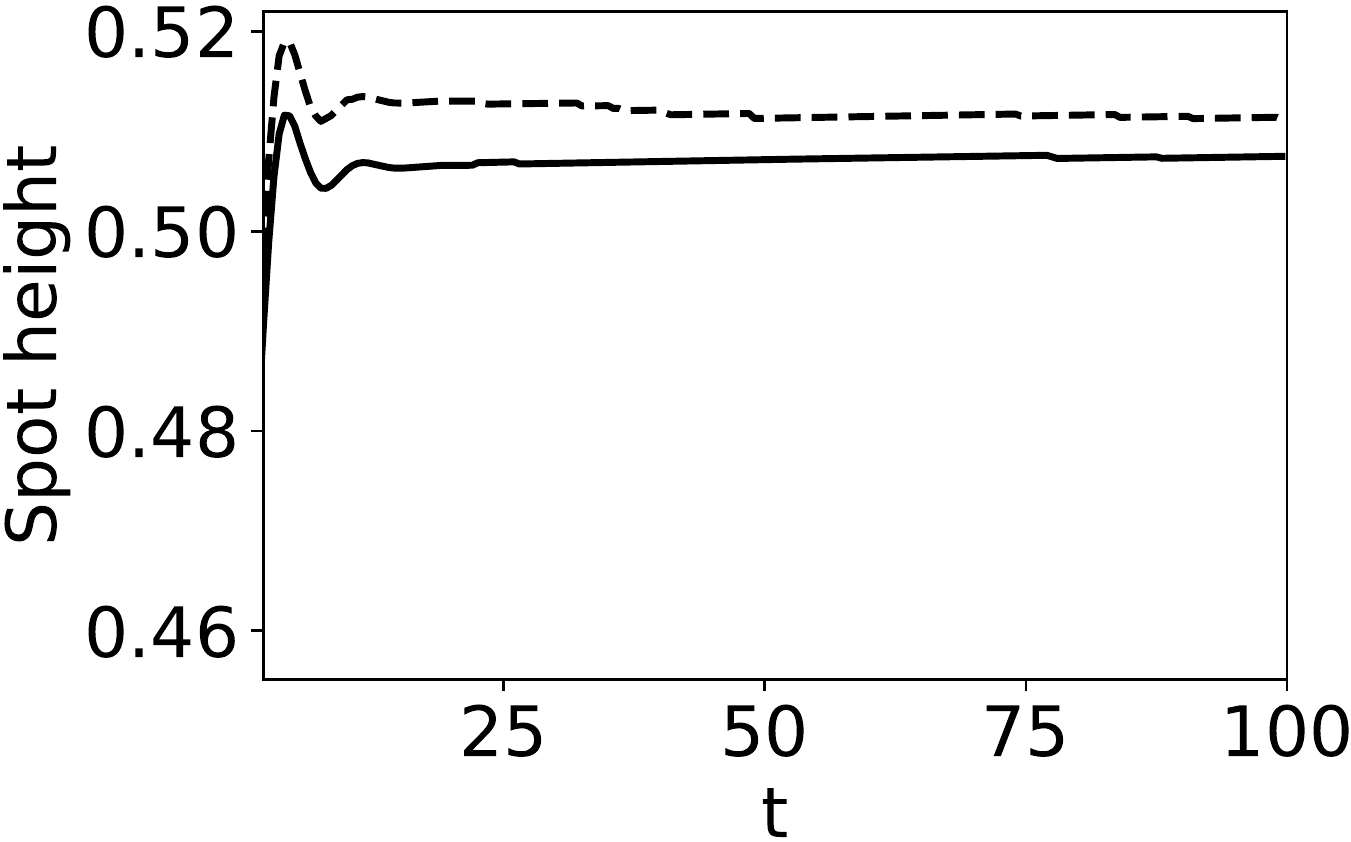}
\caption{$r_{0e} = 0.3345\,,$ $a_1/a_0 = 0.5166$.}
\label{het_two_spots_exp3}
\end{subfigure}
\caption{Full PDE simulations of \eqref{proto:model} with
  \eqref{het:localized_feed} of the spot amplitudes for three ratios
  of ${a_1/a_0}$. The initial condition for the PDE is a two-spot
  equilibrium ring pattern with spots located at $(\pm r_{0e},0)$.
  Parameters are $\eps=0.02$, $D=\tau=1$, and $a_0=4.3$. The
  competition instability occurring in (a) and (b), leads to spot
  annihilation. In (c), the two-spot equilibrium is linearly stable.}
\end{figure}

To illustrate this linear stability prediction for $a_0=4.3$ and
$D=\tau=1$, we take $\eps=0.02$ and perform full PDE simulations of
\eqref{proto:model} with \eqref{het:localized_feed} for a two-spot
equilibrium ring pattern with spots located at $(\pm r_{0e},0)$. In
Fig.~\ref{het_two_spots_exp1} and Fig.~\ref{het_two_spots_exp2} we
show full PDE results for the amplitudes of the spots for the ratios
${a_1/a_0} = 0.1040$ and ${a_1/a_0} = 0.9932$, respectively, which lie
on the unstable dashed portions in the bifurcation diagram of
Fig.~\ref{het_two_spots_r0e_vs_k}. For both values of ${a_1/a_0}$, we
confirm from these figures that a competition instability occurs,
which triggers the annihilation of a spot. In contrast, for
${a_1/a_0} = 0.5166$, Fig.~\ref{het_two_spots_r0e_vs_k} predicts that
the two-spot equilibrium solution, with spots centered at
$(\pm 0.3345,0)$, will be linearly stable to a competition
instability. This prediction is confirmed from the numerical PDE
results shown in Fig.~\ref{het_two_spots_exp3}.

\begin{figure}[htbp]
\begin{subfigure}[b]{0.2\textwidth}
\includegraphics[width=\textwidth]{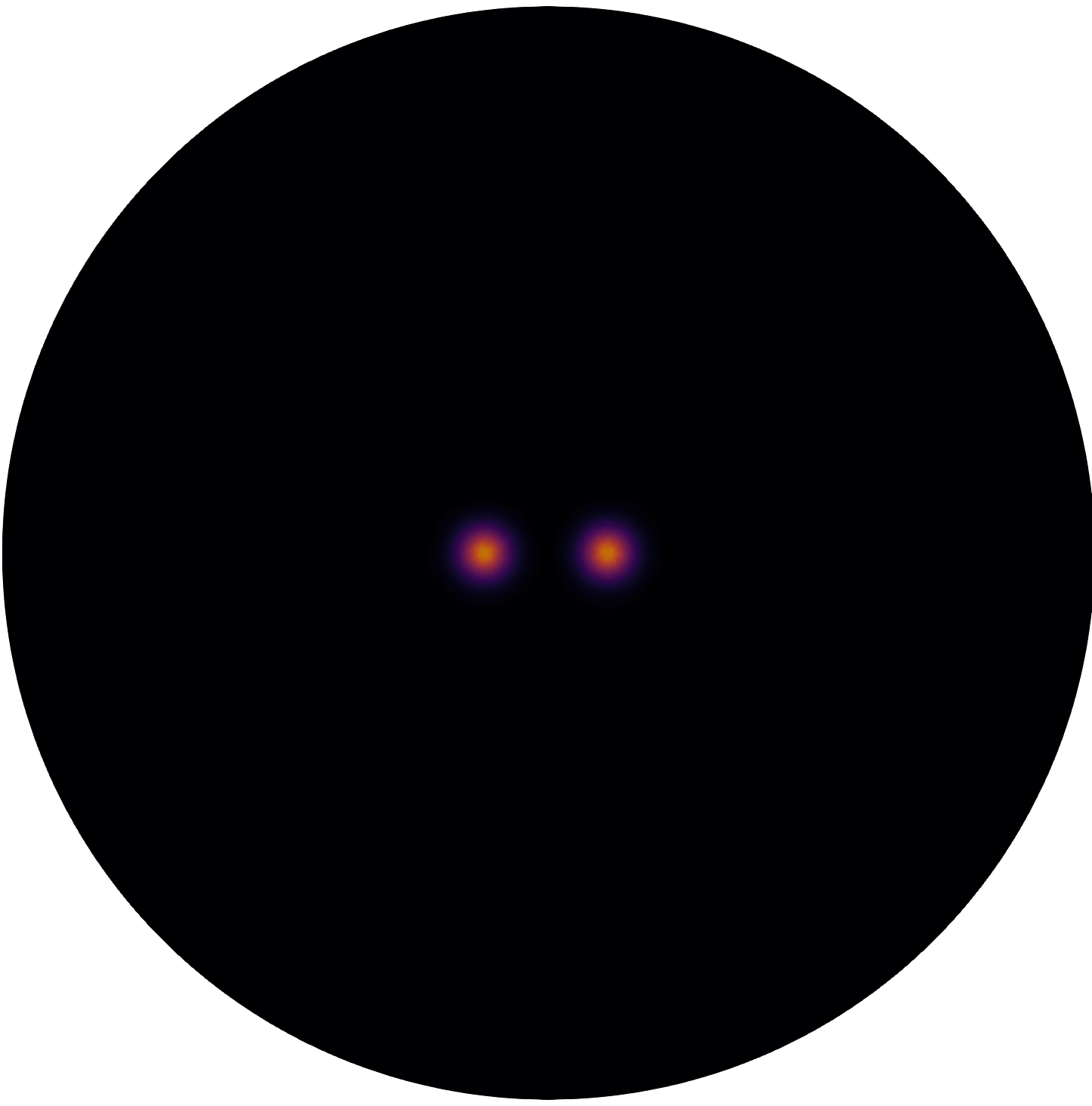}
\caption{$t=0$}
\end{subfigure}
\hfill
\begin{subfigure}[b]{0.2\textwidth}
\includegraphics[width=\textwidth]{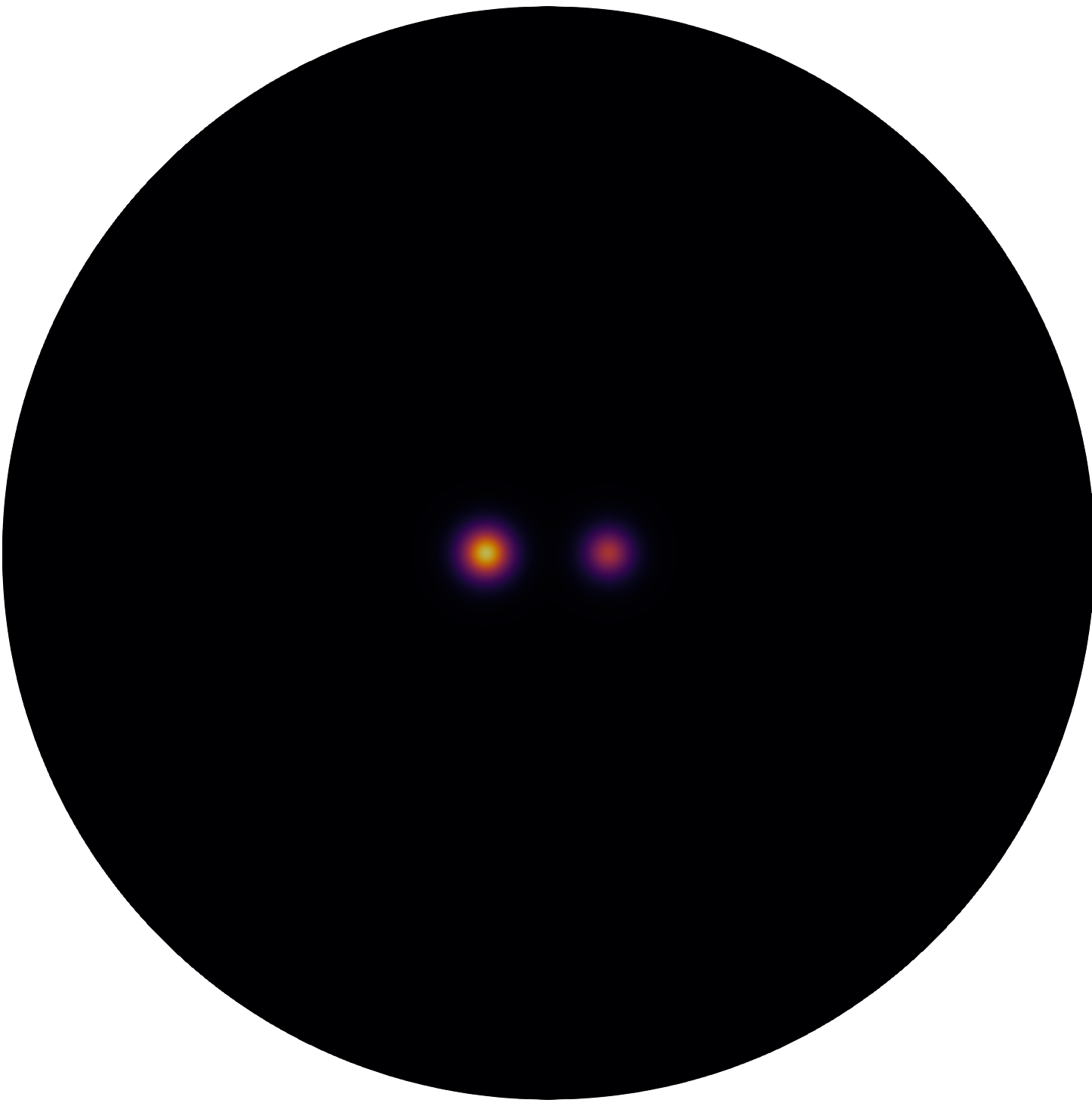}
\caption{$t=25$}
\end{subfigure}
\hfill
\begin{subfigure}[b]{0.2\textwidth}
\includegraphics[width=\textwidth]{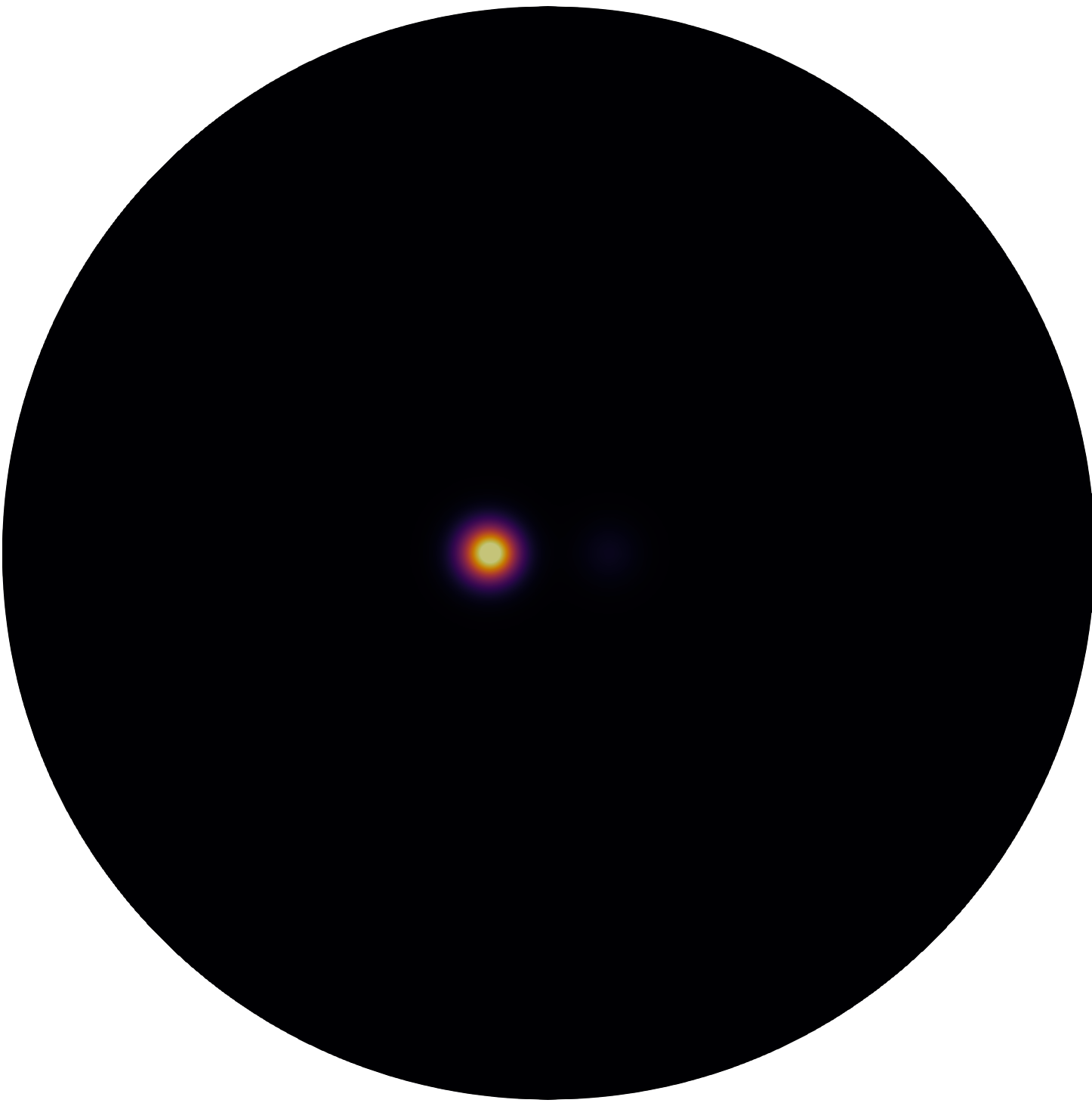}
\caption{$t=29$}
\end{subfigure}
\hfill
\begin{subfigure}[b]{0.2\textwidth}
\includegraphics[width=\textwidth]{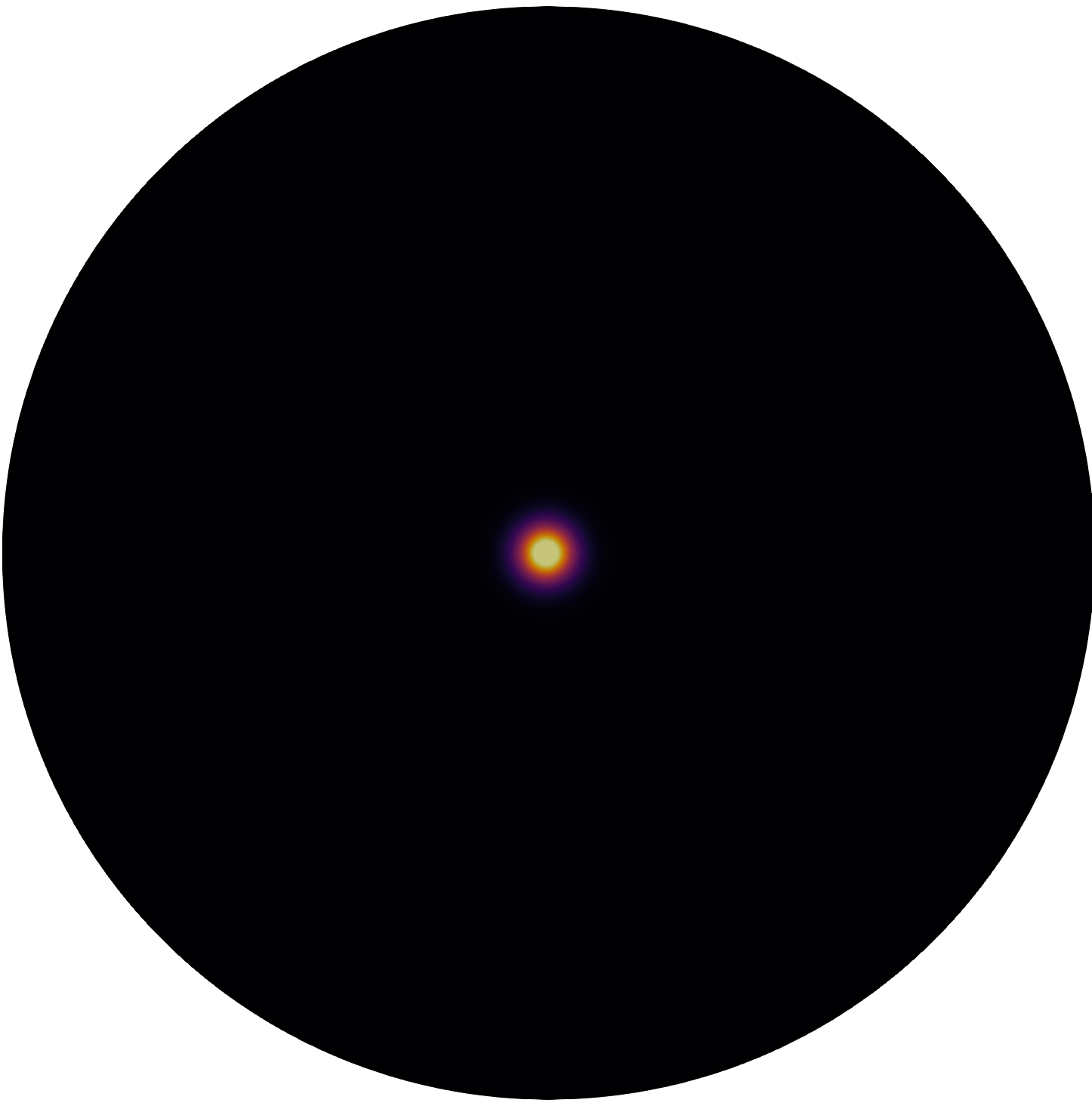}
\caption{$t=99$}
\end{subfigure}
\caption{Contour plots of $v$, from full PDE solutions of
  \eqref{proto:model} with \eqref{het:localized_feed}, corresponding
  to the parameter values shown in Fig.~\ref{het_two_spots_exp2}. A
  competition instability triggers spot annihilation, and the
  surviving spot drifts to the origin where it is pinned by the
  localized feed rate.}
\label{het_two_spots_exp2_pde}
\end{figure}

In Fig.~\ref{het_two_spots_exp2_pde}, we show some snapshots of $v$
from the full PDE numerical solution for the parameter set in
Fig.~\ref{het_two_spots_exp2}.  This figure shows that after the
competition instability triggers a spot-annihilation event, the
surviving spot ultimately get pinned at the origin where the feed rate
is concentrated. From Fig.~\ref{het_two_spots_exp2} we observe that
the spot amplitude for this pinned spot is approximately $0.8754$,
which exceeds the maximum value of approximately $0.8$, as shown in
Fig.~\ref{demo_spot_height}, for a conventional spot solution that is
not near a concentration point of the feed.  This observation
motivates the analysis in \S \ref{sec:pinned} of constructing a new
type of spot solution that is pinned at the concentration point of the
feed rate.

\section{Spot-pinning at a localized heterogeneity: A
  new type of localized structure}\label{sec:pinned}

In this section we consider the Schnakenberg model \eqref{proto:model} with
$D=\tau=1$ and with localized feed rate \eqref{het:localized_feed}, given by
\begin{equation}\label{pinned:model}
  v_t = \eps^2 \Delta v - v + uv^2 \,, \qquad u_t = \Delta u + a_0 +
  \eps^{-2} \left(a_1\Phi\left(\eps^{-1}|\v{x}-\pmb{\xi}|\right) - uv^2\right)
  \quad \mbox{in} \quad \Omega \,,
\end{equation}
with $\partial_n v = \partial_n v = 0$ on $\partial\Omega$. For the
choice $\Phi(r) \equiv {\exp(-r^2/2)/(2\pi)}$, we construct a new type
of spot solution that is pinned at the site $\pmb{\xi}\in\Omega$ of
the localization of the feed rate. Novel dynamical behaviors associated with
including this new type of spot solution in a quasi-equilibrium spot pattern
are analyzed.

\subsection{A pinned spot solution}\label{sec:pinned_one}

We construct the asymptotic profile of a pinned spot solution and we
study its linear stability properties with respect to non-radially
symmetric perturbations near the spot. We then consider the effect of
a time-varying localized concentration of the feed rate.

\subsubsection{A quasi-equilibrium one-spot pattern}

\begin{figure}[htbp]
\begin{subfigure}[b]{0.45\textwidth}
\includegraphics[width=\textwidth,height=4.3cm]{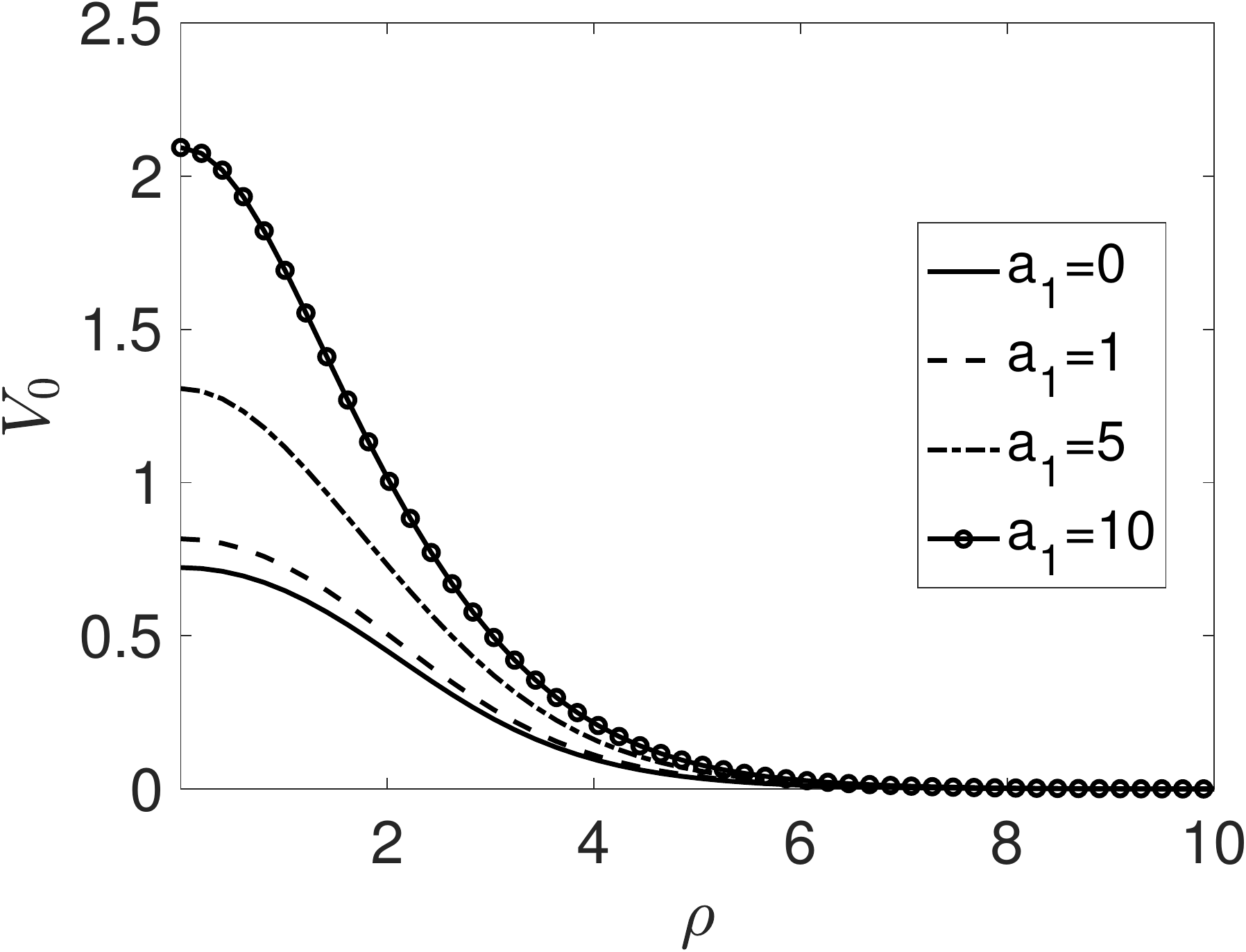}
\caption{$V_0(\rho)$ with $S=3$ for several $a_1$}
\end{subfigure}
\hfill
\begin{subfigure}[b]{0.45\textwidth}
\includegraphics[width=\textwidth,height=4.3cm]{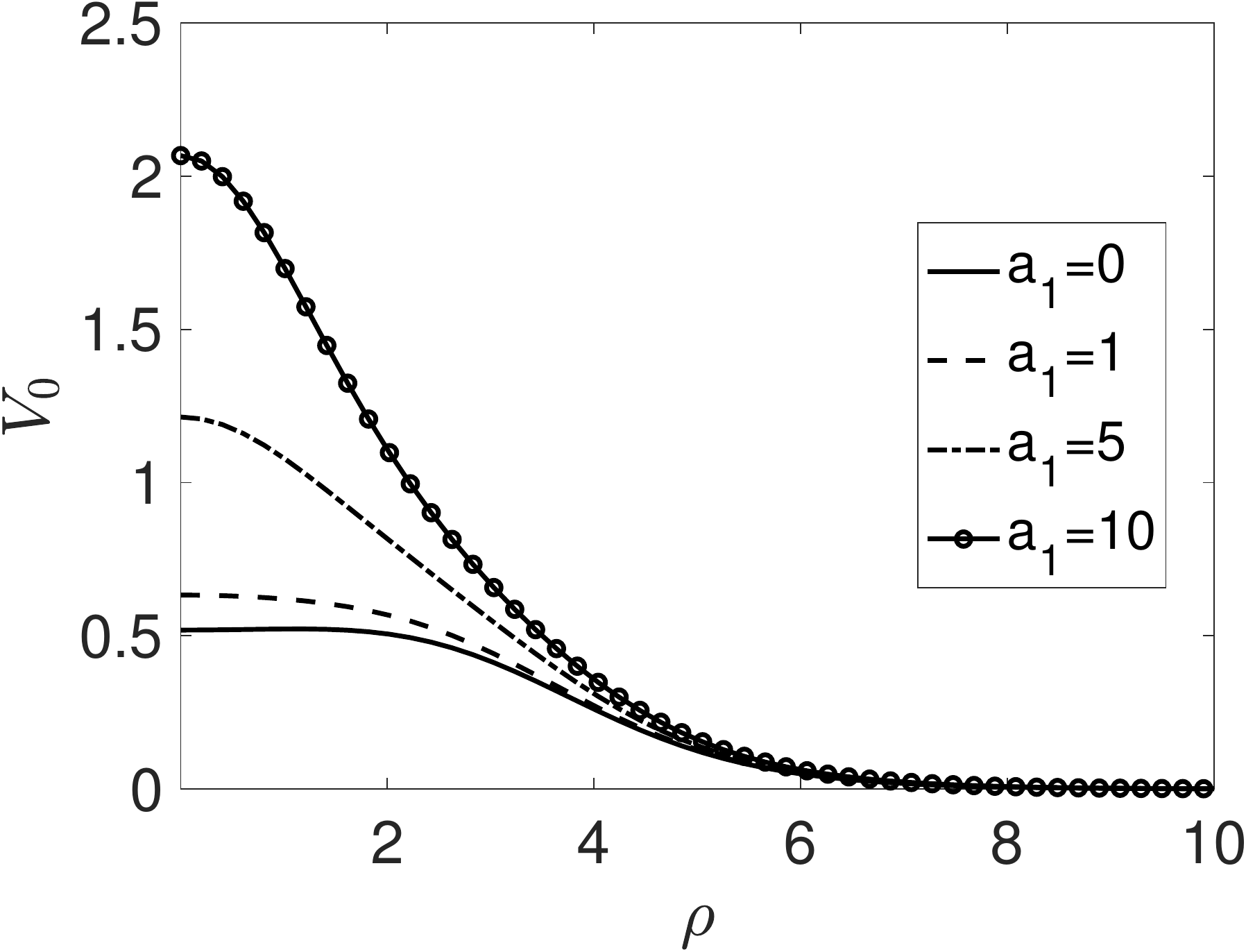}
\caption{$V_0(\rho)$ with $S=5$ for several $a_1$}
\end{subfigure}
\hfill
\caption{Solution profiles $V_0(\rho)$ with different $a_1$ for two
  values of $S_0$, as computed numerically from
  \eqref{pinned:core_full}. The spot height increases as the
  strength $a_1$ of the feed concentration increases.}
\label{fig:pinned_spot_core_solution}
\end{figure}

We begin by constructing an asymptotic quasi-equilibrium solution for
\eqref{pinned:model} corresponding to a single spot pinned at
$\pmb{\xi}$. The quasi-equilibrium problem is
\begin{equation}\label{pinned:qe}
  \eps^2\lap v_e - v_e + u_e v_e^2 = 0 \,, \quad \lap u_e + a_0 +
  \eps^{-2} \left[ a_1 \Phi\left(\eps^{-1} |\v{x}-\pmb{\xi}|/\right) -
    u_e v_e^2\right] = 0 \,,
\end{equation}
with $\partial_n v_e = \partial_n u_e = 0$ on $\partial\Omega$ and
$\Phi(r) \equiv {\exp(-r^2/2)/(2\pi)}$. In the inner region near the
pinned spot, we look for a locally radially symmetric solution of the form
$v_e \sim V_0(\rho)$ and $u_e \sim U_0(\rho)$ where
$\rho = \eps^{-1}|\v{x}-\pmb{\xi}|$. From \eqref{pinned:qe} we get
that $U_0$ and $V_0$ satisfy a new core problem
\begin{subequations}\label{pinned:core_full}
\begin{align}
  \lap_\rho V_0 - V_0 + U_0 V_0^2 &= 0 \,, \quad \lap_\rho U_0 + a_1
  \Phi(\rho) - U_0 V_0^2 = 0 \,, \qquad 0<\rho < \infty \,,
  \label{pinned:core_problem}\\ 
  V_0^{\prime}(0)=U_0^{\prime}(0)=0 \,; \qquad
  V_0 &\to 0\,, \quad U_0 \sim S_0 \log \rho + \chi(S_0 ;  a_1) \quad
  \mbox{as} \quad \rho \to \infty\,. \label{pinned:core_problem_far_field}
\end{align}
\end{subequations}
The quantity $\chi(S_0;a_1)$ is an $\bigo(1)$ nonlinear function
of $S_0$ and concentration intensity $a_1$ of the feed rate. In
Fig.~\ref{fig:pinned_spot_core_solution} we plot the numerically
computed spot profile $V_0(\rho)$ for various $S_0$ and $a_1$.

By integrating the $U_0$ equation in \eqref{pinned:core_full} on $\rho>0$,
we use $\int_{0}^{\infty} \Phi(\rho)\rho\, d\rho={1/(2\pi)}$ to
obtain the integral identity
\begin{equation}\label{pinned:S0}
S_0 + \frac{a_1}{2\pi} = \int_0^{\infty} U_0 V_0^2 \,\rho \,d\rho \,.
\end{equation}
With the identity \eqref{pinned:S0}, we derive in the sense of
distributions that, for $\eps\to 0$,
\begin{equation}\label{pinned:correspondence_rule}
  \eps^{-2}\left[ a_1 \Phi - u_e v_e^2\right] \to \left[ a_1 -
    2\pi \left(\int_0^{\infty} U_0 V_0^2 \, \rho \, d\rho \right) \right]
  \delta(\v{x}-\pmb{\xi}) = -2\pi S_0 \delta(\v{x}-\pmb{\xi}) \,.
\end{equation}

Upon using \eqref{pinned:correspondence_rule} in \eqref{pinned:qe},
we obtain that the outer problem for $u_e$, defined away from
$\pmb{\xi}$, is
\begin{equation}\label{pinned:outer_problem}
  \Delta u_e = - a_0 + 2\pi S_0 \, \delta(\v{x}-\pmb{\xi})  \quad
  \mbox{in} \quad \Omega \,, \qquad \partial_n u = 0 \quad \mbox{on} \quad
  \partial\Omega \,,
\end{equation}
which has the solution
\begin{equation}\label{pinned:one_sol}
u = -2\pi S_0 \, G(\v{x};\pmb{\xi}) + \ubar \,.
\end{equation}
Here $G$ is the Neumann Green's function satisfying
\eqref{proto:neu_green} and $\ubar$ is an undetermined constant.  By
applying the divergence theorem to \eqref{pinned:outer_problem}, we
obtain that the source strength for the pinned spot is
\begin{equation}\label{pinned:one_spot_source_strength}
S_0 = \frac{a_0|\Omega|}{2\pi} \,.
\end{equation}
To determine $\ubar$, we let $\v{x} \to \pmb{\xi}$ in
\eqref{pinned:one_sol} to obtain
$u \sim S_0 \log |\v{x} - \pmb{\xi}|- 2\pi S_0 \, R_{0,0} + \ubar$, where
$R_{0,0} = R(\pmb{\xi} ; \pmb{\xi})$. Upon matching
this expression with \eqref{pinned:core_problem_far_field} we obtain that
$\ubar = \nu^{-1} \left[S_0 + 2\pi\nu S_0 R_{0,0} + \nu\chi(S_0 ;a_1) \right]$.

We now use this construction to account for the spot height of the
pinned spot observed in the PDE simulations shown in
Fig.~\ref{het_two_spots_exp2}, in which $a_0 = 4.3$ and
$a_1 = 4.2708$. For this value of $a_0$,
\eqref{pinned:one_spot_source_strength} yields that $S_0 =
2.15$. Then, by computing the solution to the new core problem
\eqref{pinned:core_full} with $S_0 = 2.15$ and $a_1 = 4.2708$, we find
that the predicted spot height is $V_0(0) \approx 0.8755$. This value
is very close to the spot height, given approximately by $0.8754$,
observed in the full PDE simulation results shown in
Fig.~\ref{het_two_spots_exp2}.

\subsubsection{Linear stability analysis} \label{sec:pinned_linstab_one_pinned_spot}

Next, we analyze the linear stability of a pinned spot. We let
$v_e$ and $u_e$ denote the quasi-equilibrium solution and we introduce
the perturbation
\begin{equation*}
v = v_e + e^{\lambda t}\phi \,, \qquad u = u_e + e^{\lambda t}\eta \,,
\end{equation*}
into \eqref{pinned:model} and linearize. This yields the eigenvalue
problem 
\begin{equation}\label{pinned:eigenvalue_problem0}
  \eps^2 \Delta \phi - \phi + 2u_e v_e \phi + v_e^2 \eta = \lambda \phi \,,
  \qquad \Delta \eta - \eps^{2} \left(2u_e v_e \phi + v_e^2 \eta \right) =
  \lambda \eta \,.
\end{equation}
To examine the possibility of locally non-radially symmetric
instabilities near the spot, we let
$\phi \sim e^{im\theta}\Phi_0(\rho)$ and
$\eta \sim e^{im\theta}N_0(\rho)$ in
\eqref{pinned:eigenvalue_problem0} for integer modes $m\geq 2$, where
$\rho=\eps^{-1}|\v{x}-\pmb{\xi}|$. Then, upon using
$v_e \sim V_0(\rho)$ and $u_e \sim U_0(\rho)$, to leading order we obtain
an eigenvalue problem in the inner region
\begin{equation}\label{pinned:eigenvalue_problem}
  \Delta_\rho \Phi_0 - \frac{m^2}{\rho^2} \Phi_0 - \Phi_0 + 2 U_0 V_0 \Phi_0
  + V_0^2 N_0 = \lambda \Phi_0 \,, \qquad \Delta_\rho N_0 - \frac{m^2}{\rho^2} N_0
  - 2 U_0 V_0 \Phi_0 - V_0^2 N_0 = 0 \,,
\end{equation}
where $\Delta_\rho = \partial_{\rho\rho} + \rho^{-1}
\partial_\rho$. For the non-radially symmetric modes with $m\geq 2$,
we can impose that $\Phi_0 \to 0$ exponentially as $\rho \to \infty$
and impose the algebraic decay condition $N_0 \sim \mc{O}(\rho^{-m})$
as $\rho \to \infty$.  We remark that the eigenvalue problem
\eqref{pinned:eigenvalue_problem} depends on $S_0$ and $a_1$ through
the solution $V_0$ and $U_0$ to the new core problem
\eqref{pinned:core_full}.

\begin{figure}[htbp]
\begin{subfigure}{0.45\textwidth}
\includegraphics[width=\textwidth,height=4.2cm]{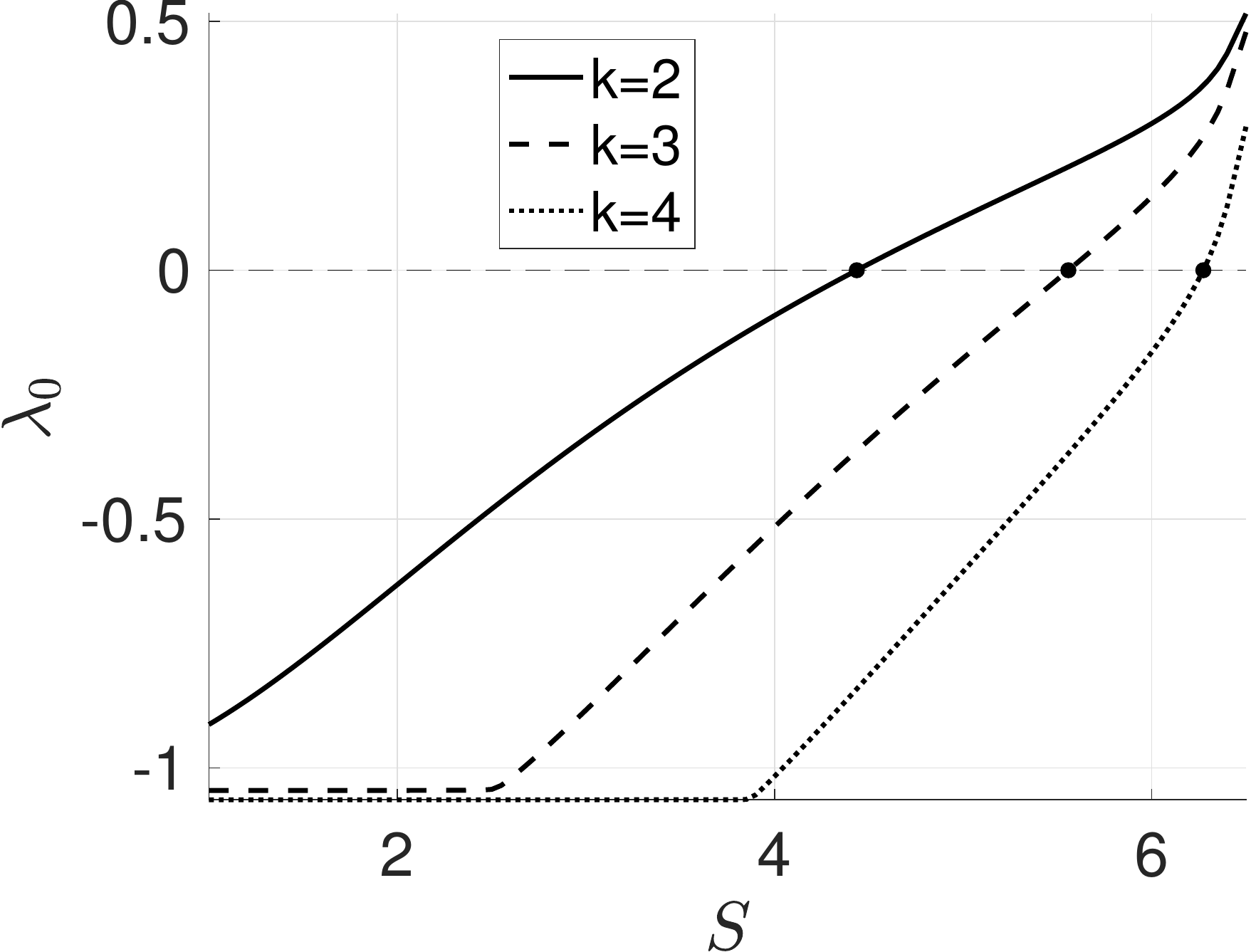}
\caption{$a_1=1$}
\end{subfigure}
\hfill
\begin{subfigure}{0.45\textwidth}
\includegraphics[width=\textwidth,height=4.2cm]{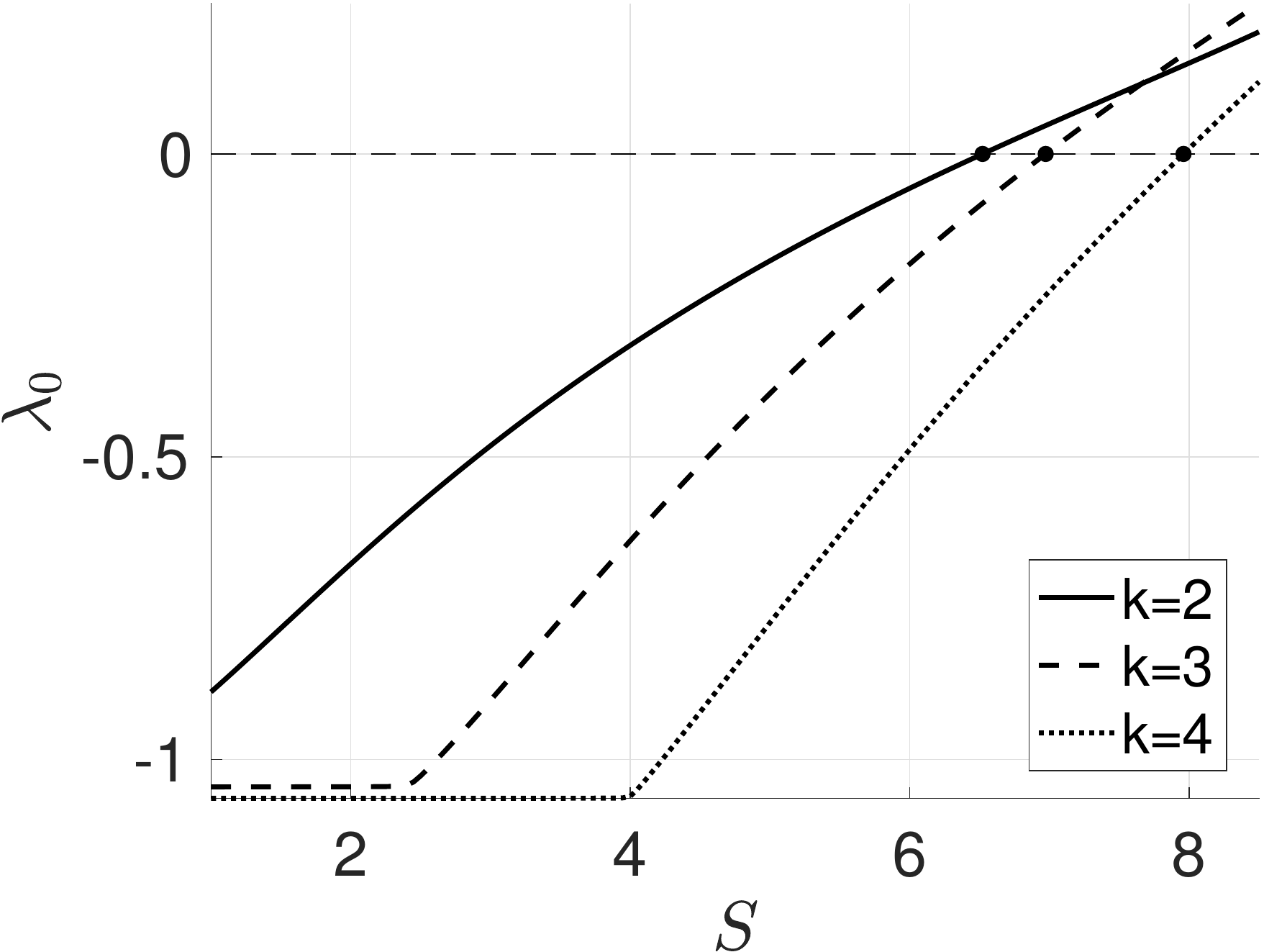}
\caption{$a_1=10$}
\end{subfigure}
\\
\begin{subfigure}{0.45\textwidth}
\includegraphics[width=\textwidth,height=4.2cm]{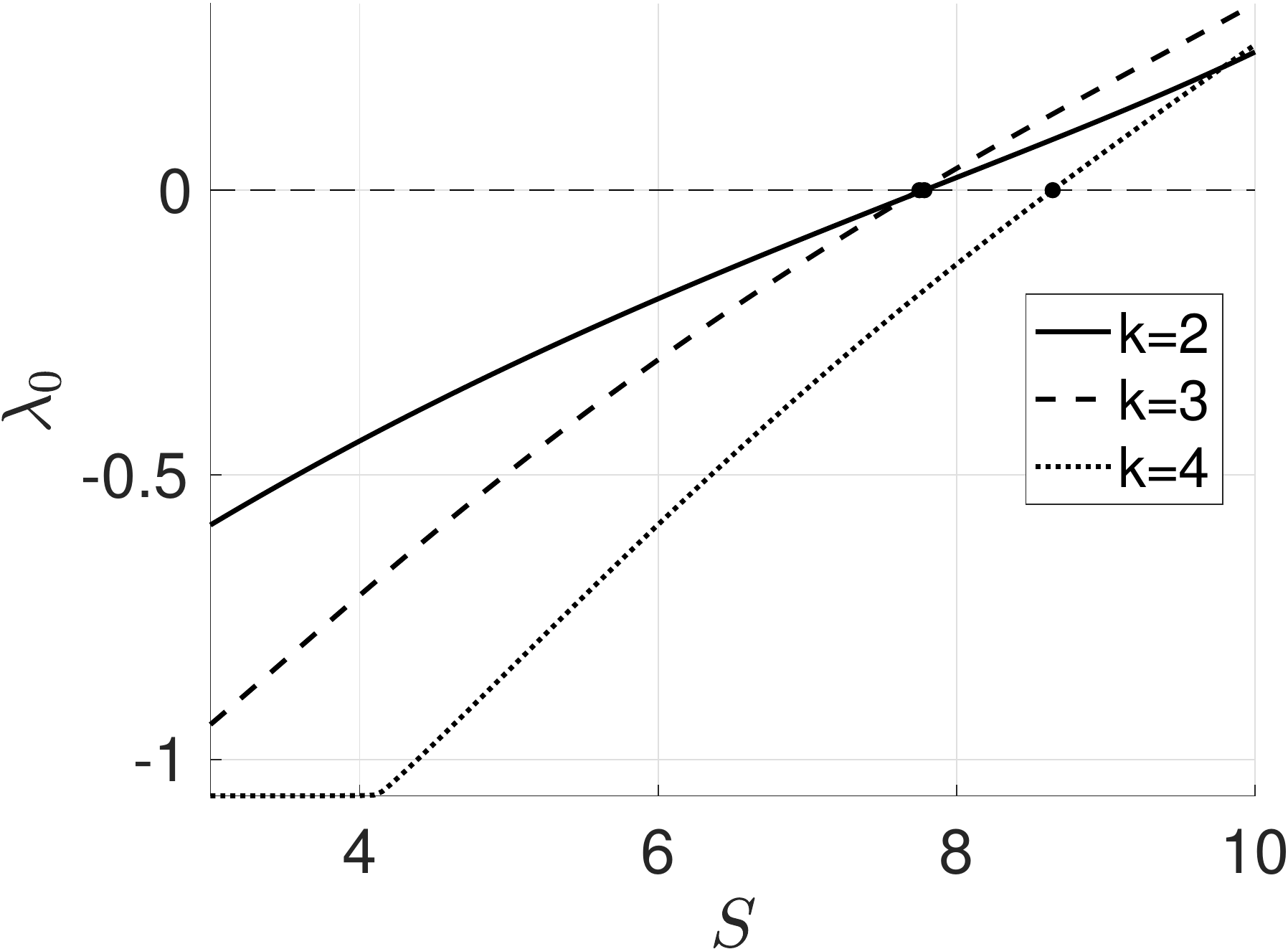}
\caption{$a_1=16$}
\end{subfigure}
\hfill
\begin{subfigure}{0.45\textwidth}
\includegraphics[width=\textwidth,height=4.2cm]{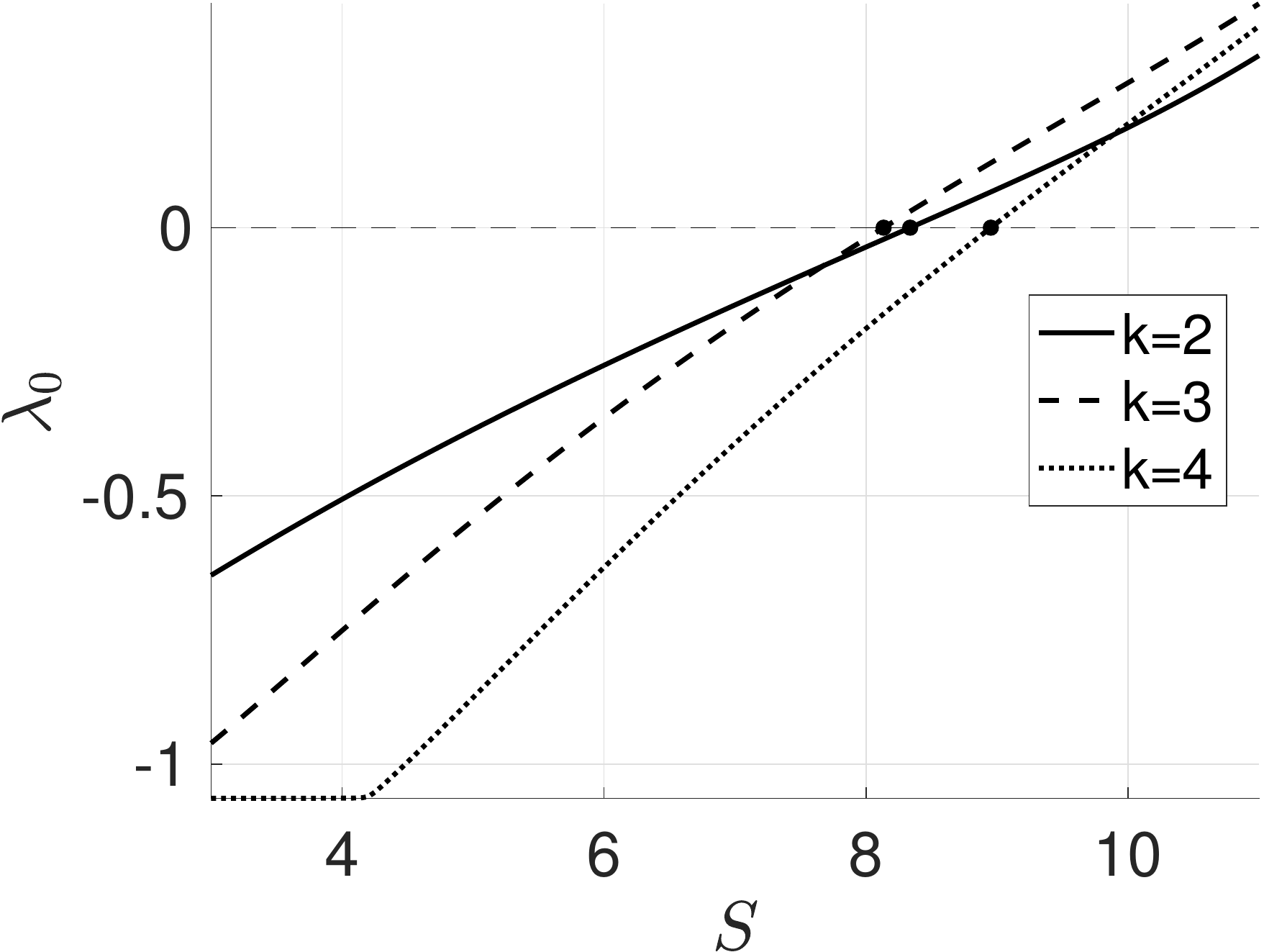}
\caption{$a_1=20$}
\label{pinned_lambda0_a1=20}
\end{subfigure}
\caption{Numerically computed eigenvalue $\lambda_0$ of
  \eqref{pinned:eigenvalue_problem} with the largest real part versus
  $S_0$ for modes $m=2,3,4$ and four different feed intensities
  $a_1$. The critical thresholds $\Sigma_m(a_1)$ are the values of
  $S_0$ where $\mbox{Re}(\lambda_0)=0$.  Top left panel: $a_1=1$,
  $\Sigma_2(1) \approx 4.4358 \,, \Sigma_3(1) \approx 5.5580 \,,
  \Sigma_4(1) \approx 6.2736$. Top right panel: $a_1=10$,
  $\Sigma_2(10) \approx 6.5219 \,, \Sigma_3(10) \approx 6.9735 \,,
  \Sigma_4(10) \approx 7.9601$. Bottom left panel: $a_1=16$,
  $\Sigma_2(16) \approx 7.7854 \,, \Sigma_3(16) \approx 7.7513 \,,
  \Sigma_4(16) \approx 8.6443$. Bottom right panel: $a_1=20$,
  $\Sigma_2(20) \approx 8.3373 \,, \Sigma_3(20) \approx 8.1346 \,,
  \Sigma_4(20) \approx 8.9531$.}
\label{pinned_lambda0}
\end{figure}

By discretizing \eqref{pinned:eigenvalue_problem}, we obtain a
generalized matrix eigenvalue problem. For each mode $m \geq 2$, we
numerically compute the eigenvalue $\lambda_0$ of the discretization
of \eqref{pinned:eigenvalue_problem} with the largest real part as a
function of $a_1$ and the source strength $S_0$. The instability
threshold occurs when $\mathrm{Re}(\lambda_0) = 0$. In
Fig.~\ref{pinned_lambda0}, we plot $\mathrm{Re}(\lambda_0)$ versus
$S_0$ for modes $m=2,3,4$ for various values of $a_1$. We define
$\Sigma_m(a_1)$ to be the spot source strength corresponding to the
stability threshold $\mbox{Re}(\lambda_0)=0$ for angular mode $m$ and
concentrated feed intensity $a_1$. When $a_1=0$, where there is no
concentration of the feed rate, we have from \cite{kww09} (see the
summary in \S \ref{sec:proto_linstab}) that there is an ordering
principle $\Sigma_2(0) < \Sigma_3(0) < \Sigma_4(0) < \ldots$ for the
mode instability thresholds. Therefore, when $a_1=0$, the
peanut-splitting mode $m=2$ is the first mode to lose stability as
$S_0$ is increased. However, a qualitatively new result for our pinned
spot solution is that this ordering principle can be violated if the
feed intensity $a_1$ is large enough. In particular, if $a_1=20$, we
observe from Fig.~\ref{pinned_lambda0_a1=20} that
$\Sigma_3(20) < \Sigma_2(20)$, which implies that the $m=3$ mode is
the first to lose stability as $S_0$ is increased.

\begin{figure}[htbp]
\begin{subfigure}[b]{0.24\textwidth}
\includegraphics[width=\textwidth]{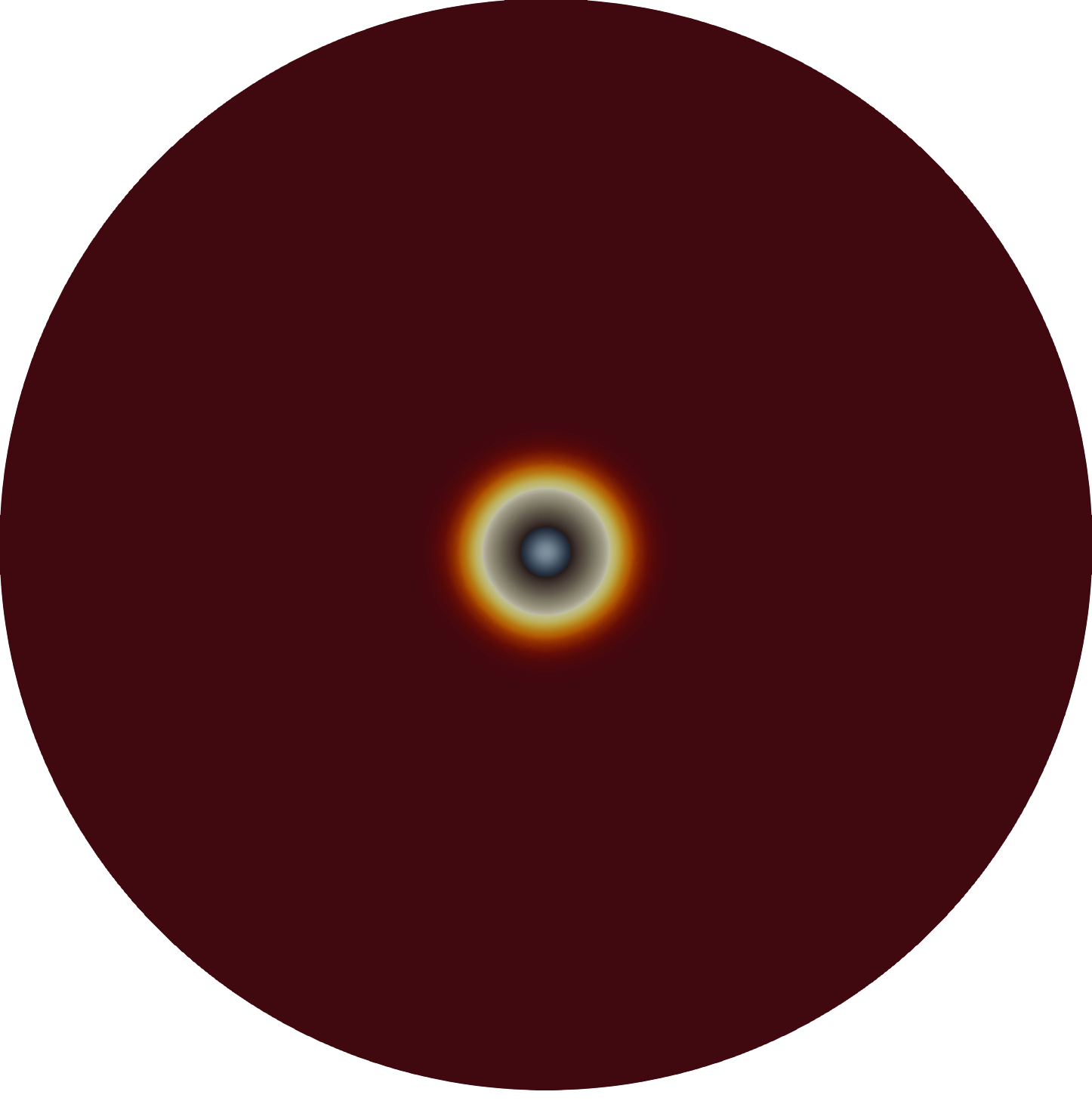}
\caption{$t=0$}
\end{subfigure}
\hfill
\begin{subfigure}[b]{0.24\textwidth}
\includegraphics[width=\textwidth]{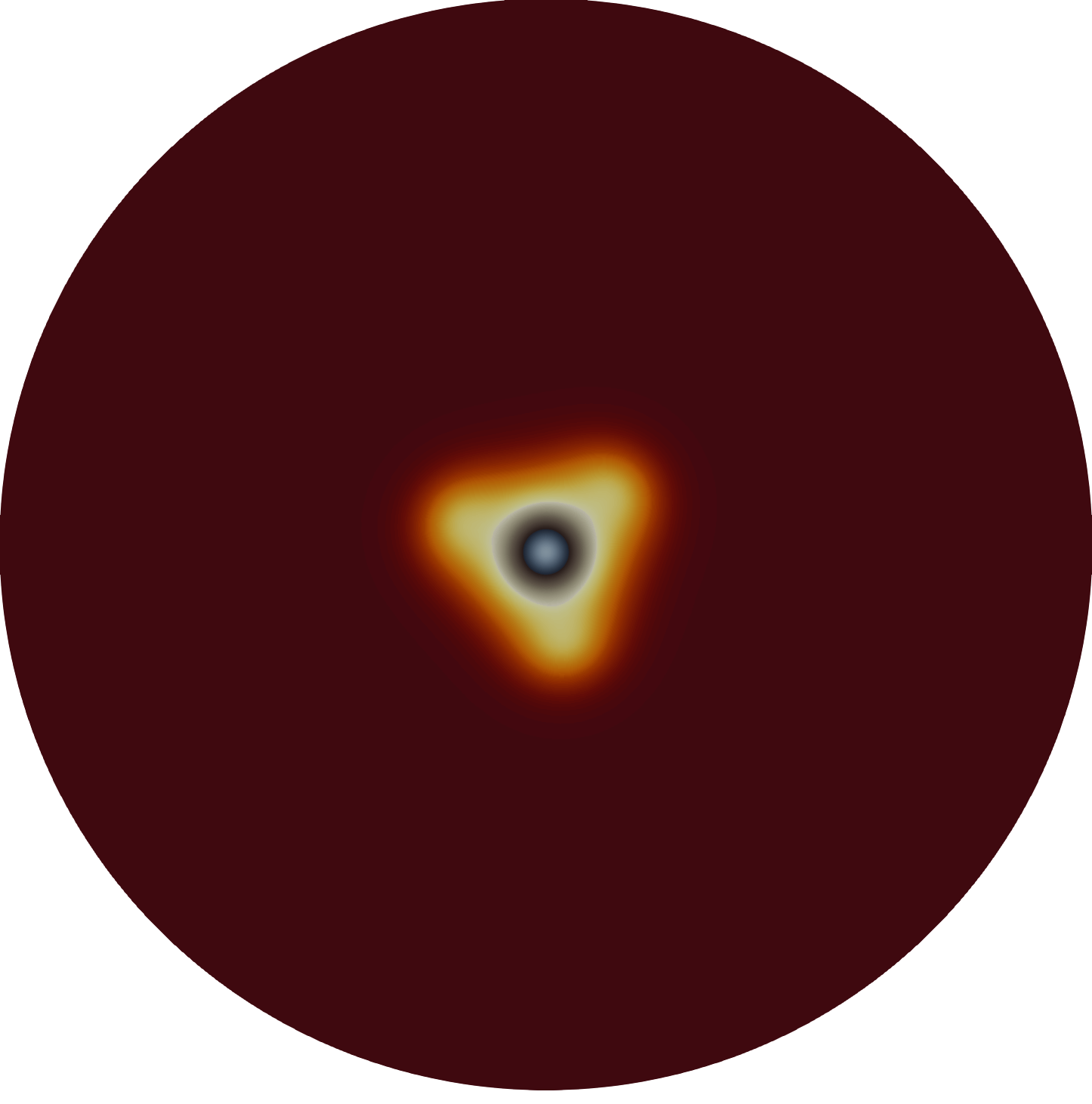}
\caption{$t=212$}
\end{subfigure}
\hfill
\begin{subfigure}[b]{0.24\textwidth}
\includegraphics[width=\textwidth]{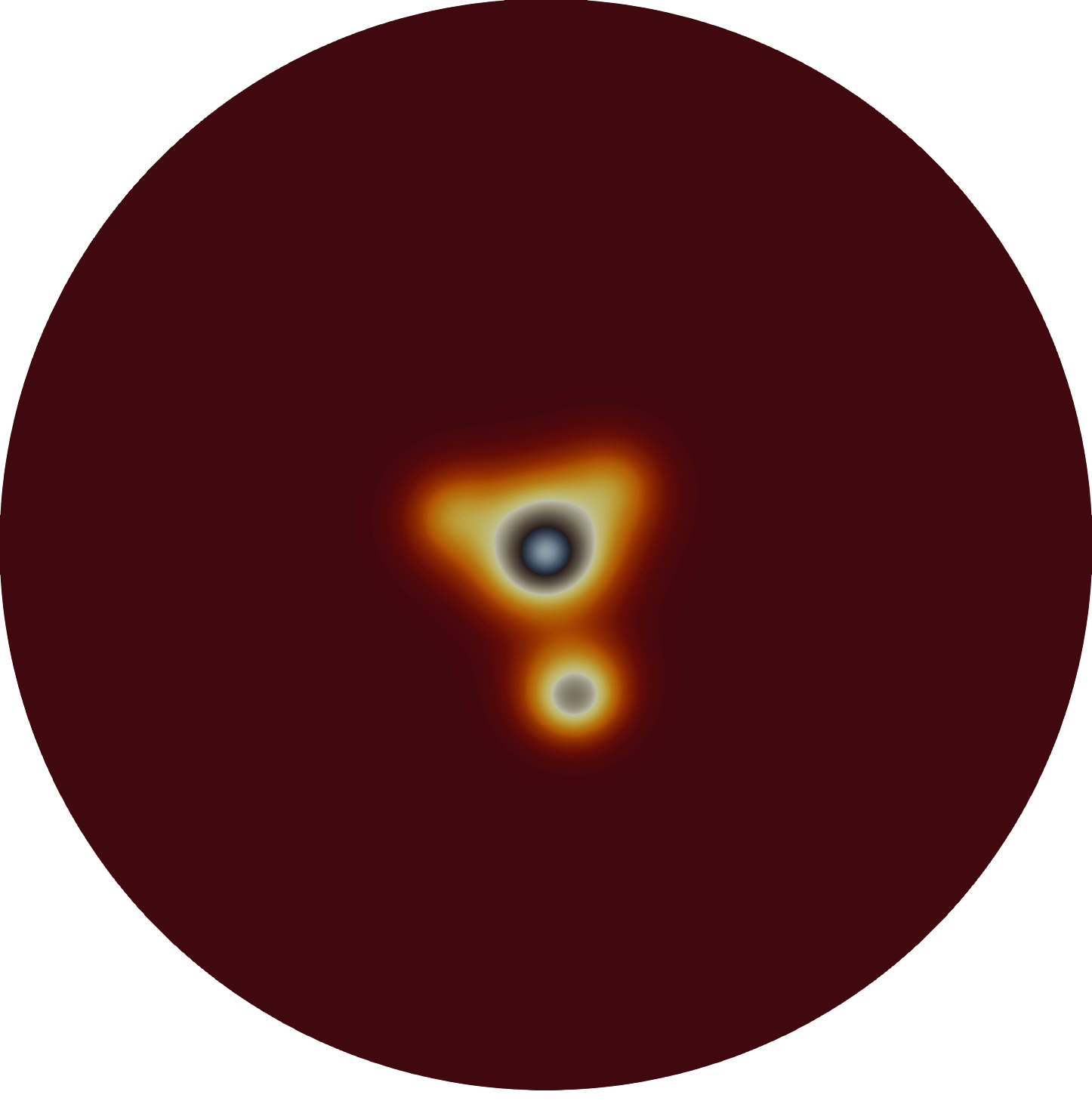}
\caption{$t=231$}
\end{subfigure}
\hfill
\begin{subfigure}[b]{0.24\textwidth}
\includegraphics[width=\textwidth]{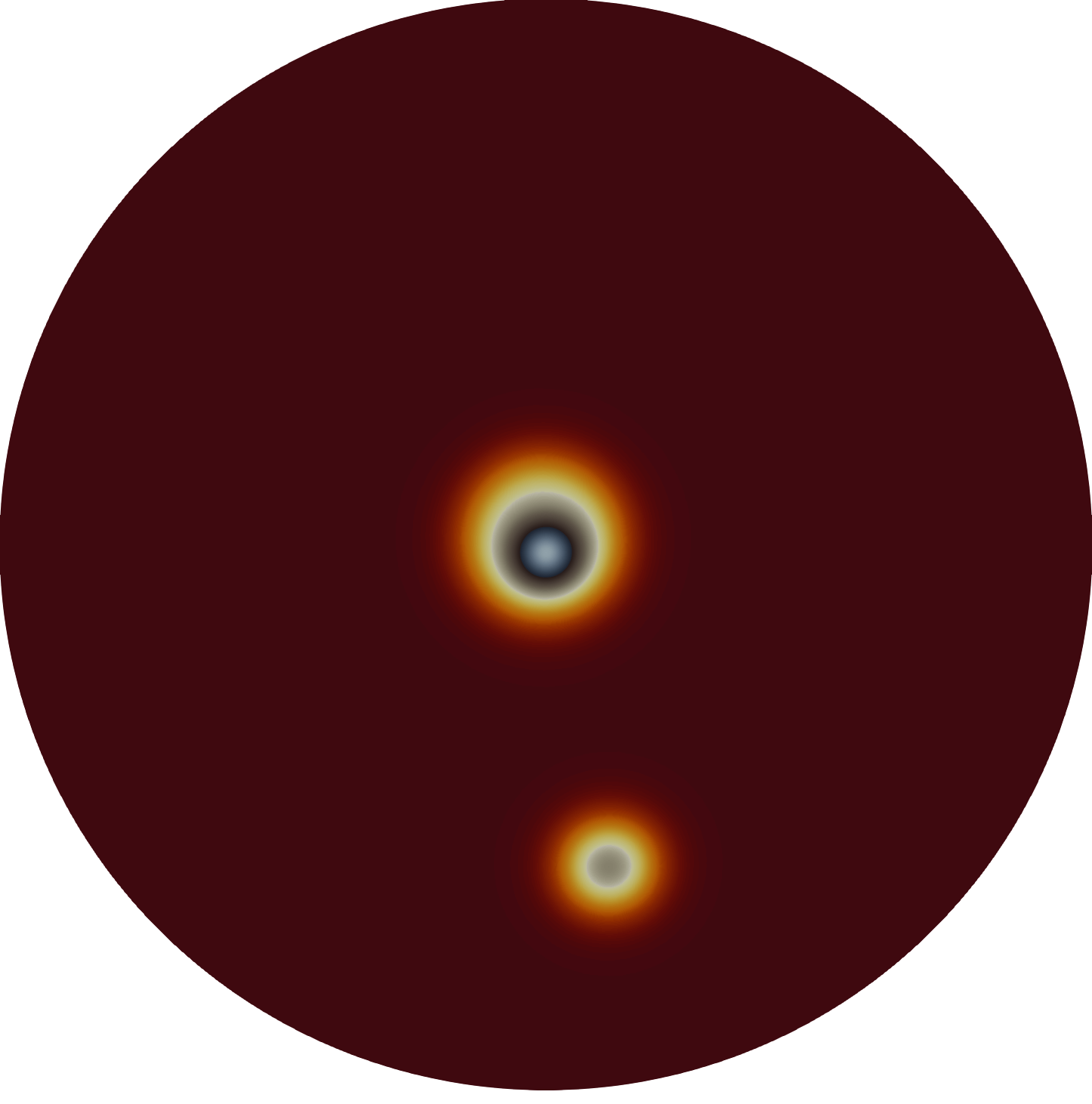}
\caption{$t=300$}
\end{subfigure}
\caption{PDE simulation results of \eqref{pinned:model} for $v$ in the
  unit disk with $\eps=0.03$, $a_1=20$, and concentrated feed rate at
  the origin $\pmb{\xi}=(0,0)$. With $a_0 = 17$ the pinned spot
  exhibits a mode $m=3$ instability by time $t=212$, but ultimately
  only one spot persists by $t=300$.}
\label{pinned_split_exp1}
\end{figure}

\begin{figure}[htbp]
\begin{subfigure}[b]{0.24\textwidth}
\includegraphics[width=\textwidth]{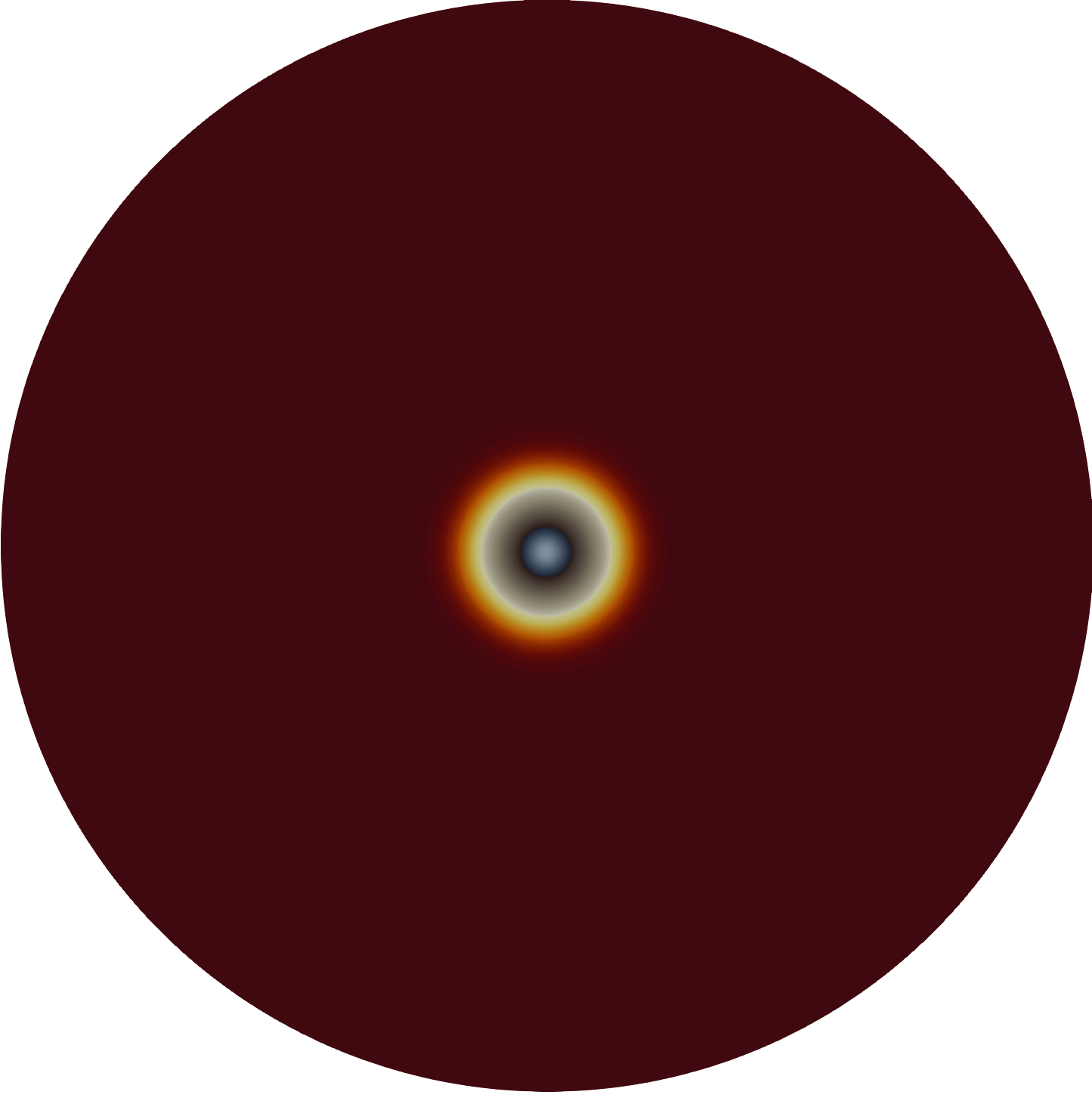}
\caption{$t=0$}
\end{subfigure}
\hfill
\begin{subfigure}[b]{0.24\textwidth}
\includegraphics[width=\textwidth]{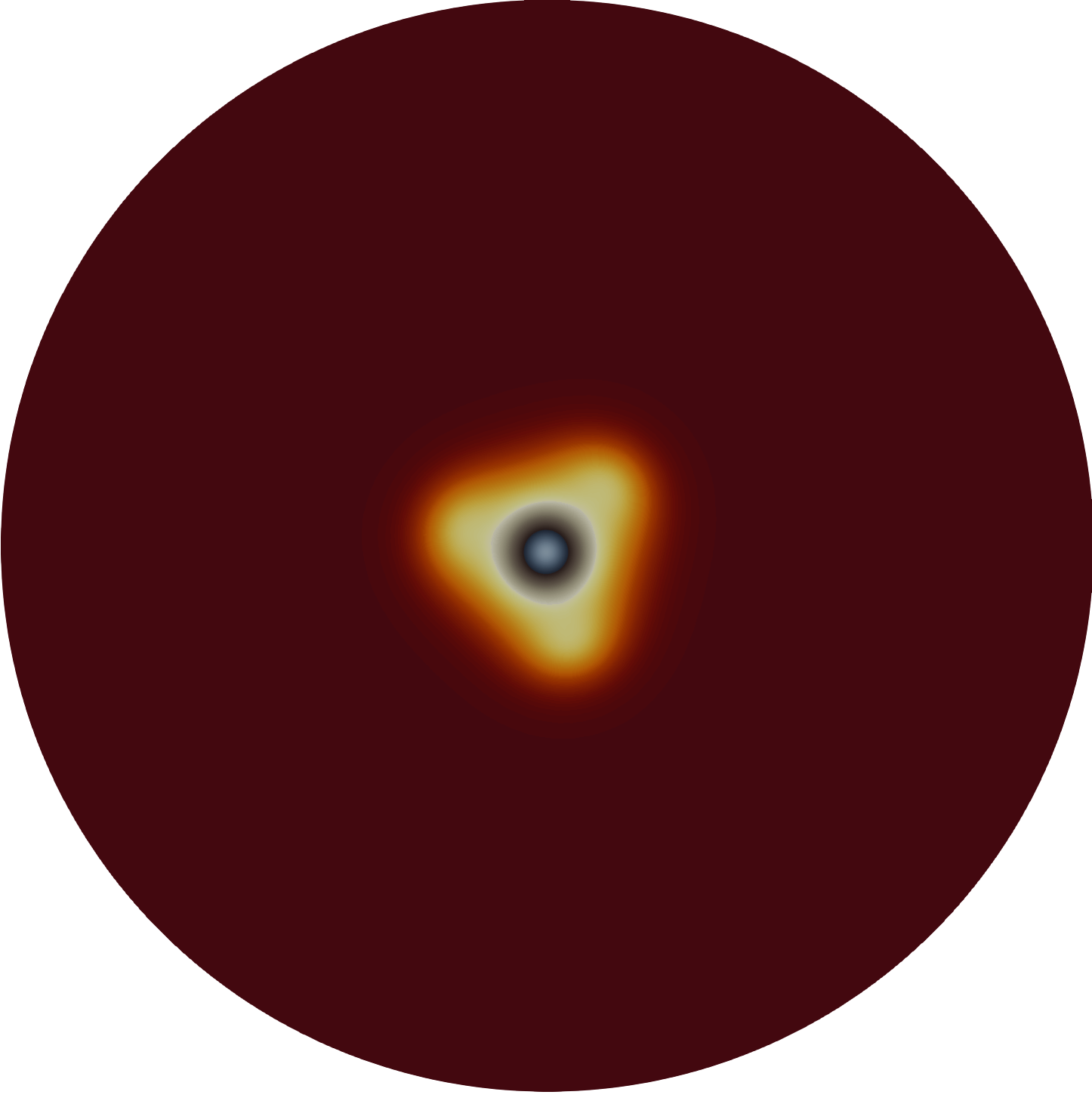}
\caption{$t=50$}
\end{subfigure}
\hfill
\begin{subfigure}[b]{0.24\textwidth}
\includegraphics[width=\textwidth]{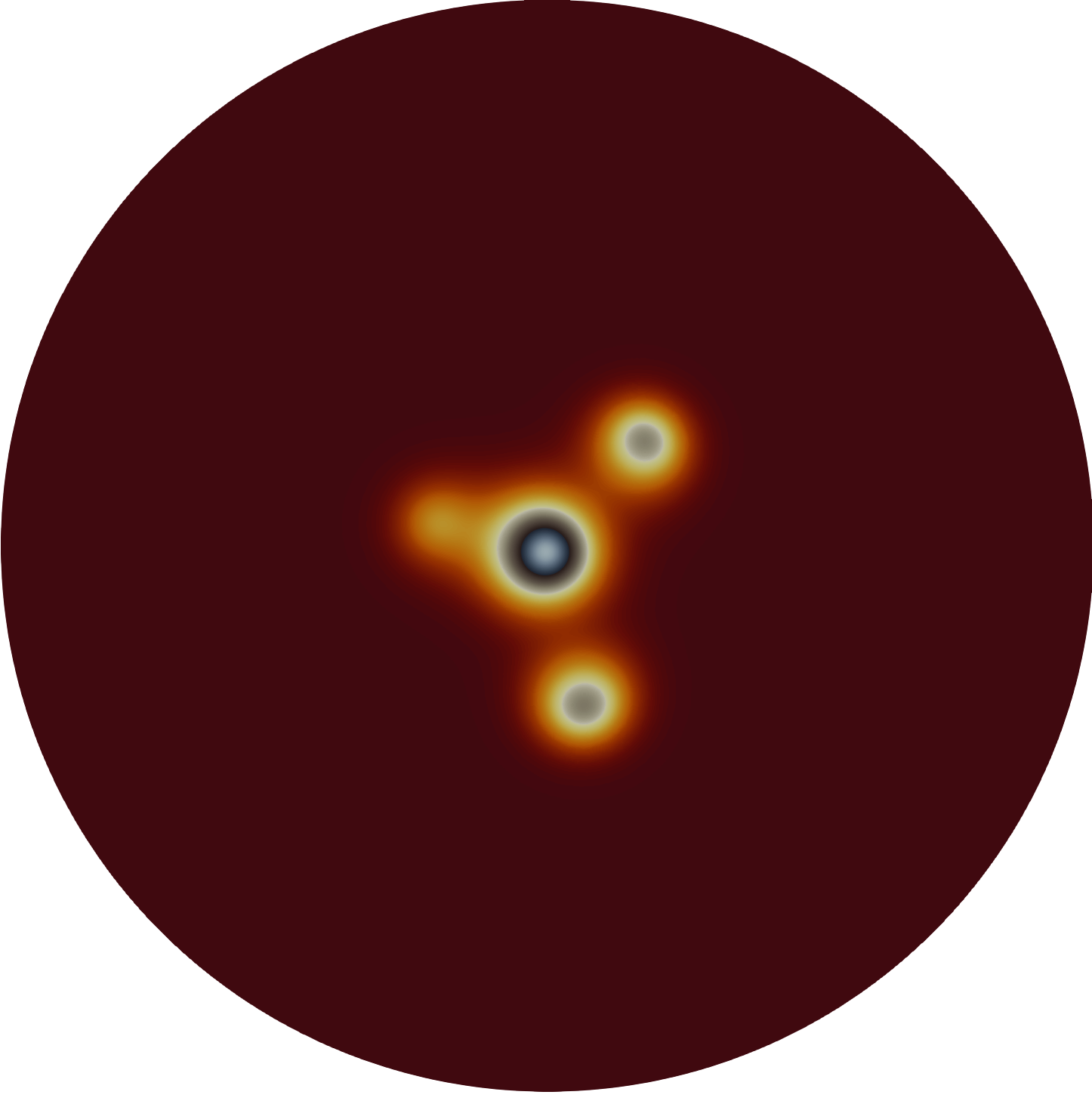}
\caption{$t=64$}
\end{subfigure}
\hfill
\begin{subfigure}[b]{0.24\textwidth}
\includegraphics[width=\textwidth]{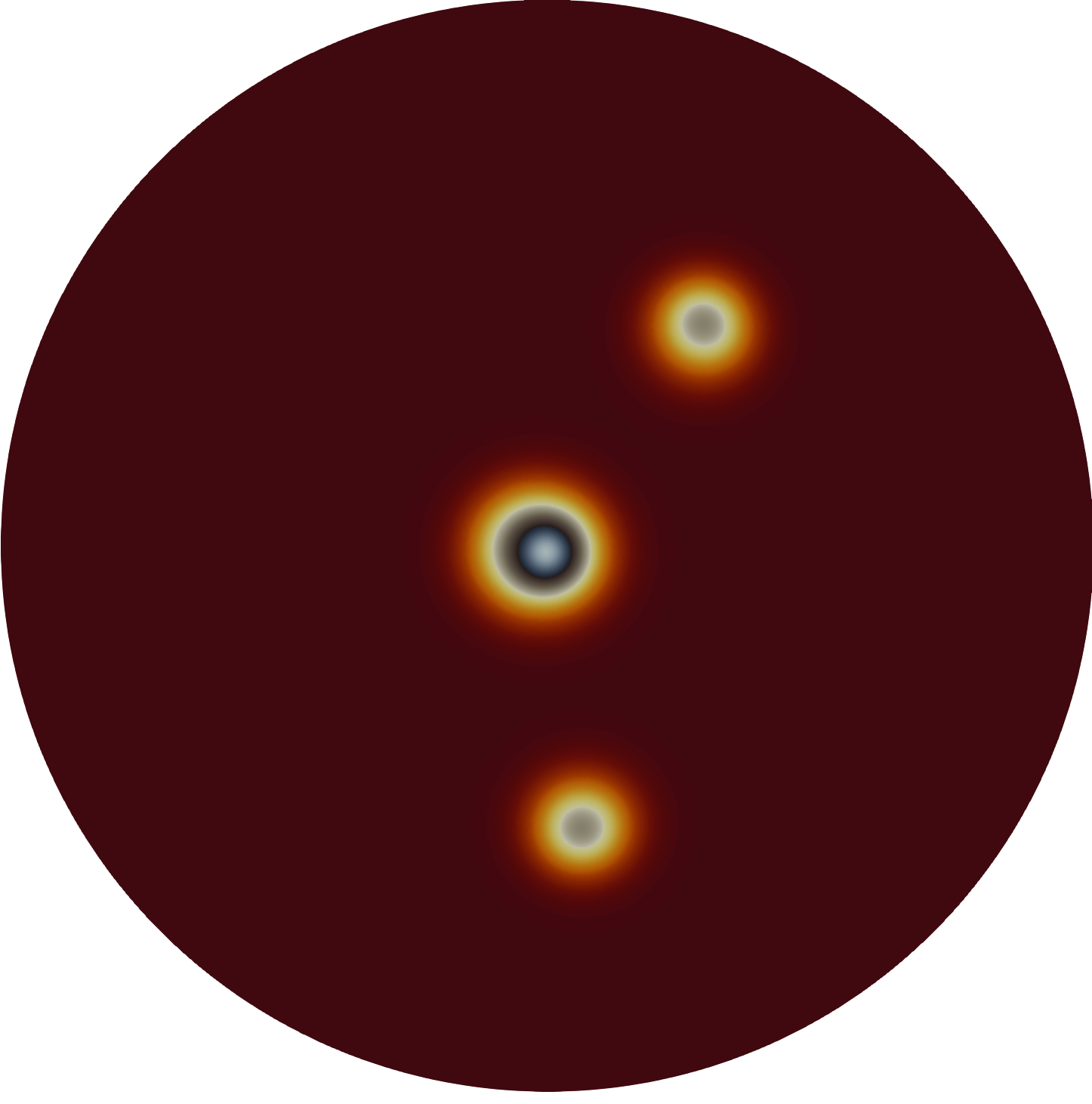}
\caption{$t=100$}
\end{subfigure}
\caption{Same caption as in Fig.~\ref{pinned_split_exp1} except that
  $a_0$ is increased to $a_0=18$. The mode $m=3$ instability of the
  pinned spot leads to two new spots.}\label{pinned_split_exp2}
\end{figure}

\begin{figure}[htbp]
\begin{subfigure}[b]{0.24\textwidth}
\includegraphics[width=\textwidth]{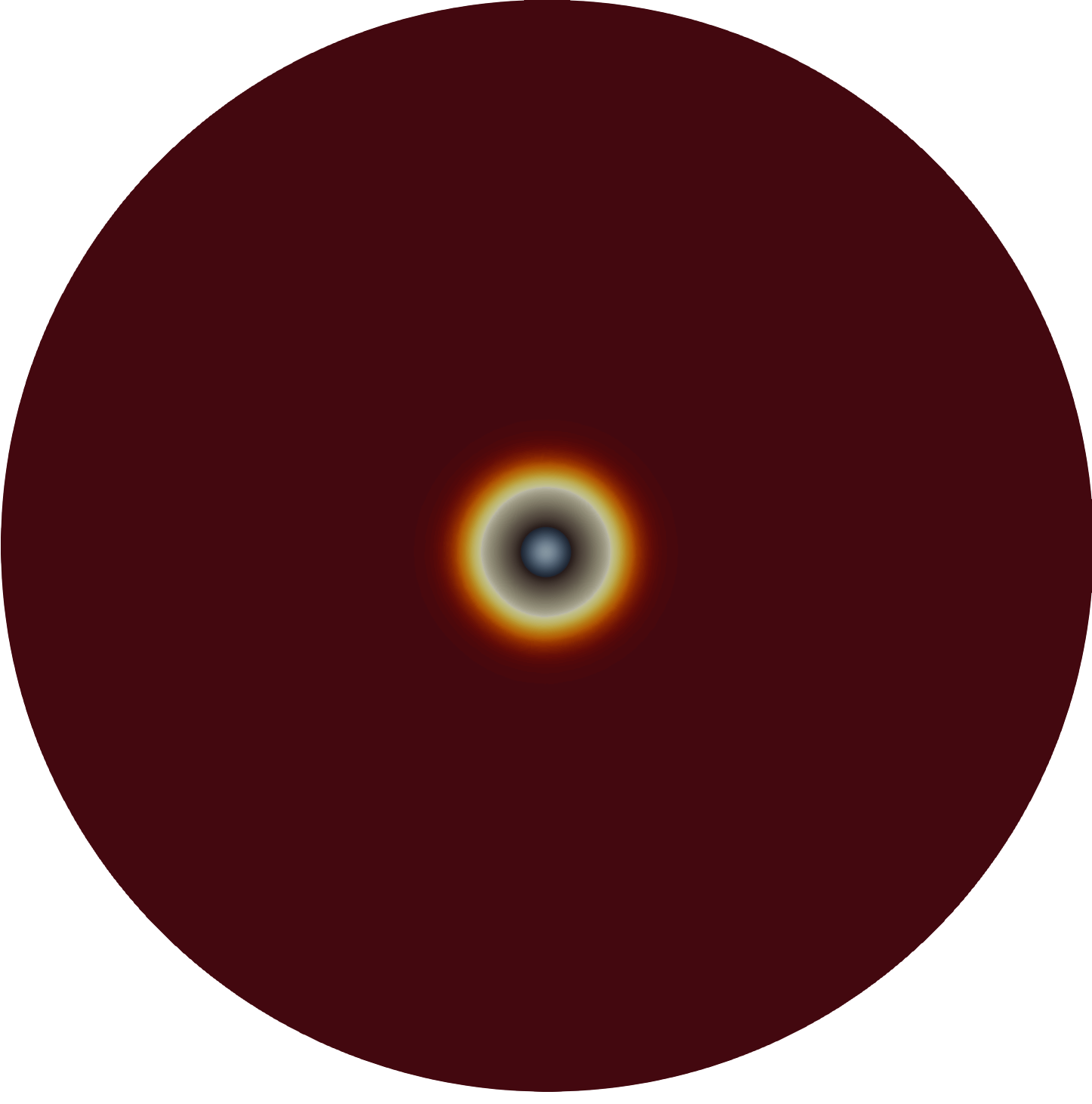}
\caption{$t=0$}
\end{subfigure}
\hfill
\begin{subfigure}[b]{0.24\textwidth}
\includegraphics[width=\textwidth]{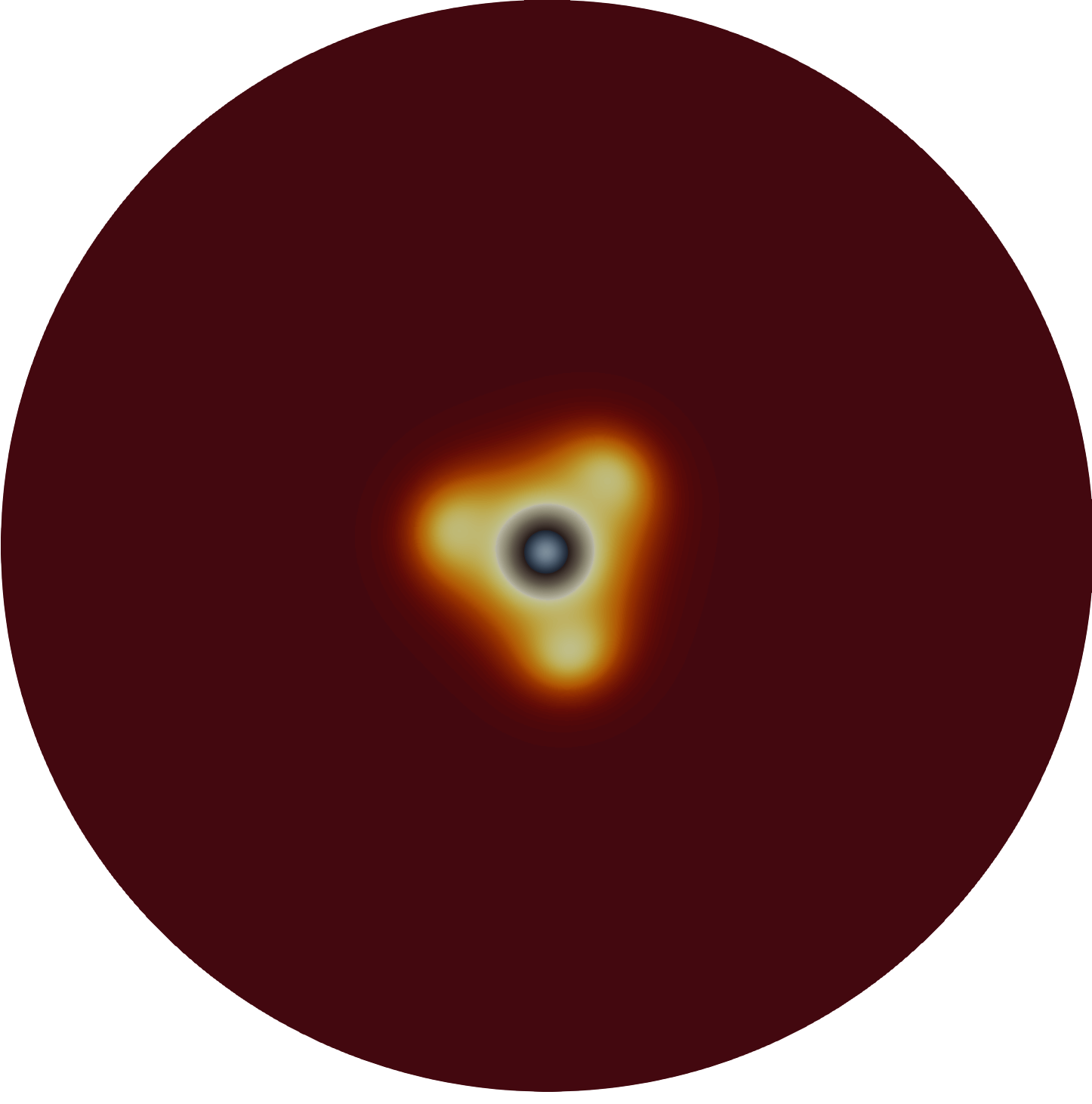}
\caption{$t=31$}
\end{subfigure}
\hfill
\begin{subfigure}[b]{0.24\textwidth}
\includegraphics[width=\textwidth]{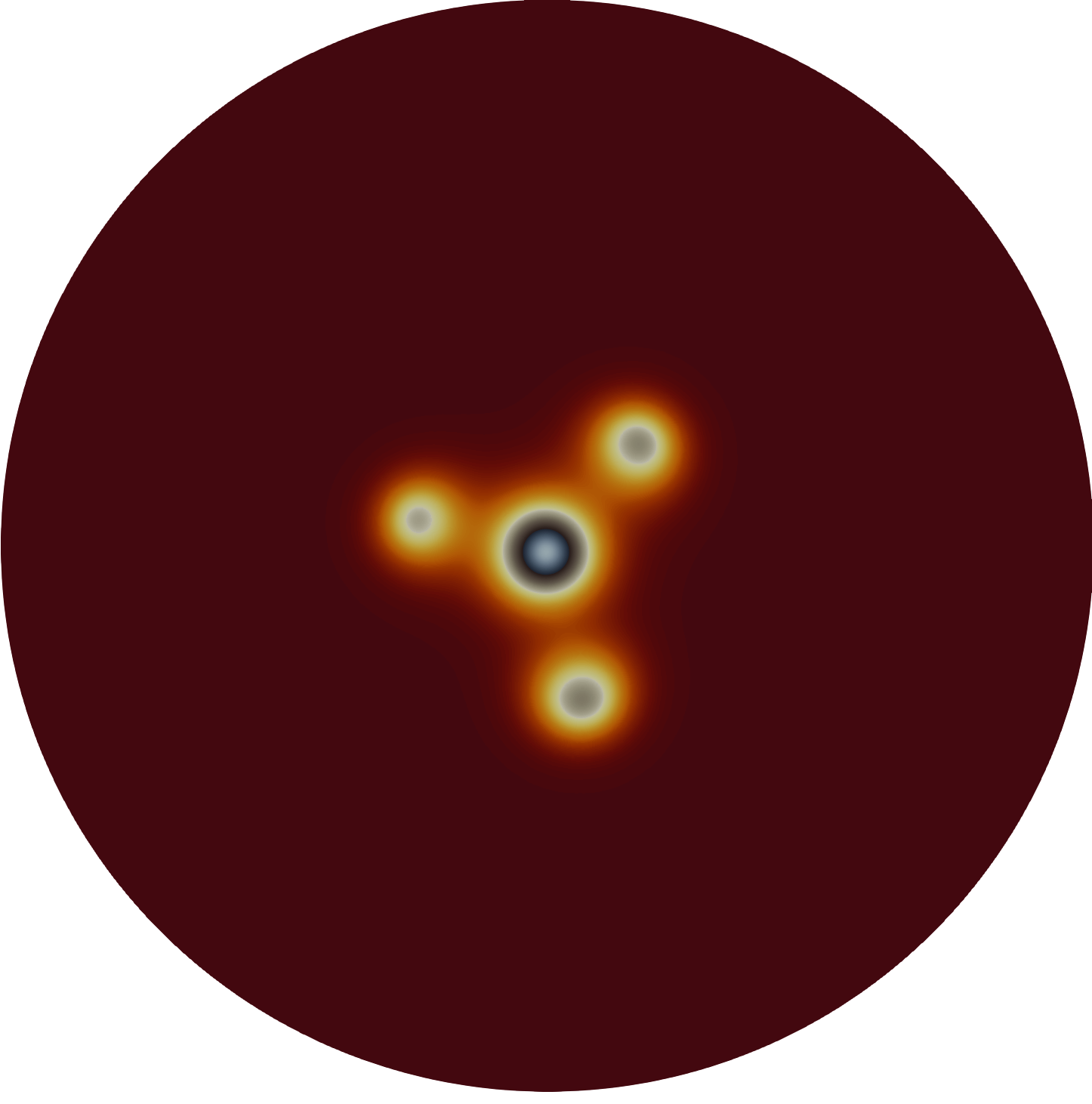}
\caption{$t=39$}
\end{subfigure}
\hfill
\begin{subfigure}[b]{0.24\textwidth}
\includegraphics[width=\textwidth]{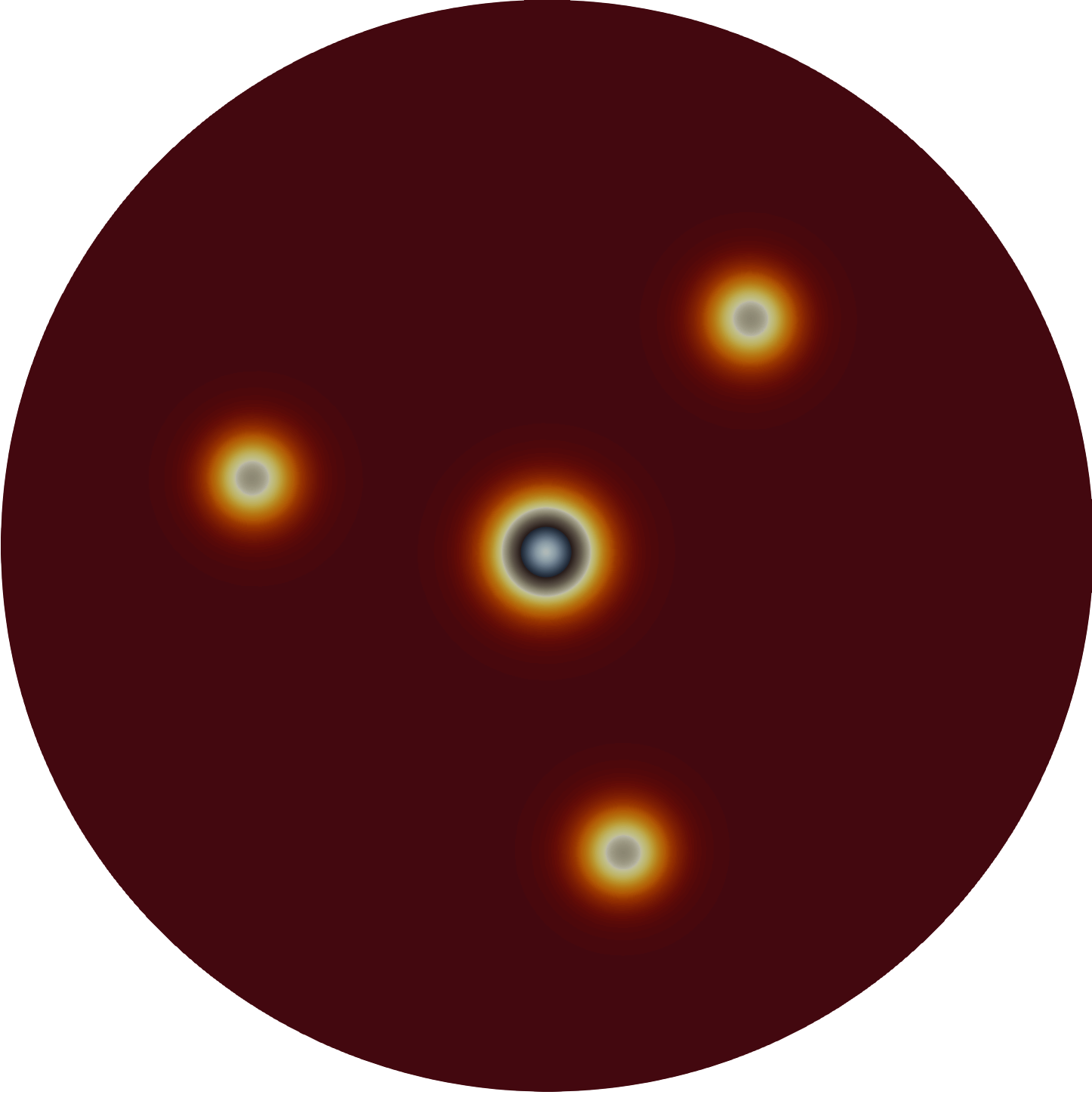}
\caption{$t=99$}
\end{subfigure}
\caption{Same caption as in Fig.~\ref{pinned_split_exp1} except that
  $a_0$ is increased further to $a_0=19$. The mode $m=3$ instability
  of the pinned spot now leads to three new
  spots.}\label{pinned_split_exp3}
\end{figure}

To illustrate this instability we compute full numerical solutions to
the PDE \eqref{pinned:model} in the unit disk with $\eps=0.03$,
$a_1=20$, and concentrated feed rate at the origin
$\pmb{\xi}=(0,0)^T$.  We choose $a_0=17$, and so from
\eqref{pinned:one_spot_source_strength} with $|\Omega|=\pi$ we get
$S_0=8.5$. From Fig.~\ref{pinned_lambda0_a1=20}, we observe that both
the $m=2$ and $m=3$ modes are unstable since
$S_0 = 8.5 > \Sigma_2(20) > \Sigma_3(20)$, with the $m=3$ mode having
the larger positive eigenvalue. In the numerical PDE results shown in
Fig.~\ref{pinned_split_exp1} at times $t=212$ and $t=231$, we observe
a mode $m=3$ instability for the pinned spot that triggers a nonlinear
spot-splitting process, but with ultimately only one new spot
surviving by time $t=300$. However, by increasing the value of $a_0$
to $a_0=18$ and $a_0=19$ for which $S_0=9$ and $S_0=9.5$, we observe
from Fig.~\ref{pinned_split_exp2} and Fig.~\ref{pinned_split_exp3},
respectively, that the most unstable mode $m=3$ mode can trigger the
creation of two or even three new spots by a nonlinear spot-splitting
event.

\subsubsection{Effect of a moving localized feed-rate}

\begin{figure}[htbp]
\includegraphics[width=0.5\textwidth,height=4.3cm]{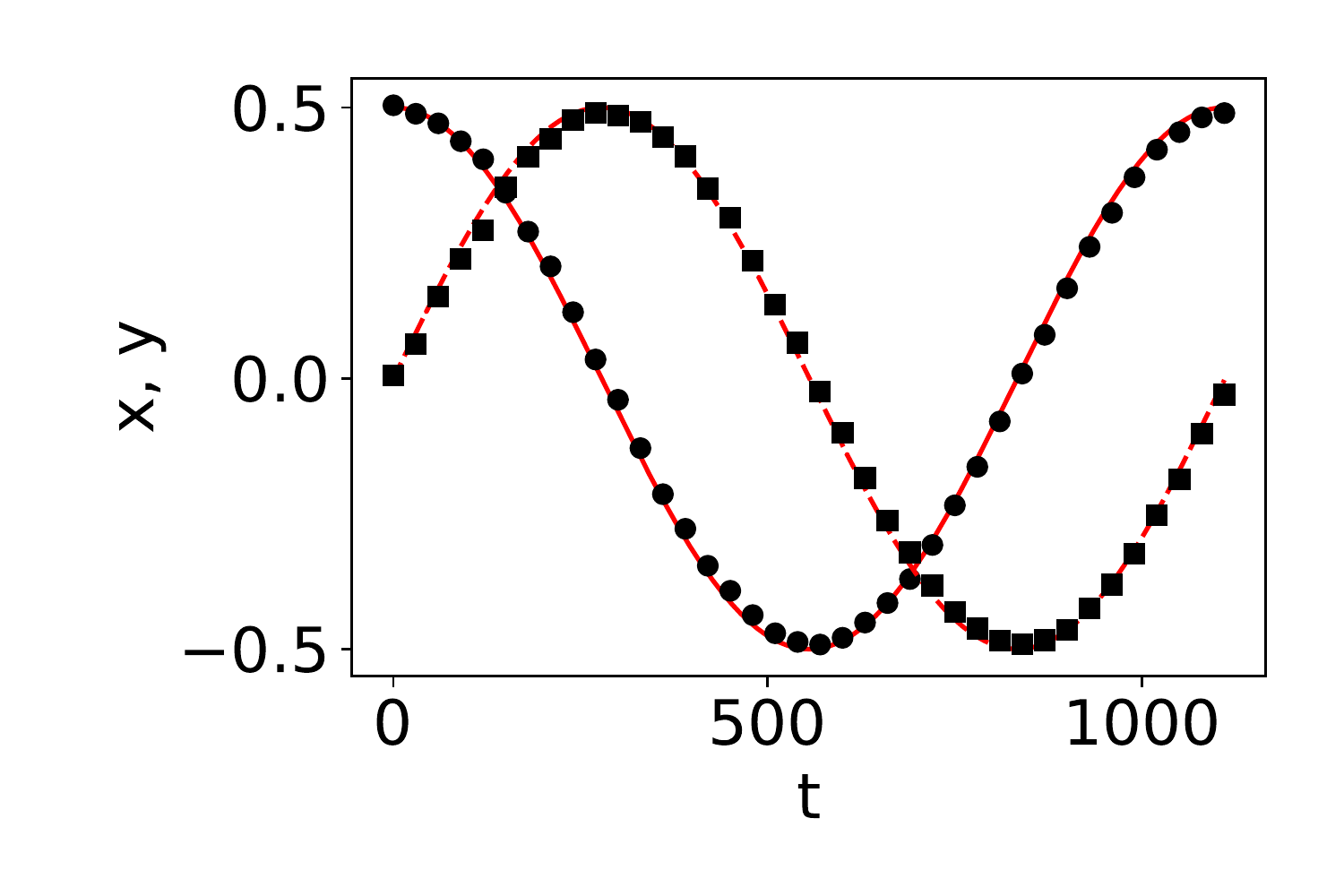}
\caption{The concentration point for the feed is rotating on the ring
  $\pmb{\xi}=(\xi_1(t),\xi_2(t))^T = 0.5(\cos(2\pi\eps^2 t),
  \sin(2\pi\eps^2 t))^T$ with $\eps=0.03$, and we choose $a_0=5$ and
  $a_1=8$. The $x$ and $y$ coordinates
  of the pinned spot, as computed numerically from the full PDE
  \eqref{pinned:model}, are shown by the black dots and square,
  respectively. We observe a close agreement between the spot
  coordinates and the coordinates $\xi_1(t)$ (solid line) and
  $\xi_2(t)$ (dashed line) of the concentration point $\pmb{\xi}$ of
  the feed rate.}
\label{fig:pinned_spot_exp1}
\end{figure}

We have shown in \S \ref{sec:het_one_spot} from the ODE
\eqref{het:DAE_one_spot} for slow spot dynamics that when a spot is
close enough to the concentration point $\pmb{\xi}$ for the feed rate,
it will get pinned to $\pmb{\xi}$ in finite time. This suggests that
if the concentration point $\pmb{\xi}$ is moving with time, the spot
will pursue $\pmb{\xi}$ and remain pinned, provided that the dynamics
of $\pmb{\xi}$ is slow enough. To examine this conjecture, we perform a
full PDE simulation of \eqref{pinned:model} for $a_1=8$, $a_0=5$, and
$\pmb{\xi} = \pmb{\xi}(\eps^2 t)$ with $\eps=0.03$, where we choose
\begin{equation}\label{pinned:trajectory}
  \pmb{\xi} = (\xi_1,\xi_2)^T \,, \quad \xi_1
  = 0.5 \cos(2\pi\eps^2 t) \,, \quad \xi_2 = 0.5 \sin(2\pi\eps^2 t) \,.
\end{equation}
In Fig.~\ref{fig:pinned_spot_exp1} we show that the trajectory of
the pinned spot aligns closely with the motion of the rotating
concentration point $\pmb{\xi}(\eps^2 t)$. This supports the
conjecture that a spot will follow the trajectory of the concentration
point of the feed rate.

\subsection{Quasi-equilibrium spot patterns with a pinned spot }\label{sec:pinned_quasi}

In this subsection we analyze the slow dynamics and linear stability
of quasi-equilibrium spot patterns that have a pinned spot, such
as shown in Fig.~\ref{pinned_split_exp1}--\ref{pinned_split_exp3}.

\subsubsection{Quasi-equilibria and slow spot dynamics}\label{sec:quasi_allpin}

We construct a quasi-equilibrium spot pattern, with $N$ spots centered
at $\v{x}_j\in\Omega$ for $j = 1,\ldots,N$, and with an additional pinned
spot at the concentration point $\pmb{\xi}\in\Omega$ of the feed
rate. We assume that the spots and the pinned-spot are well-separated
in the sense that
\begin{equation}\label{pinned:assumption}
  |\v{x}_i - \v{x}_j| = \bigo(1) \,, \quad i\neq j \,, \qquad
  |\v{x}_i - \pmb{\xi}| = \bigo(1) \,, \quad i = 1,\ldots,N \,.
\end{equation}
Near the $j^{\text{th}}$ spot centered at $\v{x}_j$, for
$j=1,\ldots,N$, we substitute the expansion
\eqref{proto:inner_expansion} with $D=1$ into
\eqref{pinned:model}. Due to the assumption \eqref{pinned:assumption},
the term $\Phi\left(|\v{x}_j-\pmb{\xi}|/\eps\right)$ is exponentially
small as $\eps\to 0$, and therefore absent to all algebraic orders in
$\eps$. We retrieve the core problem \eqref{proto:core_full} and the
integration identity \eqref{proto:Sj}. Likewise, near the the pinned
spot at $\pmb{\xi}$, we substitute $v \sim V_0(\rho)$ and
$u \sim U_0(\rho)$ into \eqref{pinned:model} to obtain the new core
problem \eqref{pinned:core_full}. Upon using the distributional limits
\eqref{proto:correspondence_rule} (with $D=1$) and
\eqref{pinned:correspondence_rule}, the outer problem for $u$, defined
away from all the spots, is
\begin{equation}\label{pinned:outer_problem2}
  \Delta u + a_0 - 2 \pi \sum \limits_{i=1}^N S_i \, \delta(\v{x}-\v{x}_j) -
  2 \pi S_0 \, \delta(\v{x}-\pmb{\xi}) = 0 \,, \quad \mbox{in} \quad
  \Omega \,, \qquad \partial_n u = 0 \,, \quad \mbox{on} \quad
  \partial\Omega \,.
\end{equation}
In terms of the Neumann Green's function of \eqref{proto:neu_green}, the
solution to \eqref{pinned:outer_problem2} is
\begin{equation}\label{pinned:outer_solution}
  u = - 2\pi S_0\, G(\v{x};\pmb{\xi}) -2\pi \sum\limits_{i=1}^N S_i \,
  G(\v{x};\v{x}_i)  + \ubar \,,
\end{equation}
where $\ubar$ is an undetermined constant. By using the divergence
theorem on \eqref{pinned:outer_problem2}, we conclude that
\begin{equation}\label{pinned:sum_Sj}
\sum\limits_{i=0}^N S_i = {a_0|\Omega|/ (2\pi)} \,.
\end{equation}

Next, we let $\v{x} \to \v{x}_j$, for $j=1,\ldots,N$, in
\eqref{pinned:outer_solution} to obtain that
\begin{equation}\label{pinned:outer_expansion2}
\begin{split}
  u &\sim S_j \log|\v{x}-\v{x}_j| - 2\pi \left(S_j R_{j,j} + 
    S_0 G(\v{x}_j;\pmb{\xi}) + \sum_{i\neq j}^{N} S_i \, G_{j,i} \right)
    + \ubar \\
    & \quad - 2\pi \left( S_j \nabla_{\v{x}} R_{j,j} +
      S_0 \nabla_{\v{x}} G(\v{x}_j;\pmb{\xi}) +
      \sum_{i\neq j}^{N} S_i \nabla_{\v{x}} G_{j,i} \right) \cdot (\v{x}-\v{x}_j)
    +  {\mathcal O}(|\v{x}-\v{x}_j|^2) \,, \qquad j=1,\ldots, N \,,
\end{split}
\end{equation}
where $R_{j,j} \equiv R(\v{x}_j; \v{x}_j)$ and
$G_{j,i} \equiv G(\v{x}_j;\v{x}_i)$. Upon matching the
${\mathcal O}(1)$ terms in \eqref{pinned:outer_expansion2} with the
far-field behavior \eqref{proto:core_problem_far_field} of the leading
order core solution, we find that
\begin{equation}\label{pinned:sj}
  S_j + 2\pi\nu\left(S_j R_{j,j} + S_0 G(\v{x}_j;\pmb{\xi}) +
    \sum\limits_{i\neq j}^N S_i \,
    G_{j,i}\right) + \nu\chi(S_j) = \nu\ubar \,, \qquad j=1\,,\ldots,N \,.
\end{equation}
Then, we expand \eqref{pinned:outer_solution} as $\v{x} \to \pmb{\xi}$
to get
\begin{equation}\label{pinned:outer_expansion3}
  u \sim S_0 \log|\v{x}-\pmb{\xi}| - 2 \pi S_0 \, R_{0,0} -
  2 \pi \sum\limits_{i=1}^N S_i \, G_{j,0} + \ubar  +
  {\mathcal O}(|\v{x}-\v{x}_j|) \,,
\end{equation}
where $R_{0,0} \equiv R(\pmb{\xi};\pmb{\xi})$ and
$G_{j,0} \equiv G(\v{x}_j;\pmb{\xi})$.  Upon matching
\eqref{pinned:outer_expansion3} with the far-field behavior
\eqref{pinned:core_problem_far_field} of the new core problem, we
conclude that
\begin{equation}\label{pinned:s0}
  S_0 + 2\pi\nu\left( S_0 R_{0,0} + \sum\limits_{i=1}^N S_i \, G_{j,0} \right) +
  \nu \chi(S_0; a_1) = \nu \ubar \,,
\end{equation}

Next, we write \eqref{pinned:sj}, \eqref{pinned:s0}, and
\eqref{pinned:sum_Sj} in matrix form as
\begin{subequations}\label{pinned:all_nas}
\begin{equation}\label{pinned:s0sj_nas}
  \v{s} + 2\pi\nu \, \Gmat\v{s} + \nu \, \pmb{\chi} = \nu\ubar\,\v{e} \,, \qquad
  \v{e}^T \v{s} = p_a \equiv \frac{a_0|\Omega|}{2\pi} \,,
\end{equation}
where we have defined
\begin{equation}\label{pinned:s0sj_def}
  \v{s} \equiv (S_0, S_1, \ldots, S_N)^T \,, \qquad \chivec \equiv
  \left(\chi(S_0;a_1), \chi(S_1), \ldots, \chi(S_N) \right)^T \,, \qquad
  \v{e} \equiv (1,\ldots,1)^T \in \mathbb{R}^{N+1} \,.
\end{equation}
\end{subequations}
Here $\chi(S_0;a_1)$ is defined by the new core problem
\eqref{pinned:core_full} for the pinned spot, while
$\Gmat \in \mathbb{R}^{(N+1)\times(N+1)}$ is the Neumann Green's
matrix of $\pmb{\xi}, \v{x_1}, \ldots, \v{x}_N$. Upon eliminating
$\ubar$ in \eqref{pinned:s0sj_def}, we obtain that the nonlinear
algebraic system for the vector $\v{s}$ of source strengths is
\begin{equation}\label{pinned:nas_all}
  \v{s} + 2\pi\nu(\Imat - \Emat)\Gmat\v{s} + \nu(\Imat - \Emat)\pmb{\chi} =
  \frac{p_a}{N+1}\v{e} \,, \qquad \mbox{with} \qquad 
  \ubar = \frac{p_a + 2\pi\nu \, \v{e}^T \Gmat \v{s} + \nu \, \v{e}^T
    \pmb{\chi}}{\nu(N+1)} \,.
\end{equation}
Here $\Emat=N^{-1}\v{e}\v{e}^T\in \mathbb{R}^{(N+1)\times(N+1)}$ and
$\Imat\in \mathbb{R}^{(N+1)\times(N+1)}$ is the identity matrix.

To derive the DAE system for slow spot dynamics we must match
\eqref{proto:matching_inner} (setting $D=1$) with
\eqref{pinned:outer_expansion2} for the $\bigo(\eps)$ gradient
terms. This matching condition yields the far-field behavior for the
inner correction term $U_{j1}$ in \eqref{proto:inner_expansion}:
\begin{equation}
  U_{j1} \sim -2 \pi \left[ S_j \nabla_\v{x} R_{j,j} +
    \sum\limits_{i\neq j}^N S_i \nabla_{\v{x}} G_{j,i} + S_0 \nabla_{\v{x}}
    G(\v{x}_j;\pmb{\xi}) \right] \cdot \v{y} \, \qquad \mbox{as}
  \quad  |\v{y}|\to \infty \,,
\end{equation}
where $\v{y} = \eps^{-1} (\v{x}-\v{x}_j)$. Following the derivation
in \S \ref{sec:proto_qe_slow}, we obtain that the DAE system for slow
spot dynamics is
\begin{equation}\label{pinned:DAE}
  \frac{d\v{x}_j}{d\sigma} = - \gamma(S_j) \left( \, \pmb{\beta}_j +
  2\pi S_0 \nabla_\v{x} G(\v{x}_j;\pmb{\xi}) \right) \,, \qquad j=1,\ldots, N\,,
\end{equation}
where $\sigma=\eps^2t$. Here, $\pmb{\beta}_j$ and $\gamma(S_j)$ are defined in
\eqref{proto:betaj} and \eqref{proto:DAE}, respectively, while
$\v{s}\equiv (S_0,S_1,\ldots,S_N)^T$ satisfies the nonlinear algebraic
system \eqref{pinned:nas_all}.

\begin{figure}[htbp]
\begin{subfigure}[b]{0.35\textwidth}
\includegraphics[width=\textwidth,height=4.2cm]{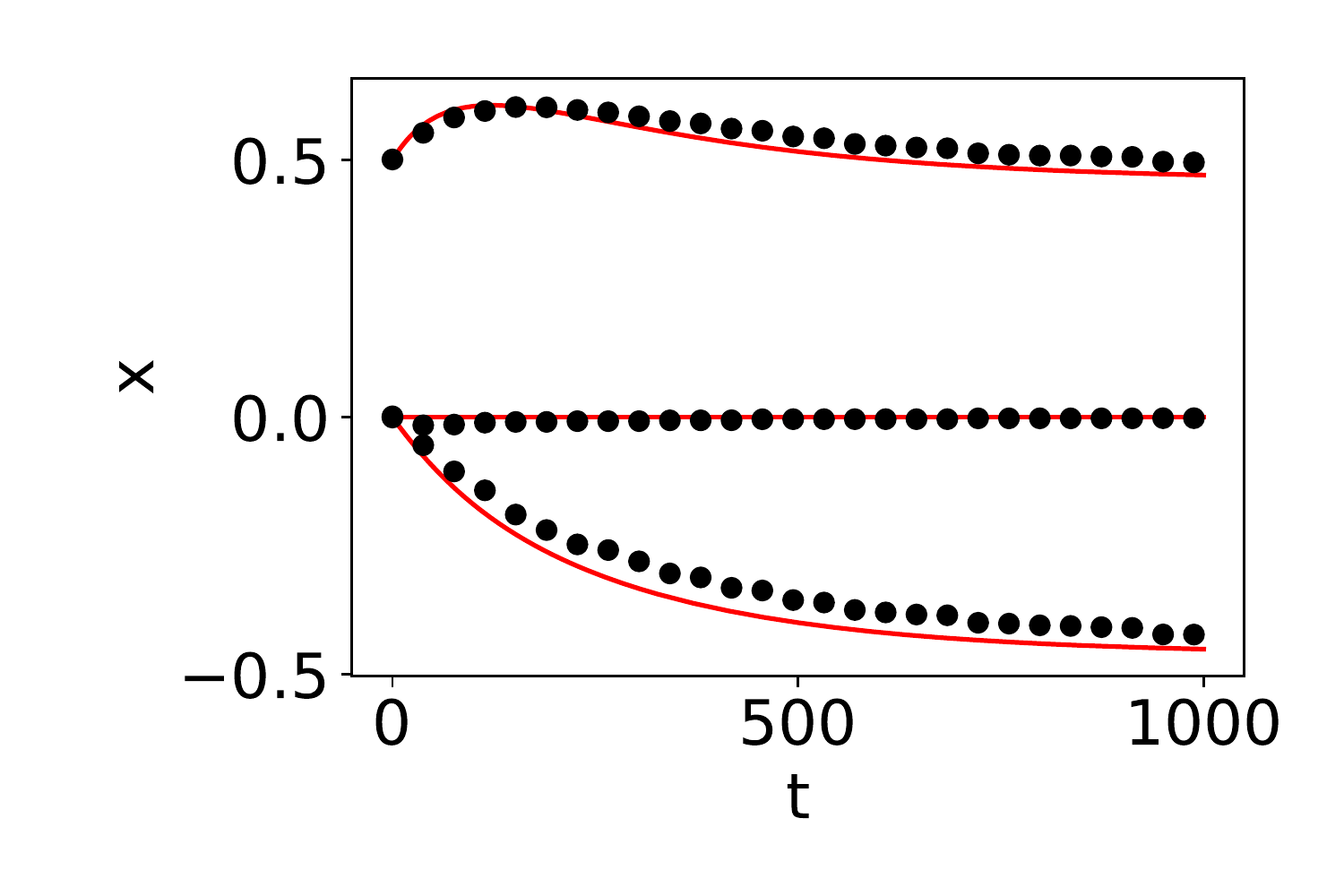}
\caption{x-coordinate}
\end{subfigure}
\hfill
\begin{subfigure}[b]{0.35\textwidth}
\includegraphics[width=\textwidth,height=4.2cm]{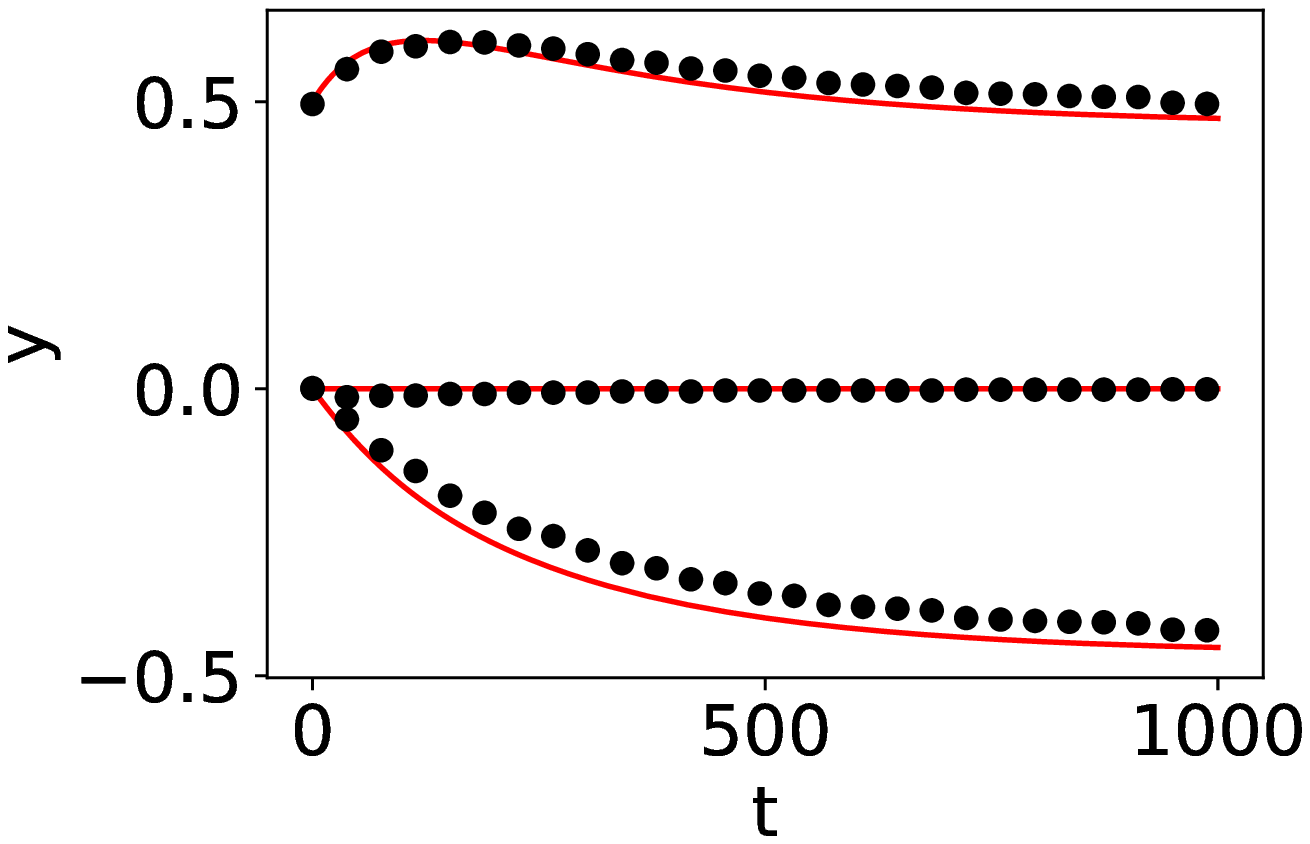}
\caption{y-coordinate}
\end{subfigure}
\hfill
\begin{subfigure}[b]{0.25\textwidth}
\includegraphics[width=\textwidth,height=4.2cm]{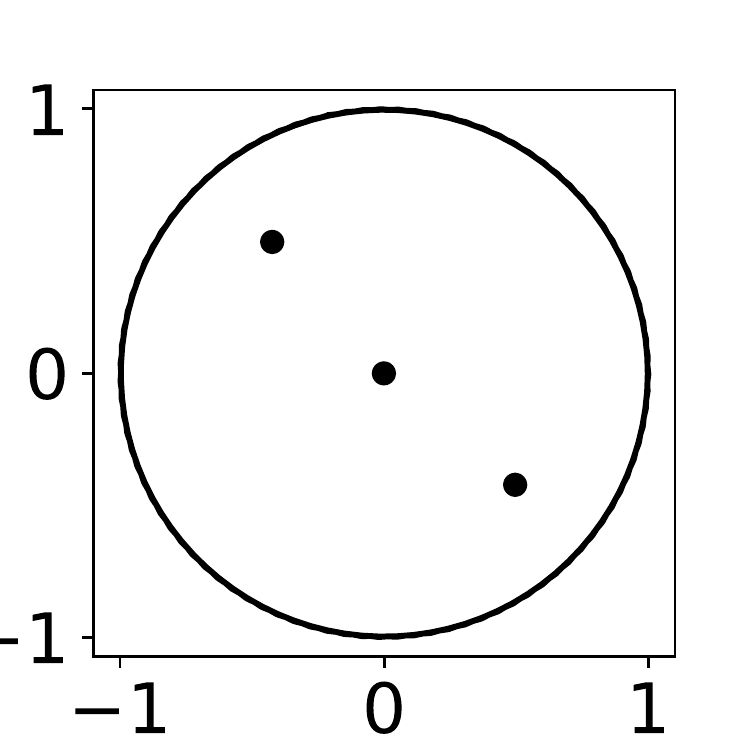}
\caption{$t=1000$}
\label{pinned_slow_exp3_t=1000}
\end{subfigure}
\caption{Left and middle panels: The spot trajectories for an initial
  quasi-equilibrium pattern with spots centered at $(0.5,0.0)^T$ and
  $(0.0,0.5)^T$, and with a pinned spot centered at the origin
  $\pmb{\xi}=\v{0}$. The full PDE results from \eqref{pinned:model}
  and DAE dynamics \eqref{pinned:DAE} and \eqref{pinned:nas_all} are
  represented by the black dots and red solid line, respectively.
  Right panel: spot locations at $t=1000$ form a colinear pattern (near
  the steady-state). Parameters are $\eps=0.03$, $a_0=15$, and $a_1=5$.}
\label{pinned_slow_exp3}
\end{figure}

We now compare the DAE dynamics \eqref{pinned:DAE} and
\eqref{pinned:nas_all} with full numerical results computed from the
PDE \eqref{pinned:model} in the unit disk. We set $\eps=0.03$, and for
the localized feed rate we choose $a_0=15$, $a_1=5$, and
$\pmb{\xi} = \v{0}$. The initial quasi-equilibrium pattern has a
pinned spot at the origin $\pmb{\xi}=\v{0}$, and two additional spots
centered at $(0.5,0)^T$ and $(0,0.5)^T$. As shown in
Fig.~\ref{pinned_slow_exp3_t=1000}, the pinned spot remains at the
origin while the other two spots move apart to form an almost colinear
pattern. The spot trajectories computed from the full PDE simulation
agree well with those from the DAE system.

\begin{figure}[htbp]
\begin{subfigure}[b]{0.45\textwidth}
\includegraphics[width=\textwidth,height=4.3cm]{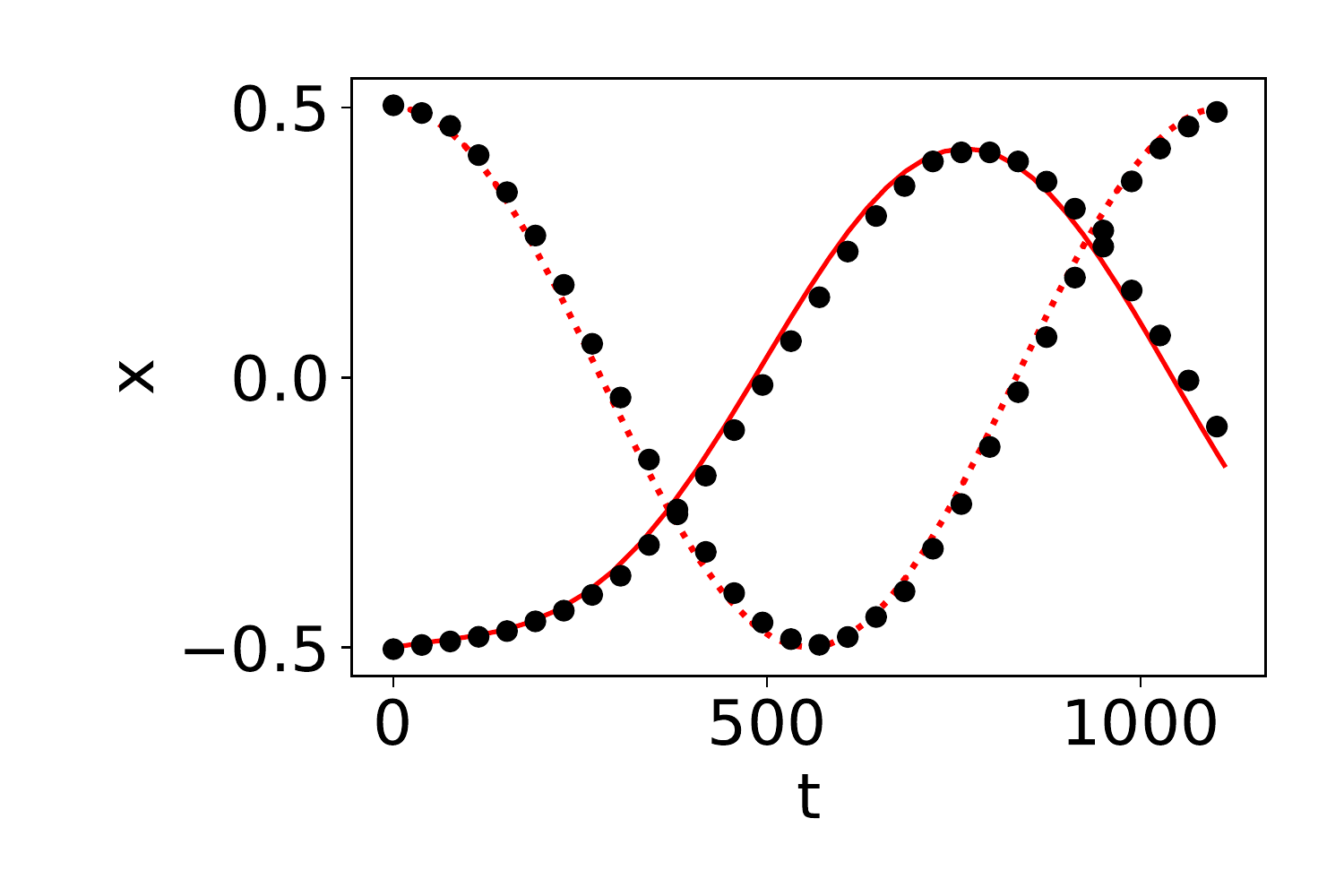}
\caption{x-coordinate}
\end{subfigure}
\hfill
\begin{subfigure}[b]{0.45\textwidth}
\includegraphics[width=\textwidth,height=4.3cm]{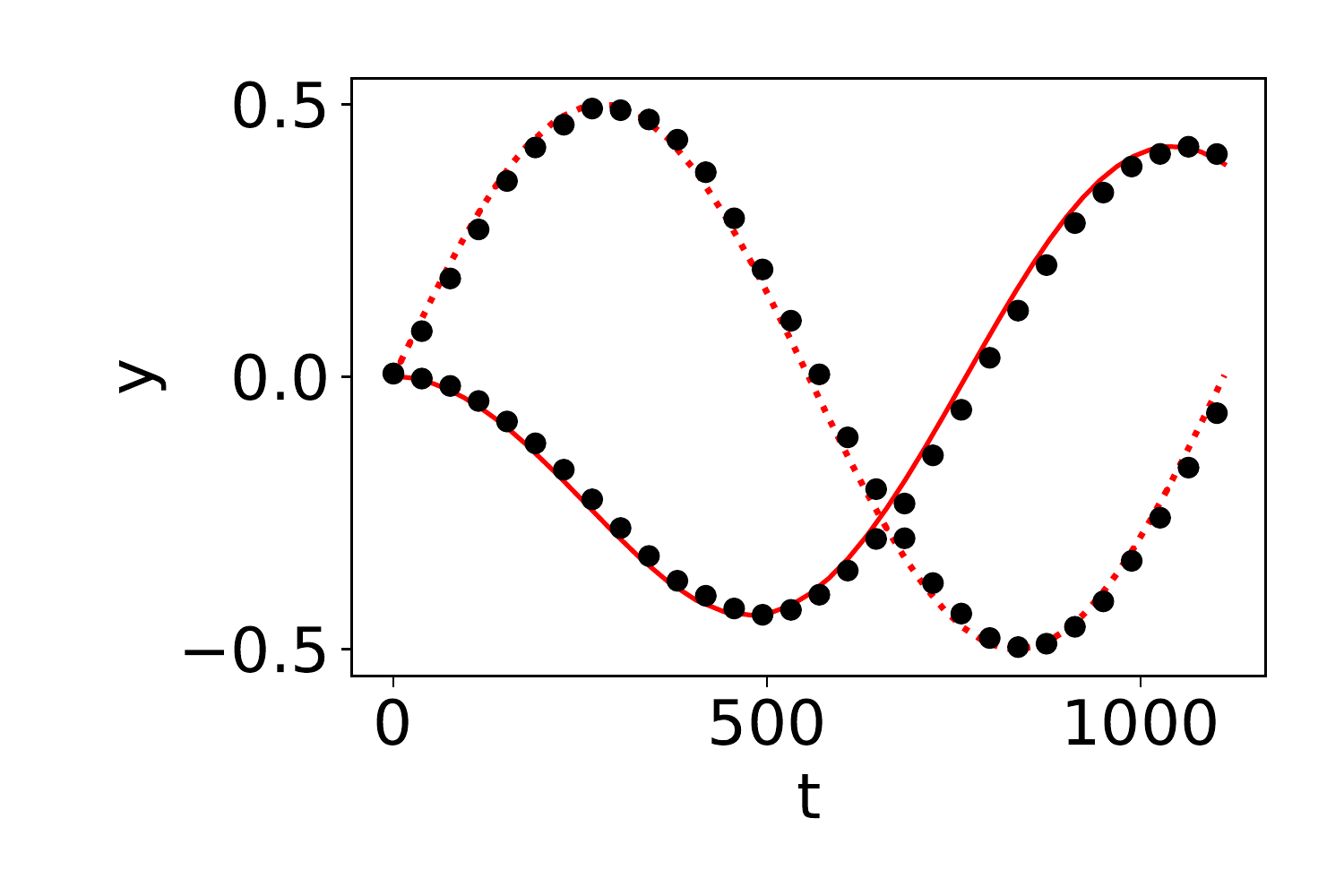}
\caption{y-coordinate}
\end{subfigure}
\caption{A two-spot quasi-equilibrium pattern with the moving
  concentration point $\pmb{\xi}(\eps^2 t)$ of the feed rate given in
  \eqref{pinned:trajectory}, and with $\eps=0.03$, $a_0=10$, and
  $a_1=5$. At time $t=0$, there is a pinned spot centered at
  $\pmb{\xi}(0)=(0.5,0.0)^T$ and an additional spot at $(-0.5,0.0)^T$. The
  solid red curve is the spot trajectory from the DAE dynamics
  \eqref{pinned:DAE} and \eqref{pinned:nas_all}. The dashed red curve
  is the moving concentration point $\pmb{\xi}(\eps^2 t)$.  The black
  dots are the locations of the spot and pinned-spot, as computed
  numerically from the PDE \eqref{pinned:model}.}
\label{pinned_slow_exp2}
\end{figure}

For our second experiment in the unit disk, we set $\eps=0.03$ and
consider a localized feed rate with $a_0 = 10$ and $a_1=5$, where the
concentration point $\pmb{\xi}$ moves slowly in time according to
\eqref{pinned:trajectory}. At time $t=0$ the quasi-equilibrium pattern
consists of the pinned spot centered at $\pmb{\xi}(0)=(0.5,0)^T$ with an
additional spot centered at $(-0.5,0)^T$. In Fig.~\ref{pinned_slow_exp2}
we show a favorable comparison between the spot trajectories obtained
from the DAE dynamics \eqref{pinned:DAE} and \eqref{pinned:nas_all}
and from the full PDE computations of \eqref{pinned:model} on
$0<t<1000$.  We observe that the initial pinned-spot remains pinned as
time increases and moves with $\pmb{\xi}(t)$ along a circular
trajectory. The other spot moves along a nearly circular trajectory in
the unit disk.

\subsubsection{Linear stability analysis}\label{sec:linstab_allpin}

We now analyze the linear stability of a quasi-equilibrium pattern
$v_e$ and $u_e$ that consists of spots centered at
$\v{x}_1,\ldots,\v{x}_N$ with an additional pinned spot at
$\pmb{\xi}$, for which the source strengths are $S_1,\ldots,S_N$ and
$S_0$, respectively. For instabilities associated with non-radially
symmetric perturbations near the spots, our previous results in \S
\ref{sec:proto_linstab} and \S \ref
{sec:pinned_linstab_one_pinned_spot} have shown that the
quasi-equilibrium pattern is linearly stable to symmetry breaking
bifurcations in the spot profiles only when
\begin{equation}\label{allpin:nonrad}
  S_j<\Sigma_2(0)\approx 4.302\,, \quad \mbox{for} \quad j=1,\ldots,N
  \qquad \mbox{and} \qquad S_0 <\min_{m\geq 2} \Sigma_m(a_1) \,.
\end{equation}
Here the symmetry-breaking stability threshold $\Sigma_m(a_1)$ for the
local angular mode $m$ was defined in \S
\ref{sec:pinned_linstab_one_pinned_spot} (see
Fig.~\ref{pinned_lambda0}).

As such, we will focus only on deriving a new GCEP associated with any
instabilities due to locally radiallly symmetric perturbations near
the spots. Upon substituting $v = v_e + e^{\lambda t} \phi$ and
$u = u_e + e^{\lambda t} \eta$ into \eqref{pinned:model}, we linearize
to get
\begin{equation}\label{allpin:full_eig}
  \eps^2 \Delta \phi - \phi + 2 u_e v_e \phi + v_e^2 \eta = \lambda \phi \,,
  \qquad  \Delta \eta - \eps^{-2}\left(2 u_e v_e \phi + v_e^2 \eta\right) =
  \lambda \eta \,, \quad \mbox{in} \quad \Omega \,,
\end{equation}
with $\partial_n \phi=\partial_n\eta=0$ on $\partial\Omega$. From the
leading-order construction of the quasi-equilibrium pattern in \S
\ref{sec:quasi_allpin}, we have that
\begin{equation}\label{pinned:qe_inner}
v_e \sim \begin{cases}
  V_{0}\left(\eps^{-1}|\v{x}-\pmb{\xi}|\right) \,, &\quad \mbox{near}\,\,
  \pmb{\xi} \,, \\[3pt]
V_{j0}\left(\eps^{-1}|\v{x}-\v{x}_j|\right) \,, &\quad \mbox{near}\,\, \v{x}_j \,,
\end{cases}
\qquad \mbox{and} \qquad
u_e \sim \begin{cases}
  U_{0}\left(\eps^{-1}|\v{x}-\pmb{\xi}|\right) \,, &\quad \mbox{near}\,\,
  \pmb{\xi} \,, \\[3pt]
U_{j0}\left(\eps^{-1}|\v{x}-\v{x}_j|\right) \,, &\quad \mbox{near}\,\, \v{x}_j \,,
\end{cases}
\end{equation}
Here $(V_{j0},U_{j0})$ is the solution to the core problem
\eqref{proto:core_full} while $(V_0,U_0)$ is the solution to the new
core problem \eqref{pinned:core_full} near the pinned spot, which
depends on the feed intensity parameter $a_1$.

In the inner region near a spot at $\v{x}_j$, for $j=1,\ldots,N$, we
let $\phi \sim c_j \tilde{\Phi}_j(\rho)$ and
$\eta \sim c_j \tilde{N}_j(\rho)$ in
\eqref{allpin:full_eig}, where $\rho = \eps^{-1}|\v{x}-\v{x}_j|$. Upon
using \eqref{pinned:qe_inner}, we retrieve the inner problem
\eqref{proto:m=0_eigenproblem} for $\tilde{\Phi}_j$ and
$\tilde{N}_j$ for each $j=1,\ldots,N$.  Similarly, by setting
$\phi \sim c_0 \tilde{\Phi}_0(\rho)$ and
$\eta \sim c_0 \tilde{N}_0(\rho)$ in
\eqref{allpin:full_eig}, where $\rho = \eps^{-1}|\v{x}-\pmb{\xi}|$,
we obtain the following inner problem for the pinned spot:
\begin{subequations}\label{pinned:m=0_eigenproblem}
\begin{align}
  \Delta_\rho \hat{\Phi}_0 &- \hat{\Phi}_0 + 2 U_0 V_0 \hat{\Phi}_0 +
    V_0^2 \hat{N}_0 = \lambda \hat{\Phi}_0 \,, \qquad
    \Delta_\rho \hat{N}_0 - 2 U_0 V_0 \hat{\Phi}_0 - V_0^2 \hat{N}_0 = 0 \,,
  \quad \rho>0 \,, \\
  &\hat{\Phi}_0^{\prime}(0)=\hat{N}_0^{\prime}(0)=0\,; \qquad \hat{\Phi}_0 \to 0
    \,, \quad \hat{N}_0 \sim \log \rho + \Bhat(S_j; a_1, \lambda) + o(1)
    \,, \quad \mbox{as} \quad \rho \to \infty \,.
\end{align}
\end{subequations}
Here $\Bhat(S_j; a_1, \lambda)$ depends on the feed intensity $a_1$
through the pinned core solution $(V_0,U_0)$. By differentiating
\eqref{pinned:core_full} with respect to $S_0$, and then comparing the
resulting system with \eqref{pinned:m=0_eigenproblem} when
$\lambda=0$, we identify
$\Bhat(S_0;a_1,0) = \partial_{S_0}\chi(S_0;a_1)$.

As in \S \ref{sec:proto_linstab}, to formulate the outer problem for
$\eta$ we first derive the distributional limit
\begin{equation*}
  \eps^{-2}\left(2 u_e v_e \phi + v_e^2 \eta\right) \to 2\pi
  c_0 \, \delta(\v{x}-\pmb{\xi}) + 2\pi \sum\limits_{i=1}^N c_i \,
  \delta(\v{x}-\v{x}_i) \,,
\end{equation*}
as $\eps\to 0$. By using this limit in \eqref{allpin:full_eig}, and by
enforcing the asymptotic matching condition to the inner solutions
near the spots, we obtain that the outer problem for $\eta$, defined
away from the spots, is
\begin{subequations}\label{pinned:linstab_allouter}
\begin{align}
  \Delta \eta - &\lambda\eta - 2\pi \left[ \, \sum\limits_{i=1}^N c_i \,
    \delta(\v{x}-\v{x}_i) + c_0 \, \delta(\v{x}-\pmb{\xi}) \, \right] = 0
  \quad \mbox{in} \quad \Omega \,, \qquad \partial_n \eta = 0 \quad
   \mbox{on} \quad \partial\Omega \,. \label{pinned:linstab_outer_problem}\\
  \eta &\sim  c_0 \left(\log |\v{x}-\pmb{\xi}| + {1/\nu} +
         \Bhat(S_0; a_1, \lambda) \right) \,, \quad \mbox{as} \quad \v{x}\to
         \pmb{\xi} \,, \label{pinned:match_0}\\
  \eta &\sim c_j \left(\log |\v{x}-\v{x}_j| + {1/\nu} +
         \Bhat(S_j; \lambda) \right) \,, \quad \mbox{as} \quad \v{x}\to
         \v{x}_j \,, \quad j = 1, \ldots, N \,. \label{pinned:match_j}
\end{align}
\end{subequations}

For $\lambda \neq 0$, the solution to
\eqref{pinned:linstab_outer_problem} is represented as
\begin{equation}\label{pinned:linstab_outer_sol}
  \eta = -2\pi c_0 \, G_\lambda(\v{x};\pmb{\xi})   - 2\pi
  \, \sum\limits_{i=1}^N c_i \, G_\lambda(\v{x};\v{x}_i) \,,
\end{equation}
where $G_\lambda$ is the eigenvalue-dependent Green's function defined
by \eqref{proto:eig_green}. By matching the near-field behavior of
\eqref{pinned:linstab_outer_sol} as $\v{x}\to\pmb{\xi}$ and as
$\v{x}\to\v{x}_j$, for $j=1,\ldots,N$, to the required singularity behavior
in \eqref{pinned:match_0} and \eqref{pinned:match_j}, respectively, we
derive a new GCEP for $\v{c} \equiv (c_0, c_1, \ldots, c_N)^T$, which we
write in matrix form as
\begin{subequations}\label{pinned:gcep_all}
\begin{equation}\label{pinned:gcep_matrix}
  \Mmat \v{c} \equiv \v{0} \qquad \mbox{where} \qquad
   \Mmat \equiv \Imat + 2\pi\nu\Gmatl + \nu\Bmat\,.
\end{equation}
Here the entries of the Green's matrix
$\mc{G}_\lambda \in \mathbb{R}^{(N+1)\times(N+1)}$ and the diagonal
matrix $\Bmat \in \mathbb{R}^{(N+1)\times(N+1)}$ are given by
\begin{equation}\label{pinned:gcep_matrix_def}
  \begin{split}
  \left(\Gmatl\right)_{i+1\, j+1} &= \begin{cases}
G_\lambda(\v{x}_i; \v{x}_j) \quad & i \neq j \,, \\
R_\lambda(\v{x}_j;\v{x}_j) \quad & i = j \,,
\end{cases}  \qquad \mbox{for} \quad i,j=0,\ldots,N \,, \\
   (\Bmat)_{1\,1} &= \Bhat(S_0; a_1, \lambda) \,, \qquad
   (\Bmat)_{j+1\,j+1} = \Bhat(S_{j};\lambda) \,, \quad \mbox{for} \quad
   j = 1, \ldots, N \,,
\end{split}
\end{equation}
\end{subequations}
where, for convenience of notation, we have defined
$\v{x}_0 \equiv \pmb{\xi}$. We conclude that the $N$-spot
quasi-equilibrium solution with an additional pinned spot at
$\pmb{\xi}$ is linearly stable on ${\mathcal O}(1)$ time-scales to
locally radially symmetric perturbations near the spots when there
is no root in $\mbox{Re}(\lambda)>0$ to
\begin{equation}\label{pinn:gcep_det}
  \det \Mmat(\lambda)=0 \,.
\end{equation}

Next, we formulate the GCEP for zero-eigenvalue crossings where
$\lambda = 0$ in \eqref{pinned:linstab_allouter}. For $\lambda=0$,
the solution to \eqref{pinned:linstab_outer_problem} is
\begin{equation}\label{pinned:linstab_outer_sol2}
  \eta = -2 \pi c_0 \, G(\v{x};\pmb{\xi})  - 2\pi \sum\limits_{i=1}^N c_i \,
  G(\v{x};\v{x}_j) + \bar{\eta} \,,
\end{equation}
where $G$ is the Neumann Green's function of \eqref{proto:neu_green}, and
$\bar{\eta}$ is an undetermined additive constant. By applying the
divergence theorem to \eqref{pinned:linstab_outer_problem} we obtain
that $\v{e}^T \v{c} = 0$. Then, by matching the near-field behavior of
\eqref{pinned:linstab_outer_sol2} as $\v{x}\to\pmb{\xi}$ and as
$\v{x}\to\v{x}_j$, for $j=1,\ldots,N$, to the required singularity behavior
in \eqref{pinned:match_0} and \eqref{pinned:match_j}, respectively, and
by recalling the identities $\Bhat(S_0;a_1,0) = \partial_{S_0}\chi(S_0;a_1)$ and
$\Bhat(S_j;0) = \chi^{\prime}(S_j)$, we obtain in matrix form that
\begin{equation}\label{pinned:c_system}
  \left( \, \Imat + 2 \pi \nu \, \Gmat  + \nu \Bmat_0 \, \right) \v{c} =
  \nu \bar{\eta} \, \v{e}  \,, \qquad \v{e}^T\v{c}=0 \,,
\end{equation}
where $\v{c}\equiv (c_0,c_1,\ldots,c_N)^T$ and
$\v{e}=(1,\ldots,1)^T\in \mathbb{R}^{N+1}$. Here $\Gmat$ is the Neumann
Green's matrix of $\pmb{\xi} \,, \v{x}_1, \ldots, \v{x}_N$, and $\Bmat_0
\in \mathbb{R}^{(N+1)\times(N+1)}$ is a diagonal matrix with diagonal entries
\begin{equation}
  \left(\Bmat_0 \right)_{1\, 1} = \partial_{S_0}\chi(S_0; a_1) \,, \qquad
  \left(\Bmat_0\right)_{j+1\, j+1} = \chi^{\prime}(S_j) \,, \quad j = 1, \ldots,
  N \,.
\end{equation}
By left-multiplying \eqref{pinned:c_system} by $\v{e}^T$, we use
$\v{e}^T\v{c}=0$ to calculate that
$\bar{\eta} = N^{-1}\left(2\pi\v{e}^T\Gmat\v{c} + \v{e}^T\Bmat_0 \v{c}\right)$.
By substituting $\bar{\eta}$ back into the first equation in
\eqref{pinned:c_system} we obtain the following GCEP for detecting
zero-eigenvalue crossings:
\begin{equation}\label{pinned:GCEP_zero_crossing}
\Mmat_0 \v{c} = \v{0} \,, \qquad \mbox{where} \qquad \Mmat_0 \equiv
\Imat + 2\pi\nu(\Imat-\Emat)\Gmat + \nu(\Imat-\Emat)\Bmat_0 \,,
\end{equation}
where $\Emat\equiv N^{-1}\v{e}\v{e}^T$. In summary, a zero-eigenvalue
crossing associated with locally radially symmetric perturbations near
the spots occurs if and only if $\det\Mmat_0=0$. Since
$\v{e}^T\v{c}=0$, this criterion detects the initiation of an
inter-spot competition instability.

\begin{figure}[htbp]
\captionsetup[subfigure]{justification=centering}
\begin{subfigure}[b]{0.25\textwidth}
\includegraphics[width=\textwidth,height=4.0cm]{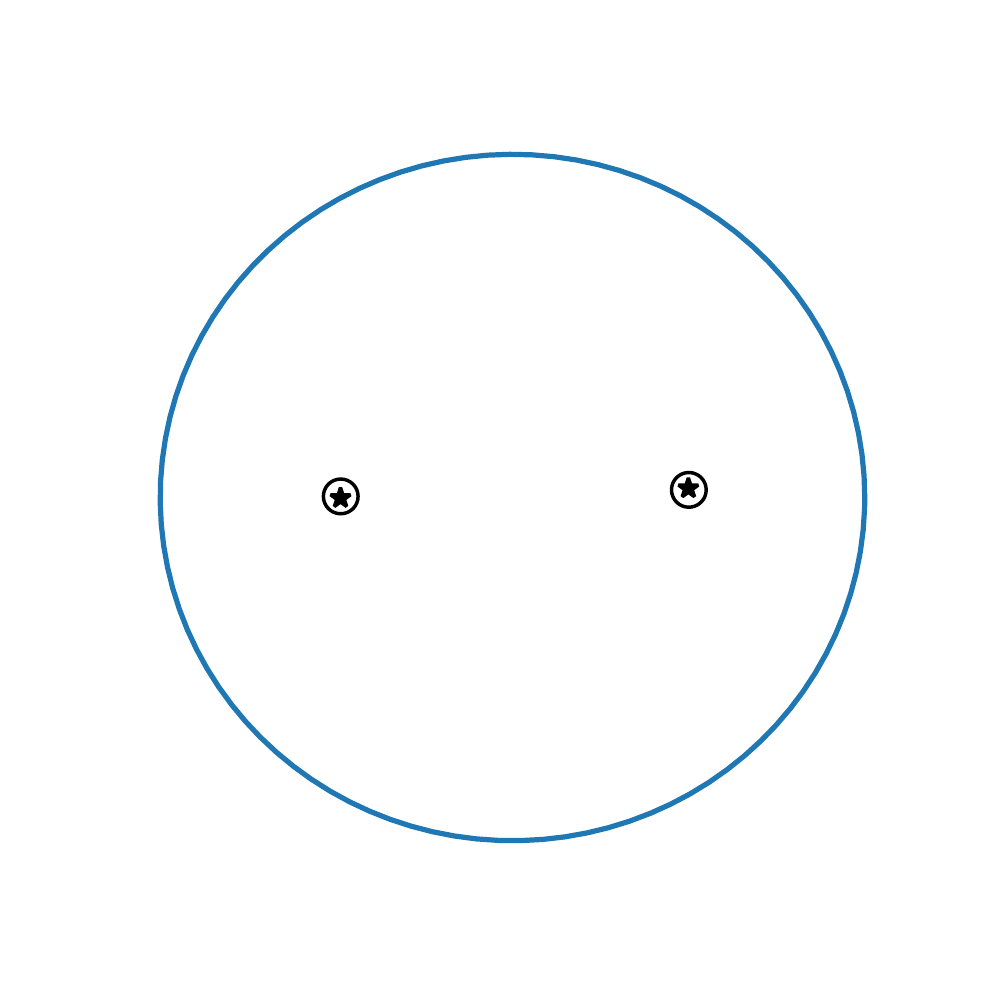}
\caption*{Before predicted \\splitting ($t=9$)}
\end{subfigure}
\qquad
\begin{subfigure}[b]{0.25\textwidth}
\includegraphics[width=\textwidth,height=4.0cm]{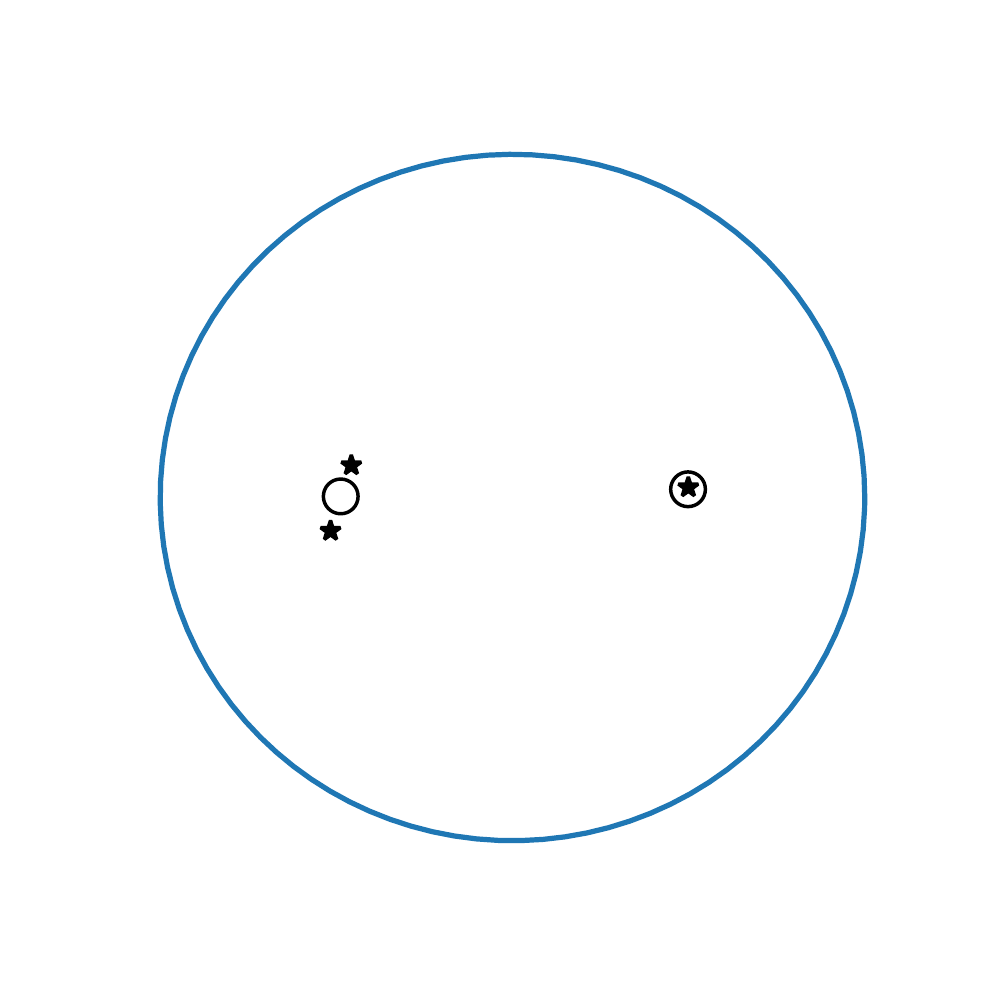}
\caption*{After predicted \\ splitting ($t=10$)}
\end{subfigure}
\qquad
\begin{subfigure}[b]{0.25\textwidth}
\includegraphics[width=\textwidth,height=4.0cm]{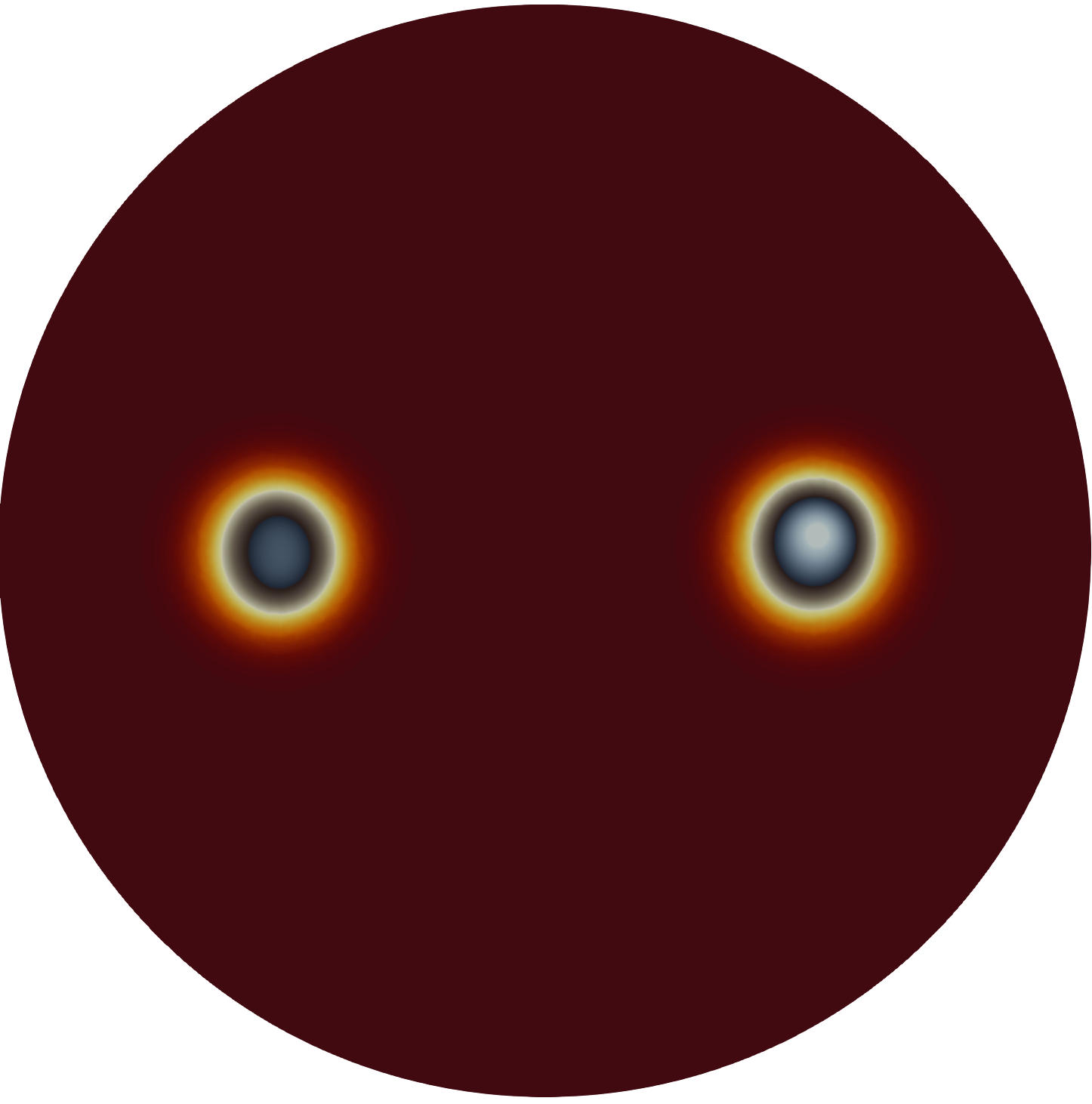}
\caption*{$v$ at $t=10$ \\ \quad}
\end{subfigure}
\\[5pt]
\begin{subfigure}[b]{0.25\textwidth}
\includegraphics[width=\textwidth,height=4.0cm]{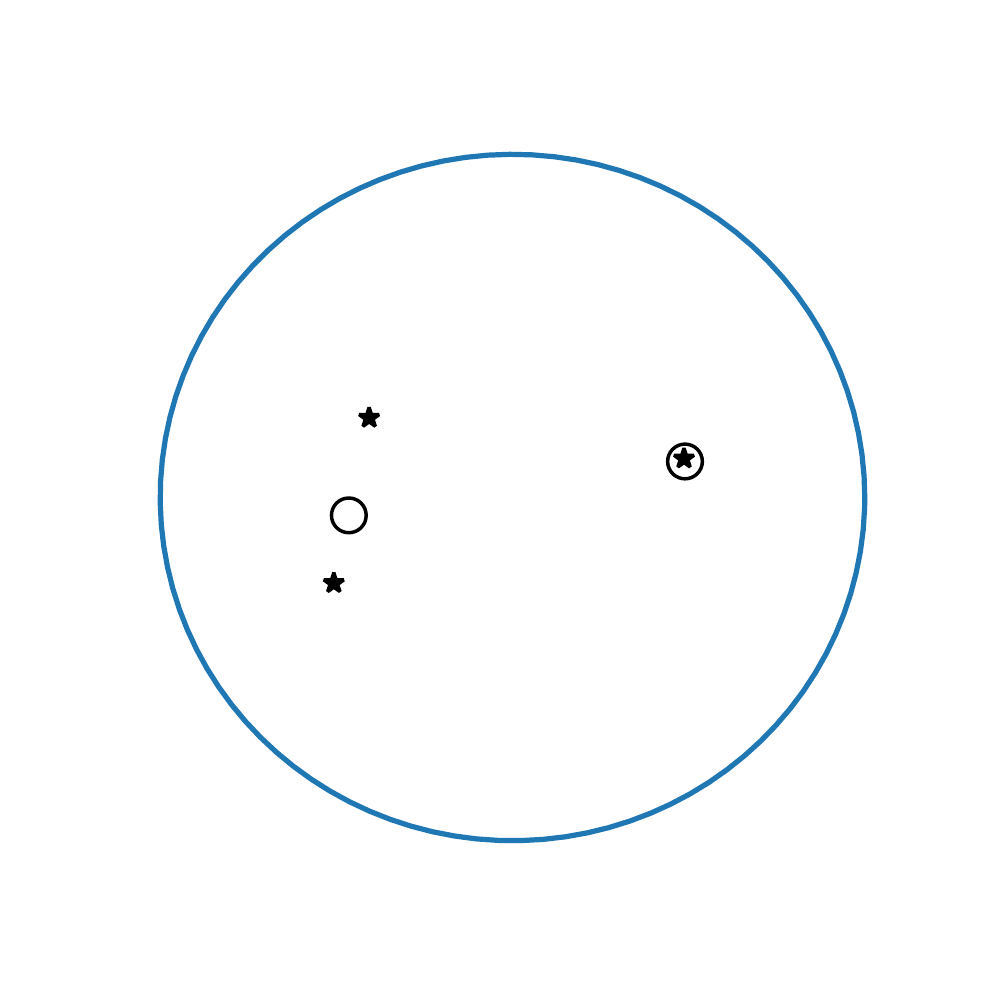}
\caption*{Before actual \\ splitting ($t=40$)}
\end{subfigure}
\qquad
\begin{subfigure}[b]{0.25\textwidth}
\includegraphics[width=\textwidth,height=4.0cm]{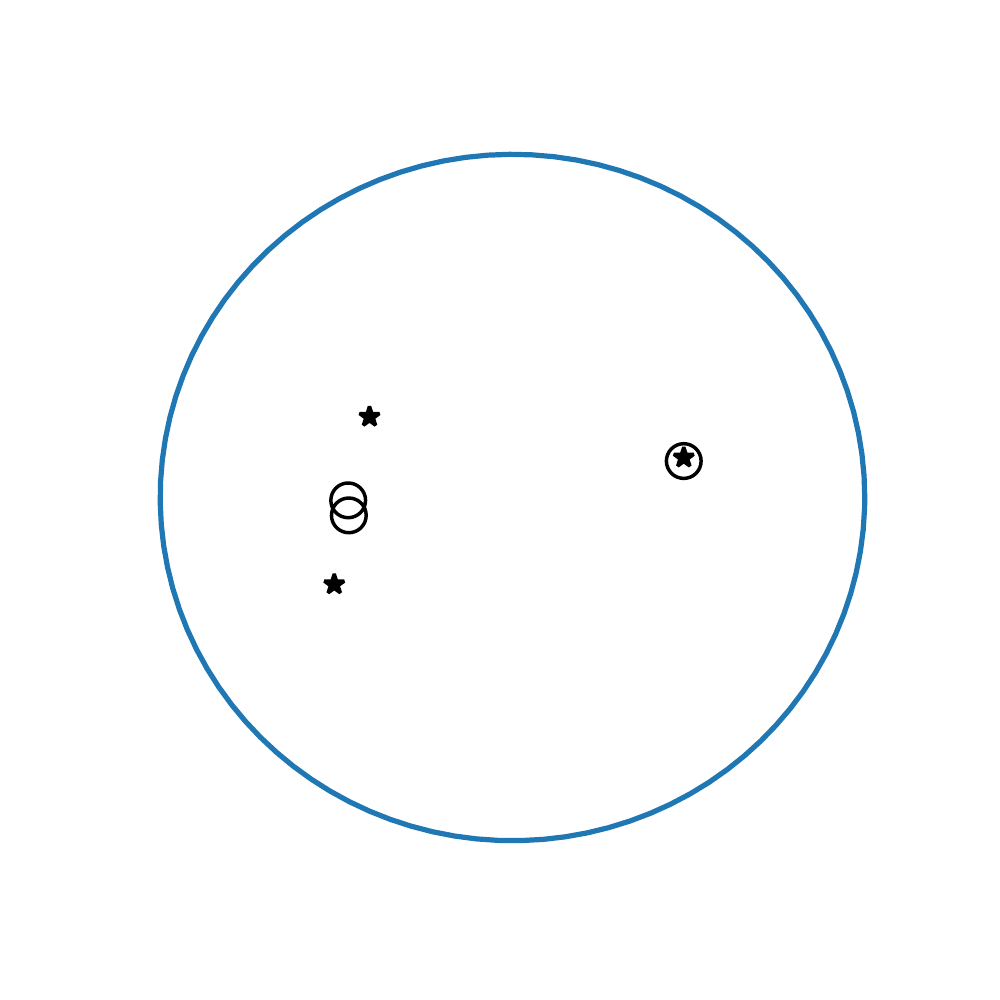}
\caption*{After actual \\ splitting ($t=41$)}
\end{subfigure}
\qquad
\begin{subfigure}[b]{0.25\textwidth}
\includegraphics[width=\textwidth,height=4.0cm]{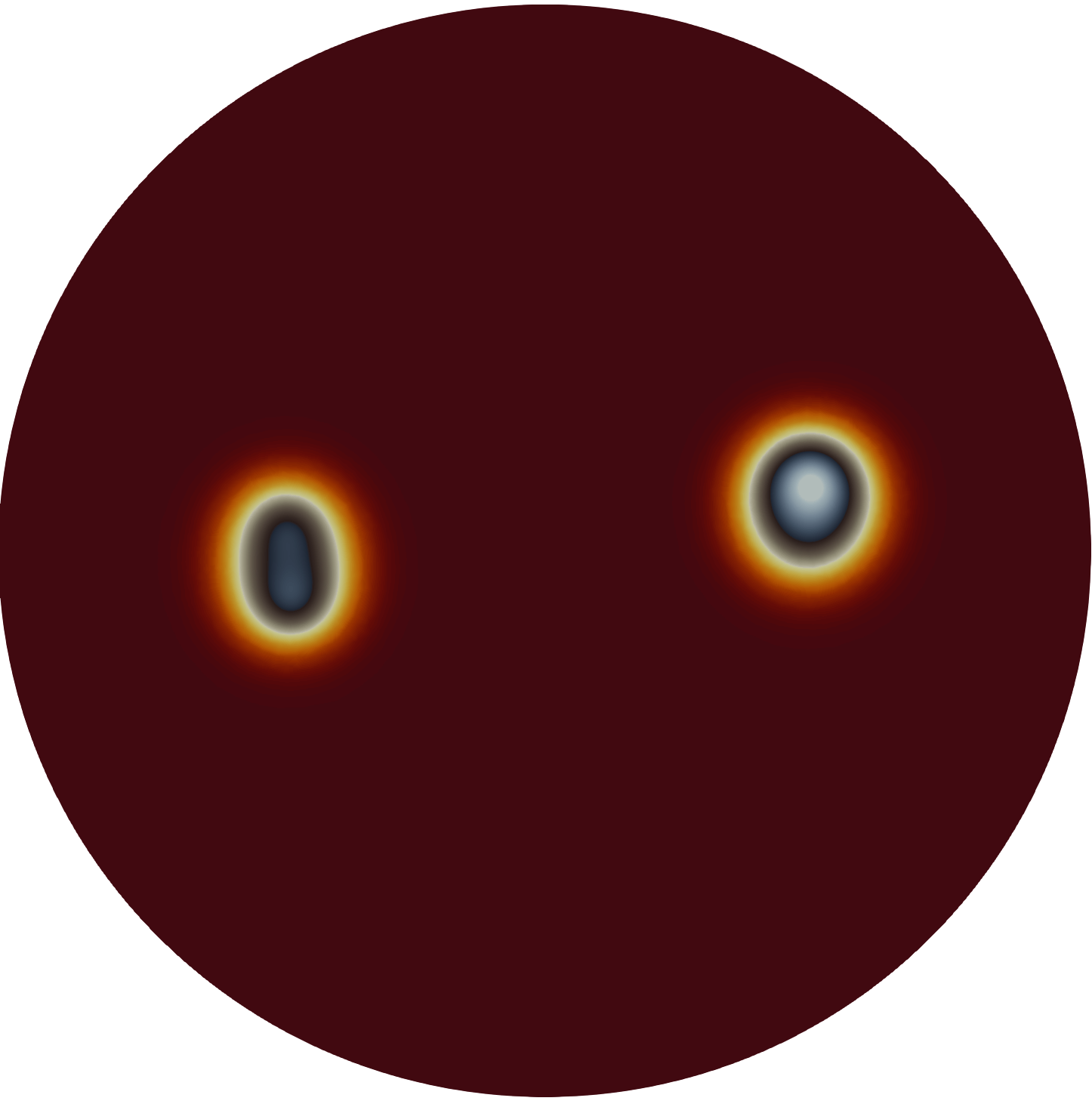}
\caption*{$v$ at $t=41$ \\ \quad}
\end{subfigure}
\caption{Left and middle panels: The spot locations obtained from PDE
  simulation of \eqref{pinned:model} are represented by the black
  circles. The star markers represent the spot locations from the DAE
  simulation of \eqref{pinned:DAE} and \eqref{pinned:nas_all}. Right
  panel: contour plot of $v$ at the indicated time, as computed
  numerically from the PDE. Parameters are $\eps=0.03$, $a_0=19$,
  $a_1=5$, with the concentration point for the feed rate evolving
  dynamically by \eqref{pinned:trajectory}.}
\label{pinned_split_comp_exp1_part1}
\end{figure}

\subsubsection{A loop of spot replication and spot annihilation}
\label{sec:loop}

\begin{figure}[htbp]
\captionsetup[subfigure]{justification=centering}	
\begin{subfigure}[b]{0.25\textwidth}
\includegraphics[width=\textwidth,height=4.0cm]{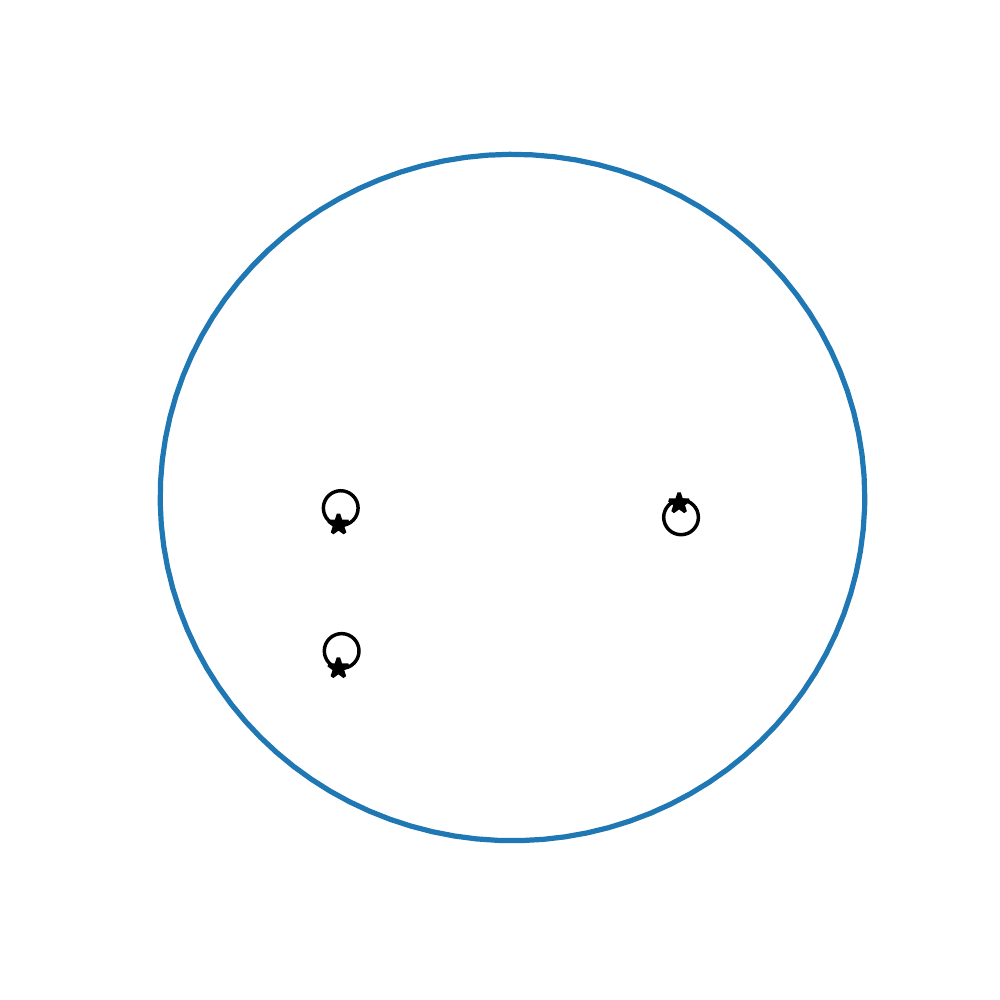}
\caption*{Before predicted killing and splitting ($t=583$)}
\end{subfigure}
\qquad
\begin{subfigure}[b]{0.25\textwidth}
\includegraphics[width=\textwidth,height=4.0cm]{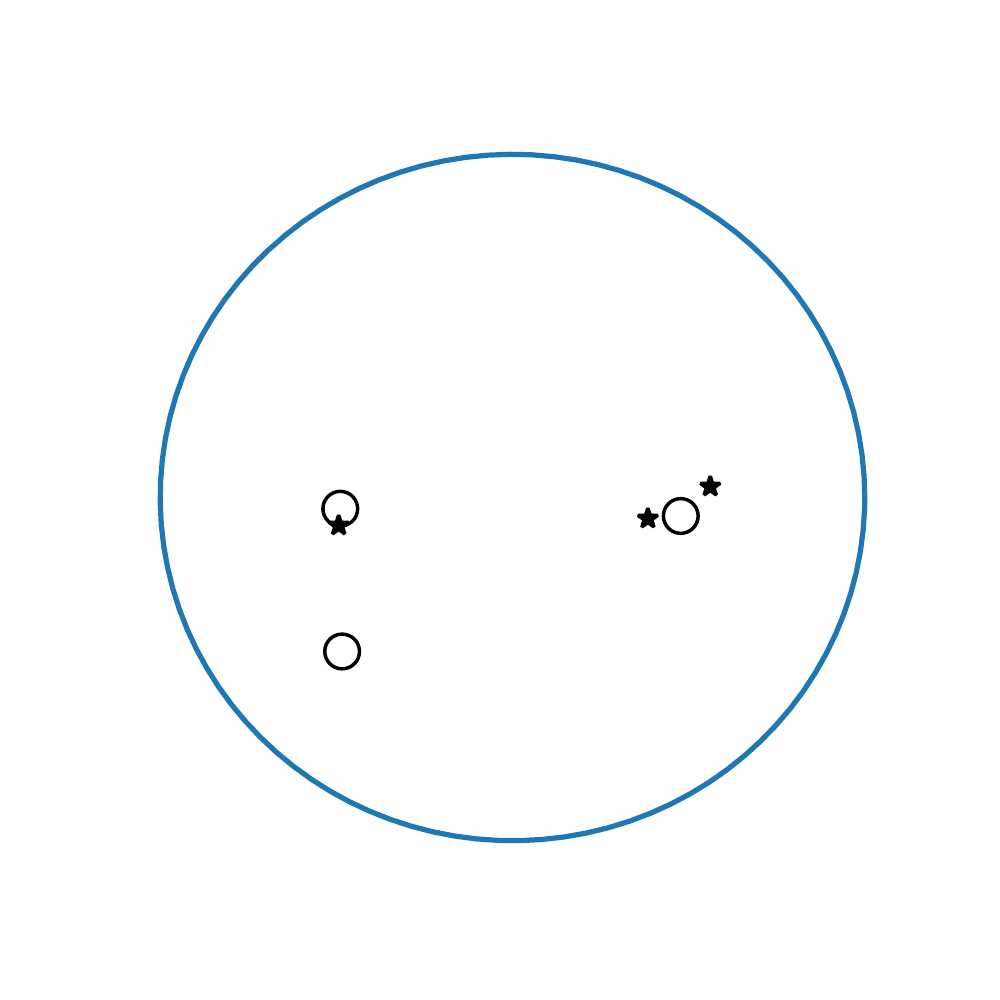}
\caption*{After predicted killing and splitting ($t=584$)}
\end{subfigure}
\qquad
\begin{subfigure}[b]{0.25\textwidth}
\includegraphics[width=\textwidth,height=4.0cm]{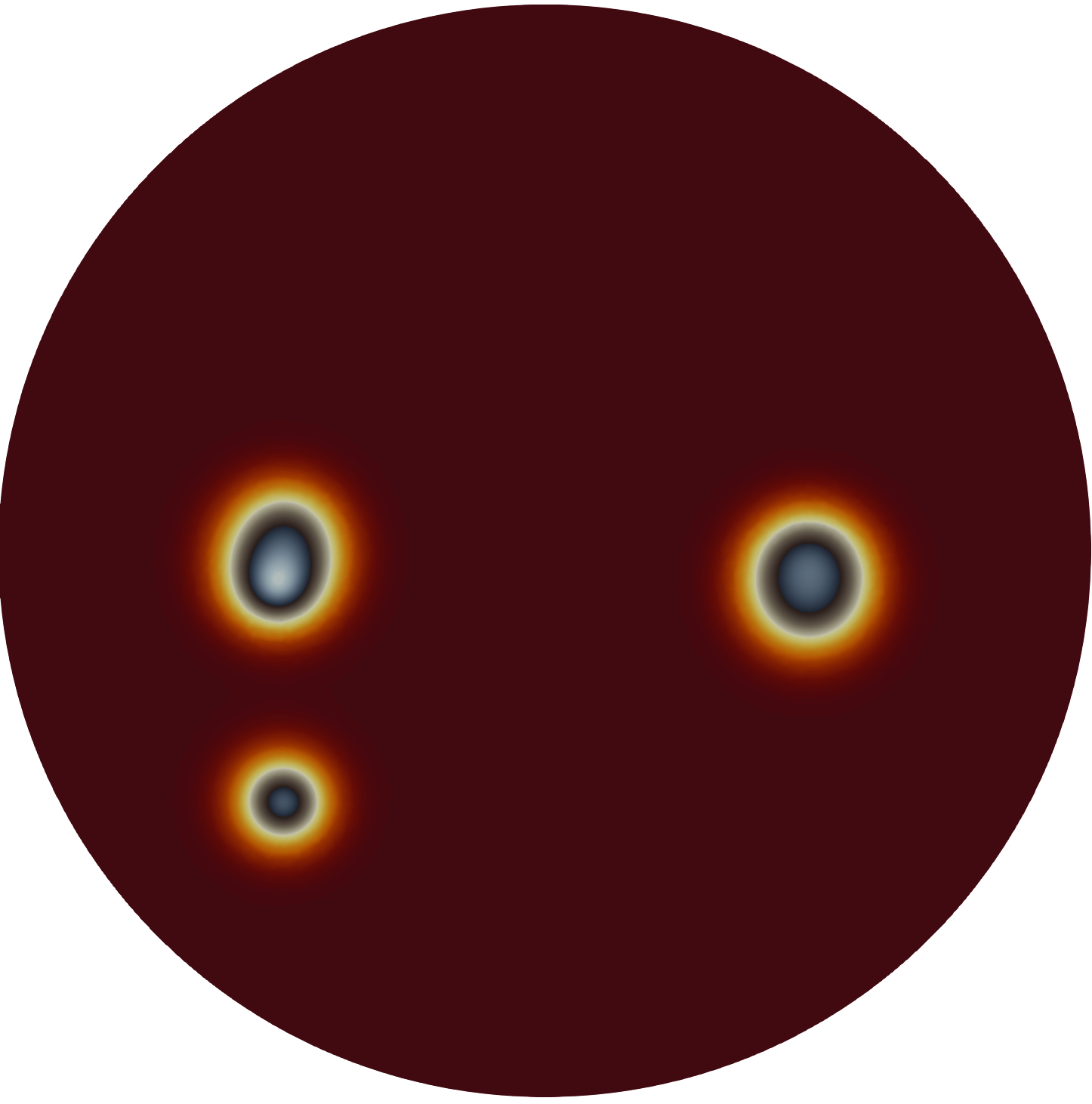}
\caption*{$v$ at $t=584$ \\ \quad }
\end{subfigure}
\\[5pt]
\begin{subfigure}[b]{0.25\textwidth}
\includegraphics[width=\textwidth,height=4.0cm]{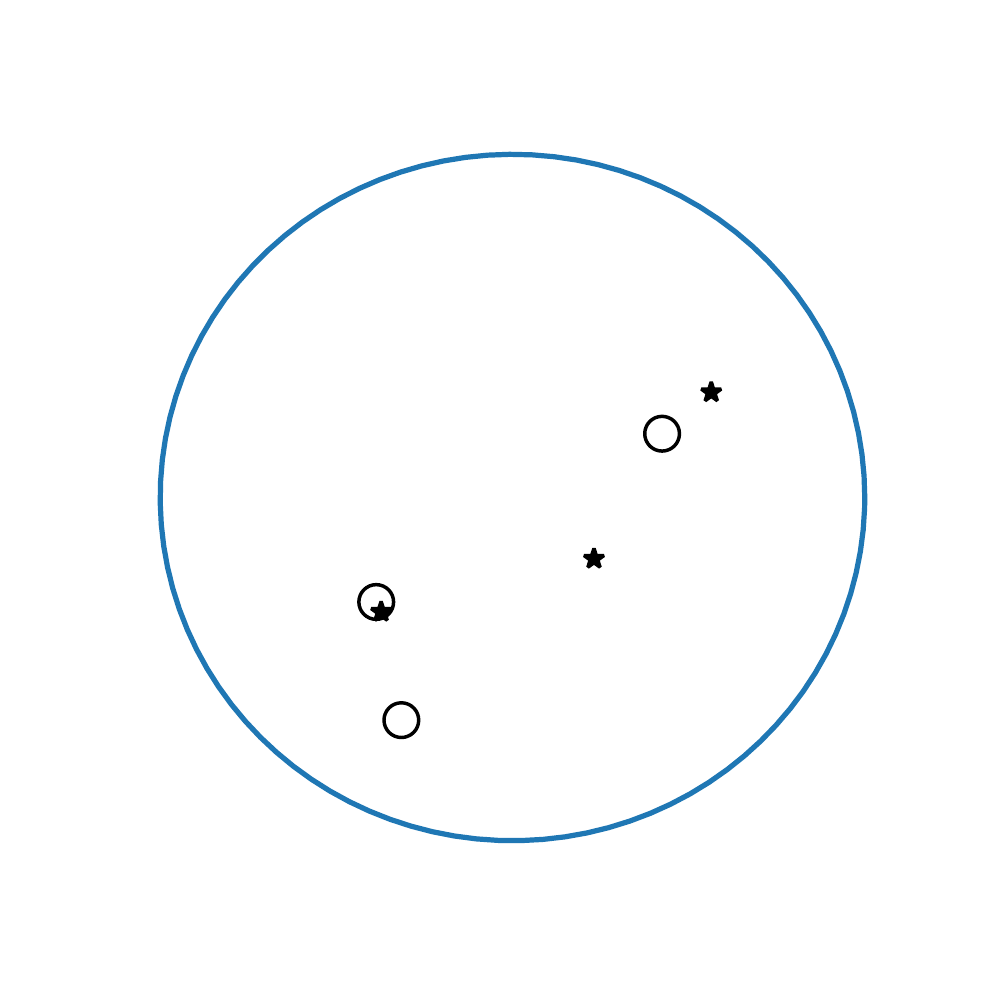}
\caption*{Before actual killing ($t=684$)}
\end{subfigure}
\qquad
\begin{subfigure}[b]{0.25\textwidth}
\includegraphics[width=\textwidth,height=4.0cm]{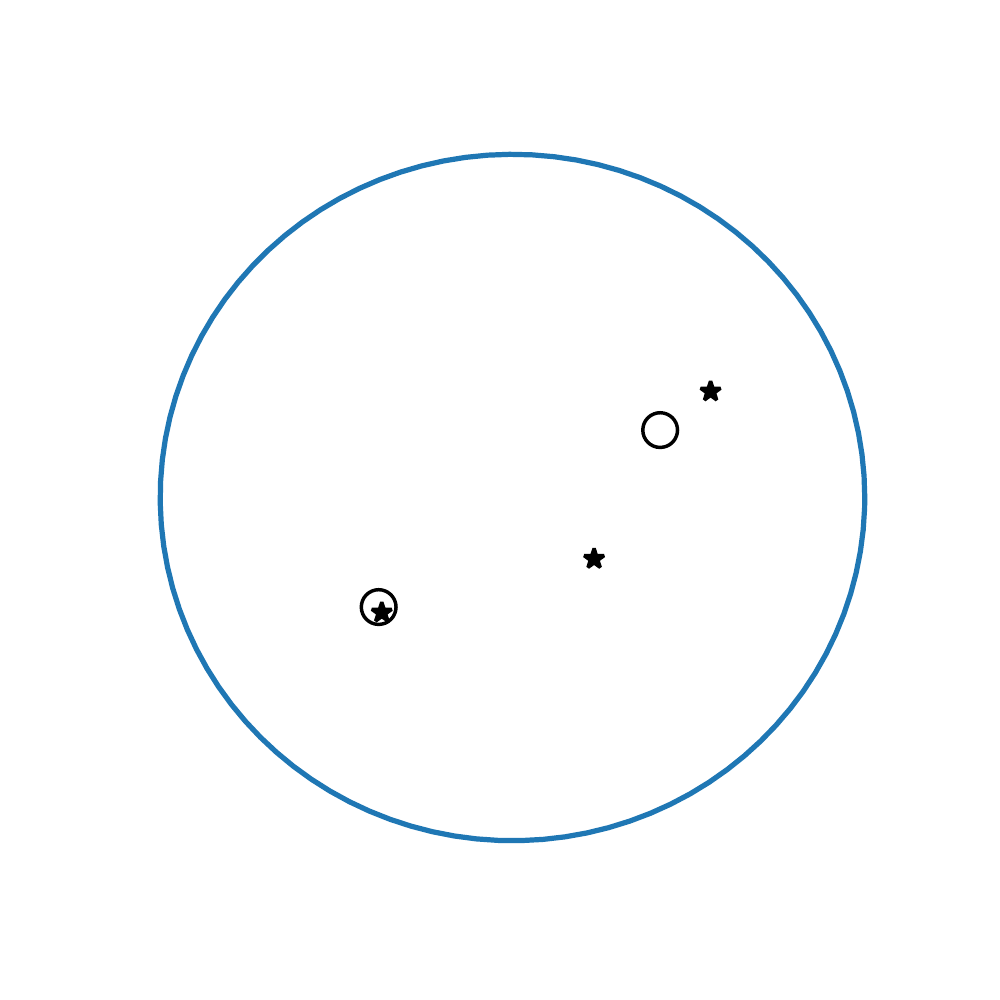}
\caption*{After actual killing ($t=685$)}
\end{subfigure}
\qquad
\begin{subfigure}[b]{0.25\textwidth}
\includegraphics[width=\textwidth,height=4.0cm]{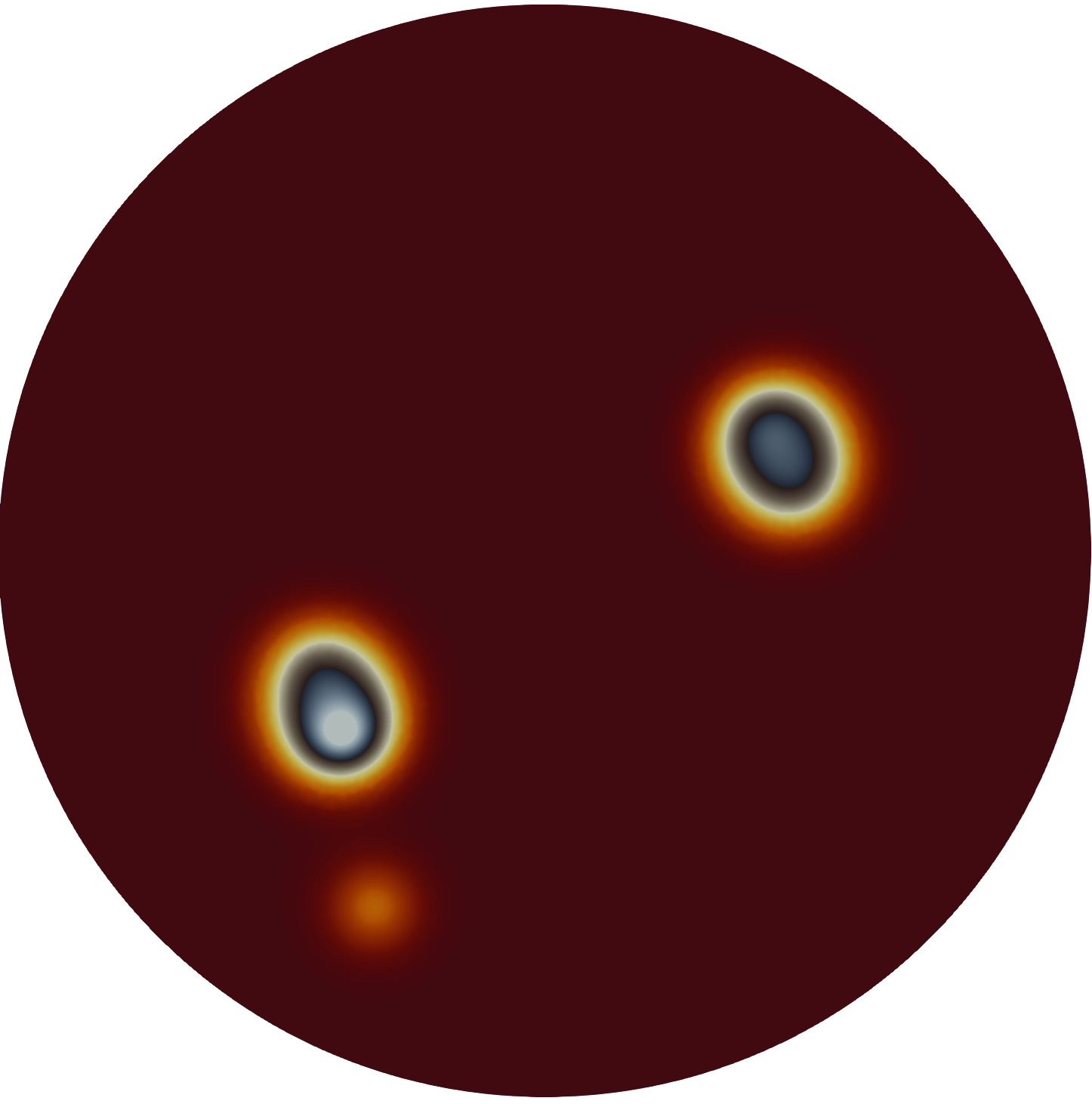}
\caption*{$v$ at $t=684$}
\end{subfigure}
\\[5pt]
\begin{subfigure}[b]{0.25\textwidth}
\includegraphics[width=\textwidth,height=4.0cm]{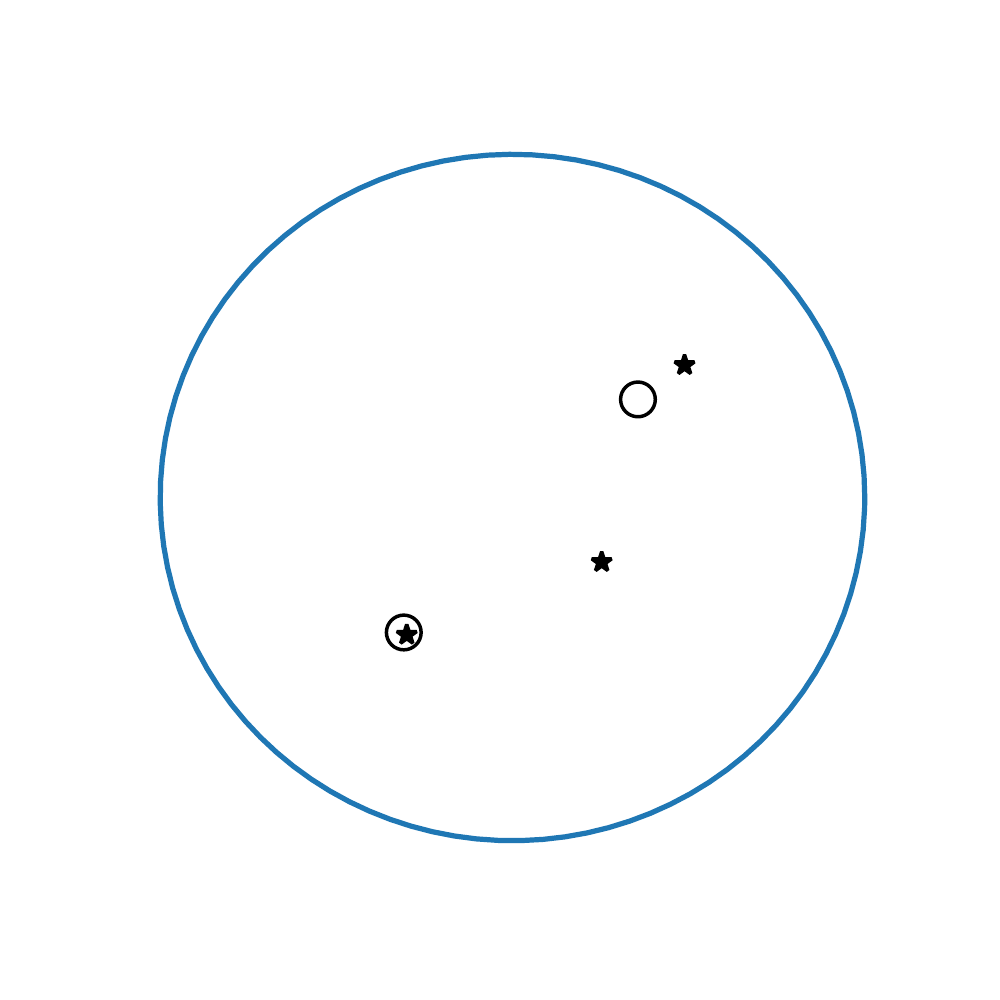}
\caption*{Before actual splitting ($t=719$)}
\end{subfigure}
\qquad
\begin{subfigure}[b]{0.25\textwidth}
\includegraphics[width=\textwidth,height=4.0cm]{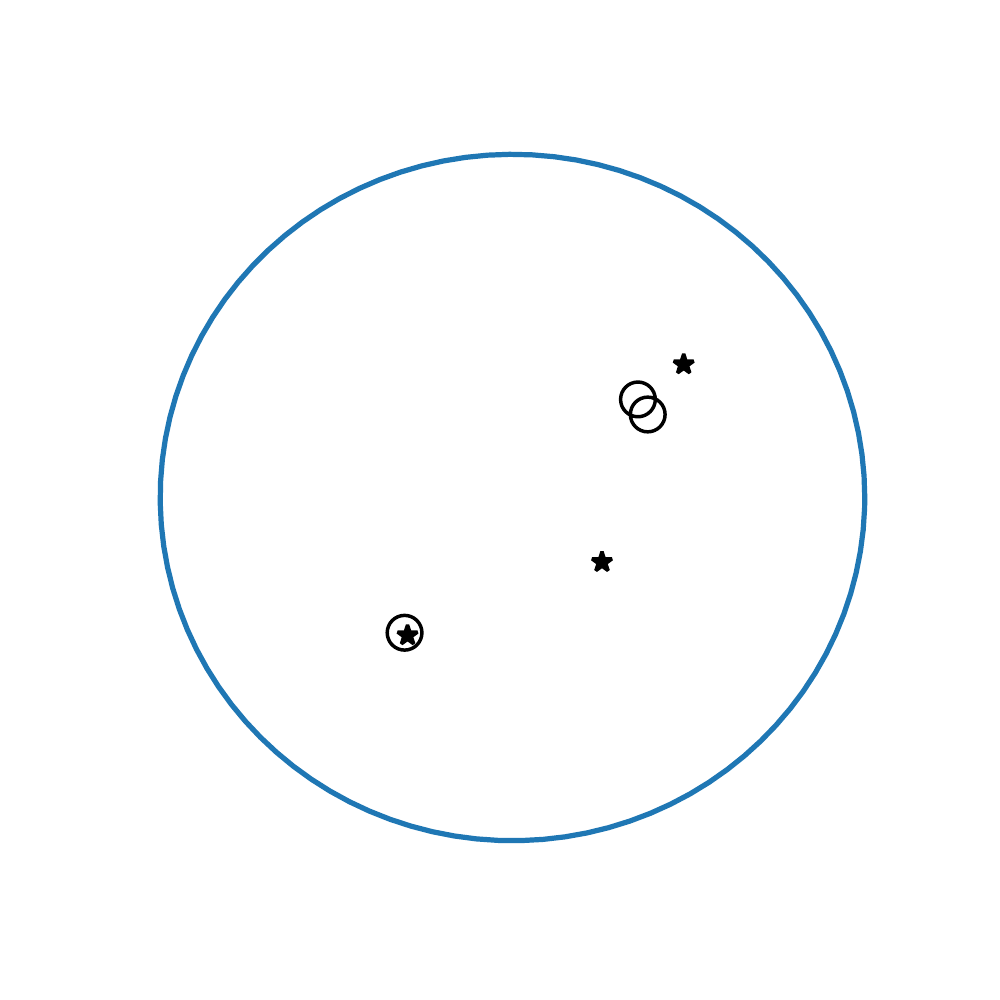}
\caption*{After actual splitting ($t=720$)}
\end{subfigure}
\qquad
\begin{subfigure}[b]{0.25\textwidth}
\includegraphics[width=\textwidth,height=4.0cm]{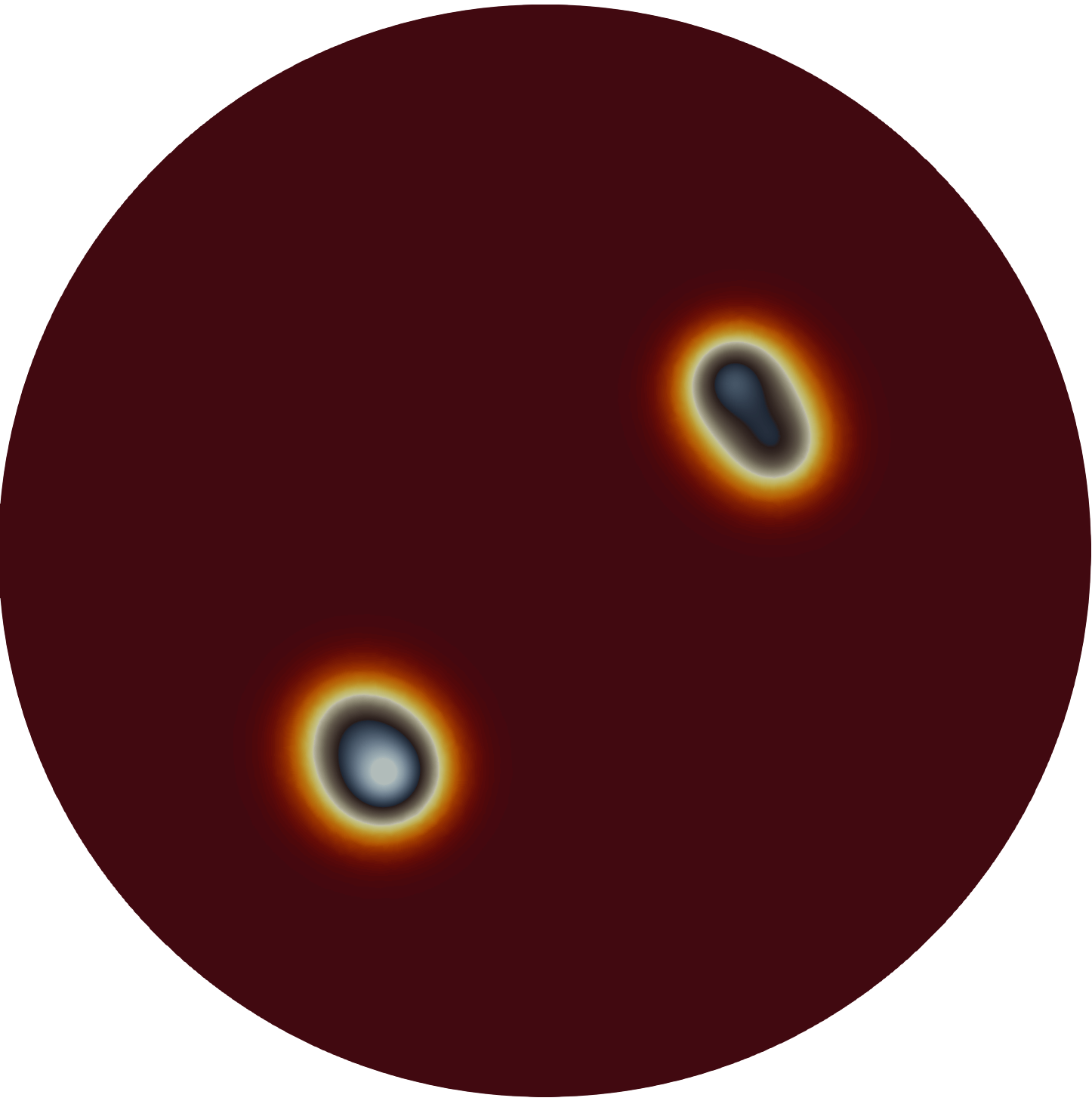}
\caption*{$v$ at $t=720$}
\end{subfigure}
\caption{Our approximation algorithm detects spot-annihilation and
  spot-splitting at the same time. On the other hand, the
  spot-tracking algorithm based on the PDE simulation detects killing
  and splitting, separately at later time.}
\label{pinned_split_comp_exp1_part2}
\end{figure}

\begin{figure}[htbp]
\captionsetup[subfigure]{justification=centering}	
\begin{subfigure}[b]{0.25\textwidth}
\includegraphics[width=\textwidth,height=4.0cm]{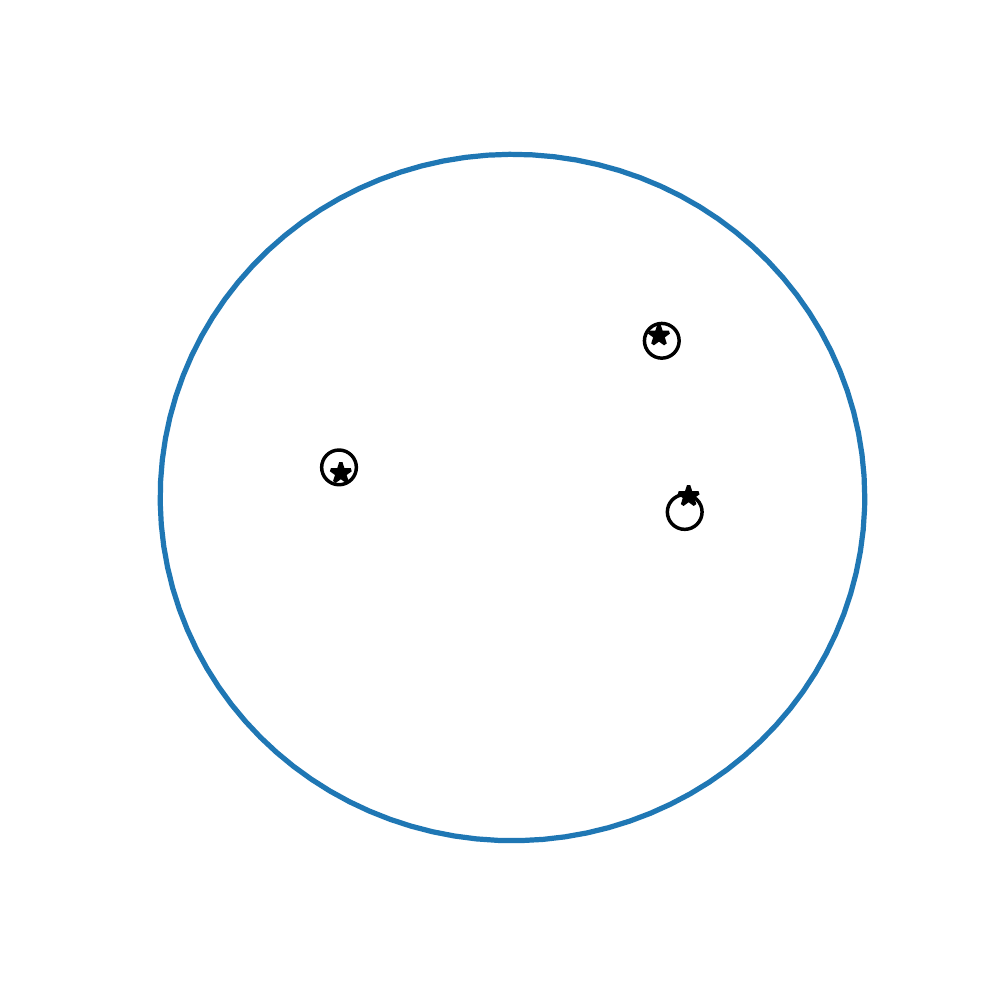}
\caption*{After one cycle ($t=1112$)}
\end{subfigure}
\qquad
\begin{subfigure}[b]{0.25\textwidth}
\includegraphics[width=\textwidth,height=4.0cm]{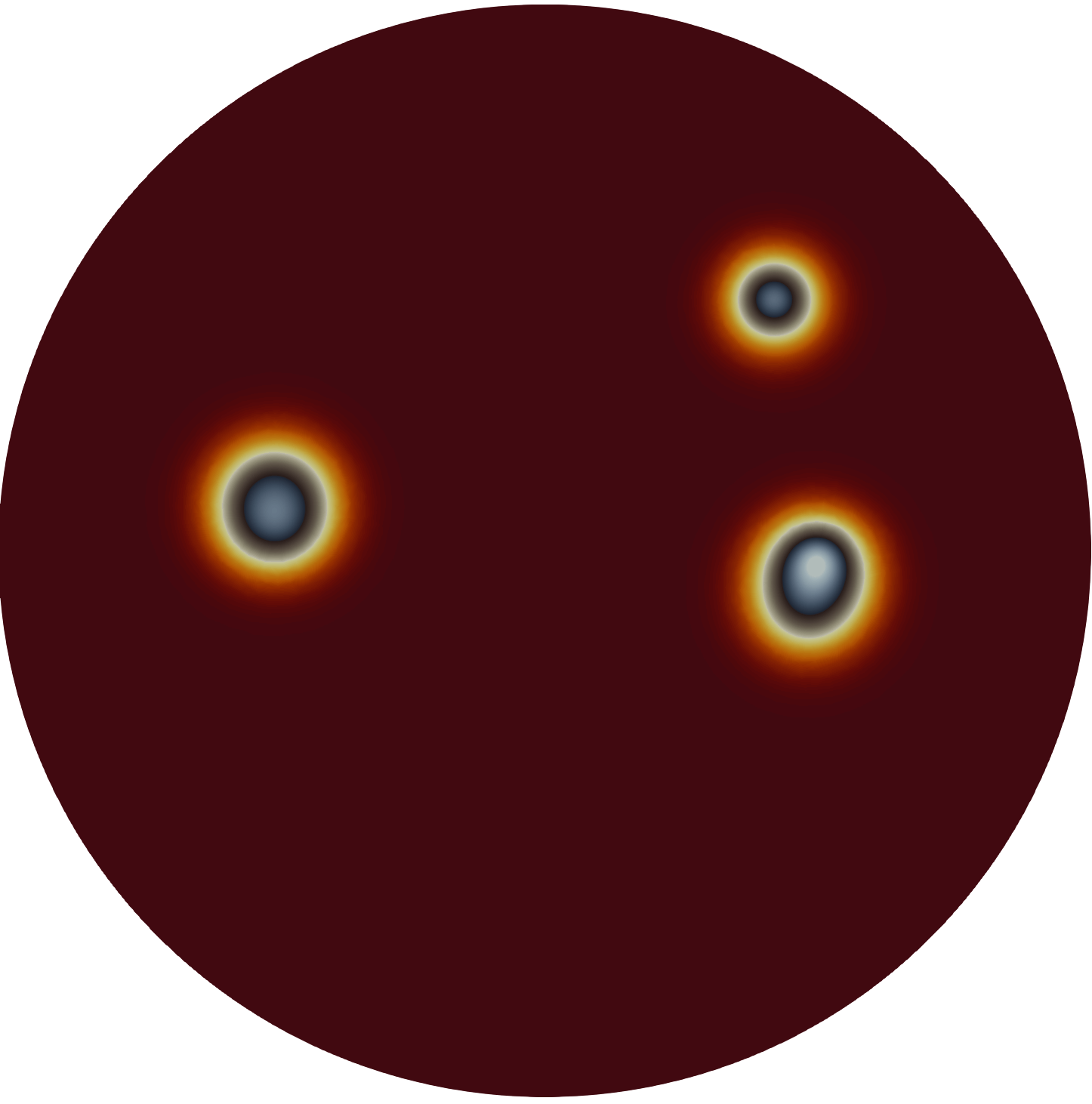}
\caption*{$v$ at $t=1112$ }
\end{subfigure}
\caption{When the pinned-spot finishes a full cycle, the predicted
  spot location from the DAE and the actual spot location (from the
  PDE spot-tracking algorithm) have a good agreement. This shows that
  the spots can catch up with the prediction from the augmented DAE
  algorithm.}
\label{pinned_split_comp_exp1_part3}
\end{figure}

\begin{figure}[htbp]
\captionsetup[subfigure]{justification=centering}	
\begin{subfigure}[b]{0.25\textwidth}
\includegraphics[width=\textwidth,height=4.0cm]{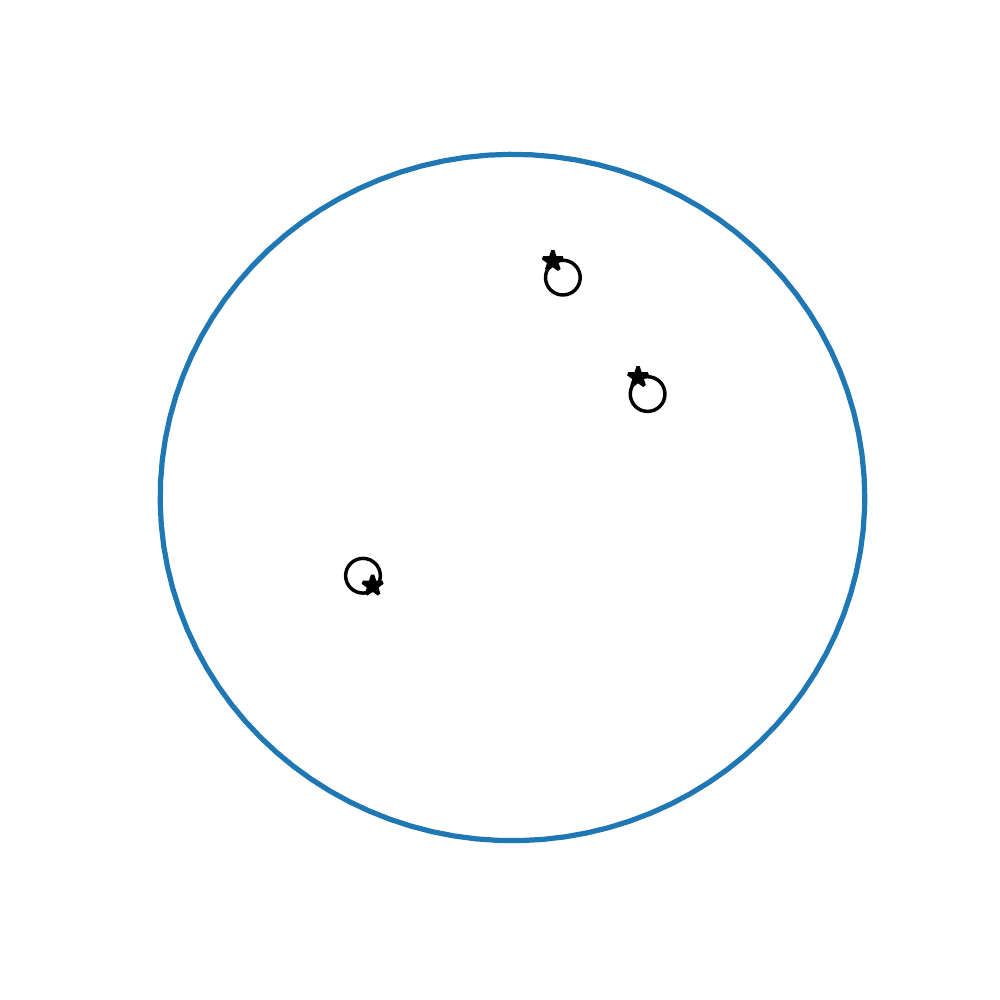}
\caption*{Before predicted killing and splitting ($t=1248$)}
\end{subfigure}
\qquad
\begin{subfigure}[b]{0.25\textwidth}
\includegraphics[width=\textwidth,height=4.0cm]{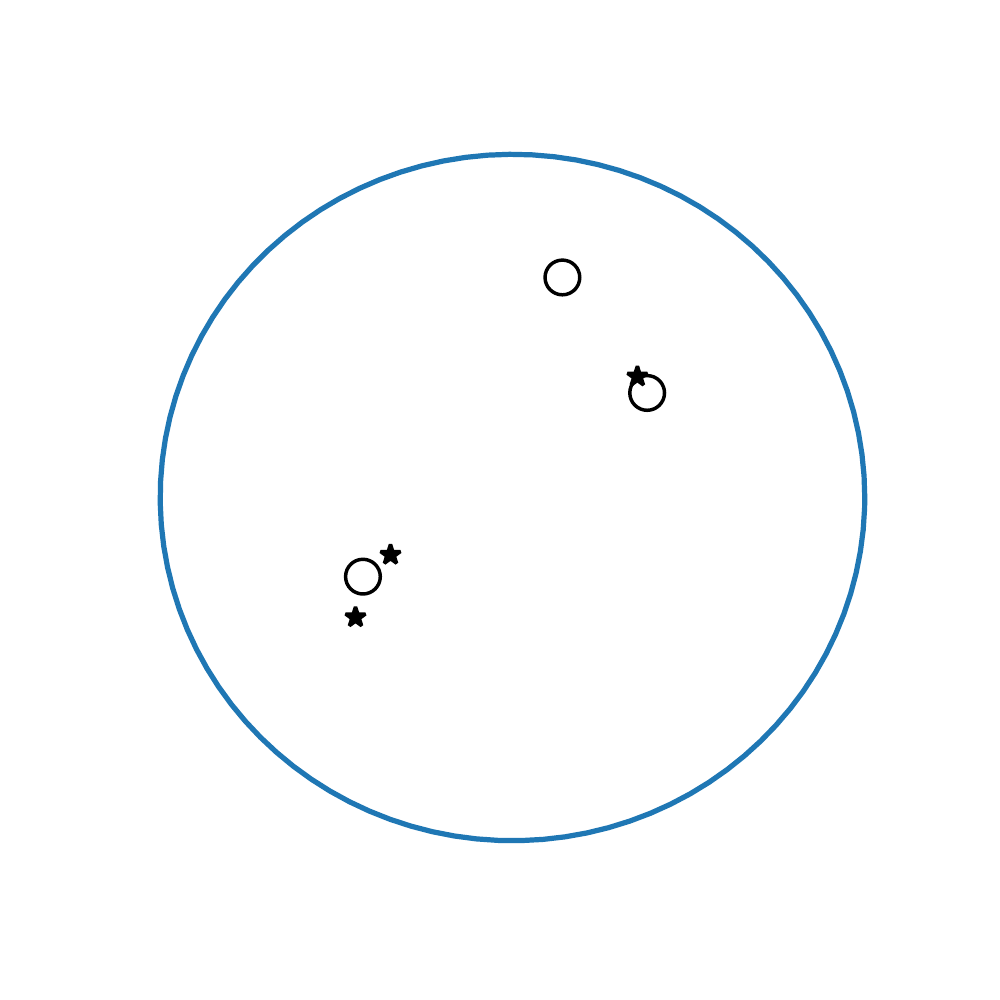}
\caption*{After predicted \\ killing and splitting \\ ($t=1249$)}
\end{subfigure}
\qquad
\begin{subfigure}[b]{0.25\textwidth}
\includegraphics[width=\textwidth,height=4.0cm]{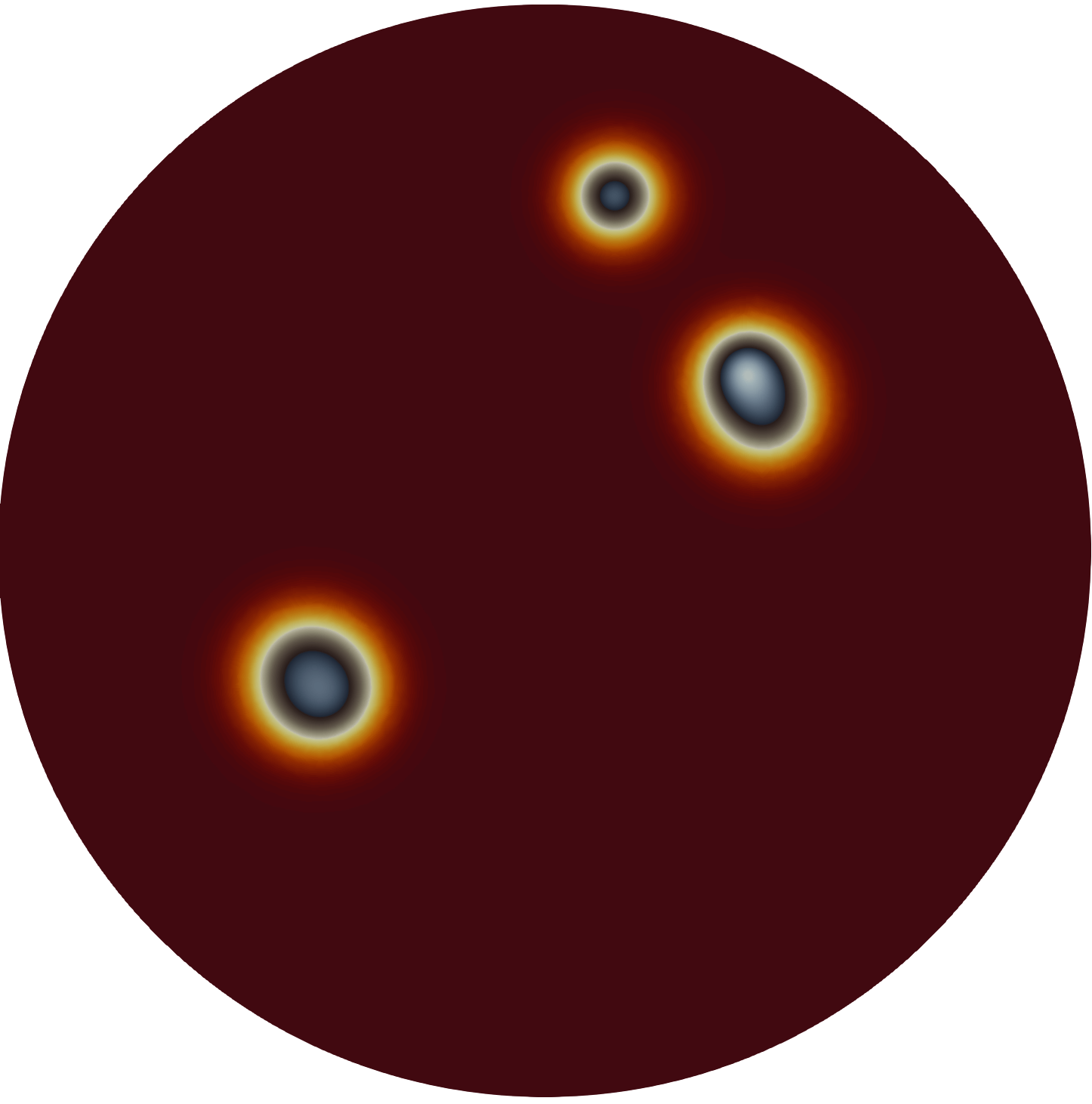}
\caption*{$v$ at $t=1249$ \\ \quad \\ \quad}
\end{subfigure}
\\[5pt]
\begin{subfigure}[b]{0.25\textwidth}
\includegraphics[width=\textwidth,height=4.0cm]{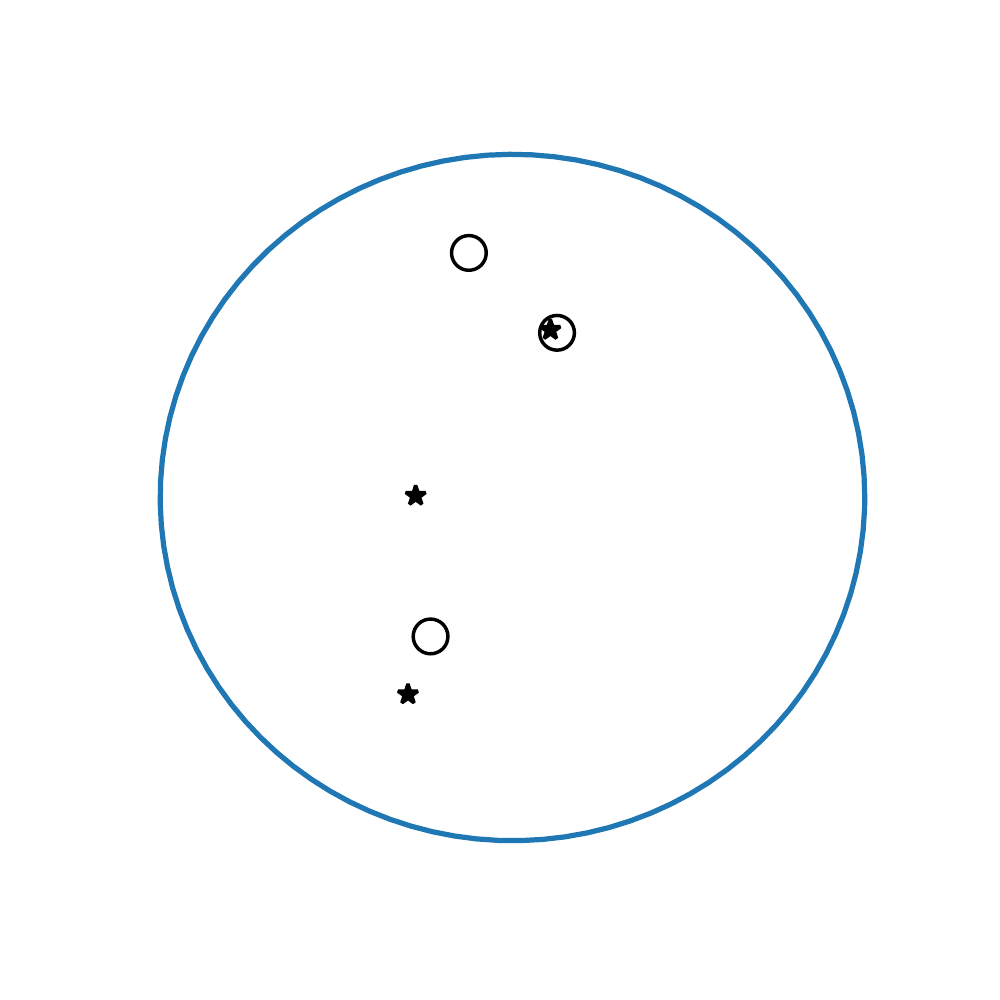}
\caption*{Before actual killing ($t=1350$)}
\end{subfigure}
\qquad
\begin{subfigure}[b]{0.25\textwidth}
\includegraphics[width=\textwidth,height=4.0cm]{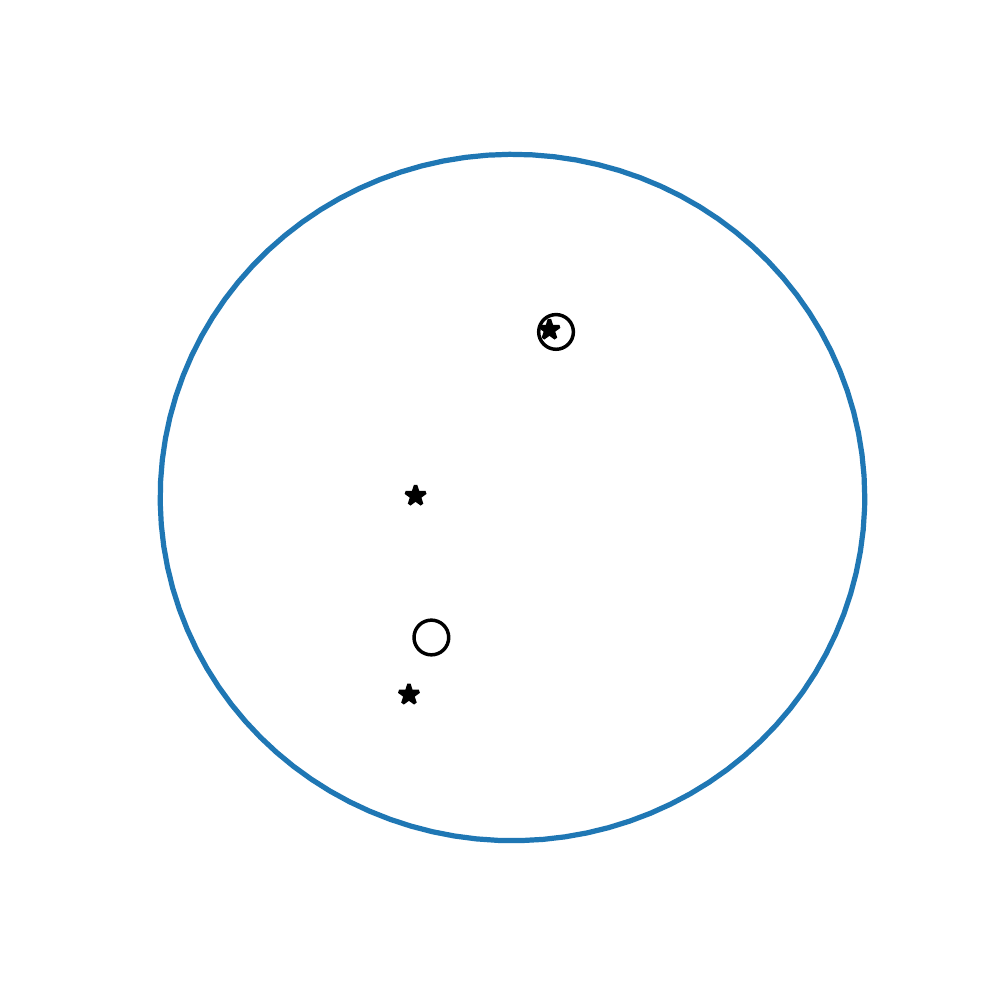}
\caption*{After actual killing ($t=1351$)}
\end{subfigure}
\qquad
\begin{subfigure}[b]{0.25\textwidth}
\includegraphics[width=\textwidth,height=4.0cm]{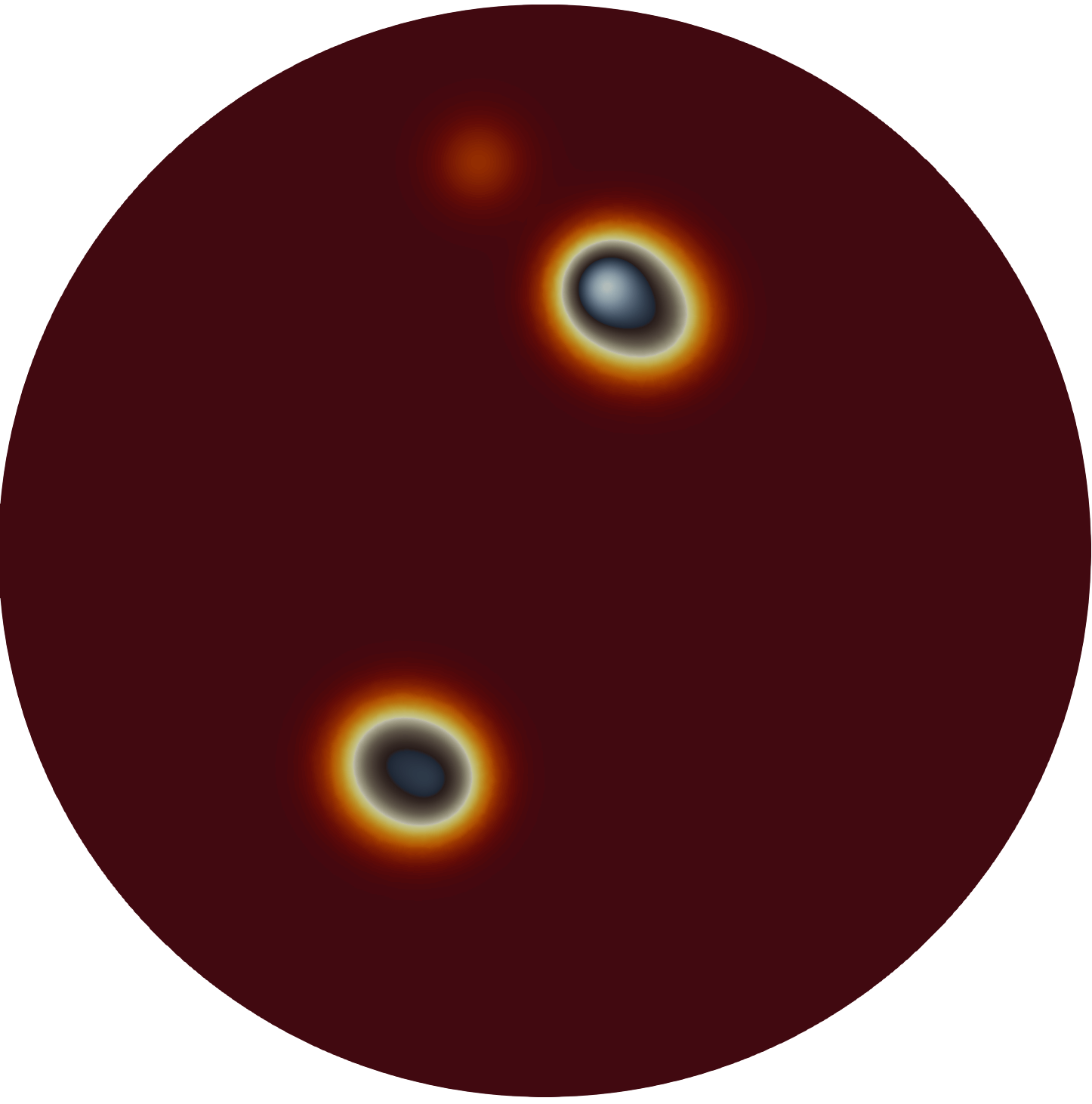}
\caption*{$v$ at $t=1350$}
\end{subfigure}
\\[5pt]
\begin{subfigure}[b]{0.25\textwidth}
\includegraphics[width=\textwidth,height=4.0cm]{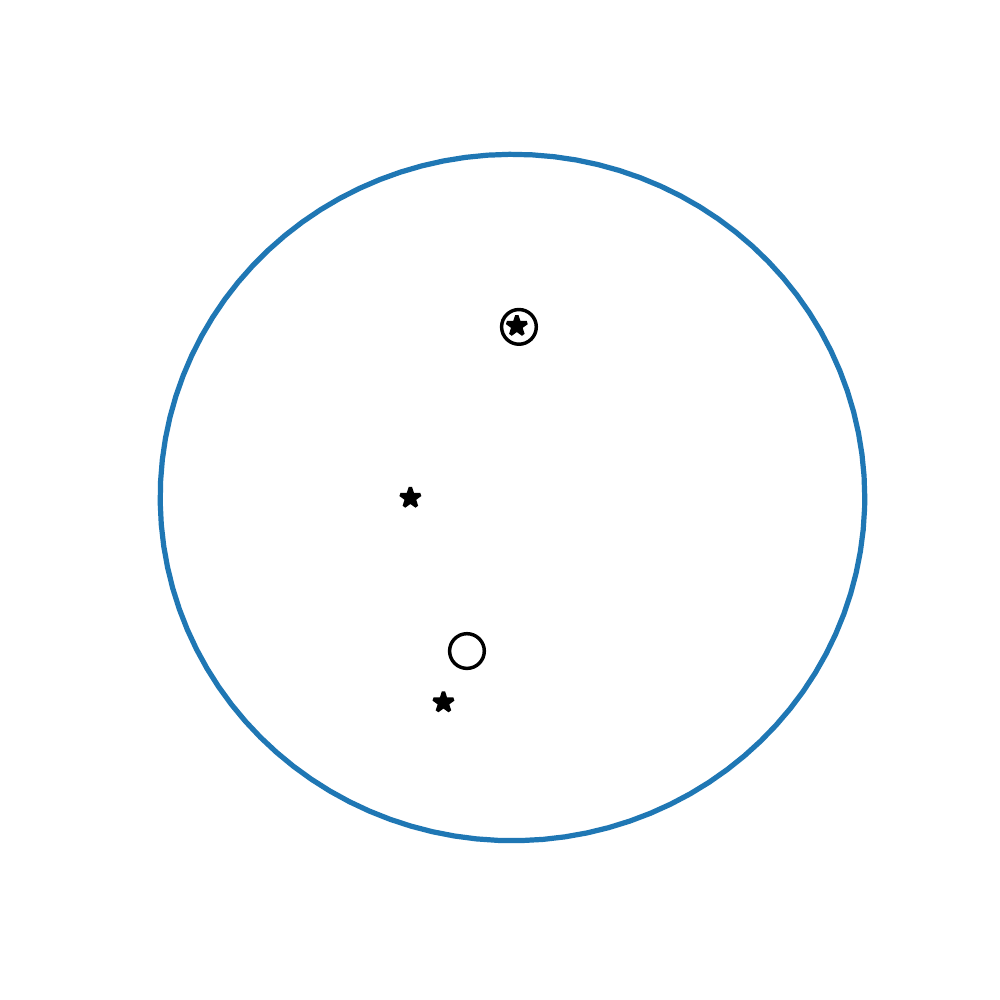}
\caption*{Before actual splitting ($t=1384$)}
\end{subfigure}
\qquad
\begin{subfigure}[b]{0.25\textwidth}
\includegraphics[width=\textwidth,height=4.0cm]{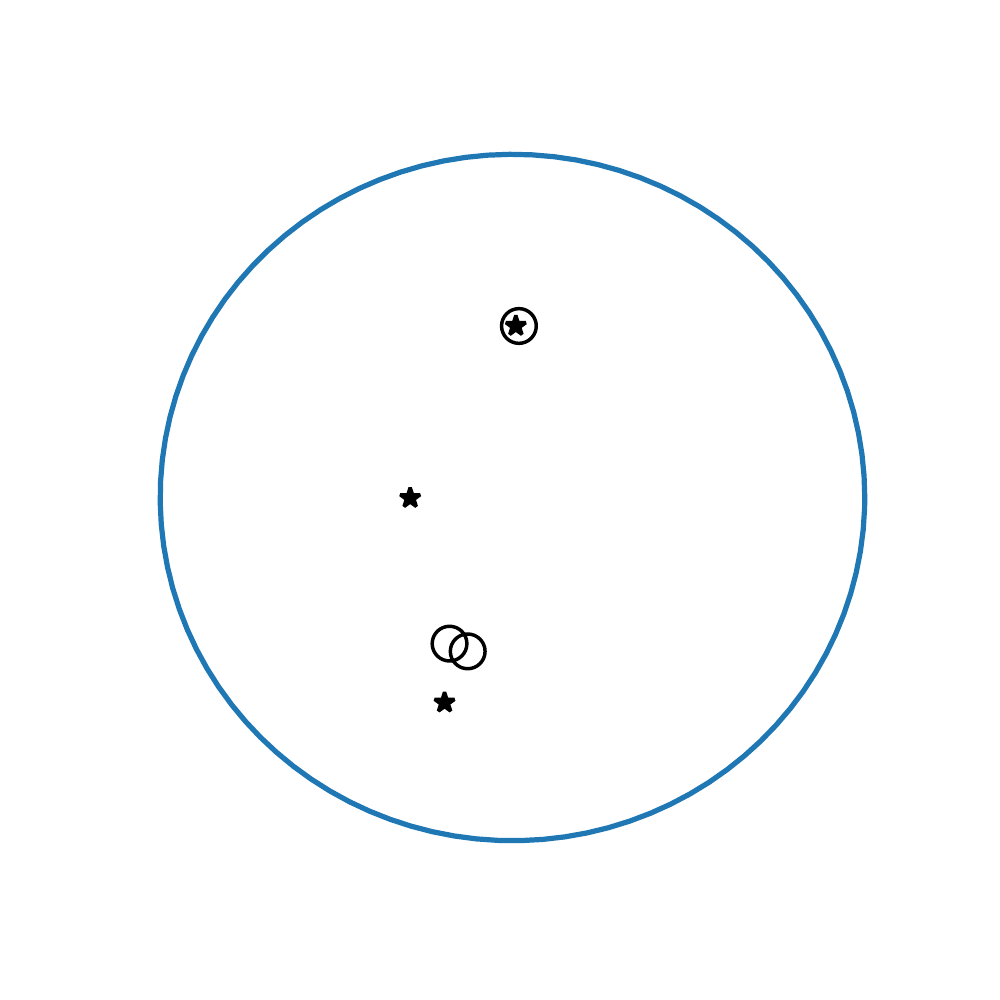}
\caption*{After actual splitting ($t=1385$)}
\end{subfigure}
\qquad
\begin{subfigure}[b]{0.25\textwidth}
\includegraphics[width=\textwidth,height=4.0cm]{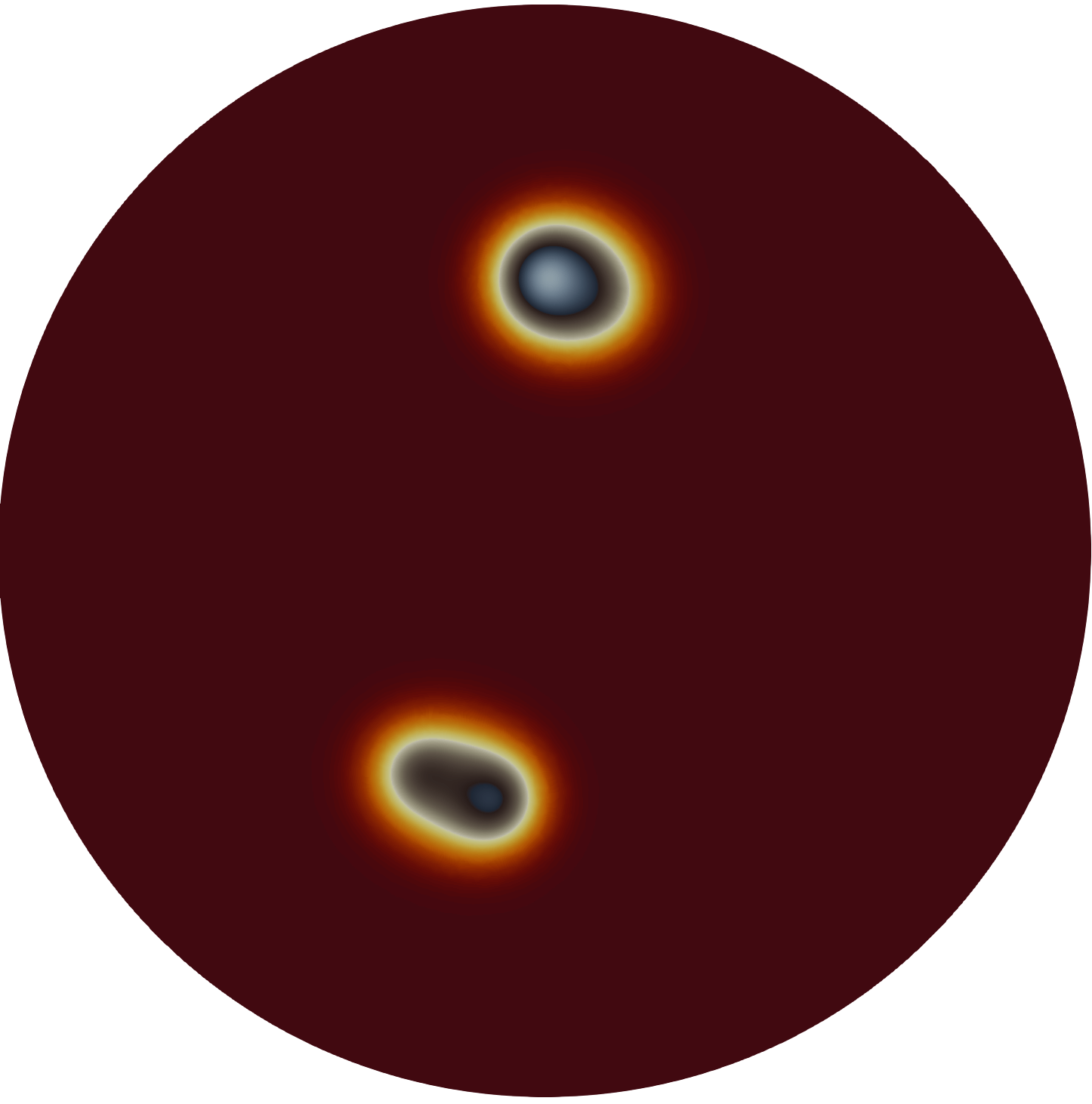}
\caption*{$v$ at $t=1385$}
\end{subfigure}
\caption{The second spot creation-annihilation event.}
\label{pinned_split_comp_exp1_part4}
\end{figure}

\begin{figure}[htbp]
\captionsetup[subfigure]{justification=centering}	
\begin{subfigure}[b]{0.25\textwidth}
\includegraphics[width=\textwidth,height=4.0cm]{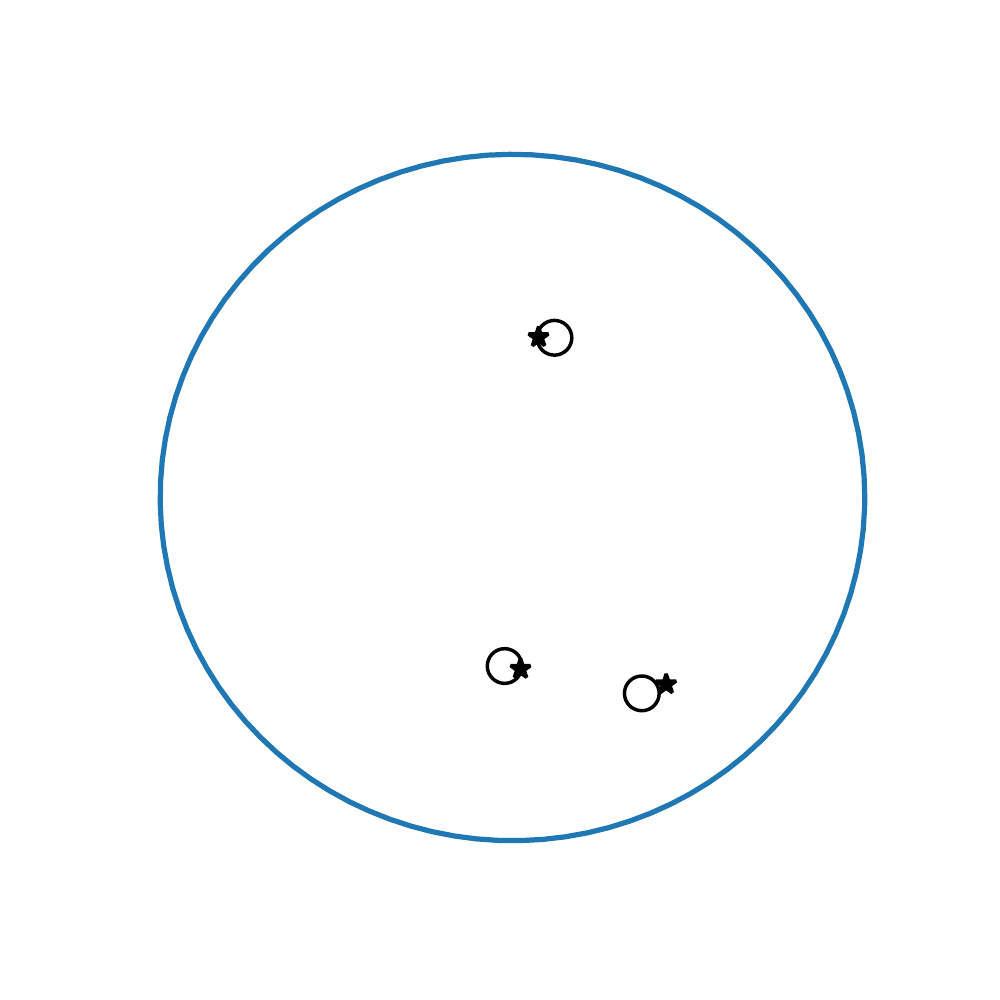}
\caption*{Before predicted killing and splitting ($t=1952$)}
\end{subfigure}
\qquad
\begin{subfigure}[b]{0.25\textwidth}
\includegraphics[width=\textwidth,height=4.0cm]{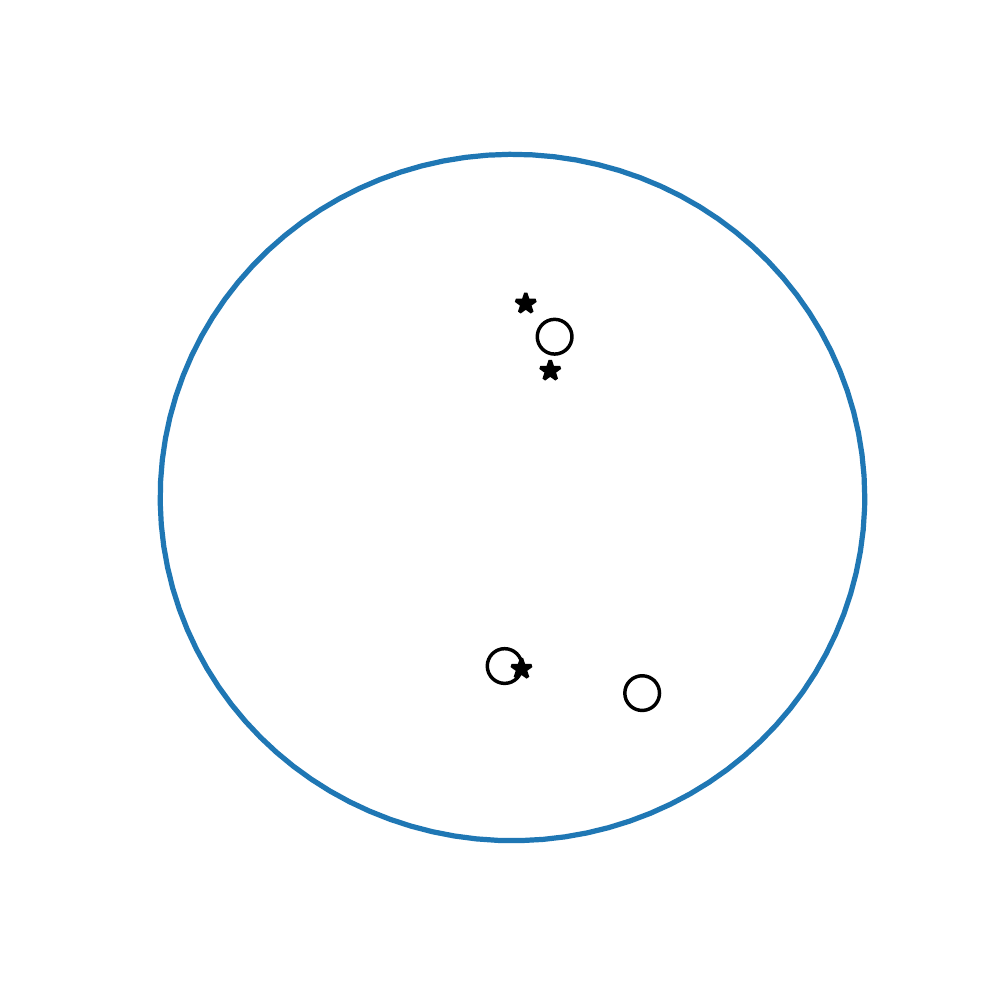}
\caption*{After predicted killing and splitting ($t=1953$)}
\end{subfigure}
\qquad
\begin{subfigure}[b]{0.25\textwidth}
\includegraphics[width=\textwidth,height=4.0cm]{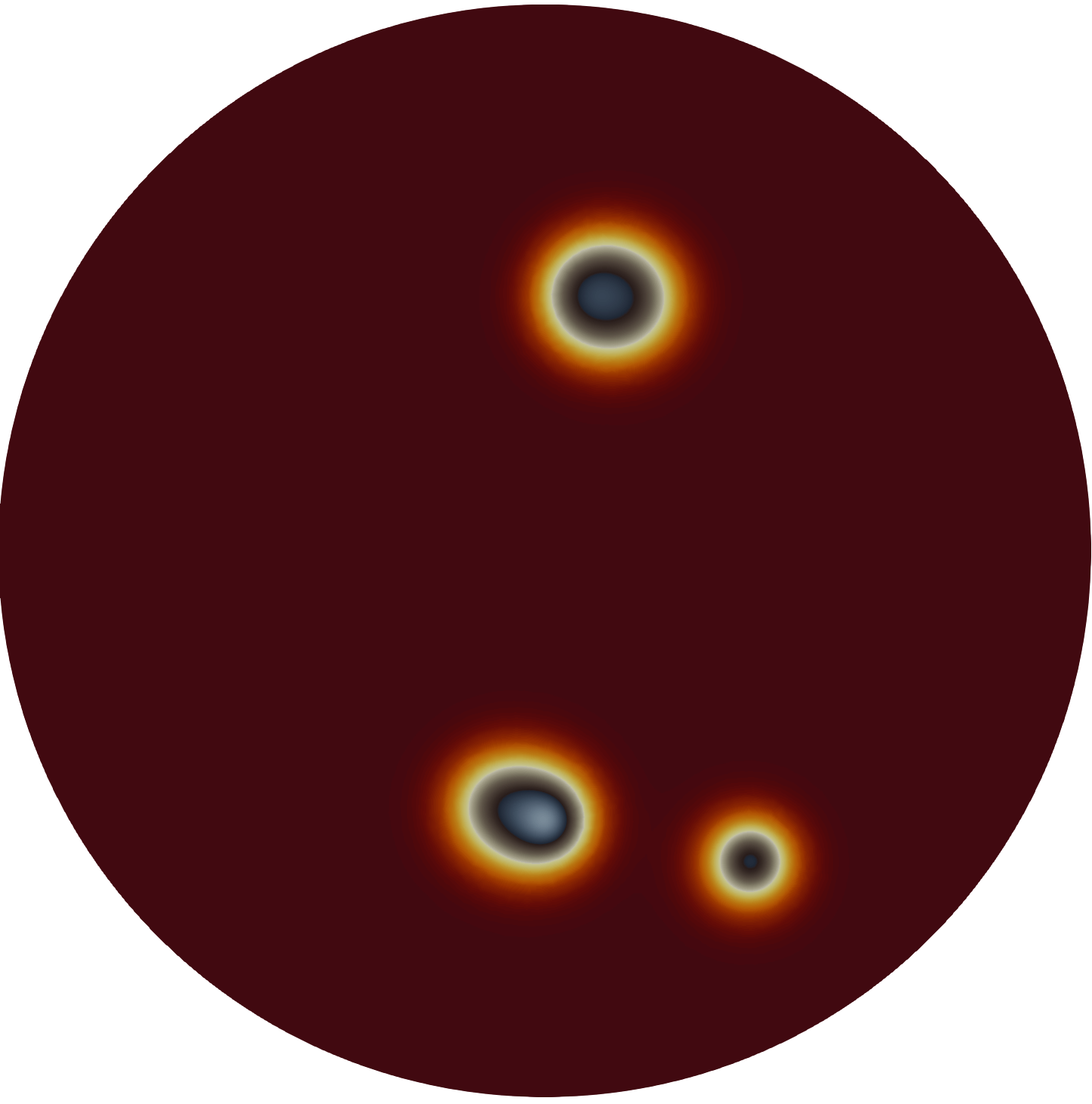}
\caption*{$v$ at $t=1953$\\ \quad }
\end{subfigure}
\\[5pt]
\begin{subfigure}[b]{0.25\textwidth}
\includegraphics[width=\textwidth,height=4.0cm]{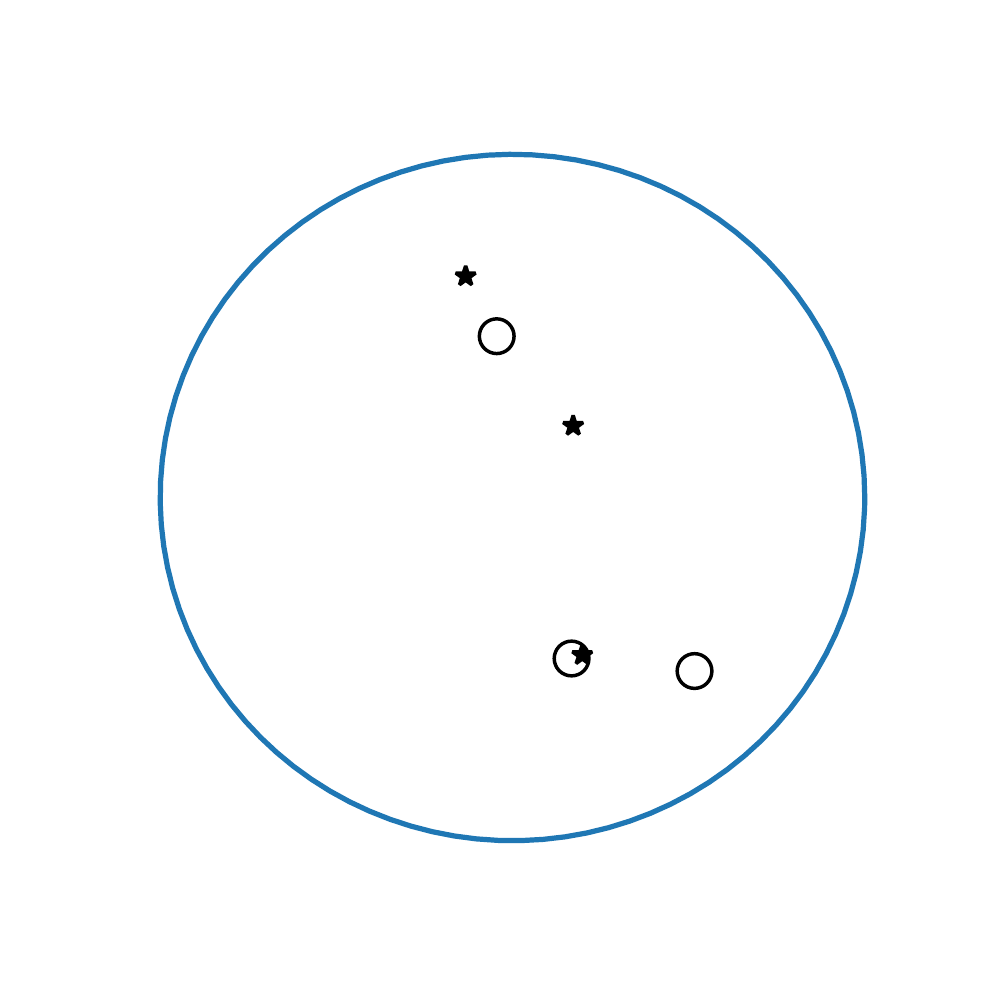}
\caption*{Before actual \\ killing ($t=2016$)}
\end{subfigure}
\qquad
\begin{subfigure}[b]{0.25\textwidth}
\includegraphics[width=\textwidth,height=4.0cm]{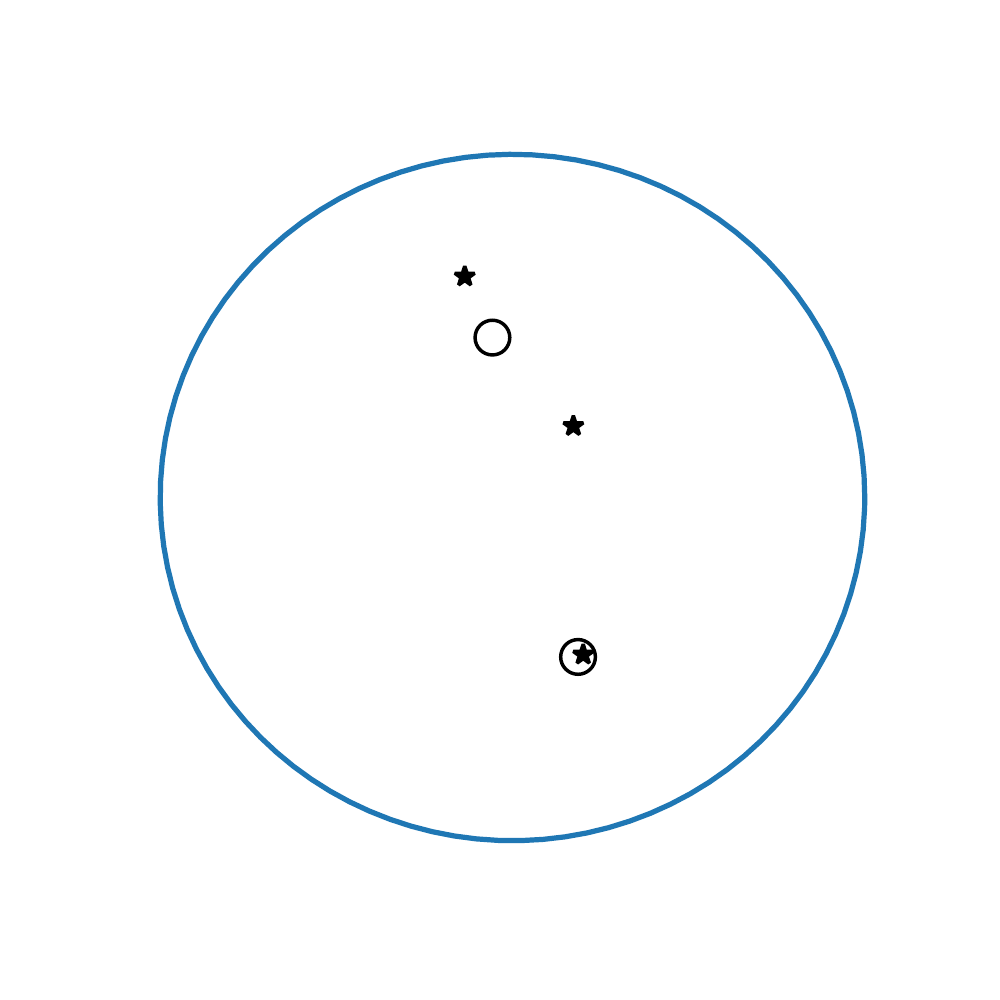}
\caption*{After actual \\ killing ($t=2017$)}
\end{subfigure}
\qquad
\begin{subfigure}[b]{0.25\textwidth}
\includegraphics[width=\textwidth,height=4.0cm]{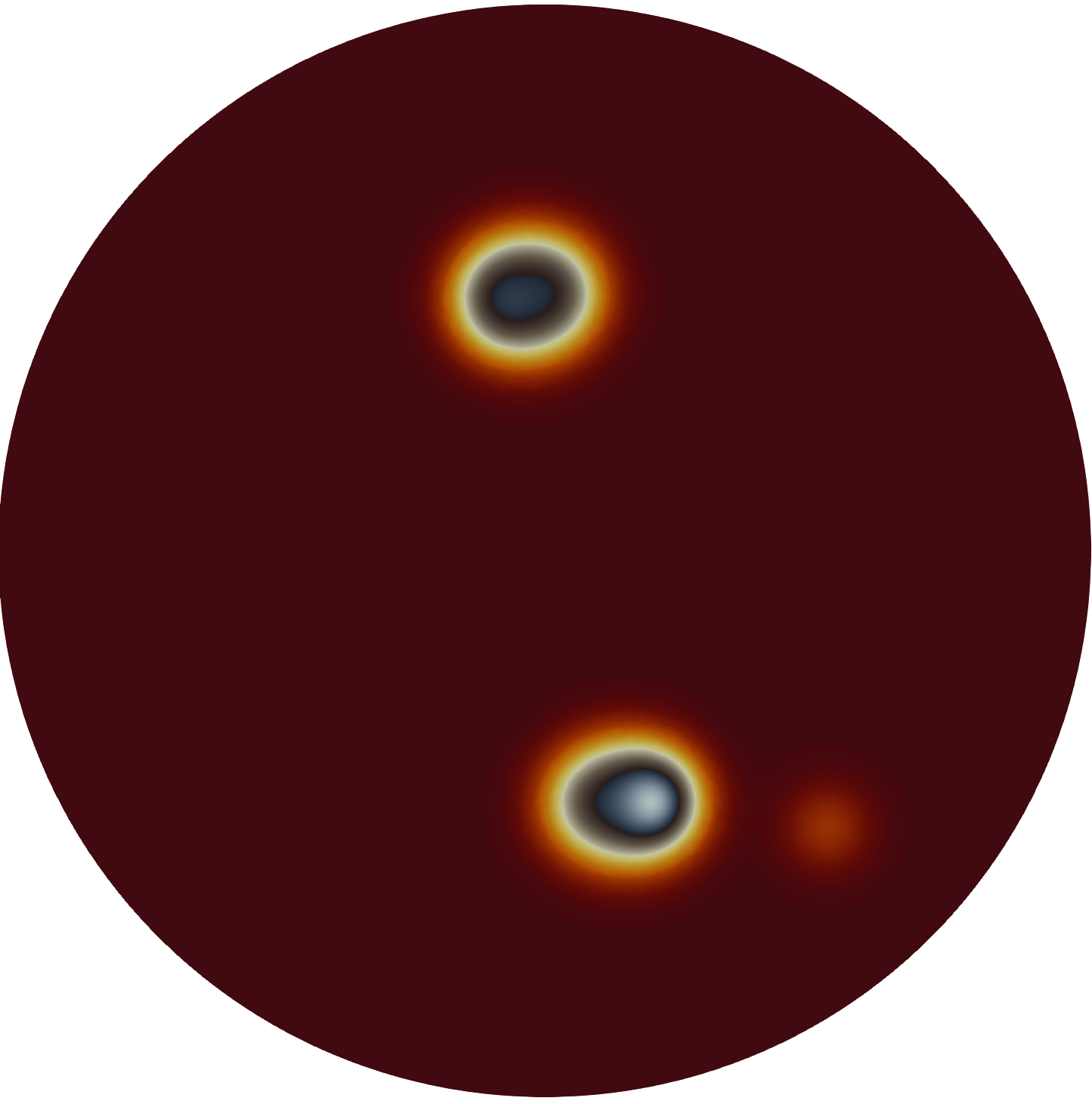}
\caption*{$v$ at $t=2016$\\ \quad }
\end{subfigure}
\\[5pt]
\begin{subfigure}[b]{0.25\textwidth}
\includegraphics[width=\textwidth,height=4.0cm]{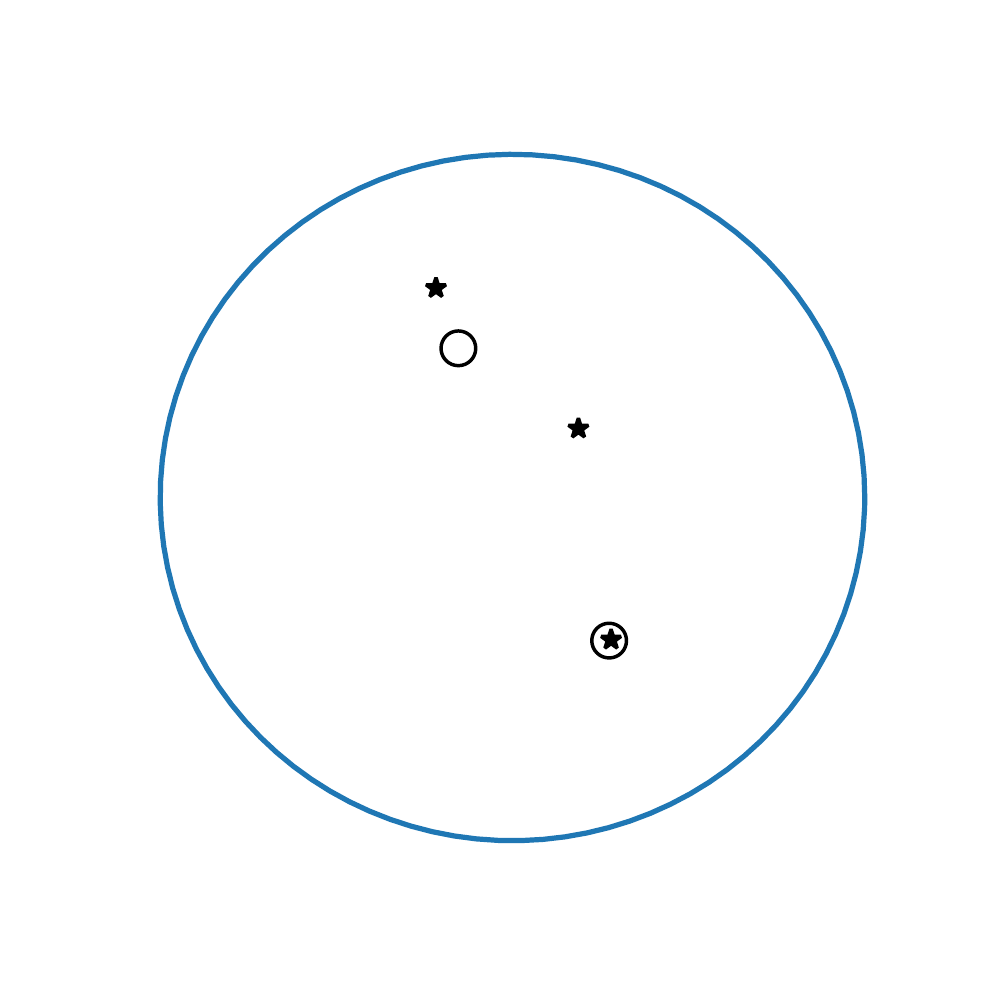}
\caption*{Before actual \\ splitting ($t=2049$)}
\end{subfigure}
\qquad
\begin{subfigure}[b]{0.25\textwidth}
\includegraphics[width=\textwidth,height=4.0cm]{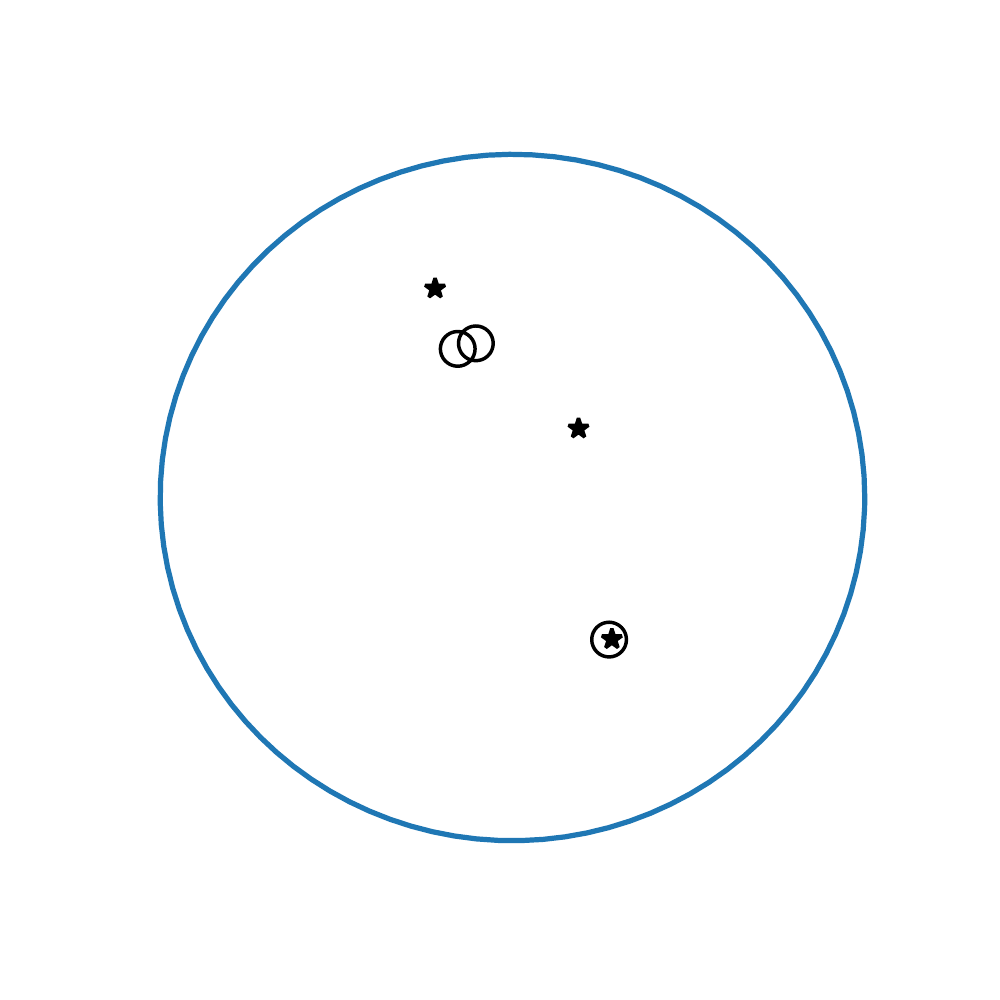}
\caption*{After actual \\ splitting ($t=2050$)}
\end{subfigure}
\qquad
\begin{subfigure}[b]{0.25\textwidth}
\includegraphics[width=\textwidth,height=4.0cm]{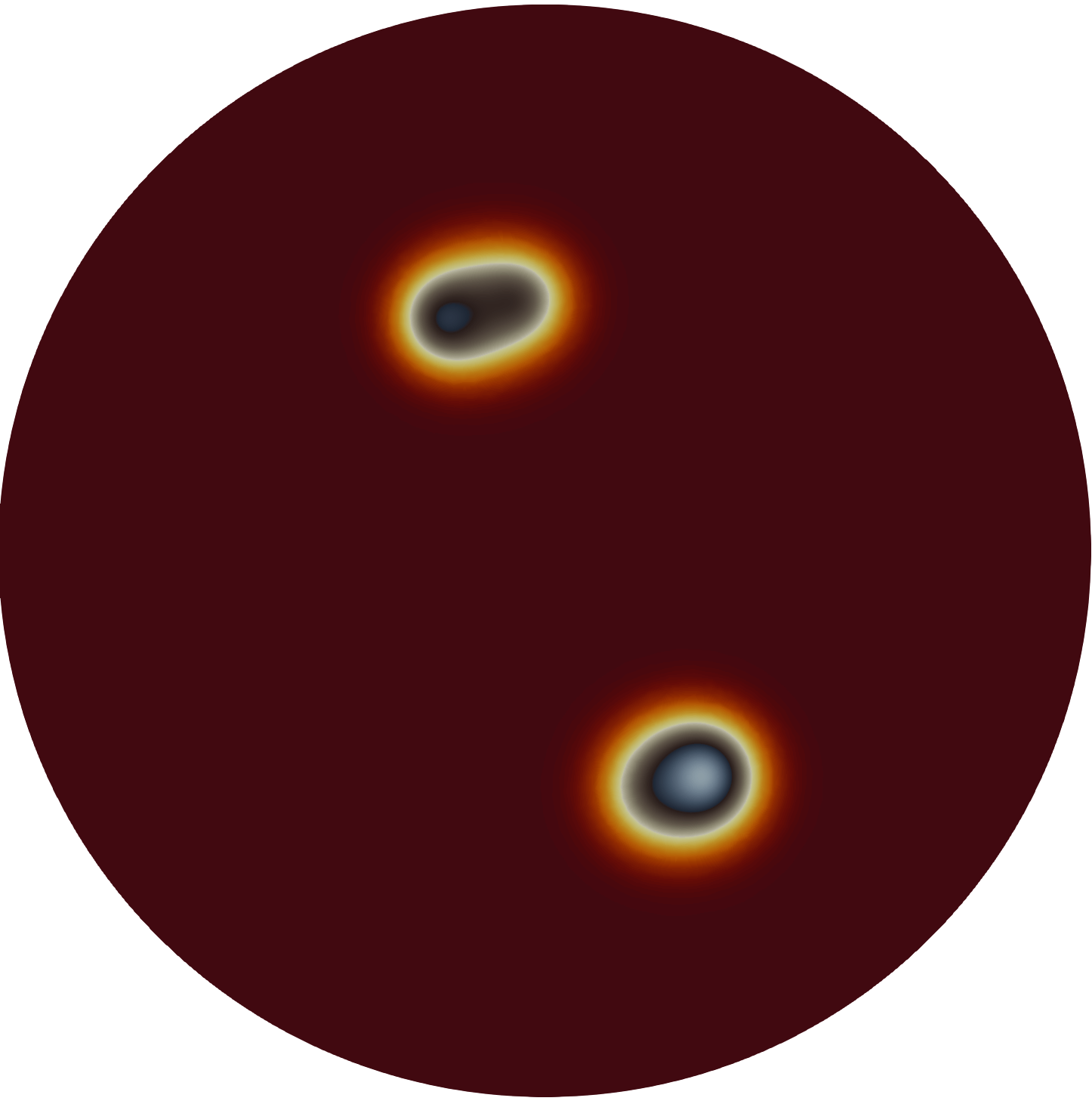}
\caption*{$v$ at $t=2050$ \\ \quad }
\end{subfigure}
\caption{The third spot creation-annihilation event}
\label{pinned_split_comp_exp1_part5}
\end{figure}

\begin{figure}[htbp]
\begin{subfigure}[b]{0.25\textwidth}
\includegraphics[width=\textwidth,height=4.0cm]{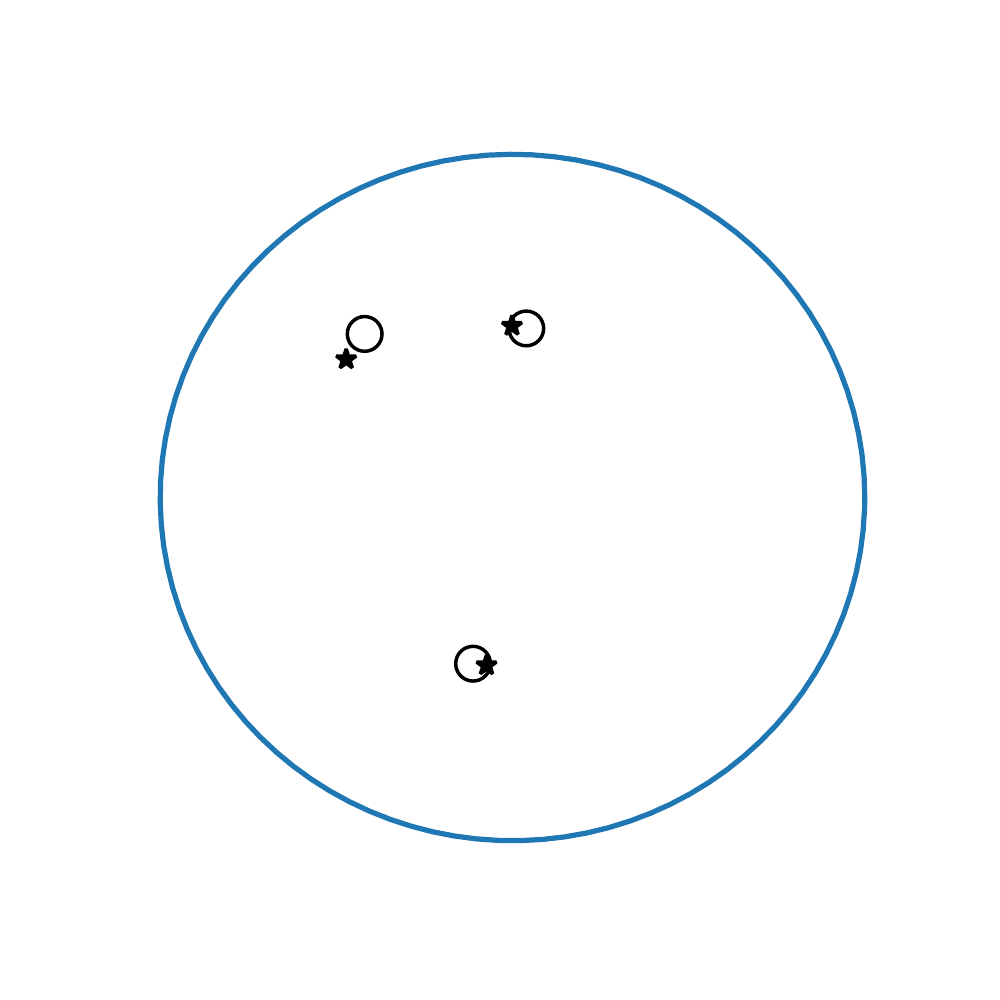}
\caption*{$t=2500$}
\end{subfigure}
\qquad
\begin{subfigure}[b]{0.25\textwidth}
\includegraphics[width=\textwidth,height=4.0cm]{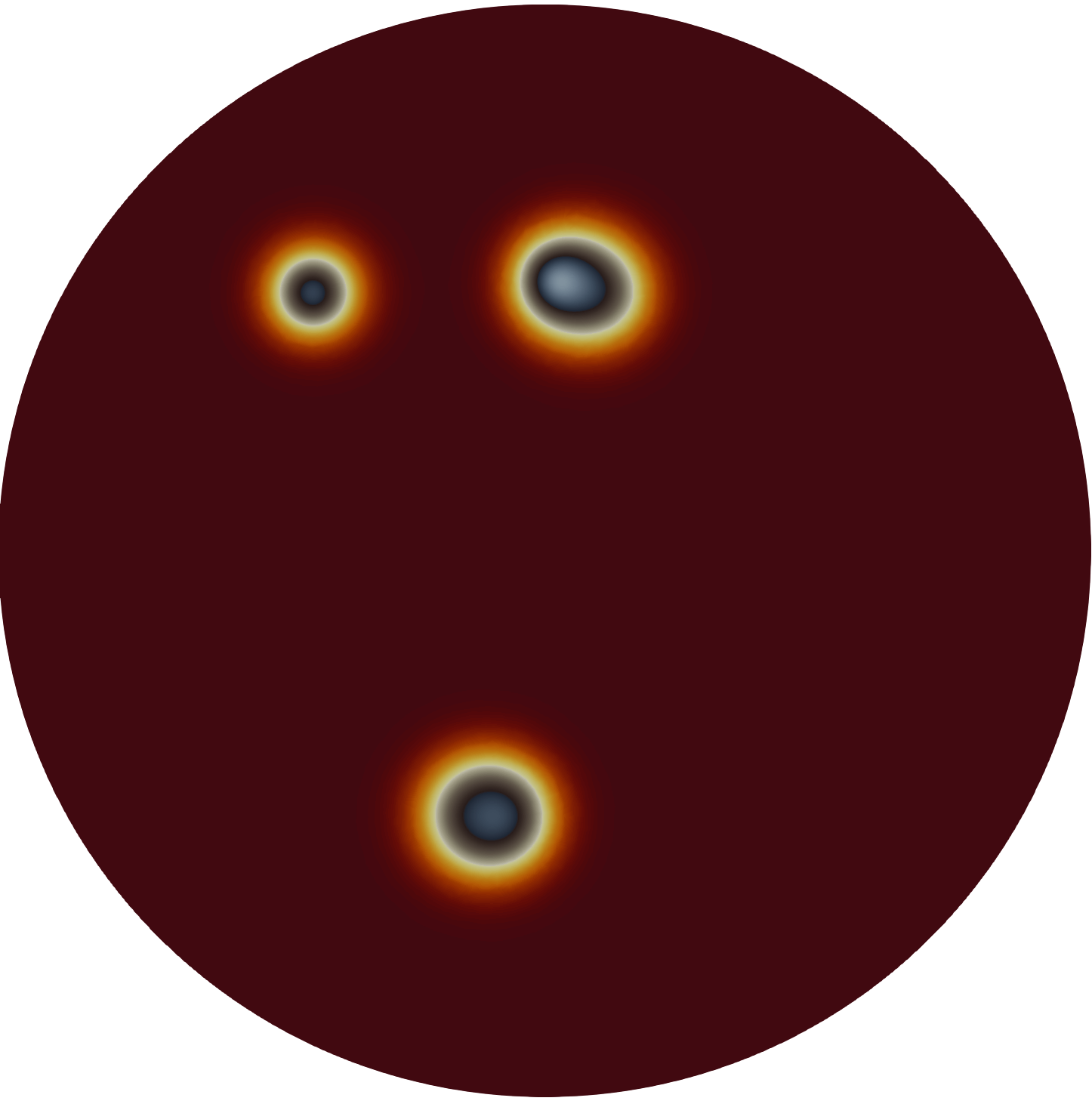}
\caption*{$v$ at $t=2500$}
\end{subfigure}
\caption{The approximation algorithm provides a reasonably close prediction
  for the unpinned spot trajectory at $t=2500$ after three spot
  creation-annihilation events.}
\label{pinned_split_comp_exp1_part6}
\end{figure}

In this subsection we show a PDE simulation of \eqref{pinned:model}
that involves a repeating loop of spot replication and
annihilation. We choose $\eps=0.03$ and consider a feed rate with
$a_0 = 19$ and $a_1=5$, where the concentration point $\pmb{\xi}$ of
the feed evolves slowly in time according to
\eqref{pinned:trajectory}.  We consider an initial two-spot
  quasi-equilibrium pattern where the pinned spot is initially at
  $\pmb{\xi}(0)=(0.5,0)^T$ and with an additional unpinned spot
  initially centered at $(-0.5,0)^T$. The PDE simulation results are
shown in the right panels of
Figs.~\ref{pinned_split_comp_exp1_part1}--\ref{pinned_split_comp_exp1_part6}
at the indicated times. In the PDE results, we observe that when
$t\approx 41$ the unpinned spot splits into two spots. The resulting
two spots evolve dynamically and remain separated from the
approaching pinned spot. However, as the pinned spot becomes close
enough to one of the two unpinned spots, at $t\approx 685$ a
competition instability is triggered and one of these spots is
annihilated, leaving a pattern with only one spot and the pinned
spot. Later at $t \approx 720$, the unpinned spot splits again, and
the spot creation-annihilation loop is repeated.  We record three
cycles of this loop in Figs.~\ref{pinned_split_comp_exp1_part1}--
\ref{pinned_split_comp_exp1_part6}.

To model this loop theoretically, we introduce an algorithim that
combines the DAE dynamics \eqref{pinned:DAE} and
\eqref{pinned:nas_all} with our linear stability theory in \S
\ref{sec:linstab_allpin} of quasi-equilibrium patterns. Since the DAE
system is valid only when there is $\mc{O}(1)$ time-scale instability
of the quasi-equilibrium pattern, we need to augment the DAE solver with
a numerical detection strategy for the initiation of spot-replication
or spot-annihilation events, and the subsequent addition or removal of
newly created or annihilated spots. At the end of each time step in the
DAE solver, we first use \eqref{pinned:GCEP_zero_crossing} and the
condition $\det\mc{M}_0=0$ to detect zero-eigenvalue crossings in GCEP.
In practice, in our algorithm we identify a zero-eigenvalue crossing
if
\begin{equation}\label{pinned:GCEP_zero_crossing_criterion}
|\det \Mmat_0| \leq \tol \,, \qquad \mbox{where} \quad \tol \ll 1\,.
\end{equation}
This zero-eigenvalue crossing corresponds to an inter-spot competition
instability, and triggers the annihilation of the spot with the
smallest source strength. Once the criterion
\eqref{pinned:GCEP_zero_crossing_criterion} is met, we eliminate that
particular spot with $S_j = \min\limits_{1\leq i \leq N} S_i$ from the
DAE system. Next, to detect a peanut-splitting instability, we choose
a number $\Sigma_2^{\mrm{eff}}$ that is slightly larger than the
peanut splitting threshold $\Sigma_2 \approx 4.302$ for the unpinned
spots. If there is an unpinned spot with $S_j > \Sigma_2^{\mrm{eff}}$,
the peanut-splitting instability triggers a nonlinear spot creation
process that divides the spot into two separate spots. To model this
process in our algorithm, we replace this spot at $\v{x}_j$, with two
new spots located at
\begin{equation}\label{pinned:add_spots}
  \v{x}_{new}^1 = \v{x}_j + \delta (v_2, -v_1) \,, \qquad
  \v{x}_{new}^2 = \v{x}_j + \delta (-v_2, v_1) \,,
\end{equation}
where $\delta \ll 1$ and $\v{v} = (v_1, v_2)$ is the normalized
velocity field ($\,|\v{v}|=1\,$) as computed from the DAE system
\eqref{pinned:DAE} and \eqref{pinned:nas_all}. The choice in
\eqref{pinned:add_spots} for this two newly created spot locations is
motivated from the result in \cite{kww09}, which showed that the
direction of spot-splitting is perpendicular to the direction of motion
of the spot.

We use this algorithm for augmenting the DAE solver with parameters
$\tol = 10^{-4} \,, \enspace \Sigma_2^{\mrm{eff}} = 4.4$ and
$\delta = 0.1$, with results shown in the left and middle panels in
Figs.~\ref{pinned_split_comp_exp1_part1}--
\ref{pinned_split_comp_exp1_part6}. At $t=9$, the algorithm detects a
peanut-splitting instability of the spot. The spot is replaced with
two spots given in \eqref{pinned:add_spots} (see
Fig.\,\ref{pinned_split_comp_exp1_part1}). Then at $t=583$, the DAE
solver detects a zero-eigenvalue crossing based on
\eqref{pinned:GCEP_zero_crossing_criterion}. The spot with the minimum
source strength is removed from the DAE system. Right after the
removal, the DAE solver detects a peanut-splitting instability, and
two new spots are created. In conclusion, the DAE solver predicts a
spot creation-annihilation event at $t=583$. The PDE simulation
confirms a spot annihilation event at $t\approx 684$, and a spot
replication event at $t\approx719$. The DAE solver also predicts the
second and third spot creation-annihilation event at $t = 1248$ and
$t = 1952$, respectively. These events are all confirmed by the full PDE
simulation (see Fig.~\ref{pinned_split_comp_exp1_part4} and
Fig.~\ref{pinned_split_comp_exp1_part5}).

\section{Discussion}\label{sec:discussion}

We have developed a hybrid asymptotic-numerical theory to analyze the
effect of several types of localized heterogeneities on the existence,
linear stability, and slow dynamics of spot patterns for the
prototypical two-component Schnakenberg model \eqref{proto:model} in a
bounded 2-D domain.  Our analysis has focused on distinct types of
localized heterogeneities: a strong localized perturbation of a
spatially uniform feed rate and the effect of removing a small hole in
the domain, through which the chemical species can leak out. Although
our overall approach relies on the theoretical framework first
introduced in \cite{kww09}, and later extended in \cite{tw18} for
analyzing the effect of heterogeneities in the Brusselator model, our
analysis of localized heterogeneities for the Schnakenberg model has
revealed a wide range of novel phenomena such as, saddle-node
bifurcations for quasi-equilibrium spot patterns that otherwise would
not occur for a homogeneous medium, a new type of spot solution pinned
at the concentration point of the feed rate, spot self-replication
behavior that generates more than two new spots, and the existence of
a creation-annihilation attractor with at most three spots. The hybrid
approach presented herein can be readily extended to other well-known
RD models such as the Gray-Scott, Gierer-Meinhardt, and Brusselator
models.

We conclude by briefly discussing a few problems that warrant further
investigation.  One interesting direction would be to extend the
algorithm, introduced in \S \ref{sec:loop}, to general $N$-spot
quasi-equilibrium patterns over much longer time scales. This would
involve coupling the ODE-DAE system for slow spot dynamics occuring on
long ${\mathcal O}(\eps^{-2})$ time scales with sudden ``surgeries'',
resulting in either spot-creation or spot-annihilation events, that
are informed through monitoring linear stability thresholds at each
time step as the quasi-equilibrium spot pattern evolves. In
particular, in this general setting, it would be interesting to
classify whether spot-annihilation events, triggered by a competition
instability due to a zero-eigenvalue crossing of the globally coupled
eigenvalue problem, can be interpreted more geometrically in terms of
crossing through a saddle-node bifurcation point of manifolds of spot
quasi-equilibria. Such manifolds depend on the instantaneous spatial
configuration of spots, and they evolve slowly in time. For a two-spot
pattern in the unit disk, and with a feed rate that is slowly ramped
in time, such a fold-point crossing was observed in
Fig.~\ref{fig:GCEP_two_all}.  In a 1-D setting, spike-annihilation
events have been recently interpreted in \cite{bast_comp} for the
extended Klausmeir RD model as arising from rapid transitions between
manifolds of spike quasi-equilibria as the pattern evolves.

Another open problem is to determine whether a creation-annihilation
attractor for spot quasi-equilibria, which involves only a few spots,
can occur for a time-independent feed rate that has a smooth
(not localized) spatial variation.  For the 1-D Schnakenberg model,
but for a very large number of spikes, such a creation-annihilation
attractor has been predicted and observed in \cite{kw_siamrev} through the
analysis of a limiting mean-field equation for the spike density. 

Finally, it would be interesting to study how variations in the domain
geometry or the domain boundary condition influence the slow spot
dynamics and the linear stability properties of quasi-equilbrium spot
patterns, leading to to new types of spot-pinning behavior. Some work
in this direction for the Brusselator model with a Robin boundary
condition is given in \cite{tw18} for the disk. For general
domains, the determination of the Neumann Green's function and the
reduced-wave Green's function would be central to this study. For an
elliptical domain of arbitrary eccentricity, spot-pinning behavior can
readily be studied using the new analytical result in
\cite{short_mfpt} for the Neumann Green's function for the ellipse.

\section*{Acknowledgements}\label{sec:ak}
Tony Wong was supported by a UBC Four-Year Graduate Fellowship.
Michael Ward gratefully acknowledges the financial support from the
NSERC Discovery Grant program. We thank Prof. Colin B. MacDonald for
helpful suggestions regarding the numerical PDE computations.

\bibliographystyle{abbrv}
\bibliography{schnak_defect}

\begin{appendix}

\section{The Green's functions for the unit disk}
\label{app:neum}

The Neumann Green's function and its regular part, satisfying
\eqref{proto:neu_green}, have explicit formulae for the unit disk
(cf.~\cite{kww09}):
\begin{subequations}\label{ring:gcomplex}
\begin{align}
  G(\v{x} ; \v{z}) &= \frac{1}{2\pi}\left( -\log|\v{x}-\v{z}| -
  \log\left| |\v{z}|\v{x} - \frac{1}{|\v{z}|}\v{z} \right| +
  \frac{1}{2} (|\v{x}|^2 + |\v{z}|^2) - \frac{3}{4}  \right) \,, \\[10pt]
  R(\v{z} ; \v{z}) &= \frac{1}{2\pi} \left( -\log\left( 1 - |\v{z}|^2
       \right) + |\v{z}|^2 - \frac{3}{4} \right) \,.
\end{align}
\end{subequations}
Their gradients are given by
\begin{equation}\label{ring:gcomplex_der}
  \nabla_{\v{x}} G =
  -\frac{1}{2\pi} \left( \frac{\left( \v{x}-\v{z} \right)
    }{|\v{x}-\v{z}|^2}  + \frac{|\v{z}|^2\left( |\v{z}|^2 \v{x} - \v{z} \right)
   }{\big| |\v{z}|^2 \v{x} - \v{z}\big|} -  \v{x} \right) \,,
  \qquad 
  \nabla_{\v{x}} R  = \frac{1}{2\pi} \left( \frac{2 - |\v{z}|^2}{1 - |\v{z}|^2}
  \right) \v{z} \,.
\end{equation}

For a ring pattern, where $\v{x}_1,\ldots,\v{x}_N$ are equally-spaced
on a ring of radius $r_0$ concentric within the unit disk as given in
\eqref{proto:ring}, we have from Proposition 4.3 of \cite{kmm05} that
\begin{equation}\label{ring:gmat_eig}
  \Gmat \v{e} = \frac{p(r_0)}{N} \v{e} \,, \qquad p(r_0) \equiv
  \frac{1}{2\pi} \left( -N \log(Nr_0^{N-1}) - N \log(1-r_0^{2N}) + r_0^2 N^2 -
    \frac{3N^2}{4}\right) \,.
\end{equation}
As such, for a ring pattern, there is a symmetric solution to the NAS
\eqref{proto:source_system} given by $S_j=S_c={p_a/N}$ for
$j=1,\ldots,N$, where $p_a$ is defined in
\eqref{proto:sum_Sj}. Then, upon defining
$\nabla_{\v{x}}R_{j,j}\equiv
\nabla_{\v{x}}R(\v{x};\v{x}_j)\vert_{\v{x}=\v{x}_j}$, and
$\nabla_{\v{x}}G_{j,i}\equiv
\nabla_{\v{x}}G(\v{x};\v{x}_i)\vert_{\v{x}=\v{x}_j}$, we then use the
reciprocity property of the Green's function to calculate
$\pmb{\beta}_j$ in \eqref{proto:betaj} as
\begin{equation}\label{proto:betaj_ring}
  \pmb{\beta}_j = 2\pi S_c \left(\nabla_{\v{x}} R_{j,j} +
    \sum\limits_{i \neq j}^N \nabla_{\v{x}} G_{j,i} \right) = 2\pi S_c \left(
   \frac{p^{\prime}(r_0)}{2N} \right) \v{e}_{\theta_j} =
    S_c
    \left( -\frac{N-1}{2r_0} + \frac{N r_0^{2N-1}}{1 - r_0^{2N}} + N r_0 \right)
    \v{e}_{\theta_j} \,,
\end{equation}
where $\v{e}_{\theta_j} \equiv (\cos\theta_j,\sin\theta_j)^T$ and
$\theta_j={2\pi (j-1)/N}$. By substituting \eqref{proto:betaj_ring} in
\eqref{proto:DAE}, and using $\v{x}_j = r_0(\sigma) \v{e}_{\theta_j}$,
we obtain the scalar ODE \eqref{proto:ring_scalar_ode} for the ring
radius $r_0$.

For the unit disk, and for $\lambda\neq 0$, the eigenvalue-dependent
Green's function $G_\lambda(\v{x};\v{x}_0)$, as defined by
\eqref{proto:eig_green}, can be expressed as an infinite series as
(cf.~Appendix A.1 of \cite{cw12})
\begin{subequations}\label{proto:eig_green_series_all}
\begin{equation}\label{proto:eig_green_series}
  G_\lambda(\v{x};\v{x}_0) = \frac{1}{2\pi} \left[K_0(\thetal|\v{x}-\v{x}_0|) -
    \frac{K_0^{\prime}(\thetal)}{I_0^{\prime}(\thetal)} I_0(\thetal r)
    I_0(\thetal r_0) \right]   - \frac{1}{\pi} \sum\limits_{n=1}^\infty
  \cos\left[n(\psi-\psi_0)\right]
  \frac{K_n^{\prime}(\thetal)}{I_n^{\prime}(\thetal)} I_n(\thetal r)
  I_n(\thetal r_0) \,.
\end{equation}
Here
$\v{x} = r \, (\cos(\psi),\sin(\psi)) \,, \, \v{x}_0 = r_0 \,
(\cos(\psi_0),\sin(\psi_0))$, $I_n$ and $K_n$ are the $n^{\text{th}}$
order modified Bessel functions of the first and second kind,
respectively, and $\thetal$ is the principal branch of
$\thetal \equiv \sqrt{\tau\lambda/D}$. The regular part of $\Gmatl$ is
\begin{equation}\label{proto:eig_green_regular_series}
  R_\lambda(\v{x}_0;\v{x}_0) = \frac{1}{2\pi} \left[ \log2 - \gamma_e -
    \frac{\log(D/\tau)}{2} - \frac{\log\lambda}{2} -
    \frac{K_0^{\prime}(\thetal)}{I_0^{\prime}(\thetal)}I_0^2(\thetal r_0)\right] -
    \frac{1}{\pi} \sum\limits_{n=1}^\infty \frac{K_n^{\prime}(\thetal)}
    {I_n^{\prime}(\thetal)} I_n^2(\thetal r_0) \,,
\end{equation}
\end{subequations}
where $\gamma_e \approx 0.5772$ is the Euler's constant.

\section{Spectrum of circulant matrices}\label{appendix:circulant}

$\mc{A} \in \mathbb{R}^{N\times N}$ is a circulant matrix if every row
is obtained by right shifting the previous row by one unit. Therefore,
$\mc{A}$ can be uniquely determined by it first row, denoted as
$\v{a} = (a_1, \ldots, a_N)$, while the second row of $\mc{A}$
is $(a_N, a_1,\ldots, a_{N-1})$. Suppose $\mc{A}$ is symmetric and
circulant. Then, the eigenvalues of $\mc{A}$ are
\begin{subequations}
\begin{equation}\label{circulant_spectrum1}
  \lambda_1 = \sum\limits_{k=1}^N a_k \,, \qquad \lambda_j =
  \sum\limits_{k=0}^{N-1} \cos\left[ \frac{2\pi(j-1)k}{N} \right] \, a_{k+1} \,,
  \quad j=2,\ldots,N \,.
\end{equation}
The corresponding eigenvectors are
$\v{q}_1 = \v{e} = (1,\ldots,1)^T \in \mathbb{R}^N$ and
\begin{equation}\label{circulant_spectrum2},
\begin{split}
  \v{q}_j &= \left( 1 \,, \cos\left( \frac{2\pi(j-1)}{N} \right) \,, \ldots \,,
    \cos\left( \frac{2\pi(j-1)(N-1)}{N} \right) \right)^T \,, \\[5pt]
  \v{q}_{N+2-j} &= \left( 0 \,, \sin\left( \frac{2\pi(j-1)}{N} \right) \,,
    \ldots \,, \sin\left( \frac{2\pi(j-1)(N-1)}{N} \right) \right)^T \,,
\end{split}
\end{equation}
for $j=2,\dots,\mrm{ceil}(N/2)$, where $\mrm{ceil}(s)$ denotes the
smallest integer that is not less than $s$. Furthermore, when $N$ is
even, we have an additional simple eigenvalue $\lambda_{N/2+1}$ with
eigenvector $\v{q}_{N/2+1} = (1, -1, \ldots, 1,-1)^T$.
\end{subequations}

\end{appendix}

\end{document}